\documentstyle[preprint,tighten,aps]{revtex}
\newtheorem{definition}{Definition}[section]
\newtheorem{theorem}{Theorem}[section]
\newtheorem{proposition}{Proposition}[section]
\begin{document}
\draft
\title{Field Interactions and Strings in\\
	 Higher Order Anisotropic Spaces}
\author{Sergiu I. Vacaru}

\address{Institute of Applied Physics,Academy of Sciences of Moldova,\\
 5 Academy str., Chi\c sin\v au 2028, Republic of Moldova\\
 e--mail: lises@cc.acad.md }
\date{29 September 1996}
\maketitle
\begin{abstract}
 We present a  geometric approach to the field theory with higher order
anisot\-rop\-ic interactions. The concepts of higher order anisotropic space
 and locally anisotropic space (in brief, h--space and la--space) are
 introduced as  general ones for various types of
 higher order  extensions  of Lagrange  and Finsler geometry and  higher
 dimension (Kaluza--Klein type) spaces. The  spinors  on h--spaces
 are defined in the framework of the geometry of Clifford
 bundles provided with compatible nonlinear and distinguished connections
 and metric structures (d--connection and d--metric).
 The spinor differential geometry of h--spaces is
 constructed. There are discussed some related issues connected with the
 geometric aspects of higher order anisotropc interactions for
 gravitational, gauge, spinor, Dirac spinor and Proca fields. Motion
 equations in higher order generalizations of Finsler spaces, of the
 mentioned type of fields, are defined in a geometric manner by using
 bundles of linear and affine  frames locally adapted to the nonlinear
 connection structure. The nearly
 autoparallel maps are introduced as maps  with deformation of connections
  extending the class of geodesic and conformal transforms. We propose two
 variants of solution of the  problem of definition of  conservation laws on
  h--spaces. A general  background of the theory of field interactions and
strings in spaces with higher order anisotro\-py is fromulated.
The conditions for consistent propagation of closed strings in higher order
anisotropic background spaces are analyzed. The connection between
the conformal invariance, the vanishing of the renormalization group
beta--function of the generalized sigma--model and field equations of
higher order anisotropic gravity are studied in detail.
\copyright{\bf \ Sergiu I. Vacaru, 1996}

\end{abstract}

\tableofcontents

\newpage

\section{Introduction}

Some of fundamental problems in physics advocate the extension to locally
anisotropic and higher order anisotropic backgrounds of physical theories
\cite{mat,ma94,am,bej,asa88,mirata,vg,v96jpa1,v96jpa2,v96jpa3}. In order to
construct physical models on higher order anisotropic spaces it is necessary
a corresponding generalization of the spinor theory. Spinor variables and
interactions of spinor fields on Finsler spaces were used in a heuristic
manner, for instance, in works \cite{asa88,ono}, where the problem of a
rigorous definition of la-spinors for la-spaces was not considered. Here we
note that, in general, the nontrivial nonlinear connection and torsion
structures and possible incompatibility of metric and connections makes the
solution of the mentioned problem very sophisticate. The geometric
definition of la-spinors and a detailed study of the relationship between
Clifford, spinor and nonlinear and distinguished connections structures in
vector bundles, generalized Lagrange and Finsler spaces are presented in
Refs. \cite{vjmp,vb295,vsp96}.

Subection II.A  contains an introduction into the geometry of higher
order anisotropic spaces, the  distinguishing of geometric objects by
N--connection structures  in such spaces is analyzed, explicit formulas for
coefficients of torsions and curvatures of N- and d--connections are presented
and the field equations for gravitational interactions with higher order
anisotropy are formulated. The distinguished Clifford algebras are introduced
in Sec. II.B and higher order anisotropic Clifford bundles are defined in
Sec. II.C. We present a study of almost complex structure for the case of
 locally anisotropic spaces modeled in the framework of the almost Hermitian
 model of generalized Lagrange spaces in Sec. II.D. The d--spinor
 techniques is analyzed in Sec. II.E and the differential
 geometry of higher order anisotropic spinors is formulated in
 Sec. II.F.
 The Sec. II.G is devoted to geometric aspects of the theory of field
 interactions with higher order anisotropy (the d--tensor and d--spinor form
 of basic field equations for gravitational, gauge and d--spinor fields are
 introduced).

Despite the charm and success of general relativity there are some
fundamental problems still unsolved in the framework of this theory. Here we
point out the undetermined status of singularities, the problem of
formulation of conservation laws in curved spaces, and the
unrenormalizability of quantum field interactions. To overcome these defects
a number of authors (see, for example, Refs. \cite{ts,wal,pon,ald}) tended
to reconsider and reformulate gravitational theories as a gauge model
similar to the theories of weak, electromagnetic, and strong forces. But, in
spite of theoretical arguments and the attractive appearance of different
proposed models of gauge gravity, the possibility and manner of
interpretation of gravity as a kind of gauge interaction remain unclear.

The work of Popov and Daikhin \cite{p,pd} is distinguished among other gauge
approaches to gravity. Popov and Dikhin did not advance a gauge extension,
or modification, of general relativity; they obtained an equivalent
reformulation (such as well-known tetrad or spinor variants) of the Einstein
equations as Yang-Mills equations for correspondingly induced Cartan
connections \cite{bis} in the affine frame bundle on the pseudo-Riemannian
space time. This result was used in solving some specific problems in
mathematical physics, for example, for formulation of a twistor-gauge
interpretation of gravity and of nearly autoparallel conservation laws on
curved spaces \cite{v87,vb12,vrjp,vk,vast}. It has also an important
conceptual role. On one hand, it points to a possible unified treatment of
gauge and gravitational fields in the language of linear connections in
corresponding vector bundles. On the other, it emphasize that the types of
fundamental interactions mentioned essentially differ one from another, even
if we admit for both of them a common gauge like formalism, because if to
Yang-Mills fields one associates semisimple gauge groups, to gauge
treatments of Einstein gravitational fields one has to introduce into
consideration nonsemisimple gauge groups.

Recent developments in theoretical physics suggest the idea that a more
adequate description of radiational, statistical, and relativistic optic
effects in classical and quantum gravity requires extensions of the
geometric background of theories \cite
{vlas,mk,am,az94,ari,bei,ikes,in,ing,ish,rox,takp,takn,tav,var,bog} by
introducing into consideration spaces with local anisotropy and formulating
corresponding variants of Lagrange and Finsler gravity and theirs extensions
 to higher order anisotropic spaces \cite{yan,mirata,v96jpa1,v96jpa2,v96jpa3}.

The presentation in the Sec. III is organized as follows:
In Sec. III.A we give a geometrical interpretation of gauge (Yang-Mills)
fields on general ha-spaces. Sec. III.B contains a geometrical definition
of anisotropic Yang-Mills equations; the variational proof of gauge field
equations is considered in connection with the "pure" geometrical method of
introducing field equations. In Sec. III.C the ha--gravity is reformulated
as a gauge theory for nonsemisimple groups. A model of nonlinear de Sitter
gauge gravity with local anisotropy is formulated in Sec. III.D. We study
gravitational gauge instantons with trivial local anisotropy in
Sec. III.E.

Theories of field interactions on locally anisotropic curved spaces form a
new branch of modern theoretical and mathematical physics. They are used for
modelling in a self--consistent manner physical processes in locally
anisotropic, stochastic and turbulent media with beak radiational reaction
and diffusion \cite{ma94,am,az94,vlas}. The first model of locally
anisotropic space was proposed by P.Finsler \cite{fin} as a generalization
of  Riemannian geometry; here we also cite the fundamental contribution made
by E. Cartan \cite{car35} and mention that in monographs \cite
{run,mat,asa,asa89,bej} detailed  bibliographies are contained. In this
work we follow R. Miron and M. Anastasiei \cite{ma87,ma94} conventions and
base our investigations on their general model of locally anisotropic (la)
gravity (in brief we shall write la-gravity) on vector bundles, v--bundles,
provided with nonlinear and distinguished connection and metric structures
(we call a such type of v--bundle as a la-space if connections and metric
are compatible).

The study of models of classical and quantum field interactions on la-spaces
is in order of the day. For instance, the problem of definition of spinors
on la-spaces is already solved (see \cite{vjmp,vod,vb295,vsp96} and Sec. II)
 and some models of locally anisotropic Yang--Mills and gauge like
gravitational interactions are analyzed (see \cite{vg,vcl96} and Sec. III
and alternative approaches in \cite{asa,bej,ish,in}). The development of
this direction entails great difficulties because of problematical character
of the possibility and manner of definition of conservation laws on
la-spaces. It will be recalled that, for instance, conservation laws of
energy--momentum type are a consequence of existence of a global group of
automorphisms of the fundamental Mikowski spaces (for (pseudo)Riemannian
spaces the tangent space' automorphisms and particular cases when there are
symmetries generated by existence of Killing vectors are considered). No
global or local automorphisms exist on generic la-spaces and in result of
this fact the formulation of la-conservation laws is sophisticate and full
of ambiguities. R. Miron and M. Anastasiei firstly pointed out the nonzero
divergence of the matter energy-momentum d--tensor, the source in Einstein
equations on la-spaces, and considered an original approach to the geometry
of time--dependent Lagrangians \cite{anask,ma87,ma94}. Nevertheless, the
rigorous definition of energy-momentum values for la-gravitational and
matter fields and the form of conservation laws for such values have not
been considered in present--day studies of the mentioned problem.

Sec. IV.A is  devoted to the formulation of the theory of nearly
autoparallel maps of  la--spaces. The classification of na--maps and
formulation of their invariant conditions are given in Sec. IV.B. In
Sec. IV.C we define the nearly  autoparallel tensor--integral on locally
anisotropic multispaces. The problem of formulation of conservation laws on
spaces with local anisotropy is studied in Sec. IV.D. We present a
definition of conservation laws for la--gravitational fields on na--images
of la--spaces in Sec. IV.E. In  Sec. IV.F, we
analyze the locally isotropic limit, to  the Einstein gravity and it
generalizations, of the na-conservation laws.

The plan of the Sec. V is as follows. We begin, Sec. V.A, with a study
of the nonlinear $\sigma$-model and ha-string propagation by developing the
d-covariant method of ha-background field. Sec. V.B is devoted to
problems of regularization and renormalization of the locally anisotropic $%
\sigma$-model and a corresponding analysis of one- and two-loop diagrams of
this model. Scattering of ha-gravitons and duality are considered in Sec. V.C.

\section{ Spinors in Higher Order Anisotropic Spaces}

The purpose of the section is to summarize and extend our investigations
\cite{vjmp,vb295,vsp96,vg,vlasg} on formulation of the theory of classical
and quantum field interactions on higher order anisotropic spaces. We
receive primary attention to the development of the necessary geometric
framework: to propose an abstract spinor formalism and formulate the
differential geometry of higher order anisotropic spaces . The next step is
the investigation of higher order anisotropic interactions of fundamental
fields on generic higher order anisotropic spaces (in brief we shall use
instead of higher order anisotropic the abbreviation ha-, for instance,
ha--spaces, ha--interactions and ha--spinors).

In order to develop the higher order anisotropic spinor theory it will be
convenient to extend the Penrose and Rindler abstract index formalism \cite
{pen,penr1,penr2} (see also the Luehr and Rosenbaum index free methods \cite
{lue}) proposed for spinors on locally isotropic spaces. We note that in
order to formulate the locally anisotropic physics usually we have
dimensions $d>4$ for the fundamental, in general higher order anisotropic
space--time and to take into account physical effects of the nonlinear
connection structure. In this case the 2-spinor calculus does not play a
 preferential role.

\subsection{ Basic Geometric Objects in Ha--Spaces}

We review some results and methods of the differential geometry of vector
bundles provided with nonlinear and distinguished connections and metric
structures \cite{ma87,ma94,mirata,v96jpa1,v96jpa2,v96jpa3}.\ This subsection
serves the twofold purpose of establishing of abstract index denotations and
starting the geometric backgrounds which are used in the next subsections of
the section.

\subsubsection{ N-connections and distinguishing of geometric objects}

Let ${\cal E}^{<z>}{\cal =}$ $\left( E^{<z>},p,M,Gr,F^{<z>}\right) $ be a
locally trivial distinguished vector bundle, dv-bundle, where $F^{<z>}={\cal %
R}^{m_1}\oplus ...\oplus {\cal R}^{m_z}$ (a real vector space of dimension $%
m=m_1+...+m_z,\dim F=m,$ ${\cal R\ }$ denotes the real number field) is the
typical fibre, the structural group is chosen to be the group of
automorphisms of ${\cal R}^m$ , i.e. $Gr=GL\left( m,{\cal R}\right) ,\,$ and
$p:E^{<z>}\rightarrow M$ (defined by intermediar projections $%
p_{<z,z-1>}:E^{<z>}\rightarrow E^{<z-1>},p_{<z-1,z-2>}:E^{<z-1>}\rightarrow
E^{<z-2>},...p:E^{<1>}\rightarrow M)$  is a differentiable surjection of
a differentiable manifold $E$ (total space, $\dim E=n+m)$ to a
differentiable manifold $M$ (base
space, $\dim M=n ).$ Local coordinates on ${\cal E}^{<z>}$ are denoted
as
$$
u^{<{\bf \alpha >}}=\left( x^{{\bf i}},y^{<{\bf a>\ }}\right) =\left( x^{%
{\bf i}}\doteq y^{{\bf a}_0},y^{{\bf a}_1},....,y^{{\bf a}_z}\right) =%
$$
$$
(...,y^{{\bf a}_{(p)}},...)=\{y_{(p)}^{{\bf a}_{(p)}}\}=\{y^{{\bf a}%
_{(p)}}\},
$$
or in brief ${\bf u=u_{<z>}=(x,y}_{(1)},...,{\bf y}_{(p)},...,{\bf y}_{(z)})$
where boldfaced indices will be considered as coordinate ones for which the
Einstein summation rule holds (Latin indices ${\bf i,j,k,...=a}_0{\bf ,b}_0%
{\bf ,c}_0{\bf ,...}=1,2,...,n$ will parametrize coordinates of geometrical
objects with respect to a base space $M,$ Latin indices ${\bf a}_p ,
{\bf b}_p,$ ${\bf c}_p,...=$ $1,2,...,m_{(p)}$ will parametrize
fibre coordinates of
geometrical objects and Greek indices ${\bf \alpha ,\beta ,\gamma ,...}$ are
considered as cumulative ones for coordinates of objects defined on the
total space of a v-bundle). We shall correspondingly use abstract indices $%
\alpha =(i,a),$ $\beta =(j,b),\gamma =(k,c),...$ in the Penrose manner \cite
{pen,penr1,penr2} in order to mark geometical objects and theirs (base,
fibre)-components or, if it will be convenient, we shall consider boldfaced
letters (in the main for pointing to the operator character of tensors and
spinors into consideration) of type ${\bf A\equiv }A=\left(
A^{(h)},A^{(v_1)},...,A^{(v_z)}\right) {\bf ,b=}\left(
b^{(h)},b^{(v_1)},...,b^{(v_z)}\right) ,...,{\bf R},$ ${\bf \omega },$
${\bf \Gamma},...$ for
geometrical objects on ${\cal E}$ and theirs splitting into horizontal (h),
or base, and vertical (v), or fibre, components. For simplicity, we shall
prefer writing out of abstract indices instead of boldface ones if this will
not give rise to ambiguities.

Coordinate trans\-forms $u^{<{\bf \alpha ^{\prime }>\ }}=u^{<{\bf \alpha
^{\prime }>\ }}\left( u^{<{\bf \alpha >}}\right) $ on ${\cal E}^{<z>}$ are
writ\-ten as
$$
\{u^{<\alpha >}=\left( x^{{\bf i}},y^{<{\bf a>}}\right) \}\rightarrow
\{u^{<\alpha ^{\prime }>}=\left( x^{{\bf i^{\prime }\ }},y^{<{\bf a^{\prime
}>}}\right) \}
$$
and written as recurrent maps
$$
x^{{\bf i^{\prime }\ }}=x^{{\bf i^{\prime }\ }}(x^{{\bf i}}),~rank\left(
\frac{\partial x^{{\bf i^{\prime }\ }}}{\partial x^{{\bf i}}}\right) =n,%
\eqno(2.1)
$$
$$
y_{(1)}^{{\bf a_1^{\prime }\ }}=K_{{\bf a}_1{\bf \ }}^{{\bf a_1^{\prime }}%
}(x^{{\bf i\ }})y_{(1)}^{{\bf a}_1},K_{{\bf a}_1{\bf \ }}^{{\bf a_1^{\prime }%
}}(x^{{\bf i\ }})\in GL\left( m_1,{\cal R}\right) ,
$$
$$
............
$$
$$
y_{(p)}^{{\bf a_p^{\prime }\ }}=K_{{\bf a}_p{\bf \ }}^{{\bf a_p^{\prime }}%
}(u_{(p-1)})y_{(p)}^{{\bf a}_p},K_{{\bf a}_p{\bf \ }}^{{\bf a_p^{\prime }}%
}(u_{(p-1)})\in GL\left( m_p,{\cal R}\right) ,
$$
$$
.............
$$
$$
y_{(z)}^{{\bf a_z^{\prime }\ }}=K_{{\bf a}_z{\bf \ }}^{{\bf a_z^{\prime }}%
}(u_{(z-1)})y_{(z-1)}^{{\bf a}_z},K_{{\bf a}_z{\bf \ }}^{{\bf a_z^{\prime }}%
}(u_{(z-1)})\in GL\left( m_z,{\cal R}\right)
$$
where matrices $K_{{\bf a}_1{\bf \ }}^{{\bf a_1^{\prime }}}(x^{{\bf i\ }%
}),...,K_{{\bf a}_p{\bf \ }}^{{\bf a_p^{\prime }}}(u_{(p-1)}),...,K_{{\bf a}%
_z{\bf \ }}^{{\bf a_z^{\prime }}}(u_{(z-1)})$ are functions of necessary
smoothness class. In brief we write transforms (2.1) in the form%
$$
x^{{\bf i^{\prime }\ }}=x^{{\bf i^{\prime }\ }}(x^{{\bf i}}),y^{<{\bf %
a^{\prime }>\ }}=K_{<{\bf a>}}^{<{\bf a^{\prime }>}}y^{<{\bf a>}}.
$$
In general form we shall write $K$--matrices $K_{<\alpha {\bf >}}^{<\alpha
{\bf ^{\prime }>}}=\left( K_{{\bf i}}^{{\bf i^{\prime }}},K_{<{\bf a>}}^{<%
{\bf a^{\prime }>}}\right) ,$ where $K_{{\bf i}}^{{\bf i^{\prime }}}=\frac{%
\partial x^{{\bf i^{\prime }\ }}}{\partial x^{{\bf i}}}.$

A local coordinate parametrization of ${\cal E}^{<z>}$ naturally defines a
coordinate basis of the module of d--vector fi\-elds
 $\Xi \left( {\cal E}%
^{<z>}\right) ,$%
$$
\partial _{<\alpha >}=(\partial _i,\partial _{<a>})=(\partial _i,\partial
_{a_1},...,\partial _{a_p},...,\partial _{a_z})=\eqno(2.2)
$$
$$
\frac \partial {\partial u^{<\alpha >}}=\left( \frac \partial {\partial
x^i},\frac \partial {\partial y^{<a>}}\right) =\left( \frac \partial
{\partial x^i},\frac \partial {\partial y^{a_1}},...\frac \partial {\partial
y^{a_p}},...,\frac \partial {\partial y^{a_z}}\right) ,
$$
and the reciprocal to (2.2) coordinate basis
$$
d^{<\alpha >}=(d^i,d^{<a>})=(d^i,d^{a_1},...,d^{a_p},...,d^{a_z})=\eqno(2.3)
$$
$$
du^{<\alpha >}=(dx^i,dy^{<a>})=(dx^i,dy^{a_1},...,dy^{a_p},...,dy^{a_z}),
$$
which is uniquely defined from the equations%
$$
d^{<\alpha >}\circ \partial _{<\beta >}=\delta _{<\beta >}^{<\alpha >},
$$
where $\delta _{<\beta >}^{<\alpha >}$ is the Kronecher symbol and by ''$%
\circ $$"$ we denote the inner (scalar) product in the tangent bundle ${\cal %
TE}^{<z>}.$

The concept of {\bf nonlinear connection,} in brief, N--connection, is
fundamental in the geometry of locally anisotropic and higher order
anisotropic spaces (see a detailed study and basic references in \cite
{ma87,ma94,mirata}). In a dv--bundle ${\cal E}^{<z>}$ it is defined
as a distribution $\{N:E_u\rightarrow H_uE,T_uE=H_uE\oplus V_u^{(1)}E\oplus
...\oplus V_u^{(p)}E...\oplus V_u^{(z)}E\}$ on $E^{<z>}$ being a global
decomposition, as a Whitney sum, into horizontal,${\cal HE,\ }$ and
vertical, ${\cal VE}^{<p>}{\cal ,}p=1,2,...,z$ subbundles of the tangent
bundle ${\cal TE:}$
$$
{\cal TE}=H{\cal E}\oplus V{\cal E}^{<1>}\oplus ...\oplus V{\cal E}%
^{<p>}\oplus ...\oplus V{\cal E}^{<z>}.\eqno(2.4)
$$

Locally a N-connection in ${\cal E}^{<z>}$ is given by it components $N_{<%
{\bf a}_f>}^{<{\bf a}_p>}({\bf u}),z\geq p>f\geq 0$ (in brief we shall write
$N_{<a_f>}^{<a_p>}(u)$ ) with respect to bases (2.2) and (2.3)):

$$
{\bf N}=N_{<a_f>}^{<a_p>}(u)\delta ^{<a_f>}\otimes \delta _{<a_p>},(z\geq
p>f\geq 0),
$$

We note that a linear connection in a dv-bundle ${\cal E}^{<z>}$ can be
considered as a particular case of a N-connection when $%
N_i^{<a>}(u)=K_{<b>i}^{<a>}\left( x\right) y^{<b>},$ where functions $%
K_{<a>i}^{<b>}\left( x\right) $ on the base $M$ are called the Christoffel
coefficients.

To coordinate locally geometric constructions with the global splitting of\\
${\cal E}^{<z>}$ defined by a N-connection structure, we have to introduce a
lo\-cal\-ly adap\-ted bas\-is ( la--basis, la--frame ):%
$$
\delta _{<\alpha >}=(\delta _i,\delta _{<a>})=(\delta _i,\delta
_{a_1},...,\delta _{a_p},...,\delta _{a_z}),\eqno(2.5)
$$
with components parametrized as

$$
\delta _i=\partial _i-N_i^{a_1}\partial _{a_1}-...-N_i^{a_z}\partial _{a_z},
$$
$$
\delta _{a_1}=\partial _{a_1}-N_{a_1}^{a_2}\partial
_{a_2}-...-N_{a_1}^{a_z}\partial _{a_z},
$$
$$
................
$$
$$
\delta _{a_p}=\partial _{a_p}-N_{a_p}^{a_{p+1}}\partial
_{a_{p+1}}-...-N_{a_p}^{a_z}\partial _{a_z},
$$
$$
...............
$$
$$
\delta _{a_z}=\partial _{a_z}
$$
and it dual la--basis%
$$
\delta ^{<\alpha >}=(\delta ^i,\delta ^{<a>})=\left( \delta ^i,\delta
^{a_1},...,\delta ^{a_p},...,\delta ^{a_z}\right) ,\eqno(2.6)
$$
$$
\delta x^i=dx^i,
$$
$$
\delta y^{a_1}=dy^{a_1}+M_i^{a_1}dx^i,
$$
$$
\delta y^{a_2}=dy^{a_2}+M_{a_1}^{a_2}dy^{a_1}+M_i^{a_2}dx^i,
$$
$$
.................
$$
$$
\delta
y^{a_p}=dy^{a_p}+M_{a_{p-1}}^{a_p}dy^{p-1}+M_{a_{p-2}}^{a_p}dy^{a_{p-2}}+...
+M_i^{a_p}dx^i,
$$
$$
...................
$$
$$
\delta
y^{a_z}=dy^{a_z}+
M_{a_{z-1}}^{a_z}dy^{z-1}+M_{a_{z-2}}^{a_z}dy^{a_{z-2}}+...+M_i^{a_z}dx^i .
$$

The{\bf \ nonholonomic coefficients }${\bf w}=\{w_{<\beta ><\gamma
>}^{<\alpha >}\left( u\right) \}$ of locally adapted to the N--connection
structure frames  are defined as%
$$
\left[ \delta _{<\alpha >},\delta _{<\beta >}\right] =\delta _{<\alpha
>}\delta _{<\beta >}-\delta _{<\beta >}\delta _{<\alpha >}=w_{<\beta
><\gamma >}^{<\alpha >}\left( u\right) \delta _{<\alpha >}.
$$

The {\bf algebra of tensorial distinguished fields} $DT\left( {\cal E}%
^{<z>}\right) $ (d--fields, d--tensors, d--objects) on ${\cal E}^{<z>}$ is
introduced as the tensor algebra\\ ${\cal T}=\{{\cal T}%
_{qs_1...s_p...s_z}^{pr_1...r_p...r_z}\}$ of the dv-bundle ${\cal E}_{\left(
d\right) }^{<z>},$
$$
p_d:{\cal HE}^{<z>}{\cal \oplus V}^1{\cal E}^{<z>}{\cal \oplus }...{\cal %
\oplus V}^p{\cal E}^{<z>}{\cal \oplus }...{\cal \oplus V}^z{\cal E}^{<z>}%
{\cal \rightarrow E}^{<z>}{\cal .\,}
$$
${\cal \ }$ An element ${\bf t}\in {\cal T}_{qs_1...s_z}^{pr_1...r_z},$
d-tensor field of type $\left(
\begin{array}{cccccc}
p & r_1 & ... & r_p & ... & r_z \\
q & s_1 & ... & s_p & ... & s_z
\end{array}
\right) ,$ can be written in local form as%
$$
{\bf t}%
=t_{j_1...j_qb_1^{(1)}...b_{r_1}^{(1)}...b_1^{(p)}...b_{r_p}^{(p)}...
b_1^{(z)}...b_{r_z}^{(z)}}^{i_1...i_pa_1^{(1)}...a_{r_1}^{(1)}...
a_1^{(p)}...a_{r_p}^{(p)}...a_1^{(z)}...a_{r_z}^{(z)}}
\left( u\right) \delta _{i_1}\otimes ...\otimes \delta _{i_p}\otimes
d^{j_1}\otimes ...\otimes d^{j_q}\otimes
$$
$$
\delta _{a_1^{(1)}}\otimes ...\otimes \delta _{a_{r_1}^{(1)}}\otimes \delta
^{b_1^{(1)}}...\otimes \delta ^{b_{s_1}^{(1)}}\otimes ...\otimes \delta
_{a_1^{(p)}}\otimes ...\otimes \delta _{a_{r_p}^{(p)}}\otimes ...\otimes
$$
$$
\delta ^{b_1^{(p)}}...\otimes \delta ^{b_{s_p}^{(p)}}\otimes \delta
_{a_1^{(z)}}\otimes ...\otimes \delta _{a_{rz}^{(z)}}\otimes \delta
^{b_1^{(z)}}...\otimes \delta ^{b_{s_z}^{(z)}}.
$$

We shall respectively use denotations $X{\cal (E}^{<z>})$ (or $X{%
\left( M\right) ),\ }\Lambda ^p\left( {\cal E}^{<z>}\right) $ (or
$\Lambda ^p\left( M\right) ) $ and ${\cal F(E}^{<z>})$ (or $%
{\cal F}$ $\left( M\right) $) for the module of d-vector fields on ${\cal E}%
^{<z>}$ (or $M$ ), the exterior algebra of p-forms on ${\cal E}^{<z>}$
(or $M)$ and the set of real functions on ${\cal E}^{<z>}$(or $%
M). $

In general, d--objects on ${\cal E}^{<z>}$ are introduced as
geometric objects with various group and coordinate transforms coordinated
with the N--connection structure on ${\cal E}^{<z>}.$ For example, a
d--connection $D$ on ${\cal E}^{<z>}$ is defined as a linear
connection $D$ on $E^{<z>}$ conserving under a parallelism the global
decomposition (2.4) into horizontal and vertical subbundles of ${\cal TE}%
^{<z>}$ .

A N-connection in ${\cal E}^{<z>}$ induces a corresponding decomposition of
d-tensors into sums of horizontal and vertical parts, for example, for every
d-vector $X\in {\cal X(E}^{<z>})$ and 1-form $\widetilde{X}\in \Lambda
^1\left( {\cal E}^{<z>}\right) $ we have respectively
$$
X=hX+v_1X+...+v_zX{\bf \ \quad }\mbox{and \quad }\widetilde{X}=h\widetilde{X}%
+v_1\widetilde{X}+...~v_z\widetilde{X}.\eqno(2.7)
$$
In consequence, we can associate to every d-covariant derivation along the
d-vector (2.7), $D_X=X\circ D,$ two new operators of h- and v-covariant
derivations defined respectively as
$$
D_X^{(h)}Y=D_{hX}Y%
$$
and
$$
D_X^{\left( v_1\right) }Y=D_{v_1X}Y{\bf ,...,D_X^{\left( v_z\right)
}Y=D_{v_zX}Y\quad }\forall Y{\bf \in }{\cal X(E}^{<z>}) ,
$$
for which the following conditions hold:%
$$
D_XY{\bf =}D_X^{(h)}Y + D_X^{(v_1)}Y+...+D_X^{(v_z)}Y, \eqno(2.8)
$$
$$
D_X^{(h)}f=(hX{\bf )}f
$$
and
$$
D_X^{(v_p)}f=(v_pX{\bf )}f,\quad X,Y{\bf \in }{\cal X\left( E\right) ,}f\in
{\cal F}\left( M\right) ,p=1,2,...z.
$$

We define a {\bf metric structure }${\bf G\ }$in the total space $E^{<z>}$
of dv-bundle ${\cal E}^{<z>}{\cal =}$ $\left( E^{<z>},p,M\right) $ over a
connected and paracompact base $M$ as a symmetric covariant tensor field of
type $\left( 0,2\right) $, $G_{<\alpha ><\beta >,}$ being nondegenerate and
of constant signature on $E^{<z>}.$

Nonlinear connection ${\bf N}$ and metric ${\bf G}$ structures on ${\cal E}%
^{<z>}$ are mutually compatible it there are satisfied the conditions:
$$
{\bf G}\left( \delta _{a_f},\delta _{a_p}\right) =0,\mbox{or equivalently, }%
G_{a_fa_p}\left( u\right) -N_{a_f}^{<b>}\left( u\right) h_{a_f<b>}\left(
u\right) =0,\eqno(2.9)
$$
where $h_{a_pb_p}={\bf G}\left( \partial _{a_p},\partial _{b_p}\right) $ and
$G_{b_fa_p}={\bf G}\left( \partial _{b_f},\partial _{a_p}\right) ,0\leq
f<p\leq z,\,$ which gives%
$$
N_{c_f}^{b_p}\left( u\right) =h^{<a>b_p}\left( u\right) G_{c_f<a>}\left(
u\right) \eqno(2.10)
$$
(the matrix $h^{a_pb_p}$ is inverse to $h_{a_pb_p}).$ In consequence one
obtains the following decomposition of metric :%
$$
{\bf G}(X,Y){\bf =hG}(X,Y)+{\bf v}_1{\bf G}(X,Y)+...+{\bf v}_z{\bf G}(X,Y)%
{\bf ,}\eqno(2.11)
$$
where the d-tensor ${\bf hG}(X,Y){\bf =G}(hX,hY)$ is of type $\left(
\begin{array}{cc}
0 & 0 \\
2 & 0
\end{array}
\right) $ and the d-tensor ${\bf v}_p{\bf G}(X,Y)={\bf G}(v_pX,v_pY)$ is of
type $\left(
\begin{array}{ccccc}
0 & ... & 0(p) & ... & 0 \\
0 & ... & 2 & ... & z
\end{array}
\right) .$ With respect to la--basis (2.6) the d--metric (2.11) is written as%
$$
{\bf G}=g_{<\alpha ><\beta >}\left( u\right) \delta ^{<\alpha >}\otimes
\delta ^{<\beta >}=g_{ij}\left( u\right) d^i\otimes d^j+h_{<a><b>}\left(
u\right) \delta ^{<a>}\otimes \delta ^{<b>},\eqno(2.12)
$$
where $g_{ij}={\bf G}\left( \delta _i,\delta _j\right) .$

A metric structure of type (2.11) (equivalently, of type (2.12)) or a metric
on $E^{<z>}$ with components satisfying constraints (2.9), equivalently
(2.10)) defines an adapted to the given N-connection inner (d--scalar)
product on the tangent bundle ${\cal TE}^{<z>}{\cal .}$

We shall say that a d-connection $\widehat{D}_X$ is compatible with the
d-scalar product on ${\cal TE}^{<z>}{\cal \ }$ (i.e. is a standard
d-connection) if
$$
\widehat{D}_X\left( {\bf X\cdot Y}\right) =\left( \widehat{D}_X{\bf Y}%
\right) \cdot {\bf Z+Y\cdot }\left( \widehat{D}_X{\bf Z}\right) ,\forall
{\bf X,Y,Z}{\bf \in }{\cal X(E}^{<z>}){\cal .}
$$
An arbitrary d--connection $D_X$ differs from the standard one $\widehat{D}_X$
by an operator $\widehat{P}_X\left( u\right) =\{X^{<\alpha >}\widehat{P}%
_{<\alpha ><\beta >}^{<\gamma >}\left( u\right) \},$ called the deformation
d-tensor with respect to $\widehat{D}_X,$ which is just a d-linear transform
of ${\cal E}_u^{<z>},$ $\forall u\in {\cal E}^{<z>}{\cal .}$ The explicit
form of $\widehat{P}_X$ can be found by using the corresponding axiom
defining linear connections \cite{lue}
$$
\left( D_X-\widehat{D}_X\right) fZ=f\left( D_X-\widehat{D}_X\right) Z{\bf ,}
$$
written with respect to la-bases (2.5) and (2.6). From the last expression
we obtain
$$
\widehat{P}_X\left( u\right) =\left[ (D_X-\widehat{D}_X)\delta _{<\alpha
>}\left( u\right) \right] \delta ^{<\alpha >}\left( u\right) ,
$$
therefore
$$
D_XZ{\bf \ }=\widehat{D}_XZ{\bf \ +}\widehat{P}_XZ.\eqno(2.13)
$$

A d-connection $D_X$ is {\bf metric (}or  compatible with met\-ric ${\bf G%
}$) on ${\cal E}^{<z>}$ if%
$$
D_X{\bf G}=0,\forall X{\bf \in }{\cal X(E}^{<z>}).\eqno(2.14)
$$

Locally adapted components $\Gamma _{<\beta ><\gamma >}^{<\alpha >}$ of a
d-connection $D_{<\alpha >}=(\delta _{<\alpha >}\circ D)$ are defined by the
equations%
$$
D_{<\alpha >}\delta _{<\beta >}=\Gamma _{<\alpha ><\beta >}^{<\gamma
>}\delta _{<\gamma >},
$$
from which one immediately follows%
$$
\Gamma _{<\alpha ><\beta >}^{<\gamma >}\left( u\right) =\left( D_{<\alpha
>}\delta _{<\beta >}\right) \circ \delta ^{<\gamma >}.\eqno(2.15)
$$

The operations of h- and v$_{(p)}$-covariant derivations, $%
D_k^{(h)}=\{L_{jk}^i,L_{<b>k\;}^{<a>}\}$ and $D_{c_p}^{(v_p)}=%
\{C_{jc_p}^i,C_{<b>c_p}^i,C_{jc_p}^{<a>},C_{<b>c_p}^{<a>}\}$ (see (2.8)),
are introduced as corresponding h- and v$_{(p)}$-paramet\-ri\-za\-ti\-ons of
(2.15):%
$$
L_{jk}^i=\left( D_k\delta _j\right) \circ d^i,\quad L_{<b>k}^{<a>}=\left(
D_k\delta _{<b>}\right) \circ \delta ^{<a>}\eqno(2.16)
$$
and%
$$
C_{jc_p}^i=\left( D_{c_p}\delta _j\right) \circ \delta ^i,\quad
C_{<b>c_p}^{<a>}=\left( D_{c_p}\delta _{<b>}\right) \circ \delta ^{<a>}%
\eqno(2.17)
$$
$$
C_{<b>c_p}^i=\left( D_{c_p}\delta _{<b>}\right) \circ \delta ^i,\quad
C_{jc_p}^{<a>}=\left( D_{c_p}\delta _j\right) \circ \delta ^{<a>}.
$$
A set of components (2.16) and (2.17), $D\Gamma =\left(
L_{jk}^i,L_{<b>k}^{<a>},C_{j<c>}^i,C_{<b><c>}^{<a>}\right) ,\,$ completely
defines the local action of a d-connection $D$ in ${\cal E}^{<z>}.$ For
instance, taken a d-tensor field of type $\left(
\begin{array}{cccc}
1 & ... & 1(p) & ... \\
1 & ... & 1(p) & ...
\end{array}
\right) ,$ ${\bf t}=t_{jb_p}^{ia_p}\delta _i\otimes \delta _{a_p}\otimes
\delta ^j\otimes \delta ^{b_p},$ and a d-vector ${\bf X}=X^i\delta
_i+X^{<a>}\delta _{<a>}$ we have%
$$
D_X{\bf t=}D_X^{(h)}{\bf t+}D_X^{(v_1)}{\bf %
t+..+.D_X^{(v_p)}t+...+D_X^{(v_z)}t=}
$$
$$
\left( X^kt_{jb_p|k}^{ia_p}+X^{<c>}t_{jb_p\perp <c>}^{ia_p}\right) \delta
_i\otimes \delta _{a_p}\otimes d^j\otimes \delta ^{b_p},
$$
where the h--covariant derivative is written as%
$$
t_{jb_p|k}^{ia_p}=\frac{\delta t_{jb_p}^{ia_p}}{\delta x^k}%
+L_{hk}^it_{jb_p}^{ha_p}+L_{c_pk}^{a_p}t_{jb_p}^{ic_p}-
L_{jk}^ht_{hb_p}^{ia_p}-L_{b_pk}^{c_p}t_{jc_p}^{ia_p}
$$
and the v--covariant derivatives are written as%
$$
t_{jb_p\perp <c>}^{ia_p}=\frac{\partial t_{jb_p}^{ia_p}}{\partial y^{<c>}}%
+C_{h<c>}^it_{jb_p}^{ha_p}+C_{d_p<c>}^{a_p}t_{jb_p}^{id_p}-
C_{j<c>}^ht_{hb_p}^{ia_p}-C_{b_p<c>}^{d_p}t_{jd_p}^{ia_p}.
$$
For a scalar function $f\in {\cal F(E}^{<z>})$ we have%
$$
D_i^{(h)}=\frac{\delta f}{\delta x^i}=\frac{\partial f}{\partial x^i}%
-N_i^{<a>}\frac{\partial f}{\partial y^{<a>}},
$$
$$
D_{a_f}^{(v_f)}=\frac{\delta f}{\delta x^{a_f}}=\frac{\partial f}{\partial
x^{a_f}}-N_{a_f}^{a_p}\frac{\partial f}{\partial y^{a_p}},1\leq f<p\leq z-1,
$$
$$
\mbox{ and }D_{c_z}^{(v_z)}f=\frac{\partial f}{\partial y^{c_z}}.
$$

We emphasize that the geometry of connections in a dv-bundle ${\cal E}^{<z>}$
is very reach. If a triple of fundamental geometric objects
$(N_{a_f}^{a_p}\left( u\right) ,$
 $\Gamma _{<\beta ><\gamma >}^{<\alpha >}\left(u\right) ,$
 $G_{<\alpha ><\beta >}\left( u\right) ) $ is fixed on ${\cal E}^{<z>},$
a multiconnection structure (with corresponding different
rules of covariant derivation, which are, or not, mutually compatible and
with the same, or not, induced d-scalar products in ${\cal TE}^{<z>}{\cal )}$
is defined on this dv-bundle. For instance, we enumerate some of connections
and covariant derivations which can present interest in investigation of
locally anisotropic gravitational and matter field interactions:

\begin{enumerate}
\item  Every N-connection in ${\cal E}^{<z>}{\cal ,}$ with coefficients $%
N_{a_f}^{a_p}\left( u\right) $ being differentiable on y-variables, induces
a structure of linear connection $\widetilde{N}_{<\beta ><\gamma >}^{<\alpha
>},$ where $\widetilde{N}_{b_pc_f}^{a_p}=\frac{\partial N_{c_f}^{a_p}}{%
\partial y^{b_p}}$ and $\widetilde{N}_{b_pc_p}^{a_p}\left( u\right) =0.$ For
some $$Y\left( u\right) =Y^i\left( u\right) \partial _i+Y^{<a>}\left(
u\right) \partial _{<a>}$$ and $$B\left( u\right) =B^{<a>}\left( u\right)
\partial _{<a>}$$ one writes%
$$
D_Y^{(\widetilde{N})}B=\left[ Y^{c_f}\left( \frac{\partial B^{a_p}}{\partial
y^{c_f}}+\widetilde{N}_{b_pi}^{a_p}B^{b_p}\right) +Y^{b_p}\frac{\partial
B^{a_p}}{\partial y^{b_p}}\right] \frac \partial {\partial y^{a_p}}~(0\leq
f<p\leq z).
$$

\item  The d--connection of Berwald type \cite{berw}

$$
\Gamma _{<\beta ><\gamma >}^{(B)<\alpha >}=\left( L_{jk}^i,\frac{\partial
N_k^{<a>}}{\partial y^{<b>}},0,C_{<b><c>}^{<a>}\right) ,\eqno(2.18)
$$
where
$$
L_{.jk}^i\left( u\right) =\frac 12g^{ir}\left( \frac{\delta g_{jk}}{\partial
x^k}+\frac{\delta g_{kr}}{\partial x^j}-\frac{\delta g_{jk}}{\partial x^r}%
\right) ,
$$

$$
C_{.<b><c>}^{<a>}\left( u\right) =\frac 12h^{<a><d>}\left( \frac{\delta
h_{<b><d>}}{\partial y^{<c>}}+\frac{\delta h_{<c><d>}}{\partial y^{<b>}}-%
\frac{\delta h_{<b><c>}}{\partial y^{<d>}}\right) ,\eqno(2.19)
$$

which is hv-metric, i.e. $D_k^{(B)}g_{ij}=0$ and $D_{<c>}^{(B)}h_{<a><b>}=0.$

\item  The canonical d--connection ${\bf \Gamma ^{(c)}}$ associated to a
metric ${\bf G}$ of type (2.12) $\Gamma _{<\beta ><\gamma >}^{(c)<\alpha
>}=\left(
L_{jk}^{(c)i},L_{<b>k}^{(c)<a>},C_{j<c>}^{(c)i},C_{<b><c>}^{(c)<a>}\right) ,$
with coefficients%
$$
L_{jk}^{(c)i}=L_{.jk}^i,C_{<b><c>}^{(c)<a>}=C_{.<b><c>}^{<a>}%
\mbox{ (see (2.19)}
$$
$$
L_{<b>i}^{(c)<a>}=\widetilde{N}_{<b>i}^{<a>}+
$$
$$
\frac 12h^{<a><c>}\left( \frac{\delta h_{<b><c>}}{\delta x^i}-\widetilde{N}%
_{<b>i}^{<d>}h_{<d><c>}-\widetilde{N}_{<c>i}^{<d>}h_{<d><b>}\right) ,
$$
$$
C_{j<c>}^{(c)i}=\frac 12g^{ik}\frac{\partial g_{jk}}{\partial y^{<c>}}.%
\eqno(2.20)
$$
This is a metric d--connection which satisfies conditions
$$
D_k^{(c)}g_{ij}=0,D_{<c>}^{(c)}g_{ij}=0,D_k^{(c)}h_{<a><b>}=0,
D_{<c>}^{(c)}h_{<a><b>}=0.
$$

\item  We can consider N-adapted Christoffel d--symbols%
$$
\widetilde{\Gamma }_{<\beta ><\gamma >}^{<\alpha >}=\frac 12G^{<\alpha
><\tau >}\left( \delta _{<\gamma >}G_{<\tau ><\beta >}+\delta _{<\beta
>}G_{<\tau ><\gamma >}-\delta _{<\tau >}G_{<\beta ><\gamma >}\right) ,%
\eqno(2.21)
$$
which have the components of d-connection $$\widetilde{\Gamma }_{<\beta
><\gamma >}^{<\alpha >}=\left( L_{jk}^i,0,0,C_{<b><c>}^{<a>}\right) $$ with $%
L_{jk}^i$ and $C_{<b><c>}^{<a>}$ as in (2.19) if $G_{<\alpha ><\beta >}$ is
taken in the form (2.12).
\end{enumerate}

Arbitrary linear connections on a dv--bundle ${\cal E}^{<z>}$ can be also
characterized by theirs deformation tensors (see (2.13)) with respect, for
instance, to d--connect\-i\-on (2.21):%
$$
\Gamma _{<\beta ><\gamma >}^{(B)<\alpha >}=\widetilde{\Gamma }_{<\beta
><\gamma >}^{<\alpha >}+P_{<\beta ><\gamma >}^{(B)<\alpha >},\Gamma _{<\beta
><\gamma >}^{(c)<\alpha >}=\widetilde{\Gamma }_{<\beta ><\gamma >}^{<\alpha
>}+P_{<\beta ><\gamma >}^{(c)<\alpha >}
$$
or, in general,%
$$
\Gamma _{<\beta ><\gamma >}^{<\alpha >}=\widetilde{\Gamma }_{<\beta ><\gamma
>}^{<\alpha >}+P_{<\beta ><\gamma >}^{<\alpha >},
$$
where $P_{<\beta ><\gamma >}^{(B)<\alpha >},P_{<\beta ><\gamma
>}^{(c)<\alpha >}$ and $P_{<\beta ><\gamma >}^{<\alpha >}$ are respectively
the deformation d--ten\-sors of d-connect\-i\-ons (2.18),\ (2.20), or of a
general one.

\subsubsection{ Torsions and curvatures of N- and d-connections}

The curvature ${\bf \Omega }$$\,$ of a nonlinear connection ${\bf N}$ in a
dv-bundle ${\cal E}^{<z>}$ can be defined as the Nijenhuis tensor field $%
N_v\left( X,Y\right) $ associated to ${\bf N\ }$ \cite{ma87,ma94}:
$$
{\bf \Omega }=N_v={\bf \left[ vX,vY\right] +v\left[ X,Y\right] -v\left[
vX,Y\right] -v\left[ X,vY\right] ,X,Y}\in {\cal X(E}^{<z>}){\cal ,}
$$
where $v=v_1\oplus ...\oplus v_z.$ In local form one has%
$$
{\bf \Omega }=\frac 12\Omega _{b_fc_f}^{a_p}\delta ^{b_f}\bigwedge \delta
^{c_f}\otimes \delta _{a_p},(0\leq f<p\leq z),
$$
where%
$$
\Omega _{b_fc_f}^{a_p}=\frac{\delta N_{c_f}^{a_p}}{\partial y^{b_f}}-\frac{%
\partial N_{b_f}^{a_p}}{\partial y^{c_f}}+N_{b_f}^{<b>}\widetilde{N}%
_{<b>c_f}^{a_p}-N_{c_f}^{<b>}\widetilde{N}_{<b>b_f}^{a_p}.\eqno(2.22)
$$

The torsion ${\bf T}$ of d--connection ${\bf D\ }$ in ${\cal E}^{<z>}$ is
defined by the equation%
$$
{\bf T\left( X,Y\right) =XY_{\circ }^{\circ }T\doteq }D_X{\bf Y-}D_Y{\bf X\
-\left[ X,Y\right] .}\eqno(2.23)
$$
One holds the following h- and v$_{(p)}-$--decompositions%
$$
{\bf T\left( X,Y\right) =T\left( hX,hY\right) +T\left( hX,vY\right) +T\left(
vX,hY\right) +T\left( vX,vY\right) .}\eqno(2.24)
$$
We consider the projections: ${\bf hT\left( X,Y\right) ,v}_{(p)}{\bf T\left(
hX,hY\right) ,hT\left( hX,hY\right) ,...}$ and say that, for instance, ${\bf %
hT\left( hX,hY\right) }$ is the h(hh)-torsion of ${\bf D}$ , ${\bf v}_{(p)}%
{\bf T\left( hX,hY\right) \ }$ is the v$_p$(hh)-torsion of ${\bf D}$ and so
on.

The torsion (2.23) is locally determined by five d-tensor fields, torsions,
defined as
$$
T_{jk}^i={\bf hT}\left( \delta _k,\delta _j\right) \cdot d^i,\quad
T_{jk}^{a_p}={\bf v}_{(p)}{\bf T}\left( \delta _k,\delta _j\right) \cdot
\delta ^{a_p},
$$
$$
P_{jb_p}^i={\bf hT}\left( \delta _{b_p},\delta _j\right) \cdot d^i,\quad
P_{jb_f}^{a_p}={\bf v}_{(p)}{\bf T}\left( \delta _{b_f},\delta _j\right)
\cdot \delta ^{a_p},
$$
$$
S_{b_fc_f}^{a_p}={\bf v}_{(p)}{\bf T}\left( \delta _{c_f},\delta
_{b_f}\right) \cdot \delta ^{a_p}.
$$
Using formulas (2.5),(2.6),(2.22) and (2.23) we can computer in explicit
form the components of torsions (2.24) for a d--connection of type (2.16)
and (2.17):
$$
T_{.jk}^i=T_{jk}^i=L_{jk}^i-L_{kj}^i,\quad
T_{j<a>}^i=C_{.j<a>}^i,T_{<a>j}^i=-C_{j<a>}^i,\eqno(2.25)
$$
$$
T_{.j<a>}^i=0,T_{.<b><c>}^{<a>}=S_{.<b><c>}^{<a>}=C_{<b><c>}^{<a>}-
C_{<c><b>}^{<a>},
$$
$$
T_{.b_fc_f}^{a_p}=\frac{\delta N_{c_f}^{a_p}}{\partial y^{b_f}}-\frac{\delta
N_{b_f}^{a_p}}{\partial y^{c_f}},T_{.<b>i}^{<a>}=P_{.<b>i}^{<a>}=\frac{%
\delta N_i^{<a>}}{\partial y^{<b>}}%
-L_{.<b>j}^{<a>},T_{.i<b>}^{<a>}=-P_{.<b>i}^{<a>}.
$$

The curvature ${\bf R}$ of d--connection in ${\cal E}^{<z>}$ is defined by
the equation
$$
{\bf R\left( X,Y\right) Z=XY_{\bullet }^{\bullet }R\bullet Z}=D_XD_Y{\bf Z}%
-D_YD_X{\bf Z-}D_{[X,Y]}{\bf Z.}\eqno(2.26)
$$
One holds the next properties for the h- and v-decompositions of curvature:%
$$
{\bf v}_{(p)}{\bf R\left( X,Y\right) hZ=0,\ hR\left( X,Y\right) v}_{(p)}{\bf %
Z=0,~v}_{(f)}{\bf R\left( X,Y\right) v}_{(p)}{\bf Z=0,}
$$
$$
{\bf R\left( X,Y\right) Z=hR\left( X,Y\right) hZ+vR\left( X,Y\right) vZ,}
$$
where ${\bf v=v}_1+...+{\bf v}_z.$ From (2.26) and the equation ${\bf %
R\left( X,Y\right) =-R\left( Y,X\right) }$ we get that the curvature of a
d-con\-nec\-ti\-on ${\bf D}$ in ${\cal E}^{<z>}$ is completely determined by
the following d-tensor fields:%
$$
R_{h.jk}^{.i}=\delta ^i\cdot {\bf R}\left( \delta _k,\delta _j\right) \delta
_h,~R_{<b>.jk}^{.<a>}=\delta ^{<a>}\cdot {\bf R}\left( \delta _k,\delta
_j\right) \delta _{<b>},\eqno(2.27)
$$
$$
P_{j.k<c>}^{.i}=d^i\cdot {\bf R}\left( \delta _{<c>},\delta _{<k>}\right)
\delta _j,~P_{<b>.<k><c>}^{.<a>}=\delta ^{<a>}\cdot {\bf R}\left( \delta
_{<c>},\delta _{<k>}\right) \delta _{<b>},
$$
$$
S_{j.<b><c>}^{.i}=d^i\cdot {\bf R}\left( \delta _{<c>},\delta _{<b>}\right)
\delta _j,~S_{<b>.<c><d>}^{.<a>}=\delta ^{<a>}\cdot {\bf R}\left( \delta
_{<d>},\delta _{<c>}\right) \delta _{<b>}.
$$
By a direct computation, using (2.5),(2.6),(2.16),(2.17) and (2.27) we get :
$$
R_{h.jk}^{.i}=\frac{\delta L_{.hj}^i}{\delta x^h}-\frac{\delta L_{.hk}^i}{%
\delta x^j}+L_{.hj}^mL_{mk}^i-L_{.hk}^mL_{mj}^i+C_{.h<a>}^iR_{.jk}^{<a>},%
\eqno(2.28)
$$
$$
R_{<b>.jk}^{.<a>}=\frac{\delta L_{.<b>j}^{<a>}}{\delta x^k}-\frac{\delta
L_{.<b>k}^{<a>}}{\delta x^j}%
+L_{.<b>j}^{<c>}L_{.<c>k}^{<a>}-L_{.<b>k}^{<c>}L_{.<c>j}^{<a>}+
C_{.<b><c>}^{<a>}R_{.jk}^{<c>},
$$
$$
P_{j.k<a>}^{.i}=\frac{\delta L_{.jk}^i}{\partial y^{<a>}}%
+C_{.j<b>}^iP_{.k<a>}^{<b>}-
$$
$$
\left( \frac{\partial C_{.j<a>}^i}{\partial x^k}%
+L_{.lk}^iC_{.j<a>}^l-L_{.jk}^lC_{.l<a>}^i-L_{.<a>k}^{<c>}C_{.j<c>}^i\right)
,
$$
$$
P_{<b>.k<a>}^{.<c>}=\frac{\delta L_{.<b>k}^{<c>}}{\partial y^{<a>}}%
+C_{.<b><d>}^{<c>}P_{.k<a>}^{<d>}-
$$
$$
\left( \frac{\partial C_{.<b><a>}^{<c>}}{\partial x^k}+L_{.<d>k}^{<c>%
\,}C_{.<b><a>}^{<d>}-L_{.<b>k}^{<d>}C_{.<d><a>}^{<c>}-
L_{.<a>k}^{<d>}C_{.<b><d>}^{<c>}\right) ,
$$
$$
S_{j.<b><c>}^{.i}=\frac{\delta C_{.j<b>}^i}{\partial y^{<c>}}-\frac{\delta
C_{.j<c>}^i}{\partial y^{<b>}}+C_{.j<b>}^hC_{.h<c>}^i-C_{.j<c>}^hC_{h<b>}^i,
$$
$$
S_{<b>.<c><d>}^{.<a>}=\frac{\delta C_{.<b><c>}^{<a>}}{\partial y^{<d>}}-%
\frac{\delta C_{.<b><d>}^{<a>}}{\partial y^{<c>}}%
+C_{.<b><c>}^{<e>}C_{.<e><d>}^{<a>}-C_{.<b><d>}^{<e>}C_{.<e><c>}^{<a>}.
$$

We note that torsions (2.25) and curvatures (2.28) can be computed by
particular cases of d-connections when d-connections (2.17), (2.20) or
(2.22) are used instead of (2.16) and (2.17).

The components of the Ricci d--tensor
$$
R_{<\alpha ><\beta >}=R_{<\alpha >.<\beta ><\tau >}^{.<\tau >}
$$
with respect to locally adapted frame (2.6) are as follows:%
$$
R_{ij}=R_{i.jk}^{.k},\quad R_{i<a>}=-^2P_{i<a>}=-P_{i.k<a>}^{.k},\eqno(2.29)
$$
$$
R_{<a>i}=^1P_{<a>i}=P_{<a>.i<b>}^{.<b>},\quad
R_{<a><b>}=S_{<a>.<b><c>}^{.<c>}.
$$
We point out that because, in general, $^1P_{<a>i}\neq ~^2P_{i<a>}$ the
Ricci d--tensor is non symmetric.

Having defined a d--metric of type (2.12) in ${\cal E}^{<z>}$ we can
introduce the scalar curvature of d--connection ${\bf D}$:
$$
{\overleftarrow{R}}=G^{<\alpha ><\beta >}R_{<\alpha ><\beta >}=R+S,%
$$
where $R=g^{ij}R_{ij}$ and $S=h^{<a><b>}S_{<a><b>}.$

For our further considerations it will be also useful to use an alternative
way of definition torsion (2.23) and curvature (2.26) by using the
commutator
$$
\Delta _{<\alpha ><\beta >}\doteq \nabla _{<\alpha >}\nabla _{<\beta
>}-\nabla _{<\beta >}\nabla _{<\alpha >}=2\nabla _{[<\alpha >}\nabla
_{<\beta >]}.\eqno(2.29)
$$
For components (2.25) of d--torsion we have
$$
\Delta _{<\alpha ><\beta >}f=T_{.<\alpha ><\beta >}^{<\gamma >}\nabla
_{<\gamma >}f\eqno(2.30)
$$
for every scalar function $f\,\,$ on ${\cal E}^{<z>}{\cal .}$ Curvature can
be introduced as an operator acting on arbitrary d--vector $V^{<\delta >}:$

$$
(\Delta _{<\alpha ><\beta >}-T_{.<\alpha ><\beta >}^{<\gamma >}\nabla
_{<\gamma >})V^{<\delta >}=R_{~<\gamma >.<\alpha ><\beta >}^{.<\delta
>}V^{<\gamma >}\eqno(2.31)
$$
(in this section we  follow conventions of Miron and
Anastasiei \cite{ma87,ma94} on d--tensors; we can obtain corresponding
Penrose and Rindler abstract index formulas \cite{penr1,penr2} just for a
trivial N-connection structure and by changing denotations for components of
torsion and curvature in this manner:\ $T_{.\alpha \beta }^\gamma
\rightarrow T_{\alpha \beta }^{\quad \gamma }$ and $R_{~\gamma .\alpha \beta
}^{.\delta }\rightarrow R_{\alpha \beta \gamma }^{\qquad \delta }).$

Here we also note that torsion and curvature of a d--connection on ${\cal E}%
^{<z>}$ satisfy generalized for ha--spaces Ricci and Bianchi identities which
in terms of components (2.30) and (2.31) are written respectively as%
$$
R_{~[<\gamma >.<\alpha ><\beta >]}^{.<\delta >}+\nabla _{[<\alpha
>}T_{.<\beta ><\gamma >]}^{<\delta >}+T_{.[<\alpha ><\beta >}^{<\nu
>}T_{.<\gamma >]<\nu >}^{<\delta >}=0\eqno(2.32)
$$
and%
$$
\nabla _{[<\alpha >}R_{|<\nu >|<\beta ><\gamma >]}^{\cdot <\sigma
>}+T_{\cdot [<\alpha ><\beta >}^{<\delta >}R_{|<\nu >|.<\gamma >]<\delta
>}^{\cdot <\sigma >}=0.
$$
Identities (2.32) can be proved similarly as in \cite{penr1} by taking into
account that indices play a distinguished character.

We can also consider a ha-generalization of the so-called conformal Weyl
tensor (see, for instance, \cite{penr1}) which can be written as a d-tensor
in this form:%
$$
C_{\quad <\alpha ><\beta >}^{<\gamma ><\delta >}=R_{\quad <\alpha ><\beta
>}^{<\gamma ><\delta >}-\frac 4{n+m_1+...+m_z-2}R_{\quad [<\alpha
>}^{[<\gamma >}~\delta _{\quad <\beta >]}^{<\delta >]}+\eqno(2.33)
$$
$$
\frac 2{(n+m_1+...m_z-1)(n+m_1+...+m_z-2)}{\overleftarrow{R}~\delta _{\quad
[<\alpha >}^{[<\gamma >}~\delta _{\quad <\beta >]}^{<\delta >]}.}
$$
This object is conformally invariant on ha--spaces provided with d--connection
generated by d--metric structures.

\subsubsection{ Field equations for ha--gravity}

The Einstein equations in some models of higher order anisotropic
supergravity have been considered in \cite{v96jpa3}. Here we note that the
Einstein equations and conservation laws on v--bundles provided with
N-connection structures were studied in detail in \cite
{ma87,ma94,ana86,ana87,vodg,voa,vcl96}. In Ref. \cite{vg} we proved that the
la-gravity can be formulated in a gauge like manner and analyzed the
conditions when the Einstein la-gravitational field equations are equivalent
to a corresponding form of Yang-Mills equations. Our aim here is to
write the higher order anisotropic gravitational field equations in a form
more convenient for theirs equivalent reformulation in ha--spinor variables.

We define d-tensor $\Phi _{<\alpha ><\beta >}$ as to satisfy conditions
$$
-2\Phi _{<\alpha ><\beta >}\doteq R_{<\alpha ><\beta >}-\frac
1{n+m_1+...+m_z}\overleftarrow{R}g_{<\alpha ><\beta >}
$$
which is the torsionless part of the Ricci tensor for locally isotropic
spaces \cite{penr1,penr2}, i.e. $\Phi _{<\alpha >}^{~~<\alpha >}\doteq 0$.\
The Einstein equations on ha--spaces
$$
\overleftarrow{G}_{<\alpha ><\beta >}+\lambda g_{<\alpha ><\beta >}=\kappa
E_{<\alpha ><\beta >},\eqno(2.34)
$$
where%
$$
\overleftarrow{G}_{<\alpha ><\beta >}=R_{<\alpha ><\beta >}-\frac 12%
\overleftarrow{R}g_{<\alpha ><\beta >}
$$
is the Einstein d--tensor, $\lambda $ and $\kappa $ are correspondingly the
cosmological and gravitational constants and by $E_{<\alpha ><\beta >}$ is
denoted the locally anisotropic energy--momentum d--tensor, can be rewritten
in equivalent form:%
$$
\Phi _{<\alpha ><\beta >}=-\frac \kappa 2(E_{<\alpha ><\beta >}-\frac
1{n+m_1+...+m_z}E_{<\tau >}^{~<\tau >}~g_{<\alpha ><\beta >}).\eqno(2.35)
$$

Because ha--spaces generally have nonzero torsions we shall add to (2.35)
(equivalently to (2.34)) a system of algebraic d--field equations with the
source $S_{~<\beta ><\gamma >}^{<\alpha >}$ being the locally anisotropic
spin density of matter (if we consider a variant of higher order anisotropic
Einstein--Cartan theory ):
$$
T_{~<\alpha ><\beta >}^{<\gamma >}+2\delta _{~[<\alpha >}^{<\gamma
>}T_{~<\beta >]<\delta >}^{<\delta >}=\kappa S_{~<\alpha ><\beta
>.}^{<\gamma >}\eqno(2.36)
$$
From (2.32 ) and (2.36) one follows the conservation law of higher order
anisot\-rop\-ic spin matter:%
$$
\nabla _{<\gamma >}S_{~<\alpha ><\beta >}^{<\gamma >}-T_{~<\delta ><\gamma
>}^{<\delta >}S_{~<\alpha ><\beta >}^{<\gamma >}=E_{<\beta ><\alpha
>}-E_{<\alpha ><\beta >}.
$$

Finally,  we remark that all presented geometric
constructions contain those elaborated for generalized Lagrange spaces \cite
{ma87,ma94} (for which a tangent bundle $TM$ is considered instead of a
v-bundle ${\cal E}^{<z>}$ ) and for constructions on the so called osculator
bundles with different prolongations and extensions of Finsler and Lagrange
metrics \cite{mirata}. We also note that the Lagrange (Finsler) geometry is
characterized by a metric of type (2.12) with components parametized as $%
g_{ij}=\frac 12\frac{\partial ^2{\cal L}}{\partial y^i\partial y^j}$ $\left(
g_{ij}=\frac 12\frac{\partial ^2\Lambda ^2}{\partial y^i\partial y^j}\right)
$ and $h_{ij}=g_{ij},$ where ${\cal L=L}$ $(x,y)$ $\left( \Lambda =\Lambda
\left( x,y\right) \right) $ is a Lagrangian $\left( \mbox{Finsler metric}%
\right) $ on $TM$ (see details in \cite{ma87,ma94,mat,bej}).

\subsection{Distinguished Clifford Algebras}

The typical fiber of dv-bundle $\xi _d\ ,\ \pi _d:\ HE\oplus V_1E\oplus
...\oplus V_zE\rightarrow E$ is a d-vector space, ${\cal F}=h{\cal F}\oplus
v_1{\cal F\oplus ...}\oplus v_z{\cal F},$ split into horizontal $h{\cal F}$
and verticals $v_p{\cal F,}p=1,...,z$ subspaces, with metric $G(g,h)$
induced by v-bundle metric (2.12). Clifford algebras (see, for example,
Refs. \cite{kar,tur,penr2}) formulated for d-vector spaces will be called
Clifford d-algebras \cite{vjmp,vb295,vod}. We shall consider
the main properties of Clifford d--algebras. The proof of theorems will be
based on the technique developed in Ref. \cite{kar} correspondingly adapted
to the distinguished character of spaces in consideration.

Let $k$ be a number field (for our purposes $k={\cal R}$ or $k={\cal C},%
{\cal R}$ and ${\cal C},$ are, respectively real and complex number fields)
and define ${\cal F},$ as a d-vector space on $k$ provided with
nondegenerate symmetric quadratic form (metric)\ $G.$ Let $C$ be an algebra
on $k$ (not necessarily commutative) and $j\ :\ {\cal F}$ $\rightarrow C$ a
homomorphism of underlying vector spaces such that $j(u)^2=\;G(u)\cdot 1\ (1$
is the unity in algebra $C$ and d-vector $u\in {\cal F}).$ We are interested
in definition of the pair $\left( C,j\right) $ satisfying the next
universitality conditions. For every $k$-algebra $A$ and arbitrary
homomorphism $\varphi :{\cal F}\rightarrow A$ of the underlying d-vector
spaces, such that $\left( \varphi (u)\right) ^2\rightarrow G\left( u\right)
\cdot 1,$ there is a unique homomorphism of algebras $\psi \ :\ C\rightarrow
A$ transforming the diagram 1 into a commutative one.

 The algebra solving this problem will be denoted as
$C\left( {\cal F},A\right) $ [equivalently as $C\left( G\right) $ or $%
C\left( {\cal F}\right) ]$ and called as Clifford d--algebra associated with
pair $\left( {\cal F},G\right) .$

\begin{theorem}
The above-presented diagram has a unique solution $\left( C,j\right) $ up to
isomorphism.
\end{theorem}

{\bf Proof:} (We adapt for d-algebras that of Ref. \cite{kar}, p. 127.) For
a universal problem the uniqueness is obvious if we prove the existence of
solution $C\left( G\right) $ . To do this we use tensor algebra ${\cal L}%
^{(F)}=\oplus {\cal L}_{qs}^{pr}\left( {\cal F}\right) $ =$\oplus
_{i=0}^\infty T^i\left( {\cal F}\right) ,$ where $T^0\left( {\cal F}\right)
=k$ and $T^i\left( {\cal F}\right) =k$ and $T^i\left( {\cal F}\right) ={\cal %
F}\otimes ...\otimes {\cal F}$ for $i>0.$ Let $I\left( G\right) $ be the
bilateral ideal generated by elements of form $\epsilon \left( u\right)
=u\otimes u-G\left( u\right) \cdot 1$ where $u\in {\cal F}$ and $1$ is the
unity element of algebra ${\cal L}\left( {\cal F}\right) .$ Every element
from $I\left( G\right) $ can be written as $\sum\nolimits_i\lambda
_i\epsilon \left( u_i\right) \mu _i,$ where $\lambda _{i},\mu _i\in {\cal L}(%
{\cal F})$ and $u_i\in {\cal F}.$ Let $C\left( G\right) $ =${\cal L}({\cal F}%
)/I\left( G\right) $ and define $j:{\cal F}\rightarrow C\left( G\right) $ as
the composition of monomorphism $i:{{\cal F}\rightarrow L}^1 ({\cal F}%
)\subset {\cal L}({\cal F})$ and projection $p:{\cal L}\left( {\cal F}%
\right) \rightarrow C\left( G\right) .$ In this case pair $\left( C\left(
G\right) ,j\right) $ is the solution of our problem. From the general
properties of tensor algebras the homomorphism $\varphi :{\cal F}\rightarrow
A$ can be extended to ${\cal L}({\cal F})$ , i.e., the diagram 2
is commutative, where $\rho $ is a monomorphism of algebras. Because $\left(
\varphi \left( u\right) \right) ^2=G\left( u\right) \cdot 1,$ then $\rho $
vanishes on ideal $I\left( G\right) $ and in this case the necessary
homomorphism $\tau $ is defined. As a consequence of uniqueness of $\rho ,$
the homomorphism $\tau $ is unique.

Tensor d-algebra ${\cal L}({\cal F )}$ can be considered as a ${\cal Z}/2$
graded algebra. Really, let us in\-tro\-duce ${\cal L}^{(0)}({\cal F}) =
\sum_{i=1}^\infty T^{2i}\left( {\cal F}\right) $ and ${\cal L}^{(1)}({\cal F}%
) =\sum_{i=1}^\infty T^{2i+1}\left( {\cal F}\right) .$ Setting $I^{(\alpha
)}\left( G\right) =I\left( G\right) \cap {\cal L}^{(\alpha )}({\cal F}).$
Define $C^{(\alpha )}\left( G\right) $ as $p\left( {\cal L}^{(\alpha )} (%
{\cal F})\right) ,$ where $p:{\cal L}\left( {\cal F}\right) \rightarrow
C\left( G\right) $ is the canonical projection. Then $C\left( G\right)
=C^{(0)}\left( G\right) \oplus C^{(1)}\left( G\right) $ and in consequence
we obtain that the Clifford d-algebra is ${\cal Z}/2$ graded.

It is obvious that Clifford d-algebra functorially depends on pair $\left(
{\cal F},G\right) .$ If $f:{\cal F}\rightarrow{\cal F}^{\prime }$ is a
homomorphism of k-vector spaces, such that $G^{\prime }\left( f(u)\right)
=G\left( u\right) ,$ where $G$ and $G^{\prime }$ are, respectively, metrics
on ${\cal F}$ and ${\cal F}^{\prime },$ then $f$ induces an homomorphism of
d-algebras%

$$
C\left( f\right) :C\left( G\right) \rightarrow C\left( G^{\prime }\right)
$$
with identities $C\left( \varphi \cdot f\right) =C\left( \varphi \right)
C\left( f\right) $ and $C\left( Id_{{\cal F}}\right) =Id_{C({\cal F)}}.$

If ${\cal A}^{\alpha}$ and ${\cal B}^{\beta}$ are ${\cal Z}/2$--graded
 d--algebras,
then their graded tensorial product $%
{\cal A}^\alpha \otimes {\cal B}^\beta $ is defined as a d-algebra for
k-vector d-space ${\cal A}^\alpha \otimes {\cal B}^\beta $ with the graded
product induced as $\left( a\otimes b\right) \left( c\otimes d\right)
=\left( -1\right) ^{\alpha \beta }ac\otimes bd,$ where $b\in {\cal B}^\alpha
$ and $c\in {\cal A}^\alpha \quad \left( \alpha ,\beta =0,1\right) .$

Now we reformulate for d--algebras the Chevalley theorem \cite{chev}:

\begin{theorem}
The Clifford d-algebra
$$
C\left( h{\cal F}\oplus v_1{\cal F\oplus ...}\oplus v_z{\cal F}%
,g+h_1+...+h_z\right)
$$
is naturally isomorphic to $C(g)\otimes C\left( h_1\right) \otimes
...\otimes C\left( h_z\right) .$
\end{theorem}

{\bf Proof. }Let $n:h{\cal F}\rightarrow C\left( g\right) $ and $%
n_{(p)}^{\prime }:v_{(p)}{\cal F}\rightarrow C\left( h_{(p)}\right) $ be
canonical maps and map
$$
m:h{\cal F}\oplus v_1{\cal F...}\oplus v_z{\cal F}\rightarrow C(g)\otimes
C\left( h_1\right) \otimes ...\otimes C\left( h_z\right)
$$
is defined as%
$$
m(x,y_{(1)},...,y_{(z)})=
$$
$$
n(x)\otimes 1\otimes ...\otimes 1+1\otimes n^{\prime }(y_{(1)})\otimes
...\otimes 1+1\otimes ...\otimes 1\otimes n^{\prime }(y_{(z)}),
$$
$x\in h{\cal F},y_{(1)}\in v_{(1)}{\cal F,...},y_{(z)}\in v_{(z)}{\cal F.}$
We have
$$
\left( m(x,y_{(1)},...,y_{(z)})\right) ^2=\left[ \left( n\left( x\right)
\right) ^2+\left( n^{\prime }\left( y_{(1)}\right) \right) ^2+...+\left(
n^{\prime }\left( y_{(z)}\right) \right) ^2\right] \cdot 1=
$$
$$
[g\left( x\right) +h\left( y_{(1)}\right) +...+h\left( y_{(z)}\right) ].
$$
\ Taking into account the universality property of Clifford d-algebras we
conclude that $m_1+...+m_z$ induces the homomorphism%
$$
\zeta :C\left( h{\cal F}\oplus v_1{\cal F}\oplus ...\oplus v_z{\cal F}%
,g+h_1+...+h_z\right) \rightarrow
$$
$$
C\left( h{\cal F},g\right) \widehat{\otimes }C\left( v_1{\cal F},h_1\right)
\widehat{\otimes }...C\left( v_z{\cal F},h_z\right) .
$$
We also can define a homomorphism%
$$
\upsilon :C\left( h{\cal F},g\right) \widehat{\otimes }C\left( v_1{\cal F}%
,h_{(1)}\right) \widehat{\otimes }...\widehat{\otimes }C\left( v_z{\cal F}%
,h_{(z)}\right) \rightarrow
$$
$$
C\left( h{\cal F}\oplus v_1{\cal F\oplus ...}\oplus v_z{\cal F}%
,g+h_{(1)}+...+h_{(z)}\right)
$$
by using formula $\upsilon \left( x\otimes y_{(1)}\otimes ...\otimes
y_{(z)}\right) =\delta \left( x\right) \delta _{(1)}^{\prime }\left(
y_{(1)}\right) ...\delta _{(z)}^{\prime }\left( y_{(z)}\right) ,$ where
homomorphysms $\delta $ and $\delta _{(1)}^{\prime },...,\delta
_{(z)}^{\prime }$ are, respectively, induced by imbeddings of $h{\cal F}$
and $v_1{\cal F}$ into $h{\cal F}\oplus v_1{\cal F\oplus ...}\oplus v_z{\cal %
F}:$%
$$
\delta :C\left( h{\cal F},g\right) \rightarrow C\left( h{\cal F}\oplus v_1%
{\cal F\oplus ...}\oplus v_z{\cal F},g+h_{(1)}+...+h_{(z)}\right) ,
$$
$$
\delta _{(1)}^{\prime }:C\left( v_1{\cal F},h_{(1)}\right) \rightarrow
C\left( h{\cal F}\oplus v_1{\cal F\oplus ...}\oplus v_z{\cal F}%
,g+h_{(1)}+...+h_{(z)}\right) ,
$$
$$
...................................
$$
$$
\delta _{(z)}^{\prime }:C\left( v_z{\cal F},h_{(z)}\right) \rightarrow
C\left( h{\cal F}\oplus v_1{\cal F\oplus ...}\oplus v_z{\cal F}%
,g+h_{(1)}+...+h_{(z)}\right) .
$$

Superpositions of homomorphisms $\zeta $ and $\upsilon $ lead to identities%
$$
\upsilon \zeta =Id_{C\left( h{\cal F},g\right) \widehat{\otimes }C\left( v_1%
{\cal F},h_{(1)}\right) \widehat{\otimes }...\widehat{\otimes }C\left( v_z%
{\cal F},h_{(z)}\right) },\eqno(2.37)
$$
$$
\zeta \upsilon =Id_{C\left( h{\cal F},g\right) \widehat{\otimes }C\left( v_1%
{\cal F},h_{(1)}\right) \widehat{\otimes }...\widehat{\otimes }C\left( v_z%
{\cal F},h_{(z)}\right) }.
$$
Really, d-algebra $C\left( h{\cal F}\oplus v_1{\cal F\oplus ...}\oplus v_z%
{\cal F},g+h_{(1)}+...+h_{(z)}\right) $ is generated by elements of type $%
m(x,y_{(1)},...y_{(z)}).$ Calculating
$$
\upsilon \zeta \left( m\left( x,y_{(1)},...y_{(z)}\right) \right) =\upsilon
(n\left( x\right) \otimes 1\otimes ...\otimes 1+1\otimes n_{(1)}^{\prime
}\left( y_{(1)}\right) \otimes ...\otimes 1+...+
$$
$$
1\otimes ....\otimes n_{(z)}^{\prime }\left( y_{(z)}\right) )=\delta \left(
n\left( x\right) \right) \delta \left( n_{(1)}^{\prime }\left(
y_{(1)}\right) \right) ...\delta \left( n_{(z)}^{\prime }\left(
y_{(z)}\right) \right) =
$$
$$
m\left( x,0,...,0\right) +m(0,y_{(1)},...,0)+...+m(0,0,...,y_{(z)})=m\left(
x,y_{(1)},...,y_{(z)}\right) ,
$$
we prove the first identity in (2.37).

On the other hand, d-algebra
$$
C\left( h{\cal F},g\right) \widehat{\otimes }C\left( v_1{\cal F}%
,h_{(1)}\right) \widehat{\otimes }...\widehat{\otimes }C\left( v_z{\cal F}%
,h_{(z)}\right)
$$
is generated by elements of type
$$
n\left( x\right) \otimes 1\otimes ...\otimes ,1\otimes n_{(1)}^{\prime
}\left( y_{(1)}\right) \otimes ...\otimes 1,...1\otimes ....\otimes
n_{(z)}^{\prime }\left( y_{(z)}\right) ,
$$
we prove the second identity in (2.37).

Following from the above--mentioned properties of homomorphisms $\zeta $ and
$\upsilon $ we can assert that the natural isomorphism is explicitly
constructed.$\Box $

In consequence of theorem 2.2 we conclude that all operations with Clifford
d-algebras can be reduced to calculations for $C\left( h{\cal F},g\right) $
and $C\left( v_{(p)}{\cal F},h_{(p)}\right) $ which are usual Clifford
algebras of dimension $2^n$ and, respectively, $2^{m_p}$ \cite{kar,ati}.

Of special interest is the case when $k={\cal R}$ and ${\cal F}$ is
isomorphic to vector space ${\cal R}^{p+q,a+b}$ provided with quadratic form
$-x_1^2-...-x_p^2+x_{p+q}^2-y_1^2-...-y_a^2+...+y_{a+b}^2.$ In this case,
the Clifford algebra, denoted as $\left( C^{p,q},C^{a,b}\right) ,\,$ is
generated by symbols $%
e_1^{(x)},e_2^{(x)},...,e_{p+q}^{(x)},e_1^{(y)},e_2^{(y)},...,e_{a+b}^{(y)}$
satisfying properties $\left( e_i\right) ^2=-1~\left( 1\leq i\leq p\right)
,\left( e_j\right) ^2=-1~\left( 1\leq j\leq a\right) ,\left( e_k\right)
^2=1~(p+1\leq k\leq p+q),$

$\left( e_j\right) ^2=1~(n+1\leq s\leq a+b),~e_ie_j=-e_je_i,~i\neq j.\,$
Explicit calculations of $C^{p,q}$ and $C^{a,b}$ are possible by using
isomorphisms \cite{kar,penr2}
$$
C^{p+n,q+n}\simeq C^{p,q}\otimes M_2\left( {\cal R}\right) \otimes
...\otimes M_2\left( {\cal R}\right) \cong C^{p,q}\otimes M_{2^n}\left(
{\cal R}\right) \cong M_{2^n}\left( C^{p,q}\right) ,
$$
where $M_s\left( A\right) $ denotes the ring of quadratic matrices of order $%
s$ with coefficients in ring $A.$ Here we write the simplest isomorphisms $%
C^{1,0}\simeq {\cal C}, C^{0,1}\simeq {\cal R}\oplus {\cal R ,}$ and $%
C^{2,0}={\cal H},$ where by ${\cal H}$ is denoted the body of quaternions.
We summarize this calculus as (as in Ref. \cite{ati})%
$$
C^{0,0}={\cal R}, C^{1,0}={\cal C}, C^{0,1}={\cal R}\oplus {\cal R}, C^{2,0}=%
{\cal H}, C^{0,2}= M_2\left( {\cal R}\right) ,
$$
$$
C^{3,0}={\cal H}\oplus {\cal H} , C^{0,3} = M_2\left( {\cal R}\right),
C^{4,0}=M_2\left( {\cal H}\right) , C^{0,4}=M_2\left( {\cal H}\right) ,
$$
$$
C^{5,0}=M_4\left( {\cal C}\right) ,~C^{0,5}=M_2\left( {\cal H}\right) \oplus
M_2\left( {\cal H}\right) ,~C^{6,0}=M_8\left( {\cal R}\right)
,~C^{0,6}=M_4\left( {\cal H}\right) ,
$$
$$
C^{7,0}=M_8\left( {\cal R}\right) \oplus M_8\left( {\cal R}\right)
,~C^{0,7}=M_8\left( {\cal C}\right) ,~C^{8,0}=M_{16}\left( {\cal R}\right)
,~C^{0,8}=M_{16}\left( {\cal R}\right) .
$$
One of the most important properties of real algebras $C^{0,p}~\left(
C^{0,a}\right) $ and $C^{p,0}~\left( C^{a,0}\right) $ is eightfold
periodicity of $p(a).$

Now, we emphasize that $H^{2n}$-spaces  admit locally a
structure of Clifford algebra on complex vector spaces. Really, by using
almost \ Hermitian structure $J_\alpha ^{\quad \beta }$ and considering
complex space ${\cal C}^n$ with nondegenarate quadratic form $%
\sum_{a=1}^n\left| z_a\right| ^2,~z_a\in {\cal C}^2$ induced locally by
metric (2.12) (rewritten in complex coordinates $z_a=x_a+iy_a)$ we define
Clifford algebra $\overleftarrow{C}^n=\underbrace{\overleftarrow{C}^1\otimes
...\otimes \overleftarrow{C}^1}_n,$ where $\overleftarrow{C}^1={\cal %
C\otimes }_R{\cal C=C\oplus C}$ or in consequence, $\overleftarrow{C}%
^n\simeq C^{n,0}\otimes _{{\cal R}}{\cal C}\approx C^{0,n}\otimes _{{\cal R}}%
{\cal C}.$ Explicit calculations lead to isomorphisms $\overleftarrow{C}%
^2=C^{0,2}\otimes _{{\cal R}}{\cal C}\approx M_2\left( {\cal R}\right)
\otimes _{{\cal R}}{\cal C}\approx M_2\left( \overleftarrow{C}^n\right)
,~C^{2p}\approx M_{2^p}\left( {\cal C}\right) $ and $\overleftarrow{C}%
^{2p+1}\approx M_{2^p}\left( {\cal C}\right) \oplus M_{2^p}\left( {\cal C}%
\right) ,$ which show that complex Clifford algebras, defined locally for $%
H^{2n}$-spaces, have periodicity 2 on $p.$

Considerations presented in the proof of theorem 2.2 show that map $j:{\cal F%
}\rightarrow C\left( {\cal F}\right) $ is monomorphic, so we can identify
space ${\cal F}$ with its image in $C\left( {\cal F},G\right) ,$ denoted as $%
u\rightarrow \overline{u},$ if $u\in C^{(0)}\left( {\cal F},G\right) ~\left(
u\in C^{(1)}\left( {\cal F},G\right) \right) ;$ then $u=\overline{u}$ (
respectively, $\overline{u}=-u).$

\begin{definition}
The set of elements $u\in C\left( G\right) ^{*},$ where $C\left( G\right)
^{*}$ denotes the multiplicative group of invertible elements of $C\left(
{\cal F},G\right) $ satisfying $\overline{u}{\cal F}u^{-1}\in {\cal F},$ is
called the twisted Clifford d-group, denoted as $\widetilde{\Gamma }\left(
{\cal F}\right) .$
\end{definition}

Let $\widetilde{\rho }:\widetilde{\Gamma }\left( {\cal F}\right) \rightarrow
GL\left( {\cal F}\right) $ be the homorphism given by $u\rightarrow \rho
\widetilde{u},$ where $\widetilde{\rho }_u\left( w\right) =\overline{u}%
wu^{-1}.$ We can verify that $\ker \widetilde{\rho }={\cal R}^{*}$is a
subgroup in $\widetilde{\Gamma }\left( {\cal F}\right) .$

Canonical map $j:{\cal F}\rightarrow C\left( {\cal F}\right) $ can be
interpreted as the linear map ${\cal F}\rightarrow C\left( {\cal F}\right)
^0 $ satisfying the universal property of Clifford d-algebras. This leads to
a homomorphism of algebras, $C\left( {\cal F}\right) \rightarrow C\left(
{\cal F}\right) ^t,$ considered by an anti-involution of $C\left( {\cal F}%
\right) $ and denoted as $u\rightarrow ~^tu.$ More exactly, if $u_1...u_n\in
{\cal F,}$ then $t_u=u_n...u_1$ and $^t\overline{u}=\overline{^tu}=\left(
-1\right) ^nu_n...u_1.$

\begin{definition}
The spinor norm of arbitrary $u\in C\left( {\cal F}\right) $ is defined as\\
$S\left( u\right) =~^t\overline{u}\cdot u\in C\left( {\cal F}\right) .$
\end{definition}

It is obvious that if $u,u^{\prime },u^{\prime \prime }\in \widetilde{\Gamma
}\left( {\cal F}\right) ,$ then $S(u,u^{\prime })=S\left( u\right) S\left(
u^{\prime }\right) $ and \\ $S\left( uu^{\prime }u^{\prime \prime }\right)
=S\left( u\right) S\left( u^{\prime }\right) S\left( u^{\prime \prime
}\right) .$ For $u,u^{\prime }\in {\cal F} S\left( u\right) =-G\left(
u\right) $ and $S\left( u,u^{\prime }\right) =S\left( u\right) S\left(
u^{\prime }\right) =S\left( uu^{\prime }\right) .$

Let us introduce the orthogonal group $O\left( G\right) \subset GL\left(
G\right) $ defined by metric $G$ on ${\cal F}$ and denote sets $SO\left(
G\right) =\{u\in O\left( G\right) ,\det \left| u\right| =1\},~Pin\left(
G\right) =\{u\in \widetilde{\Gamma }\left( {\cal F}\right) ,S\left( u\right)
=1\}$ and $Spin\left( G\right) =Pin\left( G\right) \cap C^0\left( {\cal F}%
\right) .$ For ${{\cal F}\cong {\cal R}}^{n+m}$ we write $Spin\left(
n_E\right) .$ By straightforward calculations (see similar considerations in
Ref. \cite{kar}) we can verify the exactness of these sequences:%
$$
1\rightarrow {\cal Z}/2\rightarrow Pin\left( G\right) \rightarrow O\left(
G\right) \rightarrow 1,
$$
$$
1\rightarrow {\cal Z}/2\rightarrow Spin\left( G\right) \rightarrow SO\left(
G\right) \rightarrow 0,
$$
$$
1\rightarrow {\cal Z}/2\rightarrow Spin\left( n_E\right) \rightarrow
SO\left( n_E\right) \rightarrow 1.
$$
We conclude this subsection by emphasizing that the spinor norm was defined
with respect to a quadratic form induced by a metric in dv-bundle ${\cal E}%
^{<z>}$. This approach differs from those presented in Refs. \cite{asa88}
and \cite{ono}.

\subsection{Higher Order An\-i\-sot\-rop\-ic Clifford Bundles}

We shall consider two variants of generalization of spinor constructions
defined for d-vector spaces to the case of distinguished vector bundle
spaces enabled with the structure of N-connection. The first is to use the
extension to the category of vector bundles. The second is to define the
Clifford fibration associated with compatible linear d-connection and metric
$G$ on a vector bundle. We shall analyze both variants.

\subsubsection{Clifford d--module structure in dv--bundles}

Because functor ${\cal F}\to C({\cal F})$ is smooth we can extend it to the
category of vector bundles of type $\xi ^{<z>}=\{\pi _d:HE^{<z>}\oplus
V_1E^{<z>}\oplus ...\oplus V_zE^{<z>}\rightarrow E^{<z>}\}.$ Recall that by $%
{\cal F}$ we denote the typical fiber of such bundles. For $\xi ^{<z>}$ we
obtain a bundle of algebras, denoted as $C\left( \xi ^{<z>}\right) ,\,$ such
that $C\left( \xi ^{<z>}\right) _u=C\left( {\cal F}_u\right) .$
Multiplication in every fibre defines a continuous map $C\left( \xi
^{<z>}\right) \times C\left( \xi ^{<z>}\right) \rightarrow C\left( \xi
^{<z>}\right) .$ If $\xi ^{<z>}$ is a vector bundle on number field $k$%
,\thinspace \thinspace the structure of the $C\left( \xi ^{<z>}\right) $%
-module, the d-module, the d-module, on $\xi ^{<z>}$ is given by the
continuous map $C\left( \xi ^{<z>}\right) \times _E\xi ^{<z>}\rightarrow \xi
^{<z>}$ with every fiber ${\cal F}_u$ provided with the structure of the $%
C\left( {\cal F}_u\right) -$module, correlated with its $k$-module
structure, Because ${\cal F}\subset C\left( {\cal F}\right) ,$ we have a
fiber to fiber map ${\cal F}\times _E\xi ^{<z>}\rightarrow \xi ^{<z>},$
inducing on every fiber the map ${\cal F}_u\times _E\xi
_{(u)}^{<z>}\rightarrow \xi _{(u)}^{<z>}$ (${\cal R}$-linear on the first
factor and $k$-linear on the second one ). Inversely, every such bilinear
map defines on $\xi ^{<z>}$ the structure of the $C\left( \xi ^{<z>}\right) $%
-module by virtue of universal properties of Clifford d-algebras.
Equivalently, the above-mentioned bilinear map defines a morphism of
v-bundles $m:\xi ^{<z>}\rightarrow HOM\left( \xi ^{<z>},\xi ^{<z>}\right)
\quad [HOM\left( \xi ^{<z>},\xi ^{<z>}\right) $ denotes the bundles of
homomorphisms] when $\left( m\left( u\right) \right) ^2=G\left( u\right) $
on every point.

Vector bundles $\xi ^{<z>}$ provided with $C\left( \xi ^{<z>}\right) $%
-structures are objects of the category with morphisms being morphisms of
dv-bundles, which induce on every point $u\in \xi ^{<z>}$ morphisms of $%
C\left( {\cal F}_u\right) -$modules. This is a Banach category contained in
the category of finite-dimensional d-vector spaces on filed $k.$ We shall
not use category formalism in this work, but point to its advantages in
further formulation of new directions of K-theory (see , for example, an
introduction in Ref. \cite{kar}) concerned with la-spaces.

Let us denote by $H^s\left( {\cal E}^{<z>},GL_{n_E}\left( {\cal R}\right)
\right) ,\,$ where $n_E=n+m_1+...+m_z,\,$ the s-dimensional cohomology group
of the algebraic sheaf of germs of continuous maps of dv-bundle ${\cal E}%
^{<z>}$ with group $GL_{n_E}\left( {\cal R}\right) $ the group of
automorphisms of ${\cal R}^{n_E}\,$ (for the language of algebraic topology
see, for example, Refs. \cite{kar} and \cite{god}). We shall also use the
group $SL_{n_E}\left( {\cal R}\right) =\{A\subset GL_{n_E}\left( {\cal R}%
\right) ,\det A=1\}.\,$ Here we point out that cohomologies\\ $H^s(M,Gr)$
characterize the class of a principal bundle $\pi :P\rightarrow M$ on $M$
with structural group $Gr.$ Taking into account that we deal with bundles
distinguished by an N-connection we introduce into consideration
cohomologies $H^s\left( {\cal E}^{<z>},GL_{n_E}\left( {\cal R}\right)
\right) $ as distinguished classes (d-classes) of bundles ${\cal E}^{<z>}$
provided with a global N-connection structure.

For a real vector bundle $\xi ^{<z>}$ on compact base ${\cal E}^{<z>}$ we
can define the orientation on $\xi ^{<z>}$ as an element $\alpha _d\in
H^1\left( {\cal E}^{<z>},GL_{n_E}\left( {\cal R}\right) \right) $ whose
image on map%
$$
H^1\left( {\cal E}^{<z>},SL_{n_E}\left( {\cal R}\right) \right) \rightarrow
H^1\left( {\cal E}^{<z>},GL_{n_E}\left( {\cal R}\right) \right)
$$
is the d-class of bundle ${\cal E}^{<z>}.$

\begin{definition}
The spinor structure on $\xi ^{<z>}$ is defined as an element\\ $\beta _d\in
H^1\left( {\cal E}^{<z>},Spin\left( n_E\right) \right) $ whose image in the
composition%
$$
H^1\left( {\cal E}^{<z>},Spin\left( n_E\right) \right) \rightarrow H^1\left(
{\cal E}^{<z>},SO\left( n_E\right) \right) \rightarrow H^1\left( {\cal E}%
^{<z>},GL_{n_E}\left( {\cal R}\right) \right)
$$
is the d-class of ${\cal E}^{<z>}.$
\end{definition}

The above definition of spinor structures can be reformulated in terms of
principal bundles. Let $\xi ^{<z>}$ be a real vector bundle of rank n+m on a
compact base ${\cal E}^{<z>}.$ If there is a principal bundle $P_d$ with
structural group $SO( n_E ) $ \newline  [ or $Spin( n_E ) ],$ this bundle $%
\xi ^{<z>}$ can be provided with orientation (or spinor) structure. The
bundle $P_d$ is associated with element $\alpha _d\in H^1\left( {\cal E}%
^{<z>},SO(n_{<z>})\right) $ [or $\beta _d\in H^1\left( {\cal E}%
^{<z>},Spin\left( n_E\right) \right) .$

We remark that a real bundle is oriented if and only if its first
Stiefel-Whitney d--class vanishes,
$$
w_1\left( \xi _d\right) \in H^1\left( \xi ,{\cal Z}/2\right) =0,
$$
where $H^1\left( {\cal E}^{<z>},{\cal Z}/2\right) $ is the first group of
Chech cohomology with coefficients in ${\cal Z}/2,$ Considering the second
Stiefel--Whitney class $w_2\left( \xi ^{<z>}\right) \in H^2\left( {\cal E}%
^{<z>},{\cal Z}/2\right) $ it is well known that vector bundle $\xi ^{<z>}$
admits the spinor structure if and only if $w_2\left( \xi ^{<z>}\right) =0.$
Finally,  we emphasize that taking into account that base
space ${\cal E}^{<z>}$ is also a v-bundle, $p:E^{<z>}\rightarrow M,$ we have
to make explicit calculations in order to express cohomologies $H^s\left(
{\cal E}^{<z>},GL_{n+m}\right) \,$ and $H^s\left( {\cal E}^{<z>},SO\left(
n+m\right) \right) $ through cohomologies $H^s\left( M,GL_n\right)
,H^s\left( M,SO\left( m_1\right) \right) ,$ $...H^s\left( M,SO\left(
m_z\right) \right) ,$ , which depends on global topological structures of
spaces $M$ and ${\cal E}^{<z>}$ $.$ For general bundle and base spaces this
requires a cumbersome cohomological calculus.

\subsubsection{Clifford fibration}

Another way of defining the spinor structure is to use Clifford fibrations.
Consider the principal bundle with the structural group $Gr$ being a
subgroup of orthogonal group $O\left( G\right) ,$ where $G$ is a quadratic
nondegenerate form (see(2.12)) defined on the base (also being a bundle
space) space ${\cal E}^{<z>}.$ The fibration associated to principal
fibration $P\left( {\cal E}^{<z>},Gr\right) $ with a typical fiber having
Clifford algebra $C\left( G\right) $ is, by definition, the Clifford
fibration $PC\left( {\cal E}^{<z>},Gr\right) .$ We can always define a
metric on the Clifford fibration if every fiber is isometric to $PC\left(
{\cal E}^{<z>},G\right) $ (this result is proved for arbitrary quadratic
forms $G$ on pseudo-Riemannian bases \cite{tur}). If, additionally, $%
Gr\subset SO\left( G\right) $ a global section can be defined on $PC\left(
G\right) .$

Let ${\cal P}\left( {\cal E}^{<z>},Gr\right) $ be the set of principal
bundles with differentiable base ${\cal E}^{<z>}$ and structural group $Gr.$
If $g:Gr\rightarrow Gr^{\prime }$ is an homomorphism of Lie groups and $%
P\left( {\cal E}^{<z>},Gr\right) \subset {\cal P}\left( {\cal E}%
^{<z>},Gr\right) $ (for simplicity in this subsection we shall denote mentioned
bundles and sets of bundles as $P,P^{\prime }$ and respectively, ${\cal P},%
{\cal P}^{\prime }),$ we can always construct a principal bundle with the
property that there is as homomorphism $f:P^{\prime }\rightarrow P$ of
principal bundles which can be projected to the identity map of ${\cal E}%
^{<z>}$ and corresponds to isomorphism $g:Gr\rightarrow Gr^{\prime }.$ If
the inverse statement also holds, the bundle $P^{\prime }$ is called as the
extension of $P$ associated to $g$ and $f$ is called the extension
homomorphism denoted as $\widetilde{g.}$

Now we can define distinguished spinor structures on bundle spaces (compare
with definition 2.3 ).

\begin{definition}
Let $P\in {\cal P}\left( {\cal E}^{<z>},O\left( G\right) \right) $ be a
principal bundle. A distinguished spinor structure of $P,$ equivalently a
ds-structure of ${\cal E}^{<z>}$ is an extension $\widetilde{P}$ of $P$
associated to homomorphism $h:PinG\rightarrow O\left( G\right) $ where $%
O\left( G\right) $ is the group of orthogonal rotations, generated by metric
$G,$ in bundle ${\cal E}^{<z>}.$
\end{definition}

So, if $\widetilde{P}$ is a spinor structure of the space ${\cal E}^{<z>},$
then $\widetilde{P}\in {\cal P}\left( {\cal E}^{<z>},PinG\right) .$

The definition of spinor structures on varieties was given in Ref.\cite{cru1}.
In Refs. \cite{cru2} and \cite{cru2} it is proved that a necessary and
sufficient condition for a space time to be orientable is to admit a global
field of orthonormalized frames. We mention that spinor structures can be
also defined on varieties modeled on Banach spaces \cite{ana77}. As we have
shown  similar constructions are possible for the cases
when space time has the structure of a v-bundle with an N-connection.

\begin{definition}
A special distinguished spinor structure, ds-structure, of principal bundle $%
P=P\left( {\cal E}^{<z>},SO\left( G\right) \right) $ is a principal bundle $%
\widetilde{P}=\widetilde{P}\left( {\cal E}^{<z>},SpinG\right) $ for which a
homomorphism of principal bundles $\widetilde{p}:\widetilde{P}\rightarrow P,$
projected on the identity map of ${\cal E}^{<z>}$ and corresponding to
representation%
$$
R:SpinG\rightarrow SO\left( G\right) ,
$$
is defined.
\end{definition}

In the case when the base space variety is oriented, there is a natural
bijection between tangent spinor structures with a common base. For special
ds--structures we can define, as for any spinor structure, the concepts of
spin tensors, spinor connections, and spinor covariant derivations (see
 Sec. II.F.1 and Refs. \cite{vb295,vod,vsp96}).

\subsection{Almost Complex Spinor Structures}

Almost complex structures are an important characteristic of $H^{2n}$-spaces
and of osculator bundles $Osc^{k=2k_1}(M),$ where $k_1=1,2,...$ . For
simplicity in this subsection we restrict our analysis to the case of $H^{2n}$%
-spaces. We can rewrite the almost Hermitian metric \cite{ma87,ma94}, $%
H^{2n} $-metric ( see considerations for
metrics and conditions of type (2.12) and correspondingly (2.14) ), in
complex form \cite{vjmp}:

$$
G=H_{ab}\left( z,\xi \right) dz^a\otimes dz^b,\eqno(2.38)
$$
where
$$
z^a=x^a+iy^a,~\overline{z^a}=x^a-iy^a,~H_{ab}\left( z,\overline{z}\right)
=g_{ab}\left( x,y\right) \mid _{y=y\left( z,\overline{z}\right) }^{x=x\left(
z,\overline{z}\right) },
$$
and define almost complex spinor structures. For given metric (2.38) on $%
H^{2n}$-space there is always a principal bundle $P^U$ with unitary
structural group U(n) which allows us to transform $H^{2n}$-space into
v-bundle $\xi ^U\approx P^U\times _{U\left( n\right) }{\cal R}^{2n}.$ This
statement will be proved after we introduce complex
spinor structures on oriented real vector bundles \cite{kar}.

Let us consider momentarily $k={\cal C}$ and introduce into consideration
[instead of the group $Spin(n)]$ the group $Spin^c\times _{{\cal Z}%
/2}U\left( 1\right) $ being the factor group of the product $Spin(n)\times
U\left( 1\right) $ with the respect to equivalence%
$$
\left( y,z\right) \sim \left( -y,-a\right) ,\quad y\in Spin(m).
$$
This way we define the short exact sequence%
$$
1\rightarrow U\left( 1\right) \rightarrow Spin^c\left( n\right) \stackrel{S^c%
}{\to }SO\left( n\right) \rightarrow 1,        \eqno(2.39)
$$
where $\rho ^c\left( y,a\right) =\rho ^c\left( y\right) .$ If $\lambda $ is
oriented , real, and rank $n,$ $\gamma $-bundle $\pi :E_\lambda \rightarrow
M^n,$ with base $M^n,$ the complex spinor structure, spin structure, on
$\lambda $ is given by the principal bundle $P$ with structural group $%
Spin^c\left( m\right) $ and isomorphism $\lambda \approx P\times
_{Spin^c\left( n\right) }{\cal R}^n$ (see (2.39)).
For such bundles the categorial equivalence can be defined as
$$
\epsilon ^c:{\cal E}_{{\cal C}}^T\left( M^n\right) \rightarrow {\cal E}_{%
{\cal C}}^\lambda \left( M^n\right) ,\eqno(2.40)
$$
where $\epsilon ^c\left( E^c\right) =P\bigtriangleup _{Spin^c\left( n\right)
}E^c$ is the category of trivial complex bundles on $M^n,{\cal E}_{{\cal C}%
}^\lambda \left( M^n\right) $ is the category of complex v-bundles on $M^n$
with action of Clifford bundle $C\left( \lambda \right) ,P\bigtriangleup
_{Spin^c(n)}$ and $E^c$ is the factor space of the bundle product $P\times
_ME^c$ with respect to the equivalence $\left( p,e\right) \sim \left( p%
\widehat{g}^{-1},\widehat{g}e\right) ,p\in P,e\in E^c,$ where $\widehat{g}%
\in Spin^c\left( n\right) $ acts on $E$ by via the imbedding $Spin\left(
n\right) \subset C^{0,n}$ and the natural action $U\left( 1\right) \subset
{\cal C}$ on complex v-bundle $\xi ^c,E^c=tot\xi ^c,$ for bundle $\pi
^c:E^c\rightarrow M^n.$

Now we return to the bundle $\xi ={\cal E}^{<1>}.$ A real v-bundle (not
being a spinor bundle) admits a complex spinor structure if and only if
there exist a homomorphism $\sigma :U\left( n\right) \rightarrow
Spin^c\left( 2n\right) $ making the diagram 3 commutative. The explicit
construction of $\sigma $ for arbitrary $\gamma $-bundle is given in Refs.
\cite{kar} and \cite{ati}. For $H^{2n}$-spaces it is obvious that a diagram
similar to (2.40) can be defined for the tangent bundle $TM^n,$ which
directly points to the possibility of defining the $^cSpin$-structure on $%
H^{2n}$-spaces.

Let $\lambda $ be a complex, rank\thinspace $n,$ spinor bundle with
$$
\tau :Spin^c\left( n\right) \times _{{\cal Z}/2}U\left( 1\right) \rightarrow
U\left( 1\right) \eqno(2.41)
$$
the homomorphism defined by formula $\tau \left( \lambda ,\delta \right)
=\delta ^2.$ For $P_s$ being the principal bundle with fiber $Spin^c\left(
n\right) $ we introduce the complex linear bundle $L\left( \lambda ^c\right)
=P_S\times _{Spin^c(n)}{\cal C}$ defined as the factor space of $P_S\times
{\cal C}$ on equivalence relation%

$$
\left( pt,z\right) \sim \left( p,l\left( t\right) ^{-1}z\right) ,
$$
where $t\in Spin^c\left( n\right) .$ This linear bundle is associated to
complex spinor structure on $\lambda ^c.$

If $\lambda ^c$ and $\lambda ^{c^{\prime }}$ are complex spinor bundles, the
Whitney sum $\lambda ^c\oplus \lambda ^{c^{\prime }}$ is naturally provided
with the structure of the complex spinor bundle. This follows from the
holomorphism%
$$
\omega ^{\prime }:Spin^c\left( n\right) \times Spin^c\left( n^{\prime
}\right) \rightarrow Spin^c\left( n+n^{\prime }\right) ,\eqno(2.42)
$$
given by formula $\left[ \left( \beta ,z\right) ,\left( \beta ^{\prime
},z^{\prime }\right) \right] \rightarrow \left[ \omega \left( \beta ,\beta
^{\prime }\right) ,zz^{\prime }\right] ,$ where $\omega $ is the
homomorphism making the diagram 4 commutative.%
Here, $z,z^{\prime }\in U\left( 1\right) .$ It is obvious that $L\left(
\lambda ^c\oplus \lambda ^{c^{\prime }}\right) $ is isomorphic to $L\left(
\lambda ^c\right) \otimes L\left( \lambda ^{c^{\prime }}\right) .$

We conclude this subsection by formulating our main result on complex spinor
structures for $H^{2n}$-spaces:

\begin{theorem}
Let $\lambda ^c$ be a complex spinor bundle of rank $n$ and $H^{2n}$-space
considered as a real vector bundle $\lambda ^c\oplus \lambda ^{c^{\prime }}$
provided with almost complex structure $J_{\quad \beta }^\alpha ;$
multiplication on $i$ is given by $\left(
\begin{array}{cc}
0 & -\delta _j^i \\
\delta _j^i & 0
\end{array}
\right) $. Then, the diagram 5 is commutative up to isomorphisms $\epsilon ^c
$ and $\widetilde{\epsilon }^c$ defined as in (2.40), ${\cal H}$ is functor $%
E^c\rightarrow E^c\otimes L\left( \lambda ^c\right) $ and ${\cal E}_{{\cal C}%
}^{0,2n}\left( M^n\right) $ is defined by functor ${\cal E}_{{\cal C}}\left(
M^n\right) \rightarrow {\cal E}_{{\cal C}}^{0,2n}\left( M^n\right) $ given
as correspondence $E^c\rightarrow \Lambda \left( {\cal C}^n\right) \otimes
E^c$ (which is a categorial equivalence), $\Lambda \left( {\cal C}^n\right) $
is the exterior algebra on ${\cal C}^n.$ $W$ is the real bundle $\lambda
^c\oplus \lambda ^{c^{\prime }}$ provided with complex structure.
\end{theorem}

{\bf Proof: }We use composition of homomorphisms%
$$
\mu :Spin^c\left( 2n\right) \stackrel{\pi }{\to }SO\left( n\right) \stackrel{%
r}{\to }U\left( n\right) \stackrel{\sigma }{\to }Spin^c\left( 2n\right)
\times _{{\cal Z}/2}U\left( 1\right) ,
$$
commutative diagram 6 and introduce composition of homomorphisms%
$$
\mu :Spin^c\left( n\right) \stackrel{\Delta }{\to }Spin^c\left( n\right)
\times Spin^c\left( n\right) \stackrel{{\omega }^c}{\to }Spin^c\left(
n\right) ,
$$
where $\Delta $ is the diagonal homomorphism and $\omega ^c$ is defined as
in (2.42). Using homomorphisms (2.41) and (2.42) we obtain formula $\mu
\left( t\right) =\mu \left( t\right) r\left( t\right) .$

Now consider bundle $P\times _{Spin^c\left( n\right) }Spin^c\left( 2n\right)
$ as the principal $Spin^c\left( 2n\right) $-bundle, associated to $M\oplus
M $ being the factor space of the product $P\times Spin^c\left( 2n\right) $
on the equivalence relation $\left( p,t,h\right) \sim \left( p,\mu \left(
t\right) ^{-1}h\right) .$ In this case the categorial equivalence (2.40) can
be rewritten as
$$
\epsilon ^c\left( E^c\right) =P\times _{Spin^c\left( n\right) }Spin^c\left(
2n\right) \Delta _{Spin^c\left( 2n\right) }E^c
$$
and seen as factor space of $P\times Spin^c\left( 2n\right) \times _ME^c$ on
equivalence relation
$$
\left( pt,h,e\right) \sim \left( p,\mu \left( t\right) ^{-1}h,e\right)
\mbox{and}\left( p,h_1,h_2,e\right) \sim \left( p,h_1,h_2^{-1}e\right)
$$
(projections of elements $p$ and $e$ coincides on base $M).$ Every element
of $\epsilon ^c\left( E^c\right) $ can be represented as $P\Delta
_{Spin^c\left( n\right) }E^c,$ i.e., as a factor space $P\Delta E^c$ on
equivalence relation $\left( pt,e\right) \sim \left( p,\mu ^c\left( t\right)
,e\right) ,$ when $t\in Spin^c\left( n\right) .$
The complex line bundle $L\left( \lambda ^c\right) $ can be interpreted as
the factor space of\\ $P\times _{Spin^c\left( n\right) }{\cal C}$ on
equivalence relation $\left( pt,\delta \right) \sim \left( p,r\left(
t\right) ^{-1}\delta \right) .$

Putting $\left( p,e\right) \otimes \left( p,\delta \right) \left( p,\delta
e\right) $ we introduce morphism%
$$
\epsilon ^c\left( E\right) \times L\left( \lambda ^c\right) \rightarrow
\epsilon ^c\left( \lambda ^c\right)
$$
with properties $\left( pt,e\right) \otimes \left( pt,\delta \right)
\rightarrow \left( pt,\delta e\right) =\left( p,\mu ^c\left( t\right)
^{-1}\delta e\right) ,$

$\left( p,\mu ^c\left( t\right) ^{-1}e\right) \otimes \left( p,l\left(
t\right) ^{-1}e\right) \rightarrow \left( p,\mu ^c\left( t\right) r\left(
t\right) ^{-1}\delta e\right) $ pointing to the fact that we have defined
the isomorphism correctly and that it is an isomorphism on every fiber. $%
\Box $

\subsection{ D--Spinor Techniques}

The purpose of this subsection is to show how a corresponding abstract spinor
technique entailing notational and calculations advantages can be developed
for arbitrary splits of dimensions of a d-vector space ${\cal F}=h{\cal F}%
\oplus v_1{\cal F\oplus ...}\oplus v_z{\cal F}$, where $\dim h{\cal F}=n$
and $\dim v_p{\cal F}=m_p.$ For convenience we shall also present some
necessary coordinate expressions.

The problem of a rigorous definition of spinors on la-spaces (la-spinors,
d-spinors) was posed and solved \cite{vjmp,vb295,vod} (see previous Secs.
II.B--II.D for generalizations on higher order anisotropic superspaces) in the
framework of the formalism of Clifford and spinor structures on v-bundles
provided with compatible nonlinear and distinguished connections and metric.
We introduced d-spinors as corresponding objects of the Clifford d-algebra $%
{\cal C}\left( {\cal F},G\right) $, defined for a d-vector space ${\cal F}$
in a standard manner (see, for instance, \cite{kar}) and proved that
operations with ${\cal C}\left( {\cal F},G\right) \ $ can be reduced to
calculations for ${\cal C}\left( h{\cal F},g\right) ,{\cal C}\left( v_1{\cal %
F},h_1\right) ,...$ and ${\cal C}\left( v_z{\cal F},h_z\right) ,$ which are
usual Clifford algebras of respective dimensions 2$^n,2^{m_1},...$ and 2$%
^{m_z}$ (if it is necessary we can use quadratic forms $g$ and $h_p$
correspondingly induced on $h{\cal F}$ and $v_p{\cal F}$ by a metric ${\bf G}
$ (2.12)). Considering the orthogonal subgroup $O{\bf \left( G\right) }%
\subset GL{\bf \left( G\right) }$ defined by a metric ${\bf G}$ we can
define the d-spinor norm and parametrize d-spinors by ordered pairs of
elements of Clifford algebras ${\cal C}\left( h{\cal F},g\right) $ and $%
{\cal C}\left( v_p{\cal F},h_p\right) ,p=1,2,...z.$ We emphasize that the
splitting of a Clifford d-algebra associated to a dv-bundle ${\cal E}^{<z>}$
is a straightforward consequence of the global decomposition (2.3) defining
a N-connection structure in ${\cal E}^{<z>}{\cal .}$

In this subsection we shall omit detailed proofs which in most cases are
mechanical but rather tedious. We can apply the methods developed in \cite
{pen,penr1,penr2,lue} in a straightforward manner on h- and v-subbundles in
order to verify the correctness of affirmations.

\subsubsection{Clifford d--algebra, d--spinors and d--twistors}

In order to relate the succeeding constructions with Clifford d-algebras
\cite{vjmp,vb295} we consider a la-frame decomposition of the metric (2.12):%
$$
G_{<\alpha ><\beta >}\left( u\right) =l_{<\alpha >}^{<\widehat{\alpha }%
>}\left( u\right) l_{<\beta >}^{<\widehat{\beta }>}\left( u\right) G_{<%
\widehat{\alpha }><\widehat{\beta }>},
$$
where the frame d-vectors and constant metric matrices are distinguished as

$$
l_{<\alpha >}^{<\widehat{\alpha }>}\left( u\right) =\left(
\begin{array}{cccc}
l_j^{\widehat{j}}\left( u\right) & 0 & ... & 0 \\
0 & l_{a_1}^{\widehat{a}_1}\left( u\right) & ... & 0 \\
... & ... & ... & ... \\
0 & 0 & .. & l_{a_z}^{\widehat{a}_z}\left( u\right)
\end{array}
\right) ,
$$
$$
G_{<\widehat{\alpha }><\widehat{\beta }>}=\left(
\begin{array}{cccc}
g_{\widehat{i}\widehat{j}} & 0 & ... & 0 \\
0 & h_{\widehat{a}_1\widehat{b}_1} & ... & 0 \\
... & ... & ... & ... \\
0 & 0 & 0 & h_{\widehat{a}_z\widehat{b}_z}
\end{array}
\right) ,
$$
$g_{\widehat{i}\widehat{j}}$ and $h_{\widehat{a}_1\widehat{b}_1},...,h_{%
\widehat{a}_z\widehat{b}_z}$ are diagonal matrices with $g_{\widehat{i}%
\widehat{i}}=$ $h_{\widehat{a}_1\widehat{a}_1}=...=h_{\widehat{a}_z\widehat{b%
}_z}=\pm 1.$

To generate Clifford d-algebras we start with matrix equations%
$$
\sigma _{<\widehat{\alpha }>}\sigma _{<\widehat{\beta }>}+\sigma _{<\widehat{%
\beta }>}\sigma _{<\widehat{\alpha }>}=-G_{<\widehat{\alpha }><\widehat{%
\beta }>}I,\eqno(2.43)
$$
where $I$ is the identity matrix, matrices $\sigma _{<\widehat{\alpha }%
>}\,(\sigma $-objects) act on a d-vector space ${\cal F}=h{\cal F}\oplus v_1%
{\cal F}\oplus ...\oplus v_z{\cal F}$ and theirs components are
distinguished as
$$
\sigma _{<\widehat{\alpha }>}\,=\left\{ (\sigma _{<\widehat{\alpha }>})_{%
\underline{\beta }}^{\cdot \underline{\gamma }}=\left(
\begin{array}{cccc}
(\sigma _{\widehat{i}})_{\underline{j}}^{\cdot \underline{k}} & 0 & ... & 0
\\
0 & (\sigma _{\widehat{a}_1})_{\underline{b}_1}^{\cdot \underline{c}_1} &
... & 0 \\
... & ... & ... & ... \\
0 & 0 & ... & (\sigma _{\widehat{a}_z})_{\underline{b}_z}^{\cdot \underline{c%
}_z}
\end{array}
\right) \right\} ,\eqno(2.44)
$$
indices \underline{$\beta $},\underline{$\gamma $},... refer to spin spaces
of type ${\cal S}=S_{(h)}\oplus S_{(v_1)}\oplus ...\oplus S_{(v_z)}$ and
underlined Latin indices \underline{$j$},$\underline{k},...$ and $\underline{%
b}_1,\underline{c}_1,...,\underline{b}_z,\underline{c}_z...$ refer
respectively to h-spin space ${\cal S}_{(h)}$ and v$_p$-spin space ${\cal S}%
_{(v_p)},(p=1,2,...,z)\ $which are correspondingly associated to a h- and v$%
_p$-decomposition of a dv-bundle ${\cal E}^{<z>}.$ The irreducible algebra
of matrices $\sigma _{<\widehat{\alpha }>}$ of minimal dimension $N\times N,$
where $N=N_{(n)}+N_{(m_1)}+...+N_{(m_z)},$ $\dim {\cal S}_{(h)}$=$N_{(n)}$
and $\dim {\cal S}_{(v_p)}$=$N_{(m_p)},$ has these dimensions%
$$
{N_{(n)}=\left\{
\begin{array}{rl}
{\ 2^{(n-1)/2},} & n=2k+1 \\
{2^{n/2},\ } & n=2k;
\end{array}
\right. }\quad \mbox{ and}\quad {N}_{(m_p)}{=}\left|
\begin{array}{cc}
2^{(m_p-1)/2}, & m_p=2k_p+1 \\
2^{m_p}, & m_p=2k_p
\end{array}
\right| ,
$$
where $k=1,2,...,k_p=1,2,....$

The Clifford d-algebra is generated by sums on $n+1$ elements of form%
$$
A_1I+B^{\widehat{i}}\sigma _{\widehat{i}}+C^{\widehat{i}\widehat{j}}\sigma _{%
\widehat{i}\widehat{j}}+D^{\widehat{i}\widehat{j}\widehat{k}}\sigma _{%
\widehat{i}\widehat{j}\widehat{k}}+...\eqno(2.45)
$$
and sums of $m_p+1$ elements of form%
$$
A_{2(p)}I+B^{\widehat{a}_p}\sigma _{\widehat{a}_p}+C^{\widehat{a}_p\widehat{b%
}_p}\sigma _{\widehat{a}_p\widehat{b}_p}+D^{\widehat{a}_p\widehat{b}_p%
\widehat{c}_p}\sigma _{\widehat{a}_p\widehat{b}_p\widehat{c}_p}+...
$$
with antisymmetric coefficients\\ $C^{\widehat{i}\widehat{j}}=C^{[\widehat{i}%
\widehat{j}]},C^{\widehat{a}_p\widehat{b}_p}=C^{[\widehat{a}_p\widehat{b}%
_p]},D^{\widehat{i}\widehat{j}\widehat{k}}=D^{[\widehat{i}\widehat{j}%
\widehat{k}]},D^{\widehat{a}_p\widehat{b}_p\widehat{c}_p}=D^{[\widehat{a}_p%
\widehat{b}_p\widehat{c}_p]},...$ and matrices $\sigma _{\widehat{i}\widehat{%
j}}=\sigma _{[\widehat{i}}\sigma _{\widehat{j}]},\sigma _{\widehat{a}_p%
\widehat{b}_p}=\sigma _{[\widehat{a}_p}\sigma _{\widehat{b}_p]},\sigma _{%
\widehat{i}\widehat{j}\widehat{k}}=\sigma _{[\widehat{i}}\sigma _{\widehat{j}%
}\sigma _{\widehat{k}]},...$ . Really, we have 2$^{n+1}$ coefficients $%
\left( A_1,C^{\widehat{i}\widehat{j}},D^{\widehat{i}\widehat{j}\widehat{k}%
},...\right) $ and 2$^{m_p+1}$ coefficients $( A_{2(p)},C^{\widehat{a}%
_p\widehat{b}_p},D^{\widehat{a}_p\widehat{b}_p\widehat{c}_p},...) $ of
the Clifford algebra on ${\cal F.}$

For simplicity,  we shall present the necessary geometric
constructions only for h-spin spaces ${\cal S}_{(h)}$ of dimension $N_{(n)}.$
Considerations for a v-spin space ${\cal S}_{(v)}$ are similar but with
proper characteristics for a dimension $N_{(m)}.$

In order to define the scalar (spinor) product on ${\cal S}_{(h)}$ we
introduce into consideration this finite sum (because of a finite number of
elements $\sigma _{[\widehat{i}\widehat{j}...\widehat{k}]}$ ):%
$$
^{(\pm )}E_{\underline{k}\underline{m}}^{\underline{i}\underline{j}}=\delta
_{\underline{k}}^{\underline{i}}\delta _{\underline{m}}^{\underline{j}%
}+\frac 2{1!}(\sigma _{\widehat{i}})_{\underline{k}}^{.\underline{i}}(\sigma
^{\widehat{i}})_{\underline{m}}^{.\underline{j}}+\frac{2^2}{2!}(\sigma _{%
\widehat{i}\widehat{j}})_{\underline{k}}^{.\underline{i}}(\sigma ^{\widehat{i%
}\widehat{j}})_{\underline{m}}^{.\underline{j}}+\frac{2^3}{3!}(\sigma _{%
\widehat{i}\widehat{j}\widehat{k}})_{\underline{k}}^{.\underline{i}}(\sigma
^{\widehat{i}\widehat{j}\widehat{k}})_{\underline{m}}^{.\underline{j}}+...%
\eqno(2.46)
$$
which can be factorized as
$$
^{(\pm )}E_{\underline{k}\underline{m}}^{\underline{i}\underline{j}}=N_{(n)}{%
\ }^{(\pm )}\epsilon _{\underline{k}\underline{m}}{\ }^{(\pm )}\epsilon ^{%
\underline{i}\underline{j}}\mbox{ for }n=2k\eqno(2.47)
$$
and%
$$
^{(+)}E_{\underline{k}\underline{m}}^{\underline{i}\underline{j}%
}=2N_{(n)}\epsilon _{\underline{k}\underline{m}}\epsilon ^{\underline{i}%
\underline{j}},{\ }^{(-)}E_{\underline{k}\underline{m}}^{\underline{i}%
\underline{j}}=0\mbox{ for }n=3(mod4),\eqno(2.48)
$$
$$
^{(+)}E_{\underline{k}\underline{m}}^{\underline{i}\underline{j}}=0,{\ }%
^{(-)}E_{\underline{k}\underline{m}}^{\underline{i}\underline{j}%
}=2N_{(n)}\epsilon _{\underline{k}\underline{m}}\epsilon ^{\underline{i}%
\underline{j}}\mbox{ for }n=1(mod4).
$$

Antisymmetry of $\sigma _{\widehat{i}\widehat{j}\widehat{k}...}$ and the
construction of the objects (2.45),(2.46),\\ (2.47) and (2.48) define the
properties of $\epsilon $-objects $^{(\pm )}\epsilon _{\underline{k}%
\underline{m}}$ and $\epsilon _{\underline{k}\underline{m}}$ which have an
eight-fold periodicity on $n$ (see details in \cite{penr2} and, with respect
to la-spaces, \cite{vjmp}).

For even values of $n$ it is possible the decomposition of every h-spin
space ${\cal S}_{(h)}$into irreducible h-spin spaces ${\bf S}_{(h)}$ and $%
{\bf S}_{(h)}^{\prime }$ (one considers splitting of h-indices, for
instance, \underline{$l$}$=L\oplus L^{\prime },\underline{m}=M\oplus
M^{\prime },...;$ for v$_p$-indices we shall write $\underline{a}%
_p=A_p\oplus A_p^{\prime },\underline{b}_p=B_p\oplus B_p^{\prime },...)$ and
defines new $\epsilon $-objects
$$
\epsilon ^{\underline{l}\underline{m}}=\frac 12\left( ^{(+)}\epsilon ^{%
\underline{l}\underline{m}}+^{(-)}\epsilon ^{\underline{l}\underline{m}%
}\right) \mbox{ and }\widetilde{\epsilon }^{\underline{l}\underline{m}%
}=\frac 12\left( ^{(+)}\epsilon ^{\underline{l}\underline{m}}-^{(-)}\epsilon
^{\underline{l}\underline{m}}\right) \eqno(2.49)
$$
We shall omit similar formulas for $\epsilon $-objects with lower indices.

We can verify, by using expressions (2.48) and straightforward calculations,
these para\-met\-ri\-za\-ti\-ons on symmetry properties of $\epsilon $%
-objects (2.49)
$$
\epsilon ^{\underline{l}\underline{m}}=\left(
\begin{array}{cc}
\epsilon ^{LM}=\epsilon ^{ML} & 0 \\
0 & 0
\end{array}
\right) \mbox{ and }\widetilde{\epsilon }^{\underline{l}\underline{m}%
}=\left(
\begin{array}{cc}
0 & 0 \\
0 & \widetilde{\epsilon }^{LM}=\widetilde{\epsilon }^{ML}
\end{array}
\right) \mbox{ for }n=0(mod8);\eqno(2.50)
$$
$$
\epsilon ^{\underline{l}\underline{m}}=-\frac 12{}^{(-)}\epsilon ^{%
\underline{l}\underline{m}}=\epsilon ^{\underline{m}\underline{l}},%
\mbox{ where }^{(+)}\epsilon ^{\underline{l}\underline{m}}=0,\mbox{ and }
$$
$$
\widetilde{\epsilon }^{\underline{l}\underline{m}}=-\frac 12{}^{(-)}\epsilon
^{\underline{l}\underline{m}}=\widetilde{\epsilon }^{\underline{m}\underline{%
l}}\mbox{
for }n=1(mod8);
$$
$$
\epsilon ^{\underline{l}\underline{m}}=\left(
\begin{array}{cc}
0 & 0 \\
\epsilon ^{L^{\prime }M} & 0
\end{array}
\right) \mbox{ and }\widetilde{\epsilon }^{\underline{l}\underline{m}%
}=\left(
\begin{array}{cc}
0 & \widetilde{\epsilon }^{LM^{\prime }}=-\epsilon ^{M^{\prime }L} \\ 0 & 0
\end{array}
\right) \mbox{ for }n=2(mod8);
$$
$$
\epsilon ^{\underline{l}\underline{m}}=-\frac 12{}^{(+)}\epsilon ^{%
\underline{l}\underline{m}}=-\epsilon ^{\underline{m}\underline{l}},%
\mbox{ where }^{(-)}\epsilon ^{\underline{l}\underline{m}}=0,\mbox{ and }
$$
$$
\widetilde{\epsilon }^{\underline{l}\underline{m}}=\frac 12{}^{(+)}\epsilon
^{\underline{l}\underline{m}}=-\widetilde{\epsilon }^{\underline{m}%
\underline{l}}\mbox{
for }n=3(mod8);
$$
$$
\epsilon ^{\underline{l}\underline{m}}=\left(
\begin{array}{cc}
\epsilon ^{LM}=-\epsilon ^{ML} & 0 \\
0 & 0
\end{array}
\right) \mbox{ and }\widetilde{\epsilon }^{\underline{l}\underline{m}%
}=\left(
\begin{array}{cc}
0 & 0 \\
0 & \widetilde{\epsilon }^{LM}=-\widetilde{\epsilon }^{ML}
\end{array}
\right) \mbox{ for }n=4(mod8);
$$
$$
\epsilon ^{\underline{l}\underline{m}}=-\frac 12{}^{(-)}\epsilon ^{%
\underline{l}\underline{m}}=-\epsilon ^{\underline{m}\underline{l}},%
\mbox{ where }^{(+)}\epsilon ^{\underline{l}\underline{m}}=0,\mbox{ and }
$$
$$
\widetilde{\epsilon }^{\underline{l}\underline{m}}=-\frac 12{}^{(-)}\epsilon
^{\underline{l}\underline{m}}=-\widetilde{\epsilon }^{\underline{m}%
\underline{l}}\mbox{ for }n=5(mod8);
$$
$$
\epsilon ^{\underline{l}\underline{m}}=\left(
\begin{array}{cc}
0 & 0 \\
\epsilon ^{L^{\prime }M} & 0
\end{array}
\right) \mbox{ and }\widetilde{\epsilon }^{\underline{l}\underline{m}%
}=\left(
\begin{array}{cc}
0 & \widetilde{\epsilon }^{LM^{\prime }}=\epsilon ^{M^{\prime }L} \\ 0 & 0
\end{array}
\right) \mbox{ for }n=6(mod8);
$$
$$
\epsilon ^{\underline{l}\underline{m}}=\frac 12{}^{(-)}\epsilon ^{\underline{%
l}\underline{m}}=\epsilon ^{\underline{m}\underline{l}},\mbox{ where }%
{}^{(+)}\epsilon ^{\underline{l}\underline{m}}=0,\mbox{ and }
$$
$$
\widetilde{\epsilon }^{\underline{l}\underline{m}}=-\frac 12{}^{(-)}\epsilon
^{\underline{l}\underline{m}}=\widetilde{\epsilon }^{\underline{m}\underline{%
l}}\mbox{
for }n=7(mod8).
$$

Let denote reduced and irreducible h-spinor spaces in a form pointing to the
symmetry of spinor inner products in dependence of values $n=8k+l$ ($%
k=0,1,2,...;l=1,2,...7)$ of the dimension of the horizontal subbundle (we
shall write respectively $\bigtriangleup $ and $\circ $ for antisymmetric
and symmetric inner products of reduced spinors and $\diamondsuit
=(\bigtriangleup ,\circ )$ and $\widetilde{\diamondsuit }=(\circ
,\bigtriangleup )$ for corresponding parametrizations of inner products, in
brief {\it i.p.}, of irreducible spinors; properties of scalar products of
spinors are defined by $\epsilon $-objects (2.50); we shall use $\Diamond $
for a general {\it i.p.} when the symmetry is not pointed out):%
$$
{\cal S}_{(h)}{\ }\left( 8k\right) ={\bf S}_{\circ }\oplus {\bf S}_{\circ
}^{\prime};\quad \eqno(2.51)
$$
$$
{\cal S}_{(h)}{\ }\left( 8k+1\right) ={\cal S}_{\circ }^{(-)}\
\mbox{({\it i.p.} is defined by an }^{(-)}\epsilon \mbox{-object);}
$$
$$
{\cal S}_{(h)}{\ }\left( 8k+2\right) =\{
\begin{array}{c}
{\cal S}_{\Diamond }=({\bf S}_{\Diamond },{\bf S}_{\Diamond }),\mbox{ or} \\
{\cal S}_{\Diamond }^{\prime }=({\bf S}_{\widetilde{\Diamond }}^{\prime },%
{\bf S}_{\widetilde{\Diamond }}^{\prime });
\end{array}
\qquad
$$
$$
{\cal S}_{(h)}\left( 8k+3\right) ={\cal S}_{\bigtriangleup }^{(+)}\
\mbox{({\it i.p.} is defined by an }^{(+)}\epsilon \mbox{-object);}
$$
$$
{\cal S}_{(h)}\left( 8k+4\right) ={\bf S}_{\bigtriangleup }\oplus {\bf S}%
_{\bigtriangleup }^{\prime };\quad
$$
$$
{\cal S}_{(h)}\left( 8k+5\right) ={\cal S}_{\bigtriangleup }^{(-)}\
\mbox{({\it i.p. }is defined
by an }^{(-)}\epsilon \mbox{-object),}
$$
$$
{\cal S}_{(h)}\left( 8k+6\right) =\{
\begin{array}{c}
{\cal S}_{\Diamond }=({\bf S}_{\Diamond },{\bf S}_{\Diamond }),\mbox{ or} \\
{\cal S}_{\Diamond }^{\prime }=({\bf S}_{\widetilde{\Diamond }}^{\prime },%
{\bf S}_{\widetilde{\Diamond }}^{\prime });
\end{array}
$$
\qquad
$$
{\cal S}_{(h)}\left( 8k+7\right) ={\cal S}_{\circ }^{(+)}\
\mbox{({\it i.p. } is defined by an }^{(+)}\epsilon \mbox{-object)}.
$$
We note that by using corresponding $\epsilon $-objects we can lower and
rise indices of reduced and irreducible spinors (for $n=2,6(mod4)$ we can
exclude primed indices, or inversely, see details in \cite{pen,penr1,penr2}).

The similar v-spinor spaces are denoted by the same symbols as in (2.51)
provided with a left lower mark ''$|"$ and parametrized with respect to the
values $m=8k^{\prime }+l$ (k'=0,1,...; l=1,2,...,7) of the dimension of the
vertical subbundle, for example, as
$$
{\cal S}_{(v_p)}(8k^{\prime })={\bf S}_{|\circ }\oplus {\bf S}_{|\circ
}^{\prime },{\cal S}_{(v_p)}\left( 8k+1\right) ={\cal S}_{|\circ }^{(-)},...%
\eqno(2.52)
$$
We use '' $\widetilde{}$ ''-overlined symbols,
$$
{\widetilde{{\cal S}}}_{(h)}\left( 8k\right) ={\widetilde{{\bf S}}}_{\circ
}\oplus \widetilde{S}_{\circ }^{\prime },{\widetilde{{\cal S}}}_{(h)}\left(
8k+1\right) ={\widetilde{{\cal S}}}_{\circ }^{(-)},...\eqno(2.53)
$$
and
$$
{\widetilde{{\cal S}}}_{(v_p)}(8k^{\prime })={\widetilde{{\bf S}}}_{|\circ
}\oplus {\widetilde{S}}_{|\circ }^{\prime },{\widetilde{{\cal S}}}%
_{(v_p)}\left( 8k^{\prime }+1\right) ={\widetilde{{\cal S}}}_{|\circ
}^{(-)},...\eqno(2.54)
$$
respectively for the dual to (2.50) and (2.51) spinor spaces.

The spinor spaces (2.50)-(2.54) are called the prime spinor spaces, in brief
p-spinors. They are considered as building blocks of distinguished, for
simplicity we consider $\left( n,m_1\right) $--spinor spaces constructed in
this manner:%
$$
{\cal S}(_{\circ \circ ,\circ \circ })={\bf S_{\circ }\oplus S_{\circ
}^{\prime }\oplus S_{|\circ }\oplus S_{|\circ }^{\prime },}{\cal S}(_{\circ
\circ ,\circ }\mid ^{\circ })={\bf S_{\circ }\oplus S_{\circ }^{\prime
}\oplus S_{|\circ }\oplus \widetilde{S}_{|\circ }^{\prime },}\eqno(2.55)
$$
$$
{\cal S}(_{\circ \circ ,}\mid ^{\circ \circ })={\bf S_{\circ }\oplus
S_{\circ }^{\prime }\oplus \widetilde{S}_{|\circ }\oplus \widetilde{S}%
_{|\circ }^{\prime },}{\cal S}(_{\circ }\mid ^{\circ \circ \circ })={\bf %
S_{\circ }\oplus \widetilde{S}_{\circ }^{\prime }\oplus \widetilde{S}%
_{|\circ }\oplus \widetilde{S}_{|\circ }^{\prime },}
$$
$$
...............................................
$$
$$
{\cal S}(_{\triangle },_{\triangle })={\cal S}_{\triangle }^{(+)}\oplus
S_{|\bigtriangleup }^{(+)},S(_{\triangle },^{\triangle })={\cal S}%
_{\triangle }^{(+)}\oplus \widetilde{S}_{|\triangle }^{(+)},
$$
$$
................................
$$
$$
{\cal S}(_{\triangle }|^{\circ },_\diamondsuit )={\bf S}_{\triangle }\oplus
\widetilde{S_{\circ }}^{\prime }\oplus {\cal S}_{|\diamondsuit },{\cal S}%
(_{\triangle }|^{\circ },^\diamondsuit )={\bf S}_{\triangle }\oplus
\widetilde{S_{\circ }}^{\prime }\oplus {\cal \widetilde{S}}_{|}^\diamondsuit
,
$$
$$
................................
$$
Considering the operation of dualization of prime components in (2.55) we
can generate different isomorphic variants of distinguished $\left(
n,m_1\right) $-spinor spaces. If we add anisotropic ''shalls'' with $%
m_2,...,m_z,$ we have to extend correspondingly spaces (2.55), for instance,%
$$
{\cal S}(_{\circ \circ ,\circ \circ (1)},...,_{\infty (p)},...,_{\infty
(z)})={\bf S_{\circ }\oplus S_{\circ }^{\prime }\oplus S_{|(1)\circ }\oplus
S_{|(1)\circ }^{\prime }\oplus ...}
$$
$$
{\bf \oplus S_{|(p)\circ }\oplus S_{|(p)\circ }^{\prime }\oplus ...\oplus
S_{|(z)\circ }\oplus S_{|(z)\circ ,}^{\prime }}
$$
and so on.

We define a d-spinor space ${\cal S}_{(n,m_1)}\ $ as a direct sum of a
horizontal and a vertical spinor spaces of type (2.55), for instance,
$$
{\cal S}_{(8k,8k^{\prime })}={\bf S}_{\circ }\oplus {\bf S}_{\circ }^{\prime
}\oplus {\bf S}_{|\circ }\oplus {\bf S}_{|\circ }^{\prime },{\cal S}%
_{(8k,8k^{\prime }+1)}\ ={\bf S}_{\circ }\oplus {\bf S}_{\circ }^{\prime
}\oplus {\cal S}_{|\circ }^{(-)},...,
$$
$$
{\cal S}_{(8k+4,8k^{\prime }+5)}={\bf S}_{\triangle }\oplus {\bf S}%
_{\triangle }^{\prime }\oplus {\cal S}_{|\triangle }^{(-)},...
$$
The scalar product on a ${\cal S}_{(n,m_1)}\ $ is induced by (corresponding
to fixed values of $n$ and $m_1$ ) $\epsilon $-objects (2.50) considered for
h- and v$_1$-components. We present also an example for ${\cal S}%
_{(n,m_1+...+m_z)}:$%
$$
{\cal S}_{(8k+4,8k_{(1)}+5,...,8k_{(p)}+4,...8k_{(z)})}=
$$
$$
{\bf S}_{\triangle }\oplus {\bf S}_{\triangle }^{\prime }\oplus {\cal S}%
_{|(1)\triangle }^{(-)}\oplus ...\oplus {\bf S}_{|(p)\triangle }\oplus {\bf S%
}_{|(p)\triangle }^{\prime }\oplus ...\oplus {\bf S}_{|(z)\circ }\oplus {\bf %
S}_{|(z)\circ }^{\prime }.
$$

Having introduced d-spinors for dimensions $\left( n,m_1+...+m_z\right) $ we
can write out the generalization for ha--spaces of twistor equations \cite
{penr1} by using the distinguished $\sigma $-objects (2.44):%
$$
(\sigma _{(<\widehat{\alpha }>})_{|\underline{\beta }|}^{..\underline{\gamma
}}\quad \frac{\delta \omega ^{\underline{\beta }}}{\delta u^{<\widehat{\beta
}>)}}=\frac 1{n+m_1+...+m_z}\quad G_{<\widehat{\alpha }><\widehat{\beta }%
>}(\sigma ^{\widehat{\epsilon }})_{\underline{\beta }}^{..\underline{\gamma }%
}\quad \frac{\delta \omega ^{\underline{\beta }}}{\delta u^{\widehat{%
\epsilon }}},\eqno(2.56)
$$
where $\left| \underline{\beta }\right| $ denotes that we do not consider
symmetrization on this index. The general solution of (2.56) on the d-vector
space ${\cal F}$ looks like as
$$
\omega ^{\underline{\beta }}=\Omega ^{\underline{\beta }}+u^{<\widehat{%
\alpha }>}(\sigma _{<\widehat{\alpha }>})_{\underline{\epsilon }}^{..%
\underline{\beta }}\Pi ^{\underline{\epsilon }},\eqno(2.57)
$$
where $\Omega ^{\underline{\beta }}$ and $\Pi ^{\underline{\epsilon }}$ are
constant d-spinors. For fixed values of dimensions $n$ and $m=m_1+...m_z$ we
mast analyze the reduced and irreducible components of h- and v$_p$-parts of
equations (2.56) and their solutions (2.57) in order to find the symmetry
properties of a d-twistor ${\bf Z^\alpha \ }$ defined as a pair of d-spinors%
$$
{\bf Z}^\alpha =(\omega ^{\underline{\alpha }},\pi _{\underline{\beta }%
}^{\prime }),
$$
where $\pi _{\underline{\beta }^{\prime }}=\pi _{\underline{\beta }^{\prime
}}^{(0)}\in {\widetilde{{\cal S}}}_{(n,m_1,...,m_z)}$ is a constant dual
d-spinor. The problem of definition of spinors and twistors on ha-spaces was
firstly considered in \cite{vod} (see also \cite{v87,vb12}) in connection
with the possibility to extend the equations (2.57) and theirs solutions
(2.58), by using nearly autoparallel maps, on curved, locally isotropic or
anisotropic, spaces. We note that the definition of twistors have been
extended to higher order anisotropic spaces with trivial N-- and
d--connections.

\subsubsection{ Mutual transforms of d-tensors and d-spinors}

The spinor algebra for spaces of higher dimensions can not be considered as
a real alternative to the tensor algebra as for locally isotropic spaces of
dimensions $n=3,4$ \cite{pen,penr1,penr2}. The same holds true for ha-spaces
and we emphasize that it is not quite convenient to perform a spinor
calculus for dimensions $n,m>>4$. Nevertheless, the concept of spinors is
important for every type of spaces, we can deeply understand the fundamental
properties of geometical objects on ha-spaces, and we shall consider in this
subsection some questions concerning transforms of d-tensor objects into
d-spinor ones.

\subsubsection{ Transformation of d-tensors into d-spinors}

In order to pass from d-tensors to d-spinors we must use $\sigma $-objects
(2.44) written in reduced or irreduced form \quad (in dependence of fixed
values of dimensions $n$ and $m$ ):

$$
(\sigma _{<\widehat{\alpha }>})_{\underline{\beta }}^{\cdot \underline{%
\gamma }},~(\sigma ^{<\widehat{\alpha }>})^{\underline{\beta }\underline{%
\gamma }},~(\sigma ^{<\widehat{\alpha }>})_{\underline{\beta }\underline{%
\gamma }},...,(\sigma _{<\widehat{a}>})^{\underline{b}\underline{c}%
},...,(\sigma _{\widehat{i}})_{\underline{j}\underline{k}},...,(\sigma _{<%
\widehat{a}>})^{AA^{\prime }},...,(\sigma ^{\widehat{i}})_{II^{\prime }},....%
\eqno(2.58)
$$
It is obvious that contracting with corresponding $\sigma $-objects (2.58)
we can introduce instead of d-tensors indices the d-spinor ones, for
instance,%
$$
\omega ^{\underline{\beta }\underline{\gamma }}=(\sigma ^{<\widehat{\alpha }%
>})^{\underline{\beta }\underline{\gamma }}\omega _{<\widehat{\alpha }%
>},\quad \omega _{AB^{\prime }}=(\sigma ^{<\widehat{a}>})_{AB^{\prime
}}\omega _{<\widehat{a}>},\quad ...,\zeta _{\cdot \underline{j}}^{\underline{%
i}}=(\sigma ^{\widehat{k}})_{\cdot \underline{j}}^{\underline{i}}\zeta _{%
\widehat{k}},....
$$
For d-tensors containing groups of antisymmetric indices there is a more
simple procedure of theirs transforming into d-spinors because the objects
$$
(\sigma _{\widehat{\alpha }\widehat{\beta }...\widehat{\gamma }})^{%
\underline{\delta }\underline{\nu }},\quad (\sigma ^{\widehat{a}\widehat{b}%
...\widehat{c}})^{\underline{d}\underline{e}},\quad ...,(\sigma ^{\widehat{i}%
\widehat{j}...\widehat{k}})_{II^{\prime }},\quad ...\eqno(2.59)
$$
can be used for sets of such indices into pairs of d-spinor indices. Let us
enumerate some properties of $\sigma $-objects of type (2.59) (for
simplicity we consider only h-components having q indices $\widehat{i},%
\widehat{j},\widehat{k},...$ taking values from 1 to $n;$ the properties of v%
$_p$-components can be written in a similar manner with respect to indices $%
\widehat{a}_p,\widehat{b}_p,\widehat{c}_p...$ taking values from 1 to $m$):%
$$
(\sigma _{\widehat{i}...\widehat{j}})^{\underline{k}\underline{l}}%
\mbox{
 is\ }\left\{ \
\begin{array}{c}
\mbox{symmetric on }\underline{k},\underline{l}\mbox{ for }n-2q\equiv
1,7~(mod~8); \\ \mbox{antisymmetric on }\underline{k},\underline{l}%
\mbox{
for }n-2q\equiv 3,5~(mod~8)
\end{array}
\right\} \eqno(2.60)
$$
for odd values of $n,$ and an object
$$
(\sigma _{\widehat{i}...\widehat{j}})^{IJ}~\left( (\sigma _{\widehat{i}...%
\widehat{j}})^{I^{\prime }J^{\prime }}\right)
$$
$$
\mbox{ is\ }\left\{
\begin{array}{c}
\mbox{symmetric on }I,J~(I^{\prime },J^{\prime })\mbox{ for }n-2q\equiv
0~(mod~8); \\ \mbox{antisymmetric on }I,J~(I^{\prime },J^{\prime })%
\mbox{
for }n-2q\equiv 4~(mod~8)
\end{array}
\right\} \eqno(2.61)
$$
or%
$$
(\sigma _{\widehat{i}...\widehat{j}})^{IJ^{\prime }}=\pm (\sigma _{\widehat{i%
}...\widehat{j}})^{J^{\prime }I}\{
\begin{array}{c}
n+2q\equiv 6(mod8); \\
n+2q\equiv 2(mod8),
\end{array}
\eqno(2.62)
$$
with vanishing of the rest of reduced components of the d-tensor $(\sigma _{%
\widehat{i}...\widehat{j}})^{\underline{k}\underline{l}}$ with prime/unprime
sets of indices.

\subsubsection{ Transformation of d-spinors into d-tensors; fundamental
d-spinors}

We can transform every d-spinor $\xi ^{\underline{\alpha }}=\left( \xi ^{%
\underline{i}},\xi ^{\underline{a}_1},...,\xi ^{\underline{a}_z}\right) $
into a corresponding d-tensor. For simplicity, we consider this construction
only for a h-component $\xi ^{\underline{i}}$ on a h-space being of
dimension $n$. The values%
$$
\xi ^{\underline{\alpha }}\xi ^{\underline{\beta }}(\sigma ^{\widehat{i}...%
\widehat{j}})_{\underline{\alpha }\underline{\beta }}\quad \left( n%
\mbox{ is
odd}\right) \eqno(2.63)
$$
or
$$
\xi ^I\xi ^J(\sigma ^{\widehat{i}...\widehat{j}})_{IJ}~\left( \mbox{or }\xi
^{I^{\prime }}\xi ^{J^{\prime }}(\sigma ^{\widehat{i}...\widehat{j}%
})_{I^{\prime }J^{\prime }}\right) ~\left( n\mbox{ is even}\right)
\eqno(2.64)
$$
with a different number of indices $\widehat{i}...\widehat{j},$ taken
together, defines the h-spinor $\xi ^{\underline{i}}\,$ to an accuracy to
the sign. We emphasize that it is necessary to choose only those
h-components of d-tensors (2.63) (or (2.64)) which are symmetric on pairs of
indices $\underline{\alpha }\underline{\beta }$ (or $IJ\,$ (or $I^{\prime
}J^{\prime }$ )) and the number $q$ of indices $\widehat{i}...\widehat{j}$
satisfies the condition (as a respective consequence of the properties
(2.60) and/or (2.61), (2.62))%
$$
n-2q\equiv 0,1,7~(mod~8).\eqno(2.65)
$$
Of special interest is the case when
$$
q=\frac 12\left( n\pm 1\right) ~\left( n\mbox{ is odd}\right) \eqno(2.66)
$$
or
$$
q=\frac 12n~\left( n\mbox{ is even}\right) .\eqno(2.67)
$$
If all expressions (2.63) and/or (2.64) are zero for all values of $q\,$
with the exception of one or two ones defined by the conditions (2.65),
(2.66) (or
(2.67)), the value $\xi ^{\widehat{i}}$ (or $\xi ^I$ ($\xi ^{I^{\prime }}))$
is called a fundamental h-spinor. Defining in a similar manner the
fundamental v-spinors we can introduce fundamental d-spinors as pairs of
fundamental h- and v-spinors. Here we remark that a h(v$_p$)-spinor $\xi ^{%
\widehat{i}}~(\xi ^{\widehat{a}_p})\,$ (we can also consider reduced
components) is always a fundamental one for $n(m)<7,$ which is a consequence
of (2.67)).

Finally, in this subsection, we note that the geometry of fundamental h- and
v-spinors is similar to that of usual fundamental spinors (see Appendix to
the monograph \cite{penr2}). We omit such details in this work, but
emphasize that constructions with fundamental d-spinors, for a la-space,
must be adapted to the corresponding global splitting by N-connection of the
space.

\subsection{ The Differential Geometry of D--Spinors}

This subsection is devoted to the differential geometry of d--spinors in higher
order anisotropic spaces.
We shall use denotations of type
$$
v^{<\alpha >}=(v^i,v^{<a>})\in \sigma ^{<\alpha >}=(\sigma ^i,\sigma ^{<a>})
$$
and%
$$
\zeta ^{\underline{\alpha }_p}=(\zeta ^{\underline{i}_p},\zeta ^{\underline{a%
}_p})\in \sigma ^{\alpha _p}=(\sigma ^{i_p},\sigma ^{a_p})\,
$$
for, respectively, elements of modules of d-vector and irreduced d-spinor
fields (see details in \cite{vjmp}). D-tensors and d-spinor tensors
(irreduced or reduced) will be interpreted as elements of corresponding $%
{\cal \sigma }$--modules, for instance,
$$
q_{~<\beta >...}^{<\alpha >}\in {\cal \sigma ^{<\alpha >}~_{<\beta >....}}%
,\psi _{~\underline{\beta }_p\quad ...}^{\underline{\alpha }_p\quad
\underline{\gamma }_p}\in {\cal \sigma }_{~\underline{\beta _p}\quad ...}^{%
\underline{\alpha }_p\quad \underline{\gamma }_p}~,\xi _{\quad
J_pK_p^{\prime }N_p^{\prime }}^{I_pI_p^{\prime }}\in {\cal \sigma }_{\quad
J_pK_p^{\prime }N_p^{\prime }}^{I_pI_p^{\prime }}~,...
$$

We can establish a correspondence between the la-adapted metric $g_{\alpha
\beta }$ (2.12) and d-spinor metric $\epsilon _{\underline{\alpha }%
\underline{\beta }}$ ( $\epsilon $-objects (2.50) for both h- and v$_p$%
-subspaces of ${\cal E}^{<z>}{\cal \,}$ ) of a ha-space ${\cal E}^{<z>}$ by
using the relation%
$$
g_{<\alpha ><\beta >}=-\frac 1{N(n)+N(m_1)+...+N(m_z)}\times \eqno(2.68)
$$
$$
((\sigma _{(<\alpha >}(u))^{\underline{\alpha }\underline{\beta }}(\sigma
_{<\beta >)}(u))^{\underline{\delta }\underline{\gamma }})\epsilon _{%
\underline{\alpha }\underline{\gamma }}\epsilon _{\underline{\beta }%
\underline{\delta }},
$$
where%
$$
(\sigma _{<\alpha >}(u))^{\underline{\nu }\underline{\gamma }}=l_{<\alpha
>}^{<\widehat{\alpha }>}(u)(\sigma _{<\widehat{\alpha }>})^{<\underline{\nu }%
><\underline{\gamma }>},\eqno(2.69)
$$
which is a consequence of formulas (2.43)-(2.50). In brief we can write
(2.68) as
$$
g_{<\alpha ><\beta >}=\epsilon _{\underline{\alpha }_1\underline{\alpha }%
_2}\epsilon _{\underline{\beta }_1\underline{\beta }_2}\eqno(2.70)
$$
if the $\sigma $-objects are considered as a fixed structure, whereas $%
\epsilon $-objects are treated as caring the metric ''dynamics '' , on
la-space. This variant is used, for instance, in the so-called 2-spinor
geometry \cite{penr1,penr2} and should be preferred if we have to make
explicit the algebraic symmetry properties of d-spinor objects by using
 metric decomposition (2.70). An
alternative way is to consider as fixed the algebraic structure of $\epsilon
$-objects and to use variable components of $\sigma $-objects of type (2.69)
for developing a variational d-spinor approach to gravitational and matter
field interactions on ha-spaces ( the spinor Ashtekar variables \cite{ash}
are introduced in this manner).

We note that a d--spinor metric
$$
\epsilon _{\underline{\nu }\underline{\tau }}=\left(
\begin{array}{cccc}
\epsilon _{\underline{i}\underline{j}} & 0 & ... & 0 \\
0 & \epsilon _{\underline{a}_1\underline{b}_1} & ... & 0 \\
... & ... & ... & ... \\
0 & 0 & ... & \epsilon _{\underline{a}_z\underline{b}_z}
\end{array}
\right)
$$
on the d-spinor space ${\cal S}=({\cal S}_{(h)},{\cal S}_{(v_1)},...,{\cal S}%
_{(v_z)})$ can have symmetric or antisymmetric h (v$_p$) -components $%
\epsilon _{\underline{i}\underline{j}}$ ($\epsilon _{\underline{a}_p%
\underline{b}_p})$ , see $\epsilon $-objects (2.50). For simplicity,
in order to avoid cumbersome calculations connected with eight-fold
periodicity on dimensions $n$ and $m_p$ of a ha-space ${\cal E}^{<z>},$
 we shall develop a general d-spinor formalism only by using irreduced
spinor spaces ${\cal S}_{(h)}$ and ${\cal S}_{(v_p)}.$

\subsubsection{ D-covariant derivation on ha--spaces}

Let ${\cal E}^{<z>}$ be a ha-space. We define the action on a d-spinor of a
d-covariant operator%
$$
\nabla _{<\alpha >}=\left( \nabla _i,\nabla _{<a>}\right) =(\sigma _{<\alpha
>})^{\underline{\alpha }_1\underline{\alpha }_2}\nabla _{^{\underline{\alpha
}_1\underline{\alpha }_2}}=\left( (\sigma _i)^{\underline{i}_1\underline{i}%
_2}\nabla _{^{\underline{i}_1\underline{i}_2}},~(\sigma _{<a>})^{\underline{a%
}_1\underline{a}_2}\nabla _{^{\underline{a}_1\underline{a}_2}}\right) =
$$
$$
\left( (\sigma _i)^{\underline{i}_1\underline{i}_2}\nabla _{^{\underline{i}_1%
\underline{i}_2}},~(\sigma _{a_1})^{\underline{a}_1\underline{a}_2}\nabla
_{(1)^{\underline{a}_1\underline{a}_2}},...,(\sigma _{a_p})^{\underline{a}_1%
\underline{a}_2}\nabla _{(p)^{\underline{a}_1\underline{a}_2}},...,(\sigma
_{a_z})^{\underline{a}_1\underline{a}_2}\nabla _{(z)^{\underline{a}_1%
\underline{a}_2}}\right)
$$
(in brief, we shall write
$$
\nabla _{<\alpha >}=\nabla _{^{\underline{\alpha }_1\underline{\alpha }%
_2}}=\left( \nabla _{^{\underline{i}_1\underline{i}_2}},~\nabla _{(1)^{%
\underline{a}_1\underline{a}_2}},...,\nabla _{(p)^{\underline{a}_1\underline{%
a}_2}},...,\nabla _{(z)^{\underline{a}_1\underline{a}_2}}\right) )
$$
as maps
$$
\nabla _{{\underline{\alpha }}_1{\underline{\alpha }}_2}\ :\ {\cal \sigma }^{%
\underline{\beta }}\rightarrow \sigma _{<\alpha >}^{\underline{\beta }%
}=\sigma _{{\underline{\alpha }}_1{\underline{\alpha }}_2}^{\underline{\beta
}}=
$$
$$
\left( \sigma _i^{\underline{\beta }}=\sigma _{{\underline{i}}_1{\underline{i%
}}_2}^{\underline{\beta }},\sigma _{(1)a_1}^{\underline{\beta }}=\sigma _{(1)%
{\underline{\alpha }}_1{\underline{\alpha }}_2}^{\underline{\beta }%
},...,\sigma _{(p)a_p}^{\underline{\beta }}=\sigma _{(p){\underline{\alpha }}%
_1{\underline{\alpha }}_2}^{\underline{\beta }},...,\sigma _{(z)a_z}^{%
\underline{\beta }}=\sigma _{(z){\underline{\alpha }}_1{\underline{\alpha }}%
_2}^{\underline{\beta }}\right)
$$
satisfying conditions%
$$
\nabla _{<\alpha >}(\xi ^{\underline{\beta }}+\eta ^{\underline{\beta }%
})=\nabla _{<\alpha >}\xi ^{\underline{\beta }}+\nabla _{<\alpha >}\eta ^{%
\underline{\beta }},
$$
and%
$$
\nabla _{<\alpha >}(f\xi ^{\underline{\beta }})=f\nabla _{<\alpha >}\xi ^{%
\underline{\beta }}+\xi ^{\underline{\beta }}\nabla _{<\alpha >}f
$$
for every $\xi ^{\underline{\beta }},\eta ^{\underline{\beta }}\in {\cal %
\sigma ^{\underline{\beta }}}$ and $f$ being a scalar field on ${\cal E}%
^{<z>}{\cal .\ }$ It is also required that one holds the Leibnitz rule%
$$
(\nabla _{<\alpha >}\zeta _{\underline{\beta }})\eta ^{\underline{\beta }%
}=\nabla _{<\alpha >}(\zeta _{\underline{\beta }}\eta ^{\underline{\beta }%
})-\zeta _{\underline{\beta }}\nabla _{<\alpha >}\eta ^{\underline{\beta }}
$$
and that $\nabla _{<\alpha >}\,$ is a real operator, i.e. it commuters with
the operation of complex conjugation:%
$$
\overline{\nabla _{<\alpha >}\psi _{\underline{\alpha }\underline{\beta }%
\underline{\gamma }...}}=\nabla _{<\alpha >}(\overline{\psi }_{\underline{%
\alpha }\underline{\beta }\underline{\gamma }...}).
$$

Let now analyze the question on uniqueness of action on d-spinors of an
operator $\nabla _{<\alpha >}$ satisfying necessary conditions . Denoting by
$\nabla _{<\alpha >}^{(1)}$ and $\nabla _{<\alpha >}$ two such d-covariant
operators we consider the map%
$$
(\nabla _{<\alpha >}^{(1)}-\nabla _{<\alpha >}):{\cal \sigma ^{\underline{%
\beta }}\rightarrow \sigma _{\underline{\alpha }_1\underline{\alpha }_2}^{%
\underline{\beta }}}.\eqno(2.71)
$$
Because the action on a scalar $f$ of both operators $\nabla _\alpha ^{(1)}$
and $\nabla _\alpha $ must be identical, i.e.%
$$
\nabla _{<\alpha >}^{(1)}f=\nabla _{<\alpha >}f,
$$
the action (2.71) on $f=\omega _{\underline{\beta }}\xi ^{\underline{\beta }%
} $ must be written as
$$
(\nabla _{<\alpha >}^{(1)}-\nabla _{<\alpha >})(\omega _{\underline{\beta }%
}\xi ^{\underline{\beta }})=0.
$$
In consequence we conclude that there is an element $\Theta _{\underline{%
\alpha }_1\underline{\alpha }_2\underline{\beta }}^{\quad \quad \underline{%
\gamma }}\in {\cal \sigma }_{\underline{\alpha }_1\underline{\alpha }_2%
\underline{\beta }}^{\quad \quad \underline{\gamma }}$ for which%
$$
\nabla _{\underline{\alpha }_1\underline{\alpha }_2}^{(1)}\xi ^{\underline{%
\gamma }}=\nabla _{\underline{\alpha }_1\underline{\alpha }_2}\xi ^{%
\underline{\gamma }}+\Theta _{\underline{\alpha }_1\underline{\alpha }_2%
\underline{\beta }}^{\quad \quad \underline{\gamma }}\xi ^{\underline{\beta }%
}\eqno(2.72)
$$
and%
$$
\nabla _{\underline{\alpha }_1\underline{\alpha }_2}^{(1)}\omega _{%
\underline{\beta }}=\nabla _{\underline{\alpha }_1\underline{\alpha }%
_2}\omega _{\underline{\beta }}-\Theta _{\underline{\alpha }_1\underline{%
\alpha }_2\underline{\beta }}^{\quad \quad \underline{\gamma }}\omega _{%
\underline{\gamma }}~.
$$
The action of the operator (2.71) on a d-vector $v^{<\beta >}=v^{\underline{%
\beta }_1\underline{\beta }_2}$ can be written by using formula (2.72) for
both indices $\underline{\beta }_1$ and $\underline{\beta }_2$ :%
$$
(\nabla _{<\alpha >}^{(1)}-\nabla _{<\alpha >})v^{\underline{\beta }_1%
\underline{\beta }_2}=\Theta _{<\alpha >\underline{\gamma }}^{\quad
\underline{\beta }_1}v^{\underline{\gamma }\underline{\beta }_2}+\Theta
_{<\alpha >\underline{\gamma }}^{\quad \underline{\beta }_2}v^{\underline{%
\beta }_1\underline{\gamma }}=
$$
$$
(\Theta _{<\alpha >\underline{\gamma }_1}^{\quad \underline{\beta }_1}\delta
_{\underline{\gamma }_2}^{\quad \underline{\beta }_2}+\Theta _{<\alpha >%
\underline{\gamma }_1}^{\quad \underline{\beta }_2}\delta _{\underline{%
\gamma }_2}^{\quad \underline{\beta }_1})v^{\underline{\gamma }_1\underline{%
\gamma }_2}=Q_{\ <\alpha ><\gamma >}^{<\beta >}v^{<\gamma >},
$$
where%
$$
Q_{\ <\alpha ><\gamma >}^{<\beta >}=Q_{\qquad \underline{\alpha }_1%
\underline{\alpha }_2~\underline{\gamma }_1\underline{\gamma }_2}^{%
\underline{\beta }_1\underline{\beta }_2}=\Theta _{<\alpha >\underline{%
\gamma }_1}^{\quad \underline{\beta }_1}\delta _{\underline{\gamma }%
_2}^{\quad \underline{\beta }_2}+\Theta _{<\alpha >\underline{\gamma }%
_1}^{\quad \underline{\beta }_2}\delta _{\underline{\gamma }_2}^{\quad
\underline{\beta }_1}.\eqno(2.73)
$$
The d-commutator $\nabla _{[<\alpha >}\nabla _{<\beta >]}$ defines the
d-torsion (see (2.23)-(2.25) and (2.30)). So, applying operators $\nabla
_{[<\alpha >}^{(1)}\nabla _{<\beta >]}^{(1)}$ and $\nabla _{[<\alpha
>}\nabla _{<\beta >]}$ on $f=\omega _{\underline{\beta }}\xi ^{\underline{%
\beta }}$ we can write
$$
T_{\quad <\alpha ><\beta >}^{(1)<\gamma >}-T_{~<\alpha ><\beta >}^{<\gamma
>}=Q_{~<\beta ><\alpha >}^{<\gamma >}-Q_{~<\alpha ><\beta >}^{<\gamma >}
$$
with $Q_{~<\alpha ><\beta >}^{<\gamma >}$ from (2.73).

The action of operator $\nabla _{<\alpha >}^{(1)}$ on d-spinor tensors of
type $\chi _{\underline{\alpha }_1\underline{\alpha }_2\underline{\alpha }%
_3...}^{\qquad \quad \underline{\beta }_1\underline{\beta }_2...}$ must be
constructed by using formula (2.72) for every upper index $\underline{\beta }%
_1\underline{\beta }_2...$ and formula (2.73) for every lower index $%
\underline{\alpha }_1\underline{\alpha }_2\underline{\alpha }_3...$ .

\subsubsection{Infeld--van der Waerden co\-ef\-fi\-ci\-ents and d-con\-nec\-ti\-ons}

Let
$$
\delta _{\underline{{\bf \alpha }}}^{\quad \underline{\alpha }}=\left(
\delta _{\underline{{\bf 1}}}^{\quad \underline{i}},\delta _{\underline{{\bf %
2}}}^{\quad \underline{i}},...,\delta _{\underline{{\bf N(n)}}}^{\quad
\underline{i}},\delta _{\underline{{\bf 1}}}^{\quad \underline{a}},\delta _{%
\underline{{\bf 2}}}^{\quad \underline{a}},...,\delta _{\underline{{\bf N(m)}%
}}^{\quad \underline{i}}\right)
$$
be a d--spinor basis. The dual to it basis is denoted as
$$
\delta _{\underline{\alpha }}^{\quad \underline{{\bf \alpha }}}=\left(
\delta _{\underline{i}}^{\quad \underline{{\bf 1}}},\delta _{\underline{i}%
}^{\quad \underline{{\bf 2}}},...,\delta _{\underline{i}}^{\quad \underline{%
{\bf N(n)}}},\delta _{\underline{i}}^{\quad \underline{{\bf 1}}},\delta _{%
\underline{i}}^{\quad \underline{{\bf 2}}},...,\delta _{\underline{i}%
}^{\quad \underline{{\bf N(m)}}}\right) .
$$
A d-spinor $\kappa ^{\underline{\alpha }}\in {\cal \sigma }$ $^{\underline{%
\alpha }}$ has components $\kappa ^{\underline{{\bf \alpha }}}=\kappa ^{%
\underline{\alpha }}\delta _{\underline{\alpha }}^{\quad \underline{{\bf %
\alpha }}}.$ Taking into account that
$$
\delta _{\underline{{\bf \alpha }}}^{\quad \underline{\alpha }}\delta _{%
\underline{{\bf \beta }}}^{\quad \underline{\beta }}\nabla _{\underline{%
\alpha }\underline{\beta }}=\nabla _{\underline{{\bf \alpha }}\underline{%
{\bf \beta }}},
$$
we write out the components $\nabla _{\underline{\alpha }\underline{\beta }}$
$\kappa ^{\underline{\gamma }}$ as%
$$
\delta _{\underline{{\bf \alpha }}}^{\quad \underline{\alpha }}~\delta _{%
\underline{{\bf \beta }}}^{\quad \underline{\beta }}~\delta _{\underline{%
\gamma }}^{\quad \underline{{\bf \gamma }}}~\nabla _{\underline{\alpha }%
\underline{\beta }}\kappa ^{\underline{\gamma }}=\delta _{\underline{{\bf %
\epsilon }}}^{\quad \underline{\tau }}~\delta _{\underline{\tau }}^{\quad
\underline{{\bf \gamma }}}~\nabla _{\underline{{\bf \alpha }}\underline{{\bf %
\beta }}}\kappa ^{\underline{{\bf \epsilon }}}+\kappa ^{\underline{{\bf %
\epsilon }}}~\delta _{\underline{\epsilon }}^{\quad \underline{{\bf \gamma }}%
}~\nabla _{\underline{{\bf \alpha }}\underline{{\bf \beta }}}\delta _{%
\underline{{\bf \epsilon }}}^{\quad \underline{\epsilon }}=
$$
$$
\nabla _{\underline{{\bf \alpha }}\underline{{\bf \beta }}}\kappa ^{%
\underline{{\bf \gamma }}}+\kappa ^{\underline{{\bf \epsilon }}}\gamma _{~%
\underline{{\bf \alpha }}\underline{{\bf \beta }}\underline{{\bf \epsilon }}%
}^{\underline{{\bf \gamma }}},\eqno(2.74)
$$
where the coordinate components of the d--spinor connection $\gamma _{~%
\underline{{\bf \alpha }}\underline{{\bf \beta }}\underline{{\bf \epsilon }}%
}^{\underline{{\bf \gamma }}}$ are defined as
$$
\gamma _{~\underline{{\bf \alpha }}\underline{{\bf \beta }}\underline{{\bf %
\epsilon }}}^{\underline{{\bf \gamma }}}\doteq \delta _{\underline{\tau }%
}^{\quad \underline{{\bf \gamma }}}~\nabla _{\underline{{\bf \alpha }}%
\underline{{\bf \beta }}}\delta _{\underline{{\bf \epsilon }}}^{\quad
\underline{\tau }}.\eqno(2.75)
$$
We call the Infeld - van der Waerden d-symbols a set of $\sigma $-objects ($%
\sigma _{{\bf \alpha }})^{\underline{{\bf \alpha }}\underline{{\bf \beta }}}$
paramet\-ri\-zed with respect to a coordinate d-spinor basis. Defining
$$
\nabla _{<{\bf \alpha >}}=(\sigma _{<{\bf \alpha >}})^{\underline{{\bf %
\alpha }}\underline{{\bf \beta }}}~\nabla _{\underline{{\bf \alpha }}%
\underline{{\bf \beta }}},
$$
introducing denotations
$$
\gamma ^{\underline{{\bf \gamma }}}{}_{<{\bf \alpha >\underline{\tau }}%
}\doteq \gamma ^{\underline{{\bf \gamma }}}{}_{{\bf \underline{\alpha }%
\underline{\beta }\underline{\tau }}}~(\sigma _{<{\bf \alpha >}})^{%
\underline{{\bf \alpha }}\underline{{\bf \beta }}}
$$
and using properties (2.74) we can write relations%
$$
l_{<{\bf \alpha >}}^{<\alpha >}~\delta _{\underline{\beta }}^{\quad
\underline{{\bf \beta }}}~\nabla _{<\alpha >}\kappa ^{\underline{\beta }%
}=\nabla _{<{\bf \alpha >}}\kappa ^{\underline{{\bf \beta }}}+\kappa ^{%
\underline{{\bf \delta }}}~\gamma _{~<{\bf \alpha >}\underline{{\bf \delta }}%
}^{\underline{{\bf \beta }}}\eqno(2.76)
$$
and%
$$
l_{<{\bf \alpha >}}^{<\alpha >}~\delta _{\underline{{\bf \beta }}}^{\quad
\underline{\beta }}~\nabla _{<\alpha >}~\mu _{\underline{\beta }}=\nabla _{<%
{\bf \alpha >}}~\mu _{\underline{{\bf \beta }}}-\mu _{\underline{{\bf \delta
}}}\gamma _{~<{\bf \alpha >}\underline{{\bf \beta }}}^{\underline{{\bf %
\delta }}}\eqno(2.77)
$$
for d-covariant derivations $~\nabla _{\underline{\alpha }}\kappa ^{%
\underline{\beta }}$ and $\nabla _{\underline{\alpha }}~\mu _{\underline{%
\beta }}.$

We can consider expressions similar to (2.76) and (2.77) for values having
both types of d-spinor and d-tensor indices, for instance,%
$$
l_{<{\bf \alpha >}}^{<\alpha >}~l_{<\gamma >}^{<{\bf \gamma >}}~\delta _{%
\underline{{\bf \delta }}}^{\quad \underline{\delta }}~\nabla _{<\alpha
>}\theta _{\underline{\delta }}^{~<\gamma >}=\nabla _{<{\bf \alpha >}}\theta
_{\underline{{\bf \delta }}}^{~<{\bf \gamma >}}-\theta _{\underline{{\bf %
\epsilon }}}^{~<{\bf \gamma >}}\gamma _{~<{\bf \alpha >}\underline{{\bf %
\delta }}}^{\underline{{\bf \epsilon }}}+\theta _{\underline{{\bf \delta }}%
}^{~<{\bf \tau >}}~\Gamma _{\quad <{\bf \alpha ><\tau >}}^{~<{\bf \gamma >}}
$$
(we can prove this by a straightforward calculation).

Now we shall consider some possible relations between components of
d-connec\-ti\-ons $\gamma _{~<{\bf \alpha >}\underline{{\bf \delta }}}^{%
\underline{{\bf \epsilon }}}$ and $\Gamma _{\quad <{\bf \alpha ><\tau >}}^{~<%
{\bf \gamma >}}$ and derivations of $(\sigma _{<{\bf \alpha >}})^{\underline{%
{\bf \alpha }}\underline{{\bf \beta }}}$ . According to definitions (2.12)
we can write%
$$
\Gamma _{~<{\bf \beta ><\gamma >}}^{<{\bf \alpha >}}=l_{<\alpha >}^{<{\bf %
\alpha >}}\nabla _{<{\bf \gamma >}}l_{<{\bf \beta >}}^{<\alpha >}=
$$
$$
l_{<\alpha >}^{<{\bf \alpha >}}\nabla _{<{\bf \gamma >}}(\sigma _{<{\bf %
\beta >}})^{\underline{\epsilon }\underline{\tau }}=l_{<\alpha >}^{<{\bf %
\alpha >}}\nabla _{<{\bf \gamma >}}((\sigma _{<{\bf \beta >}})^{\underline{%
{\bf \epsilon }}\underline{{\bf \tau }}}\delta _{\underline{{\bf \epsilon }}%
}^{~\underline{\epsilon }}\delta _{\underline{{\bf \tau }}}^{~\underline{%
\tau }})=
$$
$$
l_{<\alpha >}^{<{\bf \alpha >}}\delta _{\underline{{\bf \alpha }}}^{~%
\underline{\alpha }}\delta _{\underline{{\bf \epsilon }}}^{~\underline{%
\epsilon }}\nabla _{<{\bf \gamma >}}(\sigma _{<{\bf \beta >}})^{\underline{%
{\bf \alpha }}\underline{{\bf \epsilon }}}+l_{<\alpha >}^{<{\bf \alpha >}%
}(\sigma _{<{\bf \beta >}})^{\underline{{\bf \epsilon }}\underline{{\bf \tau
}}}(\delta _{\underline{{\bf \tau }}}^{~\underline{\tau }}\nabla _{<{\bf %
\gamma >}}\delta _{\underline{{\bf \epsilon }}}^{~\underline{\epsilon }%
}+\delta _{\underline{{\bf \epsilon }}}^{~\underline{\epsilon }}\nabla _{<%
{\bf \gamma >}}\delta _{\underline{{\bf \tau }}}^{~\underline{\tau }})=
$$
$$
l_{\underline{{\bf \epsilon }}\underline{{\bf \tau }}}^{<{\bf \alpha >}%
}~\nabla _{<{\bf \gamma >}}(\sigma _{<{\bf \beta >}})^{\underline{{\bf %
\epsilon }}\underline{{\bf \tau }}}+l_{\underline{{\bf \mu }}\underline{{\bf %
\nu }}}^{<{\bf \alpha >}}\delta _{\underline{\epsilon }}^{~\underline{{\bf %
\mu }}}\delta _{\underline{\tau }}^{~\underline{{\bf \nu }}}(\sigma _{<{\bf %
\beta >}})^{\underline{\epsilon }\underline{\tau }}(\delta _{\underline{{\bf %
\tau }}}^{~\underline{\tau }}\nabla _{<{\bf \gamma >}}\delta _{\underline{%
{\bf \epsilon }}}^{~\underline{\epsilon }}+\delta _{\underline{{\bf \epsilon
}}}^{~\underline{\epsilon }}\nabla _{<{\bf \gamma >}}\delta _{\underline{%
{\bf \tau }}}^{~\underline{\tau }}),
$$
where $l_{<\alpha >}^{<{\bf \alpha >}}=(\sigma _{\underline{{\bf \epsilon }}%
\underline{{\bf \tau }}})^{<{\bf \alpha >}}$ , from which it follows%
$$
(\sigma _{<{\bf \alpha >}})^{\underline{{\bf \mu }}\underline{{\bf \nu }}%
}(\sigma _{\underline{{\bf \alpha }}\underline{{\bf \beta }}})^{<{\bf \beta >%
}}\Gamma _{~<{\bf \gamma ><\beta >}}^{<{\bf \alpha >}}=(\sigma _{\underline{%
{\bf \alpha }}\underline{{\bf \beta }}})^{<{\bf \beta >}}\nabla _{<{\bf %
\gamma >}}(\sigma _{<{\bf \alpha >}})^{\underline{{\bf \mu }}\underline{{\bf %
\nu }}}+\delta _{\underline{{\bf \beta }}}^{~\underline{{\bf \nu }}}\gamma
_{~<{\bf \gamma >\underline{\alpha }}}^{\underline{{\bf \mu }}}+\delta _{%
\underline{{\bf \alpha }}}^{~\underline{{\bf \mu }}}\gamma _{~<{\bf \gamma >%
\underline{\beta }}}^{\underline{{\bf \nu }}}.
$$
Connecting the last expression on \underline{${\bf \beta }$} and \underline{$%
{\bf \nu }$} and using an orthonormalized d-spinor basis when $\gamma _{~<%
{\bf \gamma >\underline{\beta }}}^{\underline{{\bf \beta }}}=0$ (a
consequence from (2.75)) we have
$$
\gamma _{~<{\bf \gamma >\underline{\alpha }}}^{\underline{{\bf \mu }}}=\frac
1{N(n)+N(m_1)+...+N(m_z)}(\Gamma _{\quad <{\bf \gamma >~\underline{\alpha }%
\underline{\beta }}}^{\underline{{\bf \mu }}\underline{{\bf \beta }}%
}-(\sigma _{\underline{{\bf \alpha }}\underline{{\bf \beta }}})^{<{\bf \beta
>}}\nabla _{<{\bf \gamma >}}(\sigma _{<{\bf \beta >}})^{\underline{{\bf \mu }%
}\underline{{\bf \beta }}}),\eqno(2.78)
$$
where
$$
\Gamma _{\quad <{\bf \gamma >~\underline{\alpha }\underline{\beta }}}^{%
\underline{{\bf \mu }}\underline{{\bf \beta }}}=(\sigma _{<{\bf \alpha >}})^{%
\underline{{\bf \mu }}\underline{{\bf \beta }}}(\sigma _{\underline{{\bf %
\alpha }}\underline{{\bf \beta }}})^{{\bf \beta }}\Gamma _{~<{\bf \gamma
><\beta >}}^{<{\bf \alpha >}}.\eqno(2.79)
$$
We also note here that, for instance, for the canonical (see (2.18) and
(2.19)) and Berwald (see (2.20)) connections, Christoffel d-symbols (see
(2.21)) we can express d-spinor connection (2.79) through corresponding
locally adapted derivations of components of metric and N-connection by
introducing corresponding coefficients instead of $\Gamma _{~<{\bf \gamma
><\beta >}}^{<{\bf \alpha >}}$ in (2.79) and than in (2.78).

\subsubsection{ D--spinors of ha--space curvature and torsion}

The d-tensor indices of the commutator (2.29), $\Delta _{<\alpha ><\beta >},$
can be transformed into d-spinor ones:%
$$
\Box _{\underline{\alpha }\underline{\beta }}=(\sigma ^{<\alpha ><\beta >})_{%
\underline{\alpha }\underline{\beta }}\Delta _{\alpha \beta }=(\Box _{%
\underline{i}\underline{j}},\Box _{\underline{a}\underline{b}})=\eqno(2.80)
$$
$$
(\Box _{\underline{i}\underline{j}},\Box _{\underline{a}_1\underline{b}%
_1},...,\Box _{\underline{a}_p\underline{b}_p},...,\Box _{\underline{a}_z%
\underline{b}_z}),
$$
with h- and v$_p$-components,
$$
\Box _{\underline{i}\underline{j}}=(\sigma ^{<\alpha ><\beta >})_{\underline{%
i}\underline{j}}\Delta _{<\alpha ><\beta >}\mbox{ and }\Box _{\underline{a}%
\underline{b}}=(\sigma ^{<\alpha ><\beta >})_{\underline{a}\underline{b}%
}\Delta _{<\alpha ><\beta >},
$$
being symmetric or antisymmetric in dependence of corresponding values of
dimensions $n\,$ and $m_p$ (see eight-fold parametizations (2.50) and
(2.51)). Considering the actions of operator (2.80) on d-spinors $\pi ^{%
\underline{\gamma }}$ and $\mu _{\underline{\gamma }}$ we introduce the
d-spinor curvature $X_{\underline{\delta }\quad \underline{\alpha }%
\underline{\beta }}^{\quad \underline{\gamma }}\,$ as to satisfy equations%
$$
\Box _{\underline{\alpha }\underline{\beta }}\ \pi ^{\underline{\gamma }}=X_{%
\underline{\delta }\quad \underline{\alpha }\underline{\beta }}^{\quad
\underline{\gamma }}\pi ^{\underline{\delta }}\eqno(2.81)
$$
and%
$$
\Box _{\underline{\alpha }\underline{\beta }}\ \mu _{\underline{\gamma }}=X_{%
\underline{\gamma }\quad \underline{\alpha }\underline{\beta }}^{\quad
\underline{\delta }}\mu _{\underline{\delta }}.
$$
The gravitational d-spinor $\Psi _{\underline{\alpha }\underline{\beta }%
\underline{\gamma }\underline{\delta }}$ is defined by a corresponding
symmetrization of d-spinor indices:%
$$
\Psi _{\underline{\alpha }\underline{\beta }\underline{\gamma }\underline{%
\delta }}=X_{(\underline{\alpha }|\underline{\beta }|\underline{\gamma }%
\underline{\delta })}.\eqno(2.82)
$$
We note that d-spinor tensors $X_{\underline{\delta }\quad \underline{\alpha
}\underline{\beta }}^{\quad \underline{\gamma }}$ and $\Psi _{\underline{%
\alpha }\underline{\beta }\underline{\gamma }\underline{\delta }}\,$ are
transformed into similar 2-spinor objects on locally isotropic spaces \cite
{penr1,penr2} if we consider vanishing of the N-connection structure and a
limit to a locally isotropic space.

Putting $\delta _{\underline{\gamma }}^{\quad {\bf \underline{\gamma }}}$
instead of $\mu _{\underline{\gamma }}$ in (2.81) and using (2.82) we can
express respectively the curvature and gravitational d-spinors as
$$
X_{\underline{\gamma }\underline{\delta }\underline{\alpha }\underline{\beta
}}=\delta _{\underline{\delta }\underline{{\bf \tau }}}\Box _{\underline{%
\alpha }\underline{\beta }}\delta _{\underline{\gamma }}^{\quad {\bf
\underline{\tau }}}
$$
and%
$$
\Psi _{\underline{\gamma }\underline{\delta }\underline{\alpha }\underline{%
\beta }}=\delta _{\underline{\delta }\underline{{\bf \tau }}}\Box _{(%
\underline{\alpha }\underline{\beta }}\delta _{\underline{\gamma })}^{\quad
{\bf \underline{\tau }}}.
$$

The d-spinor torsion $T_{\qquad \underline{\alpha }\underline{\beta }}^{%
\underline{\gamma }_1\underline{\gamma }_2}$ is defined similarly as for
d-tensors (see (2.30)) by using the d-spinor commutator (2.80) and equations
$$
\Box _{\underline{\alpha }\underline{\beta }}f=T_{\qquad \underline{\alpha }%
\underline{\beta }}^{\underline{\gamma }_1\underline{\gamma }_2}\nabla _{%
\underline{\gamma }_1\underline{\gamma }_2}f.\eqno(2.82)
$$

The d-spinor components $R_{\underline{\gamma }_1\underline{\gamma }_2\qquad
\underline{\alpha }\underline{\beta }}^{\qquad \underline{\delta }_1%
\underline{\delta }_2}$ of the curvature d-tensor $R_{\gamma \quad \alpha
\beta }^{\quad \delta }$ can be computed by using relations (2.79), and
(2.80) and (2.82) as to satisfy the equations (the d-spinor analogous of
equations (2.31) )%
$$
(\Box _{\underline{\alpha }\underline{\beta }}-T_{\qquad \underline{\alpha }%
\underline{\beta }}^{\underline{\gamma }_1\underline{\gamma }_2}\nabla _{%
\underline{\gamma }_1\underline{\gamma }_2})V^{\underline{\delta }_1%
\underline{\delta }_2}=R_{\underline{\gamma }_1\underline{\gamma }_2\qquad
\underline{\alpha }\underline{\beta }}^{\qquad \underline{\delta }_1%
\underline{\delta }_2}V^{\underline{\gamma }_1\underline{\gamma }_2},%
$$
here d-vector $V^{\underline{\gamma }_1\underline{\gamma }_2}$ is considered
as a product of d-spinors, i.e. $V^{\underline{\gamma }_1\underline{\gamma }%
_2}=\nu ^{\underline{\gamma }_1}\mu ^{\underline{\gamma }_2}$. We find

$$
R_{\underline{\gamma }_1\underline{\gamma }_2\qquad \underline{\alpha }%
\underline{\beta }}^{\qquad \underline{\delta }_1\underline{\delta }%
_2}=\left( X_{\underline{\gamma }_1~\underline{\alpha }\underline{\beta }%
}^{\quad \underline{\delta }_1}+T_{\qquad \underline{\alpha }\underline{%
\beta }}^{\underline{\tau }_1\underline{\tau }_2}\quad \gamma _{\quad
\underline{\tau }_1\underline{\tau }_2\underline{\gamma }_1}^{\underline{%
\delta }_1}\right) \delta _{\underline{\gamma }_2}^{\quad \underline{\delta }%
_2}+
$$
$$
\left( X_{\underline{\gamma }_2~\underline{\alpha }\underline{\beta }%
}^{\quad \underline{\delta }_2}+T_{\qquad \underline{\alpha }\underline{%
\beta }}^{\underline{\tau }_1\underline{\tau }_2}\quad \gamma _{\quad
\underline{\tau }_1\underline{\tau }_2\underline{\gamma }_2}^{\underline{%
\delta }_2}\right) \delta _{\underline{\gamma }_1}^{\quad \underline{\delta }%
_1}.\eqno(2.83)
$$

It is convenient to use this d-spinor expression for the curvature d-tensor
$$
R_{\underline{\gamma }_1\underline{\gamma }_2\qquad \underline{\alpha }_1%
\underline{\alpha }_2\underline{\beta }_1\underline{\beta }_2}^{\qquad
\underline{\delta }_1\underline{\delta }_2}=\left( X_{\underline{\gamma }_1~%
\underline{\alpha }_1\underline{\alpha }_2\underline{\beta }_1\underline{%
\beta }_2}^{\quad \underline{\delta }_1}+T_{\qquad \underline{\alpha }_1%
\underline{\alpha }_2\underline{\beta }_1\underline{\beta }_2}^{\underline{%
\tau }_1\underline{\tau }_2}~\gamma _{\quad \underline{\tau }_1\underline{%
\tau }_2\underline{\gamma }_1}^{\underline{\delta }_1}\right) \delta _{%
\underline{\gamma }_2}^{\quad \underline{\delta }_2}+
$$
$$
\left( X_{\underline{\gamma }_2~\underline{\alpha }_1\underline{\alpha }_2%
\underline{\beta }_1\underline{\beta }_2}^{\quad \underline{\delta }%
_2}+T_{\qquad \underline{\alpha }_1\underline{\alpha }_2\underline{\beta }_1%
\underline{\beta }_2~}^{\underline{\tau }_1\underline{\tau }_2}\gamma
_{\quad \underline{\tau }_1\underline{\tau }_2\underline{\gamma }_2}^{%
\underline{\delta }_2}\right) \delta _{\underline{\gamma }_1}^{\quad
\underline{\delta }_1}
$$
in order to get the d--spinor components of the Ricci d-tensor%
$$
R_{\underline{\gamma }_1\underline{\gamma }_2\underline{\alpha }_1\underline{%
\alpha }_2}=R_{\underline{\gamma }_1\underline{\gamma }_2\qquad \underline{%
\alpha }_1\underline{\alpha }_2\underline{\delta }_1\underline{\delta }%
_2}^{\qquad \underline{\delta }_1\underline{\delta }_2}=
$$
$$
X_{\underline{\gamma }_1~\underline{\alpha }_1\underline{\alpha }_2%
\underline{\delta }_1\underline{\gamma }_2}^{\quad \underline{\delta }%
_1}+T_{\qquad \underline{\alpha }_1\underline{\alpha }_2\underline{\delta }_1%
\underline{\gamma }_2}^{\underline{\tau }_1\underline{\tau }_2}~\gamma
_{\quad \underline{\tau }_1\underline{\tau }_2\underline{\gamma }_1}^{%
\underline{\delta }_1}+X_{\underline{\gamma }_2~\underline{\alpha }_1%
\underline{\alpha }_2\underline{\delta }_1\underline{\gamma }_2}^{\quad
\underline{\delta }_2}+T_{\qquad \underline{\alpha }_1\underline{\alpha }_2%
\underline{\gamma }_1\underline{\delta }_2~}^{\underline{\tau }_1\underline{%
\tau }_2}\gamma _{\quad \underline{\tau }_1\underline{\tau }_2\underline{%
\gamma }_2}^{\underline{\delta }_2}\eqno(2.84)
$$
and this d-spinor decomposition of the scalar curvature:%
$$
q\overleftarrow{R}=R_{\qquad \underline{\alpha }_1\underline{\alpha }_2}^{%
\underline{\alpha }_1\underline{\alpha }_2}=X_{\quad ~\underline{~\alpha }%
_1\quad \underline{\delta }_1\underline{\alpha }_2}^{\underline{\alpha }_1%
\underline{\delta }_1~~\underline{\alpha }_2}+T_{\qquad ~~\underline{\alpha }%
_2\underline{\delta }_1}^{\underline{\tau }_1\underline{\tau }_2\underline{%
\alpha }_1\quad ~\underline{\alpha }_2}~\gamma _{\quad \underline{\tau }_1%
\underline{\tau }_2\underline{\alpha }_1}^{\underline{\delta }_1}+
$$
$$
X_{\qquad \quad \underline{\alpha }_2\underline{\delta }_2\underline{\alpha }%
_1}^{\underline{\alpha }_2\underline{\delta }_2\underline{\alpha }%
_1}+T_{\qquad \underline{\alpha }_1\quad ~\underline{\delta }_2~}^{%
\underline{\tau }_1\underline{\tau }_2~~\underline{\alpha }_2\underline{%
\alpha }_1}\gamma _{\quad \underline{\tau }_1\underline{\tau }_2\underline{%
\alpha }_2}^{\underline{\delta }_2}.\eqno(2.85)
$$

Putting (2.84) and (2.85) into (2.34) and, correspondingly, (2.35) we find
the d--spinor components of the Einstein and $\Phi _{<\alpha ><\beta >}$
d--tensors:%
$$
\overleftarrow{G}_{<\gamma ><\alpha >}=\overleftarrow{G}_{\underline{\gamma }%
_1\underline{\gamma }_2\underline{\alpha }_1\underline{\alpha }_2}=X_{%
\underline{\gamma }_1~\underline{\alpha }_1\underline{\alpha }_2\underline{%
\delta }_1\underline{\gamma }_2}^{\quad \underline{\delta }_1}+T_{\qquad
\underline{\alpha }_1\underline{\alpha }_2\underline{\delta }_1\underline{%
\gamma }_2}^{\underline{\tau }_1\underline{\tau }_2}~\gamma _{\quad
\underline{\tau }_1\underline{\tau }_2\underline{\gamma }_1}^{\underline{%
\delta }_1}+
$$
$$
X_{\underline{\gamma }_2~\underline{\alpha }_1\underline{\alpha }_2%
\underline{\delta }_1\underline{\gamma }_2}^{\quad \underline{\delta }%
_2}+T_{\qquad \underline{\alpha }_1\underline{\alpha }_2\underline{\gamma }_1%
\underline{\delta }_2~}^{\underline{\tau }_1\underline{\tau }_2}\gamma
_{\quad \underline{\tau }_1\underline{\tau }_2\underline{\gamma }_2}^{%
\underline{\delta }_2}-
$$
$$
\frac 12\varepsilon _{\underline{\gamma }_1\underline{\alpha }_1}\varepsilon
_{\underline{\gamma }_2\underline{\alpha }_2}[X_{\quad ~\underline{~\beta }%
_1\quad \underline{\mu }_1\underline{\beta }_2}^{\underline{\beta }_1%
\underline{\mu }_1~~\underline{\beta }_2}+T_{\qquad ~~\underline{\beta }_2%
\underline{\mu }_1}^{\underline{\tau }_1\underline{\tau }_2\underline{\beta }%
_1\quad ~\underline{\beta }_2}~\gamma _{\quad \underline{\tau }_1\underline{%
\tau }_2\underline{\beta }_1}^{\underline{\mu }_1}+
$$
$$
X_{\qquad \quad \underline{\beta }_2\underline{\mu }_2\underline{\nu }_1}^{%
\underline{\beta }_2\underline{\mu }_2\underline{\nu }_1}+T_{\qquad
\underline{\beta }_1\quad ~\underline{\delta }_2~}^{\underline{\tau }_1%
\underline{\tau }_2~~\underline{\beta }_2\underline{\beta }_1}\gamma _{\quad
\underline{\tau }_1\underline{\tau }_2\underline{\beta }_2}^{\underline{%
\delta }_2}]\eqno(2.86)
$$
and%
$$
\Phi _{<\gamma ><\alpha >}=\Phi _{\underline{\gamma }_1\underline{\gamma }_2%
\underline{\alpha }_1\underline{\alpha }_2}=\frac
1{2(n+m_1+...+m_z)}\varepsilon _{\underline{\gamma }_1\underline{\alpha }%
_1}\varepsilon _{\underline{\gamma }_2\underline{\alpha }_2}[X_{\quad ~%
\underline{~\beta }_1\quad \underline{\mu }_1\underline{\beta }_2}^{%
\underline{\beta }_1\underline{\mu }_1~~\underline{\beta }_2}+
$$
$$
T_{\qquad ~~\underline{\beta }_2\underline{\mu }_1}^{\underline{\tau }_1%
\underline{\tau }_2\underline{\beta }_1\quad ~\underline{\beta }_2}~\gamma
_{\quad \underline{\tau }_1\underline{\tau }_2\underline{\beta }_1}^{%
\underline{\mu }_1}+X_{\qquad \quad \underline{\beta }_2\underline{\mu }_2%
\underline{\nu }_1}^{\underline{\beta }_2\underline{\mu }_2\underline{\nu }%
_1}+T_{\qquad \underline{\beta }_1\quad ~\underline{\delta }_2~}^{\underline{%
\tau }_1\underline{\tau }_2~~\underline{\beta }_2\underline{\beta }_1}\gamma
_{\quad \underline{\tau }_1\underline{\tau }_2\underline{\beta }_2}^{%
\underline{\delta }_2}]-
$$
$$
\frac 12[X_{\underline{\gamma }_1~\underline{\alpha }_1\underline{\alpha }_2%
\underline{\delta }_1\underline{\gamma }_2}^{\quad \underline{\delta }%
_1}+T_{\qquad \underline{\alpha }_1\underline{\alpha }_2\underline{\delta }_1%
\underline{\gamma }_2}^{\underline{\tau }_1\underline{\tau }_2}~\gamma
_{\quad \underline{\tau }_1\underline{\tau }_2\underline{\gamma }_1}^{%
\underline{\delta }_1}+
$$
$$
X_{\underline{\gamma }_2~\underline{\alpha }_1\underline{\alpha }_2%
\underline{\delta }_1\underline{\gamma }_2}^{\quad \underline{\delta }%
_2}+T_{\qquad \underline{\alpha }_1\underline{\alpha }_2\underline{\gamma }_1%
\underline{\delta }_2~}^{\underline{\tau }_1\underline{\tau }_2}\gamma
_{\quad \underline{\tau }_1\underline{\tau }_2\underline{\gamma }_2}^{%
\underline{\delta }_2}].\eqno(2.87)
$$

The components of the conformal Weyl d-spinor can be computed by putting
d-spinor values of the curvature (2.83) and Ricci (2.84) d-tensors into
corresponding expression for the d-tensor (2.33). We omit this calculus in
this work.

\subsection{ Field Equations on Ha-Spaces}

The problem of formulation gravitational and gauge field equations on
different types of la-spaces is considered, for instance, in \cite
{ma94,bej,asa88} and \cite{vg}. In this subsection we shall introduce the basic
field equations for gravitational and matter field la-interactions in a
generalized form for generic higher order anisotropic spaces.

\subsubsection{ Locally anisotropic scalar field interactions}

Let $\varphi \left( u\right) =(\varphi _1\left( u\right) ,\varphi _2\left(
u\right) \dot ,...,\varphi _k\left( u\right) )$ be a complex k-component
scalar field of mass $\mu $ on ha-space ${\cal E}^{<z>}.$ The
d-covariant generalization of the conformally invariant (in the massless
case) scalar field equation \cite{penr1,penr2} can be defined by using the
d'Alambert locally anisotropic operator \cite{ana94,vst96} $\Box =D^{<\alpha
>}D_{<\alpha >}$, where $D_{<\alpha >}$ is a d-covariant derivation on $%
{\cal E}^{<z>}$ satisfying conditions (2.14) and (2.15) and constructed, for
simplicity, by using Christoffel d--symbols (2.21) (all formulas for field
equations and conservation values can be deformed by using corresponding
deformations d--tensors $P_{<\beta ><\gamma >}^{<\alpha >}$ from the
Cristoffel d--symbols, or the canonical d--connection to a general
d-connection into consideration):

$$
(\Box +\frac{n_E-2}{4(n_E-1)}\overleftarrow{R}+\mu ^2)\varphi \left(
u\right) =0,\eqno(2.88)
$$
where $n_E=n+m_1+...+m_z.$We must change d-covariant derivation $D_{<\alpha
>}$ into $^{\diamond }D_{<\alpha >}=D_{<\alpha >}+ieA_{<\alpha >}$ and take
into account the d-vector current
$$
J_{<\alpha >}^{(0)}\left( u\right) =i(\left( \overline{\varphi }\left(
u\right) D_{<\alpha >}\varphi \left( u\right) -D_{<\alpha >}\overline{%
\varphi }\left( u\right) )\varphi \left( u\right) \right)
$$
if interactions between locally anisotropic electromagnetic field ( d-vector
potential $A_{<\alpha >}$ ), where $e$ is the electromagnetic constant, and
charged scalar field $\varphi $ are considered. The equations (2.88) are
(locally adapted to the N-connection structure) Euler equations for the
Lagrangian%
$$
{\cal L}^{(0)}\left( u\right) =\sqrt{|g|}\left[ g^{<\alpha ><\beta >}\delta
_{<\alpha >}\overline{\varphi }\left( u\right) \delta _{<\beta >}\varphi
\left( u\right) -\left( \mu ^2+\frac{n_E-2}{4(n_E-1)}\right) \overline{%
\varphi }\left( u\right) \varphi \left( u\right) \right] ,\eqno(2.89)
$$
where $|g|=detg_{<\alpha ><\beta >}.$

The locally adapted variations of the action with Lagrangian (2.89) on
variables $\varphi \left( u\right) $ and $\overline{\varphi }\left( u\right)
$ leads to the locally anisotropic generalization of the energy-momentum
tensor,%
$$
E_{<\alpha ><\beta >}^{(0,can)}\left( u\right) =\delta _{<\alpha >}\overline{%
\varphi }\left( u\right) \delta _{<\beta >}\varphi \left( u\right) +$$
$$ \delta
_{<\beta >}\overline{\varphi }\left( u\right) \delta _{<\alpha >}\varphi
\left( u\right) -\frac 1{\sqrt{|g|}}g_{<\alpha ><\beta >}{\cal L}%
^{(0)}\left( u\right) ,\eqno(2.90)
$$
and a similar variation on the components of a d-metric (2.12) leads to a
symmetric metric energy-momentum d-tensor,%
$$
E_{<\alpha ><\beta >}^{(0)}\left( u\right) =E_{(<\alpha ><\beta
>)}^{(0,can)}\left( u\right) -\eqno(2.91)
$$
$$
\frac{n_E-2}{2(n_E-1)}\left[ R_{(<\alpha ><\beta >)}+D_{(<\alpha >}D_{<\beta
>)}-g_{<\alpha ><\beta >}\Box \right] \overline{\varphi }\left( u\right)
\varphi \left( u\right) .
$$
Here we note that we can obtain a nonsymmetric energy-momentum d-tensor if
we firstly vary on $G_{<\alpha ><\beta >}$ and than impose constraints of
type (2.10) in order to have a compatibility with the N-connection
structure. We also conclude that the existence of a N-connection in
dv-bundle ${\cal E}^{<z>}$ results in a nonequivalence of energy-momentum
d-tensors (2.90) and (2.91), nonsymmetry of the Ricci tensor (see (2.29)),
nonvanishing of the d-covariant derivation of the Einstein d-tensor (2.34), $%
D_{<\alpha >}\overleftarrow{G}^{<\alpha ><\beta >}\neq 0$ and, in
consequence, a corresponding breaking of conservation laws on ha-spaces when
$D_{<\alpha >}E^{<\alpha ><\beta >}\neq 0\,$ . The problem of formulation of
conservation laws on la-spaces is discussed in details and two variants of
its solution (by using nearly autoparallel maps and tensor integral
formalism on locally anisotropic and higher order multispaces) are proposed
in \cite{vst96} (see Sec. IV). In this subsection we shall present only
straightforward generalizations of field equations and necessary formulas
for energy-momentum d-tensors of matter fields on ${\cal E}^{<z>}$
considering that it is naturally that the conservation laws (usually being
consequences of global, local and/or intrinsic symmetries of the fundamental
space-time and of the type of field interactions) have to be broken on
locally anisotropic spaces.

\subsubsection{ Proca equations on ha--spaces}

Let consider a d-vector field $\varphi _{<\alpha >}\left( u\right) $ with
mass $\mu ^2$ (locally anisotropic Proca field ) interacting with exterior
la-gravitational field. From the Lagrangian
$$
{\cal L}^{(1)}\left( u\right) =\sqrt{\left| g\right| }\left[ -\frac 12{%
\overline{f}}_{<\alpha ><\beta >}\left( u\right) f^{<\alpha ><\beta >}\left(
u\right) +\mu ^2{\overline{\varphi }}_{<\alpha >}\left( u\right) \varphi
^{<\alpha >}\left( u\right) \right] ,\eqno(2.92)
$$
where $f_{<\alpha ><\beta >}=D_{<\alpha >}\varphi _{<\beta >}-D_{<\beta
>}\varphi _{<\alpha >},$ one follows the Proca equations on higher order
anisotropic spaces
$$
D_{<\alpha >}f^{<\alpha ><\beta >}\left( u\right) +\mu ^2\varphi ^{<\beta
>}\left( u\right) =0.\eqno(2.93)
$$
Equations (2.93) are a first type constraints for $\beta =0.$ Acting with $%
D_{<\alpha >}$ on (2.93), for $\mu \neq 0$ we obtain second type constraints%
$$
D_{<\alpha >}\varphi ^{<\alpha >}\left( u\right) =0.\eqno(2.94)
$$

Putting (2.94) into (2.93) we obtain second order field equations with
respect to $\varphi _{<\alpha >}$ :%
$$
\Box \varphi _{<\alpha >}\left( u\right) +R_{<\alpha ><\beta >}\varphi
^{<\beta >}\left( u\right) +\mu ^2\varphi _{<\alpha >}\left( u\right) =0.%
\eqno(2.95)
$$
The energy-momentum d-tensor and d-vector current following from the (2.95)
can be written as
$$
E_{<\alpha ><\beta >}^{(1)}\left( u\right) =-g^{<\varepsilon ><\tau >}\left(
{\overline{f}}_{<\beta ><\tau >}f_{<\alpha ><\varepsilon >}+{\overline{f}}%
_{<\alpha ><\varepsilon >}f_{<\beta ><\tau >}\right) +
$$
$$
\mu ^2\left( {\overline{\varphi }}_{<\alpha >}\varphi _{<\beta >}+{\overline{%
\varphi }}_{<\beta >}\varphi _{<\alpha >}\right) -\frac{g_{<\alpha ><\beta >}%
}{\sqrt{\left| g\right| }}{\cal L}^{(1)}\left( u\right) .
$$
and%
$$
J_{<\alpha >}^{\left( 1\right) }\left( u\right) =i\left( {\overline{f}}%
_{<\alpha ><\beta >}\left( u\right) \varphi ^{<\beta >}\left( u\right) -{%
\overline{\varphi }}^{<\beta >}\left( u\right) f_{<\alpha ><\beta >}\left(
u\right) \right) .
$$

For $\mu =0$ the d-tensor $f_{<\alpha ><\beta >}$ and the Lagrangian (2.92)
are invariant with respect to locally anisotropic gauge transforms of type
$$
\varphi _{<\alpha >}\left( u\right) \rightarrow \varphi _{<\alpha >}\left(
u\right) +\delta _{<\alpha >}\Lambda \left( u\right) ,
$$
where $\Lambda \left( u\right) $ is a d-differentiable scalar function, and
we obtain a locally anisot\-rop\-ic variant of Maxwell theory.

\subsubsection{ Higher order an\-i\-sot\-rop\-ic gravitons}

Let a massless d-tensor field $h_{<\alpha ><\beta >}\left( u\right) $ is
interpreted as a small perturbation of the locally anisotropic background
metric d-field $g_{<\alpha ><\beta >}\left( u\right) .$ Considering, for
simplicity, a torsionless background we have locally anisotropic Fierz--Pauli
equations%
$$\Box h_{<\alpha ><\beta >}\left( u\right) +2R_{<\tau ><\alpha ><\beta ><\nu
>}\left( u\right) ~h^{<\tau ><\nu >}\left( u\right) =0
$$
and d--gauge conditions%
$$
D_{<\alpha >}h_{<\beta >}^{<\alpha >}\left( u\right) =0,\quad h\left(
u\right) \equiv h_{<\beta >}^{<\alpha >}(u)=0,
$$
where $R_{<\tau ><\alpha ><\beta ><\nu >}\left( u\right) $ is curvature
d-tensor of the la-background space (these formulae can be obtained by using
a perturbation formalism with respect to $h_{<\alpha ><\beta >}\left(
u\right) $ developed in \cite{gri}; in our case we must take into account
the distinguishing of geometrical objects and operators on ha--spaces).

\subsubsection{ Higher order anisotropic Dirac equations}

Let denote the Dirac d--spinor field on ${\cal E}^{<z>}$ as $\psi \left(
u\right) =\left( \psi ^{\underline{\alpha }}\left( u\right) \right) $ and
consider as the generalized Lorentz transforms the group of automorphysm of
the metric $G_{<\widehat{\alpha }><\widehat{\beta }>}$ (see the ha-frame
decomposition of d-metric (2.12), (2.68) and (2.69) ).The d-covariant
derivation of field $\psi \left( u\right) $ is written as
$$
\overrightarrow{\nabla _{<\alpha >}}\psi =\left[ \delta _{<\alpha >}+\frac
14C_{\widehat{\alpha }\widehat{\beta }\widehat{\gamma }}\left( u\right)
~l_{<\alpha >}^{\widehat{\alpha }}\left( u\right) \sigma ^{\widehat{\beta }%
}\sigma ^{\widehat{\gamma }}\right] \psi ,\eqno(2.96)
$$
where coefficients $C_{\widehat{\alpha }\widehat{\beta }\widehat{\gamma }%
}=\left( D_{<\gamma >}l_{\widehat{\alpha }}^{<\alpha >}\right) l_{\widehat{%
\beta }<\alpha >}l_{\widehat{\gamma }}^{<\gamma >}$ generalize for ha-spaces
the corresponding Ricci coefficients on Riemannian spaces \cite{foc}. Using $%
\sigma $-objects $\sigma ^{<\alpha >}\left( u\right) $ (see (2.44) and
(2.60)--(2.62)) we define the Dirac equations on ha--spaces:
$$
(i\sigma ^{<\alpha >}\left( u\right) \overrightarrow{\nabla _{<\alpha >}}%
-\mu )\psi =0,
$$
which are the Euler equations for the Lagrangian%
$$
{\cal L}^{(1/2)}\left( u\right) =\sqrt{\left| g\right| }\{[\psi ^{+}\left(
u\right) \sigma ^{<\alpha >}\left( u\right) \overrightarrow{\nabla _{<\alpha
>}}\psi \left( u\right) -
$$
$$
(\overrightarrow{\nabla _{<\alpha >}}\psi ^{+}\left( u\right) )\sigma
^{<\alpha >}\left( u\right) \psi \left( u\right) ]-\mu \psi ^{+}\left(
u\right) \psi \left( u\right) \},\eqno(2.97)
$$
where $\psi ^{+}\left( u\right) $ is the complex conjugation and
transposition of the column$~\psi \left( u\right) .$

From (2.97) we obtain the d-metric energy-momentum d-tensor%
$$
E_{<\alpha ><\beta >}^{(1/2)}\left( u\right) =\frac i4[\psi ^{+}\left(
u\right) \sigma _{<\alpha >}\left( u\right) \overrightarrow{\nabla _{<\beta
>}}\psi \left( u\right) +\psi ^{+}\left( u\right) \sigma _{<\beta >}\left(
u\right) \overrightarrow{\nabla _{<\alpha >}}\psi \left( u\right) -
$$
$$
(\overrightarrow{\nabla _{<\alpha >}}\psi ^{+}\left( u\right) )\sigma
_{<\beta >}\left( u\right) \psi \left( u\right) -(\overrightarrow{\nabla
_{<\beta >}}\psi ^{+}\left( u\right) )\sigma _{<\alpha >}\left( u\right)
\psi \left( u\right) ]
$$
and the d-vector source%
$$
J_{<\alpha >}^{(1/2)}\left( u\right) =\psi ^{+}\left( u\right) \sigma
_{<\alpha >}\left( u\right) \psi \left( u\right) .
$$
We emphasize that la-interactions with exterior gauge fields can be
introduced by changing the higher order anisotropic partial derivation from
(2.96) in this manner:%
$$
\delta _\alpha \rightarrow \delta _\alpha +ie^{\star }B_\alpha ,
$$
where $e^{\star }$ and $B_\alpha $ are respectively the constant d-vector
potential of locally anisotropic gauge interactions on  higher order
anisotropic spaces (see \cite{vg} and the next Sec. II.G.5 ).

\subsubsection{ D--spinor locally anisotropic Yang--Mills fields}

We consider a dv-bundle ${\cal B}_E,~\pi _B:{\cal B\rightarrow E}^{<z>}{\cal %
,}$ on ha-space ${\cal E}^{<z>}{\cal .\,}$ Additionally to d-tensor and
d-spinor indices we shall use capital Greek letters, $\Phi ,\Upsilon ,\Xi
,\Psi ,...$ for fibre (of this bundle) indices (see details in \cite
{penr1,penr2} for the case when the base space of the v-bundle $\pi _B$ is a
locally isotropic space-time). Let $\underline{\nabla }_{<\alpha >}$ be, for
simplicity, a torsionless, linear connection in ${\cal B}_E$ satisfying
conditions:
$$
\underline{\nabla }_{<\alpha >}:{\em \Upsilon }^\Theta \rightarrow {\em %
\Upsilon }_{<\alpha >}^\Theta \quad \left[ \mbox{or }{\em \Xi }^\Theta
\rightarrow {\em \Xi }_{<\alpha >}^\Theta \right] ,
$$
$$
\underline{\nabla }_{<\alpha >}\left( \lambda ^\Theta +\nu ^\Theta \right) =%
\underline{\nabla }_{<\alpha >}\lambda ^\Theta +\underline{\nabla }_{<\alpha
>}\nu ^\Theta ,
$$
$$
\underline{\nabla }_{<\alpha >}~(f\lambda ^\Theta )=\lambda ^\Theta
\underline{\nabla }_{<\alpha >}f+f\underline{\nabla }_{<\alpha >}\lambda
^\Theta ,\quad f\in {\em \Upsilon }^\Theta ~[\mbox{or }{\em \Xi }^\Theta ],
$$
where by ${\em \Upsilon }^\Theta ~\left( {\em \Xi }^\Theta \right) $ we
denote the module of sections of the real (complex) v-bundle ${\cal B}_E$
provided with the abstract index $\Theta .$ The curvature of connection $%
\underline{\nabla }_{<\alpha >}$ is defined as
$$
K_{<\alpha ><\beta >\Omega }^{\qquad \Theta }\lambda ^\Omega =\left(
\underline{\nabla }_{<\alpha >}\underline{\nabla }_{<\beta >}-\underline{%
\nabla }_{<\beta >}\underline{\nabla }_{<\alpha >}\right) \lambda ^\Theta .
$$

For Yang-Mills fields as a rule one considers that ${\cal B}_E$ is enabled
with a unitary (complex) structure (complex conjugation changes mutually the
upper and lower Greek indices). It is useful to introduce instead of $%
K_{<\alpha ><\beta >\Omega }^{\qquad \Theta }$ a Hermitian matrix $%
F_{<\alpha ><\beta >\Omega }^{\qquad \Theta }=i$ $K_{<\alpha ><\beta >\Omega
}^{\qquad \Theta }$ connected with components of the Yang-Mills d-vector
potential $B_{<\alpha >\Xi }^{\quad \Phi }$ according the formula:

$$
\frac 12F_{<\alpha ><\beta >\Xi }^{\qquad \Phi }=\underline{\nabla }%
_{[<\alpha >}B_{<\beta >]\Xi }^{\quad \Phi }-iB_{[<\alpha >|\Lambda
|}^{\quad \Phi }B_{<\beta >]\Xi }^{\quad \Lambda },\eqno(2.98)
$$
where the la-space indices commute with capital Greek indices. The gauge
transforms are written in the form:

$$
B_{<\alpha >\Theta }^{\quad \Phi }\mapsto B_{<\alpha >\widehat{\Theta }%
}^{\quad \widehat{\Phi }}=B_{<\alpha >\Theta }^{\quad \Phi }~s_\Phi ^{\quad
\widehat{\Phi }}~q_{\widehat{\Theta }}^{\quad \Theta }+is_\Theta ^{\quad
\widehat{\Phi }}\underline{\nabla }_{<\alpha >}~q_{\widehat{\Theta }}^{\quad
\Theta },
$$
$$
F_{<\alpha ><\beta >\Xi }^{\qquad \Phi }\mapsto F_{<\alpha ><\beta >\widehat{%
\Xi }}^{\qquad \widehat{\Phi }}=F_{<\alpha ><\beta >\Xi }^{\qquad \Phi
}s_\Phi ^{\quad \widehat{\Phi }}q_{\widehat{\Xi }}^{\quad \Xi },
$$
where matrices $s_\Phi ^{\quad \widehat{\Phi }}$ and $q_{\widehat{\Xi }%
}^{\quad \Xi }$ are mutually inverse (Hermitian conjugated in the unitary
case). The Yang-Mills equations on torsionless la-spaces \cite{vg} (see
details in the next section) are written in this form:%
$$
\underline{\nabla }^{<\alpha >}F_{<\alpha ><\beta >\Theta }^{\qquad \Psi
}=J_{<\beta >\ \Theta }^{\qquad \Psi },\eqno(2.99)
$$
$$
\underline{\nabla }_{[<\alpha >}F_{<\beta ><\gamma >]\Theta }^{\qquad \Xi
}=0.\eqno(2.100)
$$
We must introduce deformations of connection of type  $\underline{\nabla }%
_\alpha ^{\star }~\longrightarrow \underline{\nabla }_\alpha +P_\alpha ,$
(the deformation d-tensor $P_\alpha $ is induced by the torsion in dv-bundle
${\cal B}_E)$ into the definition of the curvature of ha-gauge fields
(2.98) and motion equations (2.99) and (2.100) if interactions are modeled
on a generic ha-space.

\subsubsection{D--spinor Einstein--Cartan equ\-a\-ti\-ons }

Now we can write out the field equations of the Einstein-Cartan theory in
the d-spinor form. So, for the Einstein equations (2.34) we have

$$
\overleftarrow{G}_{\underline{\gamma }_1\underline{\gamma }_2\underline{%
\alpha }_1\underline{\alpha }_2}+\lambda \varepsilon _{\underline{\gamma }_1%
\underline{\alpha }_1}\varepsilon _{\underline{\gamma }_2\underline{\alpha }%
_2}=\kappa E_{\underline{\gamma }_1\underline{\gamma }_2\underline{\alpha }_1%
\underline{\alpha }_2},
$$
with $\overleftarrow{G}_{\underline{\gamma }_1\underline{\gamma }_2%
\underline{\alpha }_1\underline{\alpha }_2}$ from (2.86), or, by using the
d-tensor (2.87),

$$
\Phi _{\underline{\gamma }_1\underline{\gamma }_2\underline{\alpha }_1%
\underline{\alpha }_2}+(\frac{\overleftarrow{R}}4-\frac \lambda
2)\varepsilon _{\underline{\gamma }_1\underline{\alpha }_1}\varepsilon _{%
\underline{\gamma }_2\underline{\alpha }_2}=-\frac \kappa 2E_{\underline{%
\gamma }_1\underline{\gamma }_2\underline{\alpha }_1\underline{\alpha }_2},
$$
which are the d-spinor equivalent of the equations (2.35). These equations
must be completed by the algebraic equations (2.36) for the d-torsion and
d-spin density with d-tensor indices chainged into corresponding d-spinor
ones.

\section{Gauge and Gravitational Ha--Fields}

The aim of this section is twofold. The first objective is to present our
results \cite{vg,v295a,v295b} on formulation of geometrical approach to
interactions of Yang-Mills fields on spaces with higher order anisotropy in the
framework of the theory of linear connections in vector bundles (with
semisimple structural groups) on ha-spaces. The second objective is to
extend the geometrical formalism in a manner including theories with
nonsemisimple groups which permit a unique fiber bundle treatment for both
locally anisotropic Yang-Mills field and gravitational interactions. In
general lines, we shall follow the ideas and geometric methods proposed in
Refs. \cite{ts,p,pd,pon,bis}, but we shall apply them in a form convenient
for introducing into consideration geometrical constructions \cite{ma87,ma94}
and physical theories on ha-spaces.

There is a number of works on gauge models of interactions on Finsler spaces
and theirs extensions(see, for instance, \cite
{asa,asa88,asa89,bei,mirbal,ono}). One has introduced different variants of
generalized gauge transforms, postulated corresponding Lagrangians for
gravitational, gauge and matter field interactions and formulated
variational calculus (here we note the approach developed by A. Bejancu \cite
{bej89h,bej91a,bej}). The main problem of such models is the dependence of
the basic equations on chosen definition of gauge "compensation" symmetries
and on type of space and field interactions anisotropy. In order to avoid
the ambiguities connected with particular characteristics of possible
la-gauge theories we consider a "pure" geometric approach to gauge theories
(on both locally isotropic and anisotropic spaces) in the framework of the
theory of fiber bundles provided in general with different types of
nonlinear and linear multiconnection and metric structures. This way, based
on global geometric methods, holds also good for nonvariational, in the
total spaces of bundles, gauge theories (in the case of gauge gravity based
on Poincare or affine gauge groups); physical values and motion (field)
equations have adequate geometric interpretation and do not depend on the
type of local anisotropy of space-time background. It should be emphasized
here that extensions for higher order anisotropic spaces which will be
presented in this section can be realized in a straightforward manner.

\subsection{Gauge Fields on Ha-Spaces}

This subsection is devoted to formulation of the geometrical background for
gauge field theories on spaces with higher order anisotropy.

Let $\left( P,\pi ,Gr,{\cal E}^{<z>}\right) $ be a principal bundle
${\cal E}^{<z>}$ (being a ha-space) with structural group $Gr$ and surjective
map $\pi :P\rightarrow {\cal E}^{<z>}{\cal .\,}$At every point $u=\left(
x,y_{(1)},...,y_{(z)}\right) \in {\cal E}^{<z>}$ there is a vicinity ${\cal %
U\subset E}^{<z>}{\cal ,}u\in {\cal U,}$ with trivializing $P$
diffeomorphisms $f$ and $\varphi :$%
$$
f_{{\cal U}}:\pi ^{-1}\left( {\cal U}\right) \rightarrow {\cal U\times }%
Gr,\qquad f\left( p\right) =\left( \pi \left( p\right) ,\varphi \left(
p\right) \right) ,
$$
$$
\varphi _{{\cal U}}:\pi ^{-1}\left( {\cal U}\right) \rightarrow Gr,\varphi
(pq)=\varphi \left( p\right) q,\quad \forall q\in Gr,~p\in P.
$$
We remark that in the general case for two open regions%
$$
{\cal U,V}\subset {\cal E}^{<z>}{\cal ,U\cap V}\neq \emptyset ,f_{{\cal U|}%
_p}\neq f_{{\cal V|}_p},\mbox{ even }p\in {\cal U\cap V.}
$$

Transition functions $g_{{\cal UV}}$ are defined as
$$
g_{{\cal UV}}:{\cal U\cap V\rightarrow }Gr,g_{{\cal UV}}\left( u\right)
=\varphi _{{\cal U}}\left( p\right) \left( \varphi _{{\cal V}}\left(
p\right) ^{-1}\right) ,\pi \left( p\right) =u.
$$

Hereafter we shall omit, for simplicity, the specification of trivializing
regions of maps and denote, for example, $f\equiv f_{{\cal U}},\varphi
\equiv \varphi _{{\cal U}},$ $s\equiv s_{{\cal U}},$ if this will not give
rise to ambiguities.

Let $\theta \,$ be the canonical left invariant 1-form on $Gr$ with values
in algebra Lie ${\cal G}$ of group $Gr$ uniquely defined from the relation $%
\theta \left( q\right) =q,\forall q\in {\cal G,}$ and consider a 1-form $%
\omega $ on ${\cal U\subset E}^{<z>}$ with values in ${\cal G.}$ Using $%
\theta $ and $\omega ,$ we can locally define the connection form $\Omega $
in $P$ as a 1-form:%
$$
\Omega =\varphi ^{*}\theta +Ad~\varphi ^{-1}\left( \pi ^{*}\omega \right)
\eqno(3.1)
$$
where $\varphi ^{*}\theta $ and $\pi ^{*}\omega $ are, respectively, forms
induced on $\pi ^{-1}\left( {\cal U}\right) $ and $P$ by maps $\varphi $ and
$\pi $ and $\omega =s^{*}\Omega .$ The adjoint action on a form $\lambda $
with values in ${\cal G}$ is defined as
$$
\left( Ad~\varphi ^{-1}\lambda \right) _p=\left( Ad~\varphi ^{-1}\left(
p\right) \right) \lambda _p
$$
where $\lambda _p$ is the value of form $\lambda $ at point $p\in P.$

Introducing a basis $\{\Delta _{\widehat{a}}\}$ in ${\cal G}$ (index $%
\widehat{a}$ enumerates the generators making up this basis), we write the
1-form $\omega $ on ${\cal E}^{<z>}$ as
$$
\omega =\Delta _{\widehat{a}}\omega ^{\widehat{a}}\left( u\right) ,~\omega ^{%
\widehat{a}}\left( u\right) =\omega _{<\mu >}^{\widehat{a}}\left( u\right)
\delta u^{<\mu >}\eqno(3.2)
$$
where $\delta u^{<\mu >}=\left( dx^i,\delta y^{<a>}\right) $ and the
Einstein summation rule on indices $\widehat{a}$ and $<\mu >$ is used.
Functions $\omega _{<\mu >}^{\widehat{a}}\left( u\right) $ from (3.2) will
be called the components of Yang-Mills fields on ha-space ${\cal E}^{<z>}%
{\cal .}$ Gauge transforms of $\omega $ can be geometrically interpreted as
transition relations for $\omega _{{\cal U}}$ and $\omega _{{\cal V}},$ when
$u\in {\cal U\cap V,}$%
$$
\left( \omega _{{\cal U}}\right) _u=\left( g_{{\cal UV}}^{*}\theta \right)
_u+Ad~g_{{\cal UV}}\left( u\right) ^{-1}\left( \omega _{{\cal V}}\right) _u.%
\eqno(3.3)
$$

To relate $\omega _{<\mu >}^{\widehat{a}}$ with a covariant derivation we
shall consider a vector bundle $\Upsilon $ associated to $P.$ Let $\rho
:Gr\rightarrow GL\left( {\cal R}^s\right) $ and $\rho ^{\prime }:{\cal G}%
\rightarrow End\left( E^s\right) $ be, respectively, linear representations
of group $Gr$ and Lie algebra ${\cal G}$ (in a more general case we can
consider ${\cal C}^s$ instead of ${\cal R}^s).$ Map $\rho $ defines a left
action on $Gr$ and associated vector bundle $\Upsilon =P\times {\cal R}%
^s/Gr,~\pi _E:E\rightarrow {\cal E}^{<z>}{\cal .}$ Introducing the standard
basis $\xi _{\underline{i}}=\{\xi _{\underline{1}},\xi _{\underline{2}%
},...,\xi _{\underline{s}}\}$ in ${\cal R}^s,$ we can define the right
action on $P\times $ ${\cal R}^s,\left( \left( p,\xi \right) q=\left(
pq,\rho \left( q^{-1}\right) \xi \right) ,q\in Gr\right) ,$ the map induced
from $P$%
$$
p:{\cal R}^s\rightarrow \pi _E^{-1}\left( u\right) ,\quad \left( p\left( \xi
\right) =\left( p\xi \right) Gr,\xi \in {\cal R}^s,\pi \left( p\right)
=u\right)
$$
and a basis of local sections $e_{\underline{i}}:U\rightarrow \pi
_E^{-1}\left( U\right) ,~e_{\underline{i}}\left( u\right) =s\left( u\right)
\xi _{\underline{i}}.$ Every section $\varsigma :{\cal E}^{<z>}{\cal %
\rightarrow }\Upsilon $ can be written locally as $\varsigma =\varsigma
^ie_i,\varsigma ^i\in C^\infty \left( {\cal U}\right) .$ To every vector
field $X$ on ${\cal E}^{<z>}$ and Yang-Mills field $\omega ^{\widehat{a}}$
on $P$ we associate operators of covariant derivations:%
$$
\nabla _X\zeta =e_{\underline{i}}\left[ X\zeta ^{\underline{i}}+B\left(
X\right) _{\underline{j}}^{\underline{i}}\zeta ^{\underline{j}}\right]
$$
$$
B\left( X\right) =\left( \rho ^{\prime }X\right) _{\widehat{a}}\omega ^{%
\widehat{a}}\left( X\right) .\eqno(3.4)
$$
Transformation laws (3.3) and operators (3.4) are interrelated by these
transition transforms for values $e_{\underline{i}},\zeta ^{\underline{i}},$
and $B_{<\mu >}:$%
$$
e_{\underline{i}}^{{\cal V}}\left( u\right) =\left[ \rho g_{{\cal UV}}\left(
u\right) \right] _{\underline{i}}^{\underline{j}}e_{\underline{i}}^{{\cal U}%
},~\zeta _{{\cal U}}^{\underline{i}}\left( u\right) =\left[ \rho g_{{\cal UV}%
}\left( u\right) \right] _{\underline{i}}^{\underline{j}}\zeta _{{\cal V}}^{%
\underline{i}},
$$
$$
B_{<\mu >}^{{\cal V}}\left( u\right) =\left[ \rho g_{{\cal UV}}\left(
u\right) \right] ^{-1}\delta _{<\mu >}\left[ \rho g_{{\cal UV}}\left(
u\right) \right] +\left[ \rho g_{{\cal UV}}\left( u\right) \right]
^{-1}B_{<\mu >}^{{\cal U}}\left( u\right) \left[ \rho g_{{\cal UV}}\left(
u\right) \right] ,\eqno(3.5)
$$
where $B_{<\mu >}^{{\cal U}}\left( u\right) =B^{<\mu >}\left( \delta
/du^{<\mu >}\right) \left( u\right) .$

Using (3.5), we can verify that the operator $\nabla _X^{{\cal U}},$ acting
on sections of $\pi _\Upsilon :\Upsilon \rightarrow {\cal E}^{<z>}$
according to definition (3.4), satisfies the properties%
$$
\begin{array}{c}
\nabla _{f_1X+f_2Y}^{
{\cal U}}=f_1\nabla _X^{{\cal U}}+f_2\nabla _X^{{\cal U}},~\nabla _X^{{\cal U%
}}\left( f\zeta \right) =f\nabla _X^{{\cal U}}\zeta +\left( Xf\right) \zeta
, \\ \nabla _X^{{\cal U}}\zeta =\nabla _X^{{\cal V}}\zeta ,\quad u\in {\cal %
U\cap V,}f_1,f_2\in C^\infty \left( {\cal U}\right) .
\end{array}
$$

So, we can conclude that the Yang--Mills connection
 in the vector bundle $%
\pi _\Upsilon :\Upsilon \rightarrow {\cal E}^{<z>}$ is not a general one,
but is induced from the principal bundle $\pi :P\rightarrow {\cal E}^{<z>}$
with structural group $Gr.$

The curvature ${\cal K}$ of connection $\Omega $ from (3.1) is defined as%
$$
{\cal K}=D\Omega ,~D=\widehat{H}\circ d\eqno(3.6)
$$
where $d$ is the operator of exterior derivation acting on ${\cal G}$-valued
forms as\\ $d\left( \Delta _{\widehat{a}}\otimes \chi ^{\widehat{a}}\right)
=\Delta _{\widehat{a}}\otimes d\chi ^{\widehat{a}}$ and $\widehat{H}\,$ is
the horizontal projecting operator actin, for example, on the 1-form $%
\lambda $ as $\left( \widehat{H}\lambda \right) _P\left( X_p\right) =\lambda
_p\left( H_pX_p\right) ,$ where $H_p$ projects on the horizontal subspace $%
{\cal H}_p\in P_p\left[ X_p\in {\cal H}_p\mbox{ is equivalent to }\Omega
_p\left( X_p\right) =0\right] .$ We can express (3.6) locally as
$$
{\cal K}=Ad~\varphi _{{\cal U}}^{-1}\left( \pi ^{*}{\cal K}_{{\cal U}%
}\right) \eqno(3.7)
$$
where
$$
{\cal K}_{{\cal U}}=d\omega _{{\cal U}}+\frac 12\left[ \omega _{{\cal U}%
},\omega _{{\cal U}}\right] .\eqno(3.8)
$$
The exterior product of ${\cal G}$-valued form (3.8) is defined as
$$
\left[ \Delta _{\widehat{a}}\otimes \lambda ^{\widehat{a}},\Delta _{\widehat{%
b}}\otimes \xi ^{\widehat{b}}\right] =\left[ \Delta _{\widehat{a}},\Delta _{%
\widehat{b}}\right] \otimes \lambda ^{\widehat{a}}\bigwedge \xi ^{\widehat{b}%
},
$$
where the antisymmetric tensorial product is%
$$
\lambda ^{\widehat{a}}\bigwedge \xi ^{\widehat{b}}=\lambda ^{\widehat{a}}\xi
^{\widehat{b}}-\xi ^{\widehat{b}}\lambda ^{\widehat{a}}.
$$

Introducing structural coefficients $f_{\widehat{b}\widehat{c}}^{\quad
\widehat{a}}$ of ${\cal G}$ satisfying
$$
\left[ \Delta _{\widehat{b}},\Delta _{\widehat{c}}\right] =f_{\widehat{b}%
\widehat{c}}^{\quad \widehat{a}}\Delta _{\widehat{a}}
$$
we can rewrite (3.8) in a form more convenient for local considerations:%
$$
{\cal K}_{{\cal U}}=\Delta _{\widehat{a}}\otimes {\cal K}_{<\mu ><\nu >}^{%
\widehat{a}}\delta u^{<\mu >}\bigwedge \delta u^{<\nu >}\eqno(3.9)
$$
where%
$$
{\cal K}_{<\mu ><\nu >}^{\widehat{a}}=\frac{\delta \omega _{<\nu >}^{%
\widehat{a}}}{\partial u^{<\mu >}}-\frac{\delta \omega _{<\mu >}^{\widehat{a}%
}}{\partial u^{<\nu >}}+\frac 12f_{\widehat{b}\widehat{c}}^{\quad \widehat{a}%
}\left( \omega _{<\mu >}^{\widehat{b}}\omega _{<\nu >}^{\widehat{c}}-\omega
_{<\nu >}^{\widehat{b}}\omega _{<\mu >}^{\widehat{c}}\right) .
$$

This section ends by considering the problem of reduction of the local
an\-i\-sot\-rop\-ic gauge symmetries and gauge fields to isotropic ones. For
local trivial considerations we can consider that the vanishing of
dependencies on $y$ variables leads to isotropic Yang-Mills fields with the
same gauge group as in the anisotropic case, Global geometric constructions
require a more rigorous topological study of possible obstacles for
reduction of total spaces and structural groups on anisotropic bases to
their analogous on isotropic (for example, pseudo-Riemannian) base spaces.

\subsection{Yang-Mills Equations on Ha-spaces}

Interior gauge (nongravitational) symmetries are associated to semisimple
structural groups. On the principal bundle $\left( P,\pi ,Gr,{\cal E}%
^{<z>}\right) $ with nondegenerate Killing form for semisimple group $Gr$ we
can define the generalized Lagrange metric%
$$
h_p\left( X_p,Y_p\right) =G_{\pi \left( p\right) }\left( d\pi _PX_P,d\pi
_PY_P\right) +K\left( \Omega _P\left( X_P\right) ,\Omega _P\left( X_P\right)
\right) ,\eqno(3.10)
$$
where $d\pi _P$ is the differential of map $\pi :P\rightarrow {\cal E}^{<z>}%
{\cal ,}$ $G_{\pi \left( p\right) }$ is locally generated as the ha-metric
(2.12), and $K$ is the Killing form on ${\cal G:}$%
$$
K\left( \Delta _{\widehat{a}},\Delta _{\widehat{b}}\right) =f_{\widehat{b}%
\widehat{d}}^{\quad \widehat{c}}f_{\widehat{a}\widehat{c}}^{\quad \widehat{d}%
}=K_{\widehat{a}\widehat{b}}.
$$

Using the metric $G_{<\alpha ><\beta >}$ on ${\cal E}^{<z>}$ $\left[
h_P\left( X_P,Y_P\right) \mbox{ on }P\right] ,$ we can introduce operators $%
*_G$ and $\widehat{\delta }_G$ acting in the space of forms on ${\cal E}%
^{<z>}$ ($*_H$ and $\widehat{\delta }_H$ acting on forms on ${\cal E}^{<z>}%
{\cal ).}$ Let $e_{<\mu >}$ be orthonormalized frames on ${\cal U\subset E}%
^{<z>}$ and $e^{<\mu >}$ the adjoint coframes. Locally%
$$
G=\sum\limits_{<\mu >}\eta \left( <\mu >\right) e^{<\mu >}\otimes e^{<\mu
>},
$$
where $\eta _{<\mu ><\mu >}=\eta \left( <\mu >\right) =\pm 1,$ $<\mu
>=1,2,...,n_E,n_E=1,...,n+m_1+...+m_z,$ and the Hodge operator $*_G$ can be
defined as $*_G:\Lambda ^{\prime }\left( {\cal E}^{<z>}\right) \rightarrow
\Lambda ^{n+m_1+...+m_z}\left( {\cal E}^{<z>}\right) ,$ or, in explicit
form, as%
$$
*_G\left( e^{<\mu _1>}\bigwedge ...\bigwedge e^{<\mu _r>}\right) =\eta
\left( \nu _1\right) ...\eta \left( \nu _{n_E-r}\right) \times \eqno(3.11)
$$
$$
sign\left(
\begin{array}{ccccc}
1 & 2 & ...r & r+1 & ...n_E \\
<\mu _{1>} & <\mu _2> & ...<\mu _r> & <\nu _1> & ...\nu _{n_E-r}
\end{array}
\right) \times
$$
$$
e^{<\nu _1>}\bigwedge ...\bigwedge e^{<\nu _{n_E-r}>}.
$$
Next, define the operator%
$$
*_{G^{}}^{-1}=\eta \left( 1\right) ...\eta \left( n_E\right) \left(
-1\right) ^{r\left( n_E-r\right) }*_G
$$
and introduce the scalar product on forms $\beta _1,\beta _2,...\subset
\Lambda ^r\left( {\cal E}^{<z>}\right) $ with compact carrier:%
$$
\left( \beta _1,\beta _2\right) =\eta \left( 1\right) ...\eta \left(
n_E\right) \int \beta _1\bigwedge *_G\beta _2.
$$
The operator $\widehat{\delta }_G$ is defined as the adjoint to $d$
associated to the scalar product for forms, specified for $r$-forms as
$$
\widehat{\delta }_G=\left( -1\right) ^r*_{G^{}}^{-1}\circ d\circ *_G.%
\eqno(3.12)
$$

We remark that operators $*_H$ and $\delta _H$ acting in the total space of $%
P$ can be defined similarly to (3.11) and (3.12), but by using metric
(3.10). Both these operators also act in the space of ${\cal G}$-valued
forms:%
$$
*\left( \Delta _{\widehat{a}}\otimes \varphi ^{\widehat{a}}\right) =\Delta _{%
\widehat{a}}\otimes (*\varphi ^{\widehat{a}}),
$$
$$
\widehat{\delta }\left( \Delta _{\widehat{a}}\otimes \varphi ^{\widehat{a}%
}\right) =\Delta _{\widehat{a}}\otimes (\widehat{\delta }\varphi ^{\widehat{a%
}}).
$$

The form $\lambda $ on $P$ with values in ${\cal G}$ is called horizontal if
$\widehat{H}\lambda =\lambda $ and equivariant if $R^{*}\left( q\right)
\lambda =Ad~q^{-1}\varphi ,~\forall g\in Gr,R\left( q\right) $ being the
right shift on $P.$ We can verify that equivariant and horizontal forms also
satisfy the conditions%
$$
\lambda =Ad~\varphi _{{\cal U}}^{-1}\left( \pi ^{*}\lambda \right) ,\qquad
\lambda _{{\cal U}}=S_{{\cal U}}^{*}\lambda ,
$$
$$
\left( \lambda _{{\cal V}}\right) _{{\cal U}}=Ad~\left( g_{{\cal UV}}\left(
u\right) \right) ^{-1}\left( \lambda _{{\cal U}}\right) _u.
$$

Now, we can define the field equations for curvature (3.7) and connection
(3.1) :%
$$
\Delta {\cal K}=0,\eqno(3.13)
$$
$$
\nabla {\cal K}=0,\eqno(3.14)
$$
where $\Delta =\widehat{H}\circ \widehat{\delta }_H.$ Equations (3.13) are
similar to the well-known Maxwell equations and for non-Abelian gauge fields
are called Yang-Mills equations. The structural equations (3.14) are called
Bianchi identities.

The field equations (3.13) do not have a physical meaning because they are
written in the total space of bundle $\Upsilon $ and not on the base
anisotropic space-time ${\cal E}^{<z>}.$ But this dificulty may be
obviated by projecting the mentioned equations on the base. The 1-form $%
\Delta {\cal K}$ is horizontal by definition and its equivariance follows
from the right invariance of metric (3.10). So, there is a unique form $%
(\Delta {\cal K})_{{\cal U}}$ satisfying
$$
\Delta {\cal K=}Ad~\varphi _{{\cal U}}^{-1}\pi ^{*}(\Delta {\cal K})_{{\cal U%
}}.
$$
Projection of (3.13) on the base can be written as $(\Delta {\cal K})_{{\cal %
U}}=0.$ To calculate $(\Delta {\cal K})_{{\cal U}},$ we use the equality
\cite{bis,pd}%
$$
d\left( Ad~\varphi _{{\cal U}}^{-1}\lambda \right) =Ad~~\varphi _{{\cal U}%
}^{-1}~d\lambda -\left[ \varphi _{{\cal U}}^{*}\theta ,Ad~\varphi _{{\cal U}%
}^{-1}\lambda \right]
$$
where $\lambda $ is a form on $P$ with values in ${\cal G.}$ For r-forms we
have
$$
\widehat{\delta }\left( Ad~\varphi _{{\cal U}}^{-1}\lambda \right)
=Ad~\varphi _{{\cal U}}^{-1}\widehat{\delta }\lambda -\left( -1\right)
^r*_H\{\left[ \varphi _{{\cal U}}^{*}\theta ,*_HAd~\varphi _{{\cal U}%
}^{-1}\lambda \right]
$$
and, as a consequence,%
$$
\widehat{\delta }{\cal K}=Ad~\varphi _{{\cal U}}^{-1}\{\widehat{\delta }%
_H\pi ^{*}{\cal K}_{{\cal U}}+*_H^{-1}[\pi ^{*}\omega _{{\cal U}},*_H\pi ^{*}%
{\cal K}_{{\cal U}}]\}-
$$
$$
-*_H^{-1}\left[ \Omega ,Ad~\varphi _{{\cal U}}^{-1}*_H\left( \pi ^{*}{\cal K}%
\right) \right] .\eqno(3.15)
$$
By using straightforward calculations in the adapted dual basis on $\pi
^{-1}\left( {\cal U}\right) $ we can verify the equalities%
$$
\left[ \Omega ,Ad~\varphi _{{\cal U}}^{-1}~*_H\left( \pi ^{*}{\cal K}_{{\cal %
U}}\right) \right] =0,\widehat{H}\delta _H\left( \pi ^{*}{\cal K}_{{\cal U}%
}\right) =\pi ^{*}\left( \widehat{\delta }_G{\cal K}\right) ,
$$
$$
*_H^{-1}\left[ \pi ^{*}\omega _{{\cal U}},*_H\left( \pi ^{*}{\cal K}_{{\cal U%
}}\right) \right] =\pi ^{*}\{*_G^{-1}\left[ \omega _{{\cal U}},*_G{\cal K}_{%
{\cal U}}\right] \}.\eqno(3.16)
$$
From (3.15) and (3.16) it follows that
$$
\left( \Delta {\cal K}\right) _{{\cal U}}=\widehat{\delta }_G{\cal K}_{{\cal %
U}}+*_G^{-1}\left[ \omega _{{\cal U}},*_G{\cal K}_{{\cal U}}\right] .%
\eqno(3.17)
$$

Taking into account (3.17) and (3.12), we prove that projection on ${\cal E}$
of equations (3.13) and (3.14) can be expressed respectively as
$$
*_G^{-1}\circ d\circ *_G{\cal K}_{{\cal U}}+*_G^{-1}\left[ \omega _{{\cal U}%
},*_G{\cal K}_{{\cal U}}\right] =0.\eqno(3.18)
$$
$$
d{\cal K}_{{\cal U}}+\left[ \omega _{{\cal U}},{\cal K}_{{\cal U}}\right] =0.%
$$

Equations (3.18) (see (3.17)) are gauge-invariant because%
$$
\left( \Delta {\cal K}\right) _{{\cal U}}=Ad~g_{{\cal UV}}^{-1}\left( \Delta
{\cal K}\right) _{{\cal V}}.
$$

By using formulas (3.9)-(3.12) we can rewrite (3.18) in coordinate form%
$$
D_{<\nu >}\left( G^{<\nu ><\lambda >}{\cal K}_{~<\lambda ><\mu >}^{\widehat{a%
}}\right) +f_{\widehat{b}\widehat{c}}^{\quad \widehat{a}}G^{<v><\lambda
>}\omega _{<\lambda >}^{~\widehat{b}}{\cal K}_{~<\nu ><\mu >}^{\widehat{c}%
}=0,\eqno(3.19)
$$
where $D_{<\nu >}$ is, for simplicity, a compatible with metric covariant
derivation on ha-space ${\cal E}^{<z>}.$

We point out that for our bundles with semisimple structural groups the
Yang-Mills equations (3.13) (and, as a consequence, their horizontal
projections (3.18) or (3.19)) can be obtained by variation of the action%
$$
I=\int {\cal K}_{~<\mu ><\nu >}^{\widehat{a}}{\cal K}_{~<\alpha ><\beta >}^{%
\widehat{b}}G^{<\mu ><\alpha >}G^{<\nu ><\beta >}K_{\widehat{a}\widehat{b}%
}\times \eqno(3.20)
$$
$$
\left| G_{<\alpha ><\beta >}\right| ^{1/2}dx^1...dx^n\delta
y_{(1)}^1...\delta y_{(1)}^{m_1}...\delta y_{(z)}^1...\delta y_{(z)}^{m_z}.
$$
Equations for extremals of (3.20) have the form
$$
K_{\widehat{r}\widehat{b}}G^{<\lambda ><\alpha >}G^{<\kappa ><\beta
>}D_{<\alpha >}{\cal K}_{~<\lambda ><\beta >}^{\widehat{b}}-
$$
$$
K_{\widehat{a}\widehat{b}}G^{<\kappa ><\alpha >}G^{<\nu ><\beta >}f_{%
\widehat{r}\widehat{l}}^{\quad \widehat{a}}\omega _{<\nu >}^{\widehat{l}}%
{\cal K}_{~<\alpha ><\beta >}^{\widehat{b}}=0,
$$
which are equivalent to ''pure'' geometric equations (3.19) (or (3.18) due
to nondegeneration of the Killing form $K_{\widehat{r}\widehat{b}}$ for
semisimple groups.

To take into account gauge interactions with matter fields (section of
vector bundle $\Upsilon $ on ${\cal E}$ ) we have to introduce a source
1--form ${\cal J}$ in equations (3.13) and to write them as%
$$
\Delta {\cal K}={\cal J}\eqno(3.21)
$$

Explicit constructions of ${\cal J}$ require concrete definitions of the
bundle $\Upsilon ;$ for example, for spinor fields an invariant formulation
of the Dirac equations on la-spaces is necessary. We omit spinor
considerations in this section (see Sec. II.G), but we shall present the
definition of the source ${\cal J}$ for gravitational interactions (by using
the energy-momentum tensor of matter on la--space) in the next subsection.

\subsection{Gauge Higher Order Anisotropic Gravity}

A considerable body of work on the formulation of gauge gravitational models
on isotropic spaces is based on using nonsemisimple groups, for example,
Poincare and affine groups, as structural gauge groups (see critical
analysis and original results in \cite{wal,ts,lue80,pon}. The main
impediment to developing such models is caused by the degeneration of
Killing forms for nonsemisimple groups, which make it impossible to
construct consistent variational gauge field theories (functional (3.20) and
extremal equations are degenarate in these cases). There are at least two
possibilities to get around the mentioned difficulty.\ The first is to
realize a minimal extension of the nonsemisimple group to a semisimple one,
similar to the extension of the Poincare group to the de Sitter group
considered in \cite{p,pd,ts} (in the next subsection we shall use this
operation for the definition of locally anisotropic gravitational
instantons). The second possibility is to introduce into consideration the
bundle of adapted affine frames on la-space ${\cal E}^{<z>},$ to use an
auxiliary nondegenerate bilinear form $a_{\widehat{a}\widehat{b}}$ instead
of the degenerate Killing form $K_{\widehat{a}\widehat{b}}$ and to consider
a ''pure'' geometric method, illustrated in the previous subsection, of
defining gauge field equations. Projecting on the base ${\cal E}^{<z>},$ we
shall obtain gauge gravitational field equations on ha-space having a form
similar to Yang-Mills equations.

The goal of this subsection is to prove that a specific parametrization of
components of the Cartan connection in the bundle of adapted affine frames
on ${\cal E}^{<z>}$ establishes an equivalence between Yang-Mills equations
(3.21) and Einstein equations on ha-spaces.

\subsubsection{Bundles of linear ha--frames}

Let $\left( X_{<\alpha >}\right) _u=\left( X_i,X_{<a>}\right) _u=\left(
X_i,X_{a_1},...,X_{a_z}\right) _u$ be an adapted frame (see (3.4) at point $%
u\in {\cal E}^{<z>}.$ We consider a local right distinguished action of
matrices%
$$
A_{<\alpha ^{\prime }>}^{\quad <\alpha >}=\left(
\begin{array}{cccc}
A_{i^{\prime }}^{\quad i} & 0 & ... & 0 \\
0 & B_{a_1^{\prime }}^{\quad a_1} & ... & 0 \\
... & ... & ... & ... \\
0 & 0 & ... & B_{a_z^{\prime }}^{\quad a_z}
\end{array}
\right) \subset GL_{n_E}=
$$
$$
GL\left( n,{\cal R}\right) \oplus GL\left( m_1,{\cal R}\right) \oplus
...\oplus GL\left( m_z,{\cal R}\right) .
$$
Nondegenerate matrices $A_{i^{\prime }}^{\quad i}$ and $B_{j^{\prime
}}^{\quad j}$ respectively transforms linearly $X_{i|u}$ into $X_{i^{\prime
}|u}=A_{i^{\prime }}^{\quad i}X_{i|u}$ and $X_{a_p^{\prime }|u}$ into $%
X_{a_p^{\prime }|u}=B_{a_p^{\prime }}^{\quad a_p}X_{a_p|u},$ where $%
X_{<\alpha ^{\prime }>|u}=A_{<\alpha ^{\prime }>}^{\quad <\alpha
>}X_{<\alpha >}$ is also an adapted frame at the same point $u\in {\cal E}%
^{<z>}.$ We denote by $La\left( {\cal E}^{<z>}\right) $ the set of all
adapted frames $X_{<\alpha >}$ at all points of ${\cal E}^{<z>}$ and
consider the surjective map $\pi $ from $La\left( {\cal E}^{<z>}\right) $ to
${\cal E}^{<z>}$ transforming every adapted frame $X_{\alpha |u}$ and point $%
u$ into point $u.$ Every $X_{<\alpha ^{\prime }>|u}$ has a unique
representation as $X_{<\alpha ^{\prime }>}=A_{<\alpha ^{\prime }>}^{\quad
<\alpha >}X_{<\alpha >}^{\left( 0\right) },$ where $X_{<\alpha >}^{\left(
0\right) }$ is a fixed distinguished basis in tangent space $T\left( {\cal E}%
^{<z>}\right) .$ It is obvious that $\pi ^{-1}\left( {\cal U}\right) ,{\cal U%
}\subset {\cal E}^{<z>},$ is bijective to ${\cal U}\times GL_{n_E}\left(
{\cal R}\right) .$ We can transform $La\left( {\cal E}^{<z>}\right) $ in a
differentiable manifold taking $\left( u^{<\beta >},A_{<\alpha ^{\prime
}>}^{\quad <\alpha >}\right) $ as a local coordinate system on $\pi
^{-1}\left( {\cal U}\right) .$ Now, it is easy to verify that ${\cal {L}}a(%
{\cal E}^{<z>})=(La({\cal E}^{<z>},{\cal E}^{<z>},GL_{n_E}({\cal R})))$ is a
principal bundle. We call ${\cal {L}}a({\cal E}^{<z>})$ the bundle of linear
adapted frames on ${\cal E}^{<z>}.$

The next step is to identify the components of, for simplicity, compatible
d-connection $\Gamma _{<\beta ><\gamma >}^{<\alpha >}$ on ${\cal E}^{<z>}:$%
$$
\Omega _{{\cal U}}^{\widehat{a}}=\omega ^{\widehat{a}}=\{\omega _{\quad
<\lambda >}^{\widehat{\alpha }\widehat{\beta }}\doteq \Gamma _{<\beta
><\gamma >}^{<\alpha >}\}.\eqno(3.22)
$$
Introducing (3.22) in (3.17), we calculate the local 1-form%
$$
\left( \Delta {\cal R}^{(\Gamma )}\right) _{{\cal U}}=\Delta _{\widehat{%
\alpha }\widehat{\alpha }_1}\otimes (G^{<\nu ><\lambda >}D_{<\lambda >}{\cal %
R}_{\qquad <\nu ><\mu >}^{<\widehat{\alpha }><\widehat{\gamma }>}+
$$
$$
f_{\qquad <\widehat{\beta }><\widehat{\delta }><\widehat{\gamma }><\widehat{%
\varepsilon }>}^{<\widehat{\alpha }><\widehat{\gamma }>}G^{<\nu ><\lambda
>}\omega _{\qquad <\lambda >}^{<\widehat{\beta }><\widehat{\delta }>}{\cal R}%
_{\qquad <\nu ><\mu >}^{<\widehat{\gamma }><\widehat{\varepsilon }>})\delta
u^{<\mu >},\eqno(3.23)
$$
where
$$
\Delta _{\widehat{\alpha }\widehat{\beta }}=\left(
\begin{array}{cccc}
\Delta _{\widehat{i}\widehat{j}} & 0 & ... & 0 \\
0 & \Delta _{\widehat{a}_1\widehat{b}_1} & ... & 0 \\
... & ... & ... & ... \\
0 & 0 & ... & \Delta _{\widehat{a}_z\widehat{b}_z}
\end{array}
\right)
$$
is the standard distinguished basis in Lie algebra of matrices ${{\cal {G}}l}%
_{n_E}\left( {\cal R}\right) $ with $\left( \Delta _{\widehat{i}\widehat{k}%
}\right) _{jl}=\delta _{ij}\delta _{kl}$ and $\left( \Delta _{\widehat{a}_p%
\widehat{c}_p}\right) _{b_pd_p}=\delta _{a_pb_p}\delta _{c_pd_p}$ be\-ing
res\-pec\-ti\-ve\-ly the stand\-ard bas\-es in ${\cal {G}}l\left( {\cal R}%
^{n_E}\right) .$ We have denoted the curvature of connection (3.22),
considered in (3.23), as
$$
{\cal R}_{{\cal U}}^{(\Gamma )}=\Delta _{\widehat{\alpha }\widehat{\alpha }%
_1}\otimes {\cal R}_{\qquad <\nu ><\mu >}^{\widehat{\alpha }\widehat{\alpha }%
_1}X^{<\nu >}\bigwedge X^{<\mu >},
$$
where ${\cal R}_{\qquad <\nu ><\mu >}^{\widehat{\alpha }\widehat{\alpha }%
_1}=R_{<\alpha _1>\quad <\nu ><\mu >}^{\quad <\alpha >}$ (see curvatures
(2.28)).

\subsubsection{Bundles of affine ha--frames and Einstein equations}

Besides ${\cal {L}}a\left( {\cal E}^{<z>}\right) $ with ha-space ${\cal E}%
^{<z>},$ another bundle is naturally related, the bundle of adapted affine
frames with structural group $Af_{n_E}\left( {\cal R}\right) =GL_{n_E}\left(
{\cal E}^{<z>}\right)$ $\otimes {\cal R}^{n_E}.$ Because as linear space the
Lie Algebra $af_{n_E}\left( {\cal R}\right) $ is a direct sum of ${{\cal {G}}%
l}_{n_E}\left( {\cal R}\right) $ and ${\cal R}^{n_E},$ we can write forms on
${\cal {A}}a\left( {\cal E}^{<z>}\right) $ as $\Theta =\left( \Theta
_1,\Theta _2\right) ,$ where $\Theta _1$ is the ${{\cal {G}}l}_{n_E}\left(
{\cal R}\right) $ component and $\Theta _2$ is the ${\cal R}^{n_E}$
component of the form $\Theta .$ Connection (3.22), $\Omega $ in ${{\cal {L}}%
a}\left( {\cal E}^{<z>}\right) ,$ induces the Cartan connection $\overline{%
\Omega }$ in ${{\cal {A}}a}\left( {\cal E}^{<z>}\right) ;$ see the isotropic
case in \cite{p,pd,bis}. This is the unique connection on ${{\cal {A}}a}%
\left( {\cal E}^{<z>}\right) $ represented as $i^{*}\overline{\Omega }%
=\left( \Omega ,\chi \right) ,$ where $\chi $ is the shifting form and $i:{%
{\cal {A}}a}\rightarrow {{\cal {L}}a}$ is the trivial reduction of bundles.
If $s_{{\cal U}}^{(a)}$ is a local adapted frame in ${{\cal {L}}a}\left(
{\cal E}^{<z>}\right) ,$ then $\overline{s}_{{\cal U}}^{\left( 0\right)
}=i\circ s_{{\cal U}}$ is a local section in ${{\cal {A}}a}\left( {\cal E}%
^{<z>}\right) $ and
$$
\left( \overline{\Omega }_{{\cal U}}\right) =s_{{\cal U}}\Omega =\left(
\Omega _{{\cal U}},\chi _{{\cal U}}\right) ,\eqno(3.24)
$$
$$
\left( \overline{{\cal R}}_{{\cal U}}\right) =s_{{\cal U}}\overline{{\cal R}}%
=\left( {\cal R}_{{\cal U}}^{(\Gamma )},T_{{\cal U}}\right) ,
$$
where $\chi =e_{\widehat{\alpha }}\otimes \chi _{\quad <\mu >}^{\widehat{%
\alpha }}X^{<\mu >},G_{<\alpha ><\beta >}=\chi _{\quad <\alpha >}^{\widehat{%
\alpha }}\chi _{\quad <\beta >}^{\widehat{\beta }}\eta _{\widehat{\alpha }%
\widehat{\beta }}\quad (\eta _{\widehat{\alpha }\widehat{\beta }}$ is
diagonal with $\eta _{\widehat{\alpha }\widehat{\alpha }}=\pm 1)$ is a frame
decomposition of metric (2.12) on ${\cal E}^{<z>},e_{\widehat{\alpha }}$ is
the standard distinguished basis on ${\cal R}^{n_E},$ and the projection of
torsion , $T_{{\cal U}},$ on base ${\cal E}^{<z>}$ is defined as
$$
T_{{\cal U}}=d\chi _{{\cal U}}+\Omega _{{\cal U}}\bigwedge \chi _{{\cal U}%
}+\chi _{{\cal U}}\bigwedge \Omega _{{\cal U}}=\eqno(3.25)
$$
$$
e_{\widehat{\alpha }}\otimes \sum\limits_{<\mu ><<\nu >}T_{\quad <\mu ><\nu
>}^{\widehat{\alpha }}X^{<\mu >}\bigwedge X^{<\nu >}.
$$
For a fixed local adapted basis on ${\cal U}\subset {\cal E}^{<z>}$ we can
identify components $T_{\quad <\mu ><\nu >}^{\widehat{a}}$ of torsion (3.25)
with components of torsion (2.25) on ${\cal E}^{<z>},$ i.e. $T_{\quad <\mu
><\nu >}^{\widehat{\alpha }}=T_{\quad <\mu ><\nu >}^{<\alpha >}.$ By
straightforward calculation we obtain
$$
{(\Delta \overline{{\cal R}})}_{{\cal U}}=[{(\Delta {\cal R}^{(\Gamma )})}_{%
{\cal U}},\ {(R\tau )}_{{\cal U}}+{(Ri)}_{{\cal U}}],\eqno(3.26)
$$
where%
$$
\left( R\tau \right) _{{\cal U}}=\widehat{\delta }_GT_{{\cal U}%
}+*_G^{-1}\left[ \Omega _{{\cal U}},*_GT_{{\cal U}}\right] ,\quad \left(
Ri\right) _{{\cal U}}=*_G^{-1}\left[ \chi _{{\cal U}},*_G{\cal R}_{{\cal U}%
}^{(\Gamma )}\right] .
$$
Form $\left( Ri\right) _{{\cal U}}$ from (3.26) is locally constructed by
using components of the Ricci tensor (see (2.29)) as follows from
decomposition on the local adapted basis $X^{<\mu >}=\delta u^{<\mu >}:$
$$
\left( Ri\right) _{{\cal U}}=e_{\widehat{\alpha }}\otimes \left( -1\right)
^{n_E+1}R_{<\lambda ><\nu >}G^{\widehat{\alpha }<\lambda >}\delta u^{<\mu >}%
$$

We remark that for isotropic torsionless pseudo-Riemannian spaces the
requirement that $\left( \Delta \overline{{\cal R}}\right) _{{\cal U}}=0,$
i.e., imposing the connection (3.22) to satisfy Yang-Mills equations (3.13)
(equivalently (3.18) or (3.19) we obtain \cite{p,pd,ald} the equivalence of
the mentioned gauge gravitational equations with the vacuum Einstein
equations $R_{ij}=0.\,$ In the case of ha--spaces with arbitrary given
torsion, even considering vacuum gravitational fields, we have to introduce
a source for gauge gravitational equations in order to compensate for the
contribution of torsion and to obtain equivalence with the Einstein
equations.

Considerations presented in Secs. III.C.1 and III.C.2 constitute the proof of
the following result

\begin{theorem}
The Einstein equations (2.34) for la-gravity are equivalent to Yang-Mills
equations%
$$
\left( \Delta \overline{{\cal R}}\right) =\overline{{\cal J}}\eqno(3.27)
$$
for the induced Cartan connection $\overline{\Omega }$ (see (3.22), (3.24))
in the bundle of local adapted affine frames ${\cal A}a\left( {\cal E}%
\right) $ with source $\overline{{\cal J}}_{{\cal U}}$ constructed locally
by using the same formulas (3.26) (for $\left( \Delta \overline{{\cal R}}%
\right) $), where $R_{<\alpha ><\beta >}$ is changed by the matter source ${%
\tilde E}_{<\alpha ><\beta >}-\frac 12G_{<\alpha ><\beta >}{\tilde E},$
where ${\tilde E}_{<\alpha ><\beta >}=kE_{<\alpha ><\beta >}-\lambda
G_{<\alpha ><\beta >}.$
\end{theorem}

We note that this theorem is an extension for higher order anisotropic
spaces of the Popov and Dikhin results \cite{pd} with respect to a possible
gauge like treatment of the Einstein gravity. Similar theorems have been
proved for locally anisotropic gauge gravity \cite{v295a,v295b,vg} and in
the framework of some variants of locally (and higher order) anisotropic
supergravity \cite{vlasg,v96jpa3}.

\subsection{Nonlinear De Sitter Gauge Ha--Gravity}

The equivalent reexpression of the Einstein theory as a gauge like theory
implies, for both locally isotropic and anisotropic space-times, the
nonsemisimplicity of the gauge group, which leads to a nonvariational theory
in the total space of the bundle of locally adapted affine frames. A
variational gauge gravitational theory can be formulated by using a minimal
extension of the affine structural group ${{\cal A}f}_{n_E}\left( {\cal R}%
\right) $ to the de Sitter gauge group $S_{n_E}=SO\left( n_E\right) $ acting
on distinguished ${\cal R}^{n_E+1}$ space.

\subsubsection{Nonlinear gauge theories of de Sitter group}

Let us consider the de Sitter space $\Sigma ^{n_E}$ as a hypersurface given
by the equations $\eta _{AB}u^Au^B=-l^2$ in the (n+m)-dimensional spaces
enabled with diagonal metric $\eta _{AB},\eta _{AA}=\pm 1$ (in this 
subsection $%
A,B,C,...=1,2,...,n_E+1),(n_E=n+m_1+...+m_z),$ where $\{u^A\}$ are global      
Cartesian coordinates in ${\cal R}^{n_E+1};l>0$ is the curvature of de
Sitter space. The de Sitter group $S_{\left( \eta \right) }=SO_{\left( \eta
\right) }\left( n_E+1\right) $ is defined as the isometry group of $\Sigma
^{n_E}$-space with $\frac{n_E}2\left( n_E+1\right) $ generators of Lie
algebra ${{\it s}o}_{\left( \eta \right) }\left( n_E+1\right) $ satisfying
the commutation relations%
$$
\left[ M_{<AB},M_{CD}\right] =\eta _{AC}M_{BD}-\eta _{BC}M_{AD}-\eta
_{AD}M_{BC}+\eta _{BD}M_{AC}.\eqno(3.28)
$$

Decomposing indices $A,B,...$ as $A=\left( \widehat{\alpha },n_E+1\right)
,B=\left( \widehat{\beta },n_E+1\right) ,$ \\ $...,$ the metric $\eta _{AB}$
as $\eta _{AB}=\left( \eta _{\widehat{\alpha }\widehat{\beta }},\eta
_{\left( n_E+1\right) \left( n_E+1\right) }\right) ,$ and operators $M_{AB}$
as $M_{\widehat{\alpha }\widehat{\beta }}={\cal F}_{\widehat{\alpha }%
\widehat{\beta }}$ and $P_{\widehat{\alpha }}=l^{-1}M_{n_E+1,\widehat{\alpha
}},$ we can write (3.28) as
$$
\left[ {\cal F}_{\widehat{\alpha }\widehat{\beta }},{\cal F}_{\widehat{%
\gamma }\widehat{\delta }}\right] =\eta _{\widehat{\alpha }\widehat{\gamma }}%
{\cal F}_{\widehat{\beta }\widehat{\delta }}-\eta _{\widehat{\beta }\widehat{%
\gamma }}{\cal F}_{\widehat{\alpha }\widehat{\delta }}+\eta _{\widehat{\beta
}\widehat{\delta }}{\cal F}_{\widehat{\alpha }\widehat{\gamma }}-\eta _{%
\widehat{\alpha }\widehat{\delta }}{\cal F}_{\widehat{\beta }\widehat{\gamma
}},
$$
$$
\left[ P_{\widehat{\alpha }},P_{\widehat{\beta }}\right] =-l^{-2}{\cal F}_{%
\widehat{\alpha }\widehat{\beta }},\quad \left[ P_{\widehat{\alpha }},{\cal F%
}_{\widehat{\beta }\widehat{\gamma }}\right] =\eta _{\widehat{\alpha }%
\widehat{\beta }}P_{\widehat{\gamma }}-\eta _{\widehat{\alpha }\widehat{%
\gamma }}P_{\widehat{\beta }},
$$
where we have indicated the possibility to decompose ${{\it s}o}_{\left(
\eta \right) }\left( n_E+1\right) $ into a direct sum, ${{\it s}o}_{\left(
\eta \right) }\left( n_E+1\right) ={{\it s}o}_{\left( \eta \right)
}(n_E)\oplus V_{n_E},$ where $V_{n_E}$ is the vector space stretched on
vectors $P_{\widehat{\alpha }}.$ We remark that $\Sigma ^{n_E}=S_{\left(
\eta \right) }/L_{\left( \eta \right) },$ where $L_{\left( \eta \right)
}=SO_{\left( \eta \right) }\left( n_E\right) .$ For $\eta _{AB}=diag\left(
1,-1,-1,-1\right) $ and $S_{10}=SO\left( 1,4\right) ,L_6=SO\left( 1,3\right)
$ is the group of Lorentz rotations.

Let $W\left( {\cal E},{\cal R}^{n_E+1},S_{\left( \eta \right) },P\right) $
be the vector bundle associated with principal bundle $P\left( S_{\left(
\eta \right) },{\cal E}\right) $ on la-spaces. The action of the structural
group $S_{\left( \eta \right) }$ on $E\,$ can be realized by using $\left(
n_E\right) \times \left( n_E\right) $ matrices with a parametrization
distinguishing subgroup $L_{\left( \eta \right) }:$%
$$
B=bB_L,\eqno(3.29)
$$
where%
$$
B_L=\left(
\begin{array}{cc}
L & 0 \\
0 & 1
\end{array}
\right) ,
$$
$L\in L_{\left( \eta \right) }$ is the de Sitter bust matrix transforming
the vector $\left( 0,0,...,\rho \right) \in {\cal R}^{n_E+1}$ into the
arbitrary point $\left( V^1,V^2,...,V^{n_E+1}\right) \in \Sigma _\rho
^{n_E}\subset {\cal R}^{n_E+1}$ with curvature $\rho \quad \left(
V_AV^A=-\rho ^2,V^A=t^A\rho \right) .$ Matrix $b$ can be expressed as
$$
b=\left(
\begin{array}{cc}
\delta _{\quad \widehat{\beta }}^{\widehat{\alpha }}+\frac{t^{\widehat{%
\alpha }}t_{\widehat{\beta }}}{\left( 1+t^{n_E+1}\right) } & t^{
\widehat{\alpha }} \\ t_{\widehat{\beta }} & t^{n_E+1}
\end{array}
\right) .
$$

The de Sitter gauge field is associated with a linear connection in $W$,
i.e., with a ${{\it s}o}_{\left( \eta \right) }\left( n_E+1\right) $-valued
connection 1-form on ${\cal E}^{<z>}:$

$$
\widetilde{\Omega }=\left(
\begin{array}{cc}
\omega _{\quad \widehat{\beta }}^{\widehat{\alpha }} & \widetilde{\theta }^{%
\widehat{\alpha }} \\ \widetilde{\theta }_{\widehat{\beta }} & 0
\end{array}
\right) ,\eqno(3.30)
$$
where $\omega _{\quad \widehat{\beta }}^{\widehat{\alpha }}\in
so(n_E)_{\left( \eta \right) },$ $\widetilde{\theta }^{\widehat{\alpha }}\in
{\cal R}^{n_E},\widetilde{\theta }_{\widehat{\beta }}\in \eta _{\widehat{%
\beta }\widehat{\alpha }}\widetilde{\theta }^{\widehat{\alpha }}.$

Because $S_{\left( \eta \right) }$-transforms mix $\omega _{\quad \widehat{%
\beta }}^{\widehat{\alpha }}$ and $\widetilde{\theta }^{\widehat{\alpha }}$
fields in (3.30) (the introduced para\-met\-ri\-za\-ti\-on is invariant on
action on $SO_{\left( \eta \right) }\left( n_E\right) $ group we cannot
identify $\omega _{\quad \widehat{\beta }}^{\widehat{\alpha }}$ and $%
\widetilde{\theta }^{\widehat{\alpha }},$ respectively, with the connection $%
\Gamma _{~<\beta ><\gamma >}^{<\alpha >}$ and the fundamental form $\chi
^{<\alpha >}$ in ${\cal E}^{<z>}$ (as we have for (3.22) and (3.24)). To
avoid this difficulty we consider \cite{ts,pon} a nonlinear gauge
realization of the de Sitter group $S_{\left( \eta \right) },$ namely, we
introduce into consideration the nonlinear gauge field%
$$
\Omega =b^{-1}\Omega b+b^{-1}db=\left(
\begin{array}{cc}
\Gamma _{~\widehat{\beta }}^{\widehat{\alpha }} & \theta ^{
\widehat{\alpha }} \\ \theta _{\widehat{\beta }} & 0
\end{array}
\right) ,\eqno(3.31)
$$
where
$$
\Gamma _{\quad \widehat{\beta }}^{\widehat{\alpha }}=\omega _{\quad \widehat{%
\beta }}^{\widehat{\alpha }}-\left( t^{\widehat{\alpha }}Dt_{\widehat{\beta }%
}-t_{\widehat{\beta }}Dt^{\widehat{\alpha }}\right) /\left(
1+t^{n_E+1}\right) ,
$$
$$
\theta ^{\widehat{\alpha }}=t^{n_E+1}\widetilde{\theta }^{\widehat{\alpha }%
}+Dt^{\widehat{\alpha }}-t^{\widehat{\alpha }}\left( dt^{n_e+1}+\widetilde{%
\theta }_{\widehat{\gamma }}t^{\widehat{\gamma }}\right) /\left(
1+t^{n_E+1}\right) ,
$$
$$
Dt^{\widehat{\alpha }}=dt^{\widehat{\alpha }}+\omega _{\quad \widehat{\beta }%
}^{\widehat{\alpha }}t^{\widehat{\beta }}.
$$

The action of the group $S\left( \eta \right) $ is nonlinear, yielding
transforms\\ $\Gamma ^{\prime }=L^{\prime }\Gamma \left( L^{\prime }\right)
^{-1}+L^{\prime }d\left( L^{\prime }\right) ^{-1},\theta ^{\prime }=L\theta
, $ where the nonlinear matrix-valued function\\ $L^{\prime }=L^{\prime
}\left( t^{<\alpha >},b,B_T\right) $ is defined from $B_b=b^{\prime
}B_{L^{\prime }}$ (see parametrization (3.29)).

Now, we can identify components of (3.31) with components of $\Gamma
_{~<\beta ><\gamma >}^{<\alpha >}$ and $\chi _{\quad <\alpha >}^{\widehat{%
\alpha }}$ on ${\cal E}^{<z>}$ and induce in a consistent manner on the base
of bundle\\ $W\left( {\cal E},{\cal R}^{n_E+1},S_{\left( \eta \right)
},P\right) $ the la-geometry.

\subsubsection{Dynamics of the nonlinear $S\left( \eta \right) $ ha--gravity}

Instead of the gravitational potential (3.22), we introduce the
gravitational connection (similar to (3.31))
$$
\Gamma =\left(
\begin{array}{cc}
\Gamma _{\quad \widehat{\beta }}^{\widehat{\alpha }} & l_0^{-1}\chi ^{
\widehat{\alpha }} \\ l_0^{-1}\chi _{\widehat{\beta }} & 0
\end{array}
\right) \eqno(3.32)
$$
where
$$
\Gamma _{\quad \widehat{\beta }}^{\widehat{\alpha }}=\Gamma _{\quad \widehat{%
\beta }<\mu >}^{\widehat{\alpha }}\delta u^{<\mu >},
$$
$$
\Gamma _{\quad \widehat{\beta }<\mu >}^{\widehat{\alpha }}=\chi _{\quad
<\alpha >}^{\widehat{\alpha }}\chi _{\quad <\beta >}^{\widehat{\beta }%
}\Gamma _{\quad <\beta ><\gamma >}^{<\alpha >}+\chi _{\quad <\alpha >}^{%
\widehat{\alpha }}\delta _{<\mu >}\chi _{\quad \widehat{\beta }}^{<\alpha
>},
$$
$\chi ^{\widehat{\alpha }}=\chi _{\quad \mu }^{\widehat{\alpha }}\delta
u^\mu ,$ and $G_{\alpha \beta }=\chi _{\quad \alpha }^{\widehat{\alpha }%
}\chi _{\quad \beta }^{\widehat{\beta }}\eta _{\widehat{\alpha }\widehat{%
\beta }},$ and $\eta _{\widehat{\alpha }\widehat{\beta }}$ is parametrized
as
$$
\eta _{\widehat{\alpha }\widehat{\beta }}=\left(
\begin{array}{cccc}
\eta _{ij} & 0 & ... & 0 \\
0 & \eta _{a_1b_1} & ... & 0 \\
... & ... & ... & ... \\
0 & 0 & ... & \eta _{a_zb_z}
\end{array}
\right) ,
$$
$\eta _{ij}=\left( 1,-1,...,-1\right) ,...\eta _{ij}=\left( \pm 1,\pm
1,...,\pm 1\right) ,...,l_0$ is a dimensional constant.

The curvature of (3.32), ${\cal R}^{(\Gamma )}=d\Gamma +\Gamma \bigwedge
\Gamma ,$ can be written as%
$$
{\cal R}^{(\Gamma )}=\left(
\begin{array}{cc}
{\cal R}_{\quad \widehat{\beta }}^{\widehat{\alpha }}+l_0^{-1}\pi _{\widehat{%
\beta }}^{\widehat{\alpha }} & l_0^{-1}T^{
\widehat{\alpha }} \\ l_0^{-1}T^{\widehat{\beta }} & 0
\end{array}
\right) ,\eqno(3.33)
$$
where
$$
\pi _{\widehat{\beta }}^{\widehat{\alpha }}=\chi ^{\widehat{\alpha }%
}\bigwedge \chi _{\widehat{\beta }},{\cal R}_{\quad \widehat{\beta }}^{%
\widehat{\alpha }}=\frac 12{\cal R}_{\quad \widehat{\beta }<\mu ><\nu >}^{%
\widehat{\alpha }}\delta u^{<\mu >}\bigwedge \delta u^{<\nu >},
$$
and
$$
{\cal R}_{\quad \widehat{\beta }<\mu ><\nu >}^{\widehat{\alpha }}=\chi _{%
\widehat{\beta }}^{\quad <\beta >}\chi _{<\alpha >}^{\quad \widehat{\alpha }%
}R_{\quad <\beta ><\mu ><\nu >}^{<\alpha >}
$$
(see (2.27) and (2.28), the components of d-curvatures). The de Sitter gauge
group is semisimple and we are able to construct a variational gauge
gravitational locally anisotropic theory (bundle metric (3.10) is
nondegenerate). The Lagrangian of the theory is postulated as
$$
L=L_{\left( G\right) }+L_{\left( m\right) }
$$
where the gauge gravitational Lagrangian is defined as
$$
L_{\left( G\right) }=\frac 1{4\pi }Tr\left( {\cal R}^{(\Gamma )}\bigwedge *_G%
{\cal R}^{(\Gamma )}\right) ={\cal L}_{\left( G\right) }\left| G\right|
^{1/2}\delta ^{n_E}u,
$$
$$
{\cal L}_{\left( G\right) }=\frac 1{2l^2}T_{\quad <\mu ><\nu >}^{\widehat{%
\alpha }}T_{\widehat{\alpha }}^{\quad <\mu ><\nu >}+\eqno(3.34)
$$
$$
\frac 1{8\lambda }{\cal R}_{\quad \widehat{\beta }<\mu ><\nu >}^{\widehat{%
\alpha }}{\cal R}_{\quad \widehat{\alpha }}^{\widehat{\beta }\quad <\mu
><\nu >}-\frac 1{l^2}\left( {\overleftarrow{R}}\left( \Gamma \right)
-2\lambda _1\right) ,
$$
$T_{\quad <\mu ><\nu >}^{\widehat{\alpha }}=\chi _{\quad <\alpha >}^{%
\widehat{\alpha }}T_{\quad <\mu ><\nu >}^{<\alpha >}$ (the gravitational
constant $l^2$ in (3.34) satisfies the relations $l^2=2l_0^2\lambda ,\lambda
_1=-3/l_0],\quad Tr$ denotes the trace on $\widehat{\alpha },\widehat{\beta }
$ indices, and the matter field Lagrangian is defined as%
$$
L_{\left( m\right) }=-1\frac 12Tr\left( \Gamma \bigwedge *_G{\cal I}\right) =%
{\cal L}_{\left( m\right) }\left| G\right| ^{1/2}\delta ^{n_E}u,
$$
$$
{\cal L}_{\left( m\right) }=\frac 12\Gamma _{\quad \widehat{\beta }<\mu >}^{%
\widehat{\alpha }}S_{\quad <\alpha >}^{\widehat{\beta }\quad <\mu
>}-t_{\quad \widehat{\alpha }}^{<\mu >}l_{\quad <\mu >}^{\widehat{\alpha }}.%
\eqno(3.35)
$$
The matter field source ${\cal I}$ is obtained as a variational derivation
of ${\cal L}_{\left( m\right) }$ on $\Gamma $ and is parametrized as
$$
{\cal I}=\left(
\begin{array}{cc}
S_{\quad \widehat{\beta }}^{\widehat{\alpha }} & -l_0t^{
\widehat{\alpha }} \\ -l_0t_{\widehat{\beta }} & 0
\end{array}
\right) \eqno(3.36)
$$
with $t^{\widehat{\alpha }}=t_{\quad <\mu >}^{\widehat{\alpha }}\delta
u^{<\mu >}$ and $S_{\quad \widehat{\beta }}^{\widehat{\alpha }}=S_{\quad
\widehat{\beta }<\mu >}^{\widehat{\alpha }}\delta u^{<\mu >}$ being
respectively the canonical tensors of energy-momentum and spin density.
Because of the contraction of the ''interior'' indices $\widehat{\alpha },%
\widehat{\beta }$ in (3.34) and (3.35) we used the Hodge operator $*_G$
instead of $*_H$ (hereafter we consider $*_G=*).$

Varying the action
$$
S=\int \left| G\right| ^{1/2}\delta ^{n_E}u\left( {\cal L}_{\left( G\right)
}+{\cal L}_{\left( m\right) }\right)
$$
on the $\Gamma $-variables (3.26), we obtain the gauge-gravitational field
equations:%
$$
d\left( *{\cal R}^{(\Gamma )}\right) +\Gamma \bigwedge \left( *{\cal R}%
^{(\Gamma )}\right) -\left( *{\cal R}^{(\Gamma )}\right) \bigwedge \Gamma
=-\lambda \left( *{\cal I}\right) .\eqno(3.37)
$$

Specifying the variations on $\Gamma _{\quad \widehat{\beta }}^{\widehat{%
\alpha }}$ and $l^{\widehat{\alpha }}$-variables, we rewrite (3.37) as
$$
\widehat{{\cal D}}\left( *{\cal R}^{(\Gamma )}\right) +\frac{2\lambda }{l^2}%
\left( \widehat{{\cal D}}\left( *\pi \right) +\chi \bigwedge \left(
*T^T\right) -\left( *T\right) \bigwedge \chi ^T\right) =-\lambda \left(
*S\right) ,\eqno(3.38)
$$
$$
\widehat{{\cal D}}\left( *T\right) -\left( *{\cal R}^{(\Gamma )}\right)
\bigwedge \chi -\frac{2\lambda }{l^2}\left( *\pi \right) \bigwedge \chi =%
\frac{l^2}2\left( *t+\frac 1\lambda *\tau \right) ,\eqno(3.39)
$$
where
$$
T^t=\{T_{\widehat{\alpha }}=\eta _{\widehat{\alpha }\widehat{\beta }}T^{%
\widehat{\beta }},~T^{\widehat{\beta }}=\frac 12T_{\quad <\mu ><\nu >}^{%
\widehat{\beta }}\delta u^{<\mu >}\bigwedge \delta u^{<\nu >}\},
$$
$$
\chi ^T=\{\chi _{\widehat{\alpha }}=\eta _{\widehat{\alpha }\widehat{\beta }%
}\chi ^{\widehat{\beta }},~\chi ^{\widehat{\beta }}=\chi _{\quad <\mu >}^{%
\widehat{\beta }}\delta u^{<\mu >}\},\qquad \widehat{{\cal D}}=d+\widehat{%
\Gamma }
$$
($\widehat{\Gamma }$ acts as $\Gamma _{\quad \widehat{\beta }<\mu >}^{%
\widehat{\alpha }}$ on indices $\widehat{\gamma },\widehat{\delta },...$ and
as $\Gamma _{\quad <\beta ><\mu >}^{<\alpha >}$ on indices $<\gamma
>,<\delta >,...).$ In (3.39), $\tau $ defines the energy-momentum tensor of
the $S_{\left( \eta \right) }$-gauge gravitational field $\widehat{\Gamma }:$%
$$
\tau _{<\mu ><\nu >}\left( \widehat{\Gamma }\right) =\frac 12Tr\left( {\cal R%
}_{<\mu ><\alpha >}{\cal R}_{\quad <\nu >}^{<\alpha >}-\frac 14{\cal R}%
_{<\alpha ><\beta >}{\cal R}^{<\alpha ><\beta >}G_{<\mu ><\nu >}\right) .%
\eqno(3.40)
$$

Equations (3.37) (or equivalently (3.38),(3.39)) make up the complete system
of variational field equations for nonlinear de Sitter gauge gravity with
higher order  anisotropy. They can be interpreted as a generalization of
gauge like  equations for la-gravity \cite{vg} (equivalently, of gauge
gravitational equations (3.27)] to a system of gauge field equations with
dynamical torsion and corresponding spin-density source.

A. Tseytlin \cite{ts} presented a quantum analysis of the isotropic version
of equations (3.38) and (3.39). Of course, the problem of quantizing
gravitational interactions is unsolved for both variants of locally
anisotropic and isotropic gauge de Sitter gravitational theories, but we
think that the generalized Lagrange version of $S_{\left( \eta \right) }$%
-gravity is more adequate for studying quantum radiational and statistical
gravitational processes. This is a matter for further investigations.

Finally, we remark that we can obtain a nonvariational Poincare gauge
gravitational theory on la-spaces if we consider the contraction of the
gauge potential (3.32) to a potential with values in the Poincare Lie
algebra
$$
\Gamma =\left(
\begin{array}{cc}
\Gamma _{\quad \widehat{\beta }}^{\widehat{\alpha }} & l_0^{-1}\chi ^{
\widehat{\alpha }} \\ l_0^{-1}\chi _{\widehat{\beta }} & 0
\end{array}
\right) \rightarrow \Gamma =\left(
\begin{array}{cc}
\Gamma _{\quad \widehat{\beta }}^{\widehat{\alpha }} & l_0^{-1}\chi ^{
\widehat{\alpha }} \\ 0 & 0
\end{array}
\right) .
$$
Isotropic Poincare gauge gravitational theories are studied in a number of
papers (see, for example, \cite{wal,ts,lue80,pon}). In a manner similar to
considerations presented in this work, we can generalize Poincare gauge
models for spaces with local anisotropy.

\subsection{La--Gravitational Gauge Instantons}

The existence of self-dual, or instanton, topologically nontrivial solutions
of Yang-Mills equations is a very important physical consequence of gauge
theories. All known instanton-type Yang-Mills and gauge gravitational
solutions (see, for example, \cite{ts,pon}) are locally isotropic. A
variational gauge-gravitational extension of la-gravity makes possible a
straightforward application of techniques of constructing solutions for
first order gauge equations for the definition of locally anisotropic
gravitational instantons. This subsection is devoted to the study of some
particular instanton solutions of the gauge gravitational theory on la-space.

Let us consider the Euclidean formulation of the $S_{\left( \eta \right) }$%
-gauge gravitational theory by changing gauge structural groups and flat
metric:%
$$
SO_{(\eta )}(n_E+1)\rightarrow SO(n_E+1),SO_{(\eta )}(n_E)\rightarrow
SO(n_E),\eta _{AB}\rightarrow -\delta _{AB}.
$$
Self-dual (anti-self-dual) conditions for the curvature (3.33)
$$
{\cal R}^{(\Gamma )}=*{\cal R}^{(\Gamma )}\quad (-*{\cal R}^{(\Gamma )})%
$$
can be written as a system of equations
$$
\left( {\cal R}_{\quad \widehat{\beta }}^{\widehat{\alpha }}-l_0^{-2}\pi
_{\quad \widehat{\beta }}^{\widehat{\alpha }}\right) =\pm *\left( {\cal R}%
_{\quad \widehat{\beta }}^{\widehat{\alpha }}-l_0^{-2}\pi _{\quad \widehat{%
\beta }}^{\widehat{\alpha }}\right) \eqno(3.41)
$$
$$
T^{\widehat{\alpha }}=\pm *T^{\widehat{\alpha }}\eqno(3.42)
$$
(the ''-'' refers to the anti-self-dual case), where the ''-'' before $%
l_0^{-2}$ appears because of the transition of the Euclidean negatively
defined metric $-\delta _{<\alpha ><\beta >},$ which leads to $\chi _{\quad
<\alpha >}^{\widehat{\alpha }}\rightarrow i\chi _{\quad <\alpha >|(E)}^{%
\widehat{\alpha }},\pi \rightarrow -\pi _{(E)}$ (we shall omit the index $(E)
$ for Euclidean values).

For solutions of (3.41) and (3.42) the energy-momentum tensor (3.40) is
identically equal to zero. Vacuum equations (3.37) and (3.38), when source
is ${\cal I\equiv 0}$ (see (3.36)), are satisfied as a consequence of
generalized Bianchi identities for the curvature (3.33). The mentioned
solutions of (3.41) and (3.42) realize a local minimum of the Euclidean
action%
$$
S=\frac 1{8\lambda }\int \left| G^{1/2}\right| \delta ^{n_E}\{(R\left(
\Gamma \right) -l_0^{-2}\pi )^2+2T^2\},
$$
where $T^2=T_{\quad <\mu ><\nu >}^{\widehat{\alpha }}T_{\widehat{\alpha }%
}^{\quad <\mu ><\nu >}$ is extremal on the topological invariant (Pontryagin
index)%
$$
p_2=-\frac 1{8\pi ^2}\int Tr\left( {\cal R}^{(\Gamma )}\bigwedge {\cal R}%
^{(\Gamma )}\right) =--\frac 1{8\pi ^2}\int Tr\left( \widehat{{\cal R}}%
\bigwedge \widehat{{\cal R}}\right) .
$$

For the Euclidean de Sitter spaces, when%
$$
{\cal R}=0\quad \{T=0,R_{\quad <\mu ><\nu >}^{\widehat{\alpha }\widehat{%
\beta }}=-\frac 2{l_0^2}\chi _{\quad <\mu >}^{[\alpha }\chi _{\quad <\nu
>}^{\beta ]}\}\eqno(3.43)
$$
we obtain the absolute minimum, $S=0.$

We emphasize that for $R_{<\beta >\quad <\mu ><\beta >}^{\quad <\alpha
>}=\left( 2/l_0^2\right) \delta _{[<\mu >}^{<\alpha >}G_{<\nu >]<\beta >}$
torsion vanishes. Torsionless instantons also have another interpretation.\
For $T_{\quad <\beta ><\gamma >}^{<\alpha >}=0$ contraction of equations
(3.41) leads to Einstein equations with cosmological $\lambda $-term (as a
consequence of generalized Ricci identities):%
$$
R_{<\alpha ><\beta ><\mu ><\nu >}-R_{<\mu ><\nu ><\alpha ><\beta >}=\frac
32\{R_{[<\alpha ><\beta ><\mu >]<\nu >}-
$$
$$
R_{[<\alpha ><\beta ><\nu >]<\mu >}+R_{[<\mu ><\nu ><\alpha >]<\beta
>}-R_{[<\mu ><\nu ><\beta >]<\alpha >}\}.
$$
So, in the Euclidean case the locally anisotropic vacuum Einstein equations
are a subset of instanton solutions.

Now, let us study the $SO\left( n_E\right) $ solution of equations (3.41)
and\ (3.42). We consider the spherically symmetric ansatz (in order to point
out the connection between high-dimensional gravity and la-gravity the
N-connection structure is chosen to be trivial, i.e. ${N}_j^a\left( u\right)
\equiv 0):$%
$$
\Gamma _{\quad \widehat{\beta }<\mu >}^{\widehat{\alpha }}=a\left( u\right)
\left( u^{\widehat{\alpha }}\delta _{\widehat{\beta }<\mu >}-u_{\widehat{%
\beta }}\delta _{<\mu >}^{\widehat{\alpha }}\right) +q\left( u\right)
\epsilon _{\quad \widehat{\beta }<\mu ><\nu >}^{\widehat{\alpha }}u^{<\mu
>},
$$
$$
\chi _{<\alpha >}^{\widehat{\alpha }}=f\left( u\right) \delta _{\quad
<\alpha >}^{\widehat{\alpha }}+n\left( u\right) u^{\widehat{\alpha }%
}u_\alpha ,\eqno(3.44)
$$
where $u=u^{<\alpha >}u^{<\beta >}G_{<\alpha ><\beta >}=x^{\widehat{i}}x_{%
\widehat{i}}+y^{\widehat{a}}y_{\widehat{a}},$ and $a\left( u\right) ,q\left(
u\right) ,f\left( u\right) $ and $n\left( u\right) $ are some scalar
functions. Introducing (3.44) into (3.41) and (3.42), we obtain,
respectively,%
$$
u\left( \pm \frac{dq}{du}-a^2-q^2\right) +2\left( a\pm q\right) +l_0^{-1}f^2,%
\eqno(3.45)
$$
$$
2d\left( a\mp q\right) /du+\left( a\mp q\right) ^2-l_0^{-1}fn=0,\eqno(3.46)
$$
$$
2\frac{df}{du}+f\left( a\mp 2q\right) +n\left( au-1\right) =0.\eqno(3.47)
$$

The traceless part of the torsion vanishes because of the parametrization
(3.44), but in the general case the trace and pseudo-trace of the torsion
are not identical to zero:%
$$
T^\mu =q^{\left( 0\right) }u^\mu \left( -2df/du+n-a\left( f+un\right)
\right) ,
$$
$$
\overbrace{T}^\mu =q^{\left( 1\right) }u^\mu \left( 2qf\right) ,
$$
$q^{\left( 0\right) }$ and $q^{\left( 0\right) }$ are constant. Equation
(3.42) or (3.47) establishes the proportionality of $T^\mu $ and $\overbrace{%
T}^\mu .$ As a consequence we obtain that the $SO\left( n+m\right) $
solution of (3.52) is torsionless if $q\left( u\right) =0$ of $f\left(
u\right) =0.$

Let first analyze the torsionless instantons, $T_{\quad <\alpha ><\beta
>}^{<\mu >}=0.$ If $f=0,$ then from (3.46) one has two possibilities: (a) $%
n=0$ leads to nonsense because $\chi _\alpha ^{\widehat{\alpha }}=0$ or $%
G_{\alpha \beta }=0.$ b) $a=u^{-1}$ and $n\left( u\right) $ is an arbitrary
scalar function; we have from (3.46) $a\mp q=2/\left( a+C^2\right) $ or $%
q=\pm 2/u\left( u+C^2\right) ,$ where $C=const.$ If $q\left( u\right) =0,$
we obtain the de Sitter space (3.43) because equations (3.45) and (3.46)
impose vanishing of both self-dual and anti-self-dual parts of $\left( {\cal %
R}_{\quad \widehat{\beta }}^{\widehat{\alpha }}-l_0^2\pi _{\quad \widehat{%
\beta }}^{\widehat{\alpha }}\right) ,$ so, as a consequence, ${\cal R}%
_{\quad \widehat{\beta }}^{\widehat{\alpha }}-l_0^2\pi _{\quad \widehat{%
\beta }}^{\widehat{\alpha }}\equiv 0.$ There is an infinite number of $%
SO\left( n_E\right) $-symmetric solutions of (3.43):%
$$
f=l_0\left[ a\left( 2-au\right) \right] ^{1/2},\quad n=l_0\{2\frac{da}{du}+%
\frac{a^2}{\left[ a\left( 2-au\right) \right] ^{1/2}}\},
$$
$a(u)$ is a scalar function.

To find instantons with torsion, $T_{\quad \beta \gamma }^\alpha \neq 0,$ is
also possible. We present the $SO\left( 4\right) $ one-instanton solution,
obtained in \cite{pon} (which in the case of $H^4$-space parametrized by
local coordinates $\left( x^{\widehat{1}},x^{\widehat{2}},y^{\widehat{1}},y^{%
\widehat{2}}\right) ,$ with $u=x^{\widehat{1}}x_{\widehat{1}}+x^{\widehat{2}%
}x_{\widehat{2}}+y^{\widehat{1}}y_{\widehat{1}}+y^{\widehat{2}}y_{\widehat{2}%
}):$%
$$
a=a_0\left( u+c^2\right) ^{-1},q=\mp q_0\left( u+c^2\right) ^{-1}
$$
$$
f=l_0\left( \alpha u+\beta \right) ^{1/2}/\left( u+c^2\right) ,n=c_0/\left(
u+c^2\right) \left( \gamma u+\delta \right) ^{1/2}
$$
where%
$$
a_0=-1/18,q_0=5/6,\alpha =266/81,\beta =8/9,\gamma =10773/11858,\delta
=1458/5929.
$$
We suggest that local regions with $T_{\ \beta \gamma }^\alpha \neq 0$ are
similarly to Abrikosv vortexes in superconductivity and the appearance of
torsion is a possible realization of the Meisner effect in gravity (for
details and discussions on the superconducting or Higgs-like interpretation
of gravity see \cite{ts,pon}). In our case we obtain a locally anisotropic
superconductivity and we think that the formalism of gauge locally
anisotropic theories may even work for some models of anisotropic low- and
high-temperature superconductivity \cite{v91}.

\section{Na-Maps and Conservation Laws}

The aim of this section is to develop a necessary geometric background (the
theory of nearly autoparallel maps, in brief na-maps,  and tensor integral
formalism on la-multispaces) for formulation and a detailed investigation of
conservation laws on locally isotropic and anisotropic curved spaces. We
shall summarize our results on formulation of na-maps for generalized affine
spaces (GAM-spaces) \cite{vk,vth,vob12}, Einstein-Cartan and Einstein spaces
\cite{vrjp,vod,voarm}, bundle spaces \cite{vb12,vb13,vog} and different
classes of la-spaces \cite{vodg,voa,gv,vcl96} and present an extension of
the na-map theory for superspaces. For simplicity we shall restrict our
considerations only with the "first" order anisotropy.

The question of definition of tensor integration as the inverse operation of
covariant derivation was posed and studied by A.Mo\'or \cite{moo}.
Tensor--integral and bitensor formalisms turned out to be very useful in
solving certain problems connected with conservation laws in general
relativity \cite{goz,vb13}.  In order to extend tensor--integral
constructions we have proposed \cite{vob12,vog} to  take into consideration
nearly autoparallel \cite{voarm,vk,vb13,vb12} and nearly geodesic \cite{sin}
maps, ng--maps, which forms a subclass of local 1--1 maps of  curved spaces
with deformation of the connection and metric structures. A generalization
of the Sinyukov's ng--theory for spaces with local anisotropy was proposed
by considering maps with deformation of connection for Lagrange spaces (on
Lagrange spaces see \cite{ker,ma87,ma94}) and generalized Lagrange spaces
\cite{vcl96,vodg,voa,vod,gvd}. Tensor integration formalism for  generalized
Lagrange spaces was developed in \cite{v295a,gv,vcl96}. One of the main
purposes of this section is to synthesize the results obtained in the
mentioned  works and to formulate them for a very general class of
la--spaces. As the next step the la--gravity and analysis of
la--conservation laws are considered.

We note that proofs of our theorems are mechanical, but, in most cases, they
are rather tedious calculations similar to those presented in \cite
{sin,vrjp,voarm,vcl96}. Some of them, on la-spaces, will be given in detail the
rest, being similar, or consequences, will be only sketched or omitted.

\subsection{Nearly Autoparallel Maps of La--Spaces}

The aim of the subsection is to present a generalization of
 the ng-- \cite{sin} and  na--map \cite
{vk,vb12,vrjp,vob12,vob13,vog} theory by introducing into consideration maps
of vector bundles provided with compatible N--connection, d--connection and
metric structures.

Our geometric arena consists from pairs of open regions $( U, {\underline U})
$ of la--spaces, $U{\subset}{\xi},\, {\underline U}{\subset}{\underline {\xi}%
}$, and 1--1 local maps $f : U{\to}{\underline U}$ given by functions $%
f^{a}(u)$ of smoothly class $C^r( U) \, (r>2, $ or $r={\omega}$~ for
analytic functions) and their inverse functions $f^{\underline a}({%
\underline u})$ with corresponding non--zero Jacobians in every point $u{\in}
U$ and ${\underline u}{\in}{\underline U}.$

We consider that two open regions $U$~ and ${\underline U}$~ are attributed
to a common for f--map coordinate system if this map is realized on the
principle of coordinate equality $q(u^{\alpha}) {\to} {\underline q}
(u^{\alpha})$~ for every point $q {\in} U$~ and its f--image ${\underline q}
{\in} {\underline U}.$ We note that all calculations included in this work
will be local in nature and taken to refer to open subsets of mappings of
type $\frac{{\xi} {\supset} U \buildrel f }{\longrightarrow {\underline U} {%
\subset} {\underline {\xi}}.}$ For simplicity, we suppose that in a fixed
common coordinate system for $U$ and ${\underline U}$ spaces $\xi$ and ${%
\underline {\xi}}$ are characterized by a common N--connection structure (in
consequence of (2.10) by a corresponding concordance of d--metric structure),
i.e.
$$
N^{a}_{j}(u)={\underline N}^{a}_{j}(u)={\underline N}^{a}_{j} ({\underline u}%
),%
$$
which leads to the possibility to establish common local bases, adapted to a
given N--connection, on both regions $U$ and ${\underline U.}$ We consider
that on $\xi$ it is defined the linear d--connection structure with
components ${\Gamma}^{\alpha}_{{.}{\beta}{\gamma}}.$ On the space $%
\underline {\xi}$ the linear d--connection is considered to be a  general
one with torsion
$$
{\underline T}^{\alpha}_{{.}{\beta}{\gamma}}={\underline {\Gamma}}^{\alpha}_
{{.}{\beta}{\gamma}}-{\underline {\Gamma}}^{\alpha}_{{.}{\gamma}{\beta}}+
w^{\alpha}_{{.}{\beta}{\gamma}}
$$
and nonmetricity
$$
{\underline K}_{{\alpha}{\beta}{\gamma}}={{\underline D}_{\alpha}} {%
\underline G}_{{\beta}{\gamma}}. \eqno(4.1)
$$

Geometrical objects on ${\underline {\xi}}$ are specified by underlined
symbols (for example, ${\underline A}^{\alpha}, {\underline B}^{{\alpha}{%
\beta}})$~ or underlined indices (for example, $A^{\underline a}, B^{{%
\underline a}{\underline b}}).$

For our purposes it is convenient to introduce auxiliary sym\-met\-ric
d--con\-nec\-ti\-ons, ${\gamma}^{\alpha}_{{.}{\beta}{\gamma}}={\gamma}%
^{\alpha}_{{.}{\gamma}{\beta}} $~ on $\xi$ and ${\underline {\gamma}}%
^{\alpha}_{.{\beta}{\gamma}}= {\underline {\gamma}}^{\alpha}_{{.}{\gamma}{%
\beta}}$ on ${\underline {\xi}}$ defined, correspondingly, as
$$
{\Gamma}^{\alpha}_{{.}{\beta}{\gamma}}= {\gamma}^{\alpha}_{{.}{\beta}{\gamma}%
}+  T^{\alpha}_{{.}{\beta}{\gamma}}\quad {\rm \; and}\quad {\underline
{\Gamma}}^{\alpha}_{{.}{\beta}{\gamma}}= {\underline {\gamma}}^{\alpha}_{{.}{%
\beta}{\gamma}}+ {\underline T}^{\alpha}_{{.}{\beta}{\gamma}}.%
$$

We are interested in definition of local 1--1 maps from $U$ to ${\underline U%
}$ characterized by symmetric, $P^{\alpha}_{{.}{\beta}{\gamma}},$ and
antisymmetric, $Q^{\alpha}_{{.}{\beta}{\gamma}}$,~ deformations:
$$
{\underline {\gamma}}^{\alpha}_{{.}{\beta}{\gamma}} ={\gamma}^{\alpha}_{{.}{%
\beta}{\gamma}}+ P^{\alpha}_{{.}{\beta}{\gamma}} \eqno(4.2)%
$$
and
$$
{\underline T}^{\alpha}_{{.}{\beta}{\gamma}}= T^{\alpha}_{{.}{\beta}{\gamma}%
}+ Q^{\alpha}_{{.}{\beta}{\gamma}}. \eqno(4.3)
$$
The auxiliary linear covariant derivations induced by ${\gamma}^{\alpha}_{{.}%
{\beta}{\gamma}}$ and ${\underline {\gamma}}^{\alpha}_{{.}{\beta}{\gamma}}$~
are denoted respectively as $^{({\gamma})}D$~ and $^{({\gamma})}{\underline D%
}.$~

Let introduce this local coordinate parametrization of curves on $U$~:
$$
u^{\alpha}=u^{\alpha}({\eta})=(x^{i}({\eta}), y^{i}({\eta}),~{\eta}_{1}<{\eta%
}<{\eta}_{2},%
$$
where corresponding tangent vector field is defined as
$$
v^{\alpha}={\frac{{du^{\alpha}} }{d{\eta}}}= ({\frac{{dx^{i}({\eta})} }{{d{%
\eta}}}}, {\frac{{dy^{j}({\eta})} }{d{\eta}}}).%
$$

\begin{definition}
Curve $l$~ is called auto parallel, a--parallel, on $\xi $ if its tangent
vector field $v^\alpha $~ satisfies a--parallel equations:
$$
vDv^\alpha =v^\beta {^{({\gamma })}D}_\beta v^\alpha ={\rho }({\eta }%
)v^\alpha ,\eqno(4.4)
$$
where ${\rho }({\eta })$~ is a scalar function on $\xi $.
\end{definition}

Let curve ${\underline l} {\subset} {\underline {\xi}}$ is given in
parametric form as $u^{\alpha}=u^{\alpha}({\eta}),~{\eta}_1 < {\eta} <{\eta}%
_2$ with tangent vector field $v^{\alpha} = {\frac{{du^{\alpha}} }{{d{\eta}}}%
} {\ne} 0.$ We suppose that a 2--dimensional distribution $E_2({\underline l}%
)$ is defined along ${\underline l} ,$ i.e. in every point $u {\in} {%
\underline l}$ is fixed a 2-dimensional vector space $E_{2}({\underline l}) {%
\subset} {\underline {\xi}}.$ The introduced distribution $E_{2}({\underline
l})$~ is coplanar along ${\underline l}$~ if every vector ${\underline p}%
^{\alpha}(u^{b}_{(0)}) {\subset} E_{2}({\underline l}), u^{\beta}_{(0)} {%
\subset} {\underline l}$~ rests contained in the same distribution after
parallel transports along ${\underline l},$~ i.e. ${\underline p}%
^{\alpha}(u^{\beta}({\eta})) {\subset} E_{2} ({\underline l}).$

\begin{definition}
A curve ${\underline{l}}$~ is called nearly autoparallel, or in brief an
na--parallel, on space ${\underline{\xi }}$~ if a coplanar along ${%
\underline{l}}$~ distribution $E_2({\underline{l}})$ containing tangent to ${%
\underline{l}}$~ vector field $v^\alpha ({\eta })$,~ i.e. $v^\alpha ({\eta })%
{\subset }E_2({\underline{l}}),$~ is defined.
\end{definition}
We can define nearly autoparallel maps of la--spaces as an anisotropic
generalization (see also \cite{vodg,voa} ng--\cite{sin}  and na--maps \cite
{vk,vob12,vog,vob13,vb13}:

\begin{definition}
Nearly autoparallel maps, na--maps, of la--spaces are defined as local 1--1
mappings of v--bundles, $\xi {\to }{\underline{\xi }},$ changing every
a--parallel on $\xi $ into a na--parallel on ${\underline{\xi }}.$
\end{definition}

Now we formulate the general conditions when deformations (4.2) and (4.3)
charac\-ter\-ize na-maps : Let a-parallel $l{\subset }U$~ is given by
func\-ti\-ons\\ $u^\alpha =u^{({\alpha })}({\eta }),v^\alpha ={\frac{{%
du^\alpha }}{d{\eta }}}$, ${\eta }_1<{\eta }<{\eta }_2$, satisfying
equations (4.4). We suppose that to this a--parallel corresponds a
na--parallel ${\underline{l}}\subset {\underline{U}}$ given by the same
parameterization in a common for a chosen na--map coordinate system on $U$~
and ${\underline{U}}.$ This condition holds for vectors ${\underline{v}}%
_{(1)}^\alpha =v{\underline{D}}v^\alpha $~ and $v_{(2)}^\alpha =v{\underline{%
D}}v_{(1)}^\alpha $ satisfying equality
$$
{\underline{v}}_{(2)}^\alpha ={\underline{a}}({\eta })v^\alpha +{\underline{b%
}}({\eta }){\underline{v}}_{(1)}^\alpha \eqno(4.5)
$$
for some scalar functions ${\underline{a}}({\eta })$~ and ${\underline{b}}({%
\eta })$~ (see Definitions IV.4 and IV.5). Putting splittings (4.2) and (4.3)
into expressions for $${\underline{v}}_{(1)}^\alpha \mbox{ and }{%
\underline{v}}_{(2)}^\alpha $$ in (4.5) we obtain:
$$
v^\beta v^\gamma v^\delta (D_\beta P_{{.}{\gamma }{\delta }}^\alpha +P_{{.}{%
\beta }{\tau }}^\alpha P_{{.}{\gamma }{\delta }}^\tau +Q_{{.}{\beta }{\tau }%
}^\alpha P_{{.}{\gamma }{\delta }}^\tau )=bv^\gamma v^\delta P_{{.}{\gamma }{%
\delta }}^\alpha +av^\alpha ,\eqno(4.6)
$$
where
$$
b({\eta },v)={\underline{b}}-3{\rho },\qquad \mbox{and}\qquad a({\eta },v)={%
\underline{a}}+{\underline{b}}{\rho }-v^b{\partial }_b{\rho }-{\rho }^2%
\eqno(4.7)
$$
are called the deformation parameters of na--maps.

The algebraic equations for the deformation of torsion $Q_{{.}{\beta }{\tau }%
}^\alpha $ should be written as the compatibility conditions for a given
nonmetricity tensor ${\underline{K}}_{{\alpha }{\beta }{\gamma }}$~ on ${%
\underline{\xi }}$ ( or as the metricity conditions if d--connection ${%
\underline{D}}_\alpha $~ on ${\underline{\xi }}$~ is required to be metric) :%
$$
D_\alpha G_{{\beta }{\gamma }}-P_{{.}{\alpha }({\beta }}^\delta G_{{{\gamma }%
)}{\delta }}-{\underline{K}}_{{\alpha }{\beta }{\gamma }}=Q_{{.}{\alpha }({%
\beta }}^\delta G_{{\gamma }){\delta }},\eqno(4.8)
$$
where $({\quad })$ denotes the symmetric alternation.

So, we have proved this

\begin{theorem}
The na--maps from la--space $\xi $ to la--space ${\underline{\xi }}$~ with a
fixed common nonlinear connection $N_j^a(u)={\underline{N}}_j^a(u)$ and
given d--connections, ${\Gamma }_{{.}{\beta }{\gamma }}^\alpha $~ on $\xi $~
and ${\underline{\Gamma }}_{{.}{\beta }{\gamma }}^\alpha $~ on ${\underline{%
\xi }}$ are locally parametrized by the solutions of equations (4.6) and
(4.8) for every point $u^\alpha $~ and direction $v^\alpha $~ on $U{\subset }%
{\xi }.$
\end{theorem}

We call (4.6) and (4.8) the basic equations for na--maps of la--spaces. They
generalize the corresponding Sinyukov's equations \cite{sin} for isotropic
spaces provided with symmetric affine connection structure.

\subsection{ Classification of Na--Maps of La--Spaces}

Na--maps are classed on possible polynomial parametrizations on variables $%
v^{\alpha}$~ of deformations parameters $a$ and $b$ (see (4.6) and (4.7) ).

\begin{theorem}
There are four classes of na--maps characterized by corresponding
deformation parameters and tensors and basic equations:

\begin{enumerate}
\item  for $na_{(0)}$--maps, ${\pi }_{(0)}$--maps,
$$
P_{{\beta }{\gamma }}^\alpha (u)={\psi }_{{(}{\beta }}{\delta }_{{\gamma }%
)}^\alpha
$$
(${\delta }_\beta ^\alpha $~ is Kronecker symbol and ${\psi }_\beta ={\psi }%
_\beta (u)$~ is a covariant vector d--field) ;

\item  for $na_{(1)}$--maps
$$
a(u,v)=a_{{\alpha }{\beta }}(u)v^\alpha v^\beta ,\quad b(u,v)=b_\alpha
(u)v^\alpha
$$
and $P_{{.}{\beta }{\gamma }}^\alpha (u)$~ is the solution of equations
$$
D_{({\alpha }}P_{{.}{\beta }{\gamma })}^\delta +P_{({\alpha }{\beta }}^\tau
P_{{.}{\gamma }){\tau }}^\delta -P_{({\alpha }{\beta }}^\tau Q_{{.}{\gamma })%
{\tau }}^\delta =b_{({\alpha }}P_{{.}{\beta }{\gamma })}^{{\delta }}+a_{({%
\alpha }{\beta }}{\delta }_{{\gamma })}^\delta ;\eqno(4.9)
$$

\item  for $na_{(2)}$--maps
$$
a(u,v)=a_\beta (u)v^\beta ,\quad b(u,v)={\frac{{b_{{\alpha }{\beta }%
}v^\alpha v^\beta }}{{{\sigma }_\alpha (u)v^\alpha }}},\quad {\sigma }%
_\alpha v^\alpha {\neq }0,
$$
$$
P_{{.}{\alpha }{\beta }}^\tau (u)={{\psi }_{({\alpha }}}{\delta }_{{\beta }%
)}^\tau +{\sigma }_{({\alpha }}F_{{\beta })}^\tau
$$
and $F_\beta ^\alpha (u)$~ is the solution of equations
$$
{D}_{({\gamma }}F_{{\beta })}^\alpha +F_\delta ^\alpha F_{({\gamma }}^\delta
{\sigma }_{{\beta })}-Q_{{.}{\tau }({\beta }}^\alpha F_{{\gamma })}^\tau ={%
\mu }_{({\beta }}F_{{\gamma })}^\alpha +{\nu }_{({\beta }}{\delta }_{{\gamma
})}^\alpha \eqno(4.10)
$$
$({\mu }_\beta (u),{\nu }_\beta (u),{\psi }_\alpha (u),{\sigma }_\alpha (u)$%
~ are covariant d--vectors) ;

\item  for $na_{(3)}$--maps
$$
b(u,v)={\frac{{{\alpha }_{{\beta }{\gamma }{\delta }}v^\beta v^\gamma
v^\delta }}{{{\sigma }_{{\alpha }{\beta }}v^\alpha v^\gamma }}},
$$
$$
P_{{.}{\beta }{\gamma }}^\alpha (u)={\psi }_{({\beta }}{\delta }_{{\gamma }%
)}^\alpha +{\sigma }_{{\beta }{\gamma }}{\varphi }^\alpha ,
$$
where ${\varphi }^\alpha $~ is the solution of equations

$$
D_\beta {\varphi }^\alpha ={\nu }{\delta }_\beta ^\alpha +{\mu }_\beta {%
\varphi }^\alpha +{\varphi }^\gamma Q_{{.}{\gamma }{\delta }}^\alpha ,%
\eqno(4.11)
$$
${\alpha }_{{\beta }{\gamma }{\delta }}(u),{\sigma }_{{\alpha }{\beta }}(u),{%
\psi }_\beta (u),{\nu }(u)$~ and ${\mu }_\beta (u)$~ are d--tensors.
\end{enumerate}
\end{theorem}

{\it Proof.} We sketch the proof respectively for every point in the theorem:

\begin{enumerate}
\item  It is easy to verify that a--parallel equations (4.4) on $\xi $
transform into similar ones on $\underline{\xi }$ if and only if
deformations (4.2) with deformation d--tensors of type ${P^\alpha }_{\beta
\gamma }(u)={\psi }_{(\beta }{\delta }_{\gamma )}^\alpha $ are considered.

\item  Using corresponding to $na_{(1)}$--maps parametrizations of $a(u,v)$
and $b(u,v)$ (see conditions of the theorem) for arbitrary $v^\alpha \neq 0$
on $U\in \xi $ and after a redefinition of deformation parameters we obtain
that equations (4.6) hold if and only if ${P^\alpha }_{\beta \gamma }$
satisfies (4.3).

\item  In a similar manner we obtain basic $na_{(2)}$--map equations (4.10)
from (4.6) by considering $na_{(2)}$--parametrizations of deformation
parameters and d--tensor.

\item  For $na_{(3)}$--maps we mast take into consideration deformations of
torsion (4.3) and introduce $na_{(3)}$--parametrizations for $b(u,v)$ and ${%
P^\alpha }_{\beta \gamma }$ into the basic na--equations (4.6). The last
ones for $na_{(3)}$--maps are equivalent to equations (4.11) (with a
corresponding redefinition of deformation parameters). \qquad $\Box $
\end{enumerate}

We point out that for ${\pi}_{(0)}$-maps we have not differential equations
on $P^{\alpha}_{{.}{\beta}{\gamma}}$ (in the  isotropic case one considers a
first order system of differential equations on metric \cite{sin}; we omit
constructions  with deformation of metric in this subsection).\

To formulate invariant conditions for reciprocal na--maps (when every
a-parallel on ${\underline {\xi}}$~ is also transformed into na--parallel on
$\xi$ ) it is convenient to introduce into consideration the curvature and
Ricci tensors defined for auxiliary connection ${\gamma}^{\alpha}_{{.}{\beta}%
{\gamma}}$~ :
$$
r^{{.}{\delta}} _{{\alpha}{.}{\beta}{\tau}}={\partial}_{[{\beta}}{\gamma}%
^{\delta}_ {{.}{\tau}]{\alpha}}+{\gamma}^{\delta}_{{.}{\rho}[{\beta}}{\gamma}%
^{\rho}_ {{.}{\tau}]{\alpha}} + {{\gamma}^{\delta}}_{\alpha \phi} {w^{\phi}}%
_{\beta \tau}%
$$
and, respectively, $r_{{\alpha}{\tau}}=r^{{.}{\gamma}} _{{\alpha}{.}{\gamma}{%
\tau}} $, where $[\quad ]$ denotes antisymmetric alternation of indices, and
to define values:
$$
^{(0)}T^{\mu}_{{.}{\alpha}{\beta}}= {\Gamma}^{\mu}_{{.}{\alpha}{\beta}} -
T^{\mu}_{{.}{\alpha}{\beta}}- {\frac{1 }{(n+m + 1)}}({\delta}^{\mu}_{({\alpha%
}}{\Gamma}^{\delta}_ {{.}{\beta}){\delta}}-{\delta}^{\mu}_{({\alpha}%
}T^{\delta}_ {{.}{\beta}){\gamma}}),
$$
$$
{}^{(0)}{W}^{\cdot \tau}_{\alpha \cdot \beta \gamma} = {r}^{\cdot
\tau}_{\alpha \cdot \beta \gamma} + {\frac{1}{n+m+1}} [ {\gamma}%
^{\tau}_{\cdot \varphi \tau} {\delta}^{\tau}_{( \alpha} {w^{\varphi}}_{\beta
) \gamma} - ( {\delta}^{\tau}_{\alpha}{r}_{[ \gamma \beta ]} + {\delta}%
^{\tau}_{\gamma} {r}_{[ \alpha \beta ]} - {\delta}^{\tau}_{\beta} {r}_{[
\alpha \gamma ]} )] -%
$$
$$
{\frac{1}{{(n+m+1)}^2}} [ {\delta}^{\tau}_{\alpha}  ( 2 {\gamma}%
^{\tau}_{\cdot \varphi \tau} {w^{\varphi}}_{[ \gamma \beta ] } - {\gamma}%
^{\tau}_{\cdot \tau [ \gamma } {w^{\varphi}}_{\beta ] \varphi} ) + {\delta}%
^{\tau}_{\gamma}  ( 2 {\gamma}^{\tau}_{\cdot \varphi \tau} {w^{\varphi}}%
_{\alpha \beta} -{\gamma}^{\tau}_{\cdot \alpha \tau} {w^{\varphi}}_{\beta
\varphi}) -
$$
$$
{\delta}^{\tau}_{\beta}  ( 2 {\gamma}^{\tau}_{\cdot \varphi \tau} {%
w^{\varphi}}_{\alpha \gamma} - {\gamma}^{\tau}_{\cdot \alpha \tau} {%
w^{\varphi}}_{\gamma \varphi} ) ],%
$$

$$
{^{(3)}T}^{\delta}_{{.}{\alpha}{\beta}}= {\gamma}^{\delta}_{{.}{\alpha}{\beta%
}}+ {\epsilon}{\varphi}^{\tau}{^{({\gamma})}D}_{\beta}q_{\tau}+ {\frac{1 }{%
n+m}}({\delta}^{\gamma}_{\alpha}- {\epsilon}{\varphi}^{\delta}q_{\alpha})[{%
\gamma}^{\tau}_{{.}{\beta}{\tau}}+ {\epsilon}{\varphi}^{\tau}{^{({\gamma})}D}%
_{\beta}q_{\tau}+%
$$
$$
{\frac{1 }{{n+m -1}}}q_{\beta}({\epsilon}{\varphi}^{\tau}{\gamma}^{\lambda}_
{{.}{\tau}{\lambda}}+ {\varphi}^{\lambda}{\varphi}^{\tau}{^{({\gamma})}D}%
_{\tau}q_{\lambda})]- {\frac{1 }{n+m}}({\delta}^{\delta}_{\beta}-{\epsilon}{%
\varphi}^{\delta} q_{\beta})[{\gamma}^{\tau}_{{.}{\alpha}{\tau}}+ {\epsilon}{%
\varphi}^{\tau} {^{({\gamma})}D}_{\alpha}q_{\tau}+%
$$
$$
{\frac{1 }{{n+m -1}}}q_{\alpha}({\epsilon}{\varphi}^{\tau}{\gamma}%
^{\lambda}_ {{.}{\tau}{\lambda}}+ {\varphi}^{\lambda}{\varphi}^{\tau} {%
^{(\gamma)}D} _{\tau}q_{\lambda})],%
$$

$$
{^{(3)}W}^\alpha {{.}{\beta }{\gamma }{\delta }}={\rho }_{{\beta }{.}{\gamma
}{\delta }}^{{.}{\alpha }}+{\epsilon }{\varphi }^\alpha q_\tau {\rho }_{{%
\beta }{.}{\gamma }{\delta }}^{{.}{\tau }}+({\delta }_\delta ^\alpha -
$$
$$
{\epsilon }{\varphi }^\alpha q_\delta )p_{{\beta }{\gamma }}-({\delta }%
_\gamma ^\alpha -{\epsilon }{\varphi }^\alpha q_\gamma )p_{{\beta }{\delta }%
}-({\delta }_\beta ^\alpha -{\epsilon }{\varphi }^\alpha q_\beta )p_{[{%
\gamma }{\delta }]},
$$
$$
(n+m-2)p_{{\alpha }{\beta }}=-{\rho }_{{\alpha }{\beta }}-{\epsilon }q_\tau {%
\varphi }^\gamma {\rho }_{{\alpha }{.}{\beta }{\gamma }}^{{.}{\tau }}+{\frac
1{n+m}}[{\rho }_{{\tau }{.}{\beta }{\alpha }}^{{.}{\tau }}-{\epsilon }q_\tau
{\varphi }^\gamma {\rho }_{{\gamma }{.}{\beta }{\alpha }}^{{.}{\tau }}+{%
\epsilon }q_\beta {\varphi }^\tau {\rho }_{{\alpha }{\tau }}+
$$
$$
{\epsilon }q_\alpha (-{\varphi }^\gamma {\rho }_{{\tau }{.}{\beta }{\gamma }%
}^{{.}{\tau }}+{\epsilon }q_\tau {\varphi }^\gamma {\varphi }^\delta {\rho }%
_{{\gamma }{.}{\beta }{\delta }}^{{.}{\tau }}]),
$$
where $q_\alpha {\varphi }^\alpha ={\epsilon }=\pm 1,$
$$
{{\rho }^\alpha }_{\beta \gamma \delta }=r_{\beta \cdot \gamma \delta
}^{\cdot \alpha }+{\frac 12}({\psi }_{(\beta }{\delta }_{\varphi )}^\alpha +{%
\sigma }_{\beta \varphi }{\varphi }^\tau ){w^\varphi }_{\gamma \delta }
$$
( for a similar value on $\underline{\xi }$ we write ${\quad }{\underline{%
\rho }}_{\cdot \beta \gamma \delta }^\alpha ={\underline{r}}_{\beta \cdot
\gamma \delta }^{\cdot \alpha }-{\frac 12}({\psi }_{(\beta }{\delta }_{{%
\varphi })}^\alpha -{\sigma }_{\beta \varphi }{\varphi }^\tau ){w^\varphi }%
_{\gamma \delta }{\quad })$ and ${\rho }_{\alpha \beta }={\rho }_{\cdot
\alpha \beta \tau }^\tau .$

Similar values,
$$
^{(0)}{\underline T}^{\alpha}_{{.}{\beta}{\gamma}}, ^{(0)}{\underline W}%
^{\nu}_{{.}{\alpha}{\beta}{\gamma}}, {\hat T}^{\alpha} _{{.}{\beta}{\gamma}%
}, {\check T}^{\alpha}_ {{.}{\beta}{\tau}}, {\hat W}^{\delta}_{{.}{\alpha}{%
\beta}{\gamma}}, {\check W}^{\delta}_ {{.}{\alpha}{\beta}{\gamma}}, ^{(3)}{%
\underline T}^{\delta} _{{.}{\alpha}{\beta}},%
$$
and $^{(3)}{\underline W}^ {\alpha}_{{.}{\beta}{\gamma}{\delta}} $ are
given, correspondingly, by auxiliary connections ${\quad}{\underline {\Gamma}%
}^{\mu}_{{.}{\alpha}{\beta}},$~
$$
{\star {\gamma}}^{\alpha}_{{.}{\beta}{\lambda}}={\gamma}^{\alpha} _{{.}{\beta%
}{\lambda}} + {\epsilon}F^{\alpha}_{\tau}{^{({\gamma})}D}_{({\beta}}
F^{\tau}_{{\lambda})}, \quad {\check {\gamma}}^{\alpha}_{{.}{\beta}{\lambda}%
}= {\widetilde {\gamma}}^{\alpha}_{{.}{\beta}{\lambda}} + {\epsilon}%
F^{\lambda} _{\tau} {\widetilde D}_{({\beta}}F^{\tau}_{{\lambda})},%
$$
$$
{\widetilde {\gamma}}^{\alpha}_{{.}{\beta}{\tau}}={\gamma}^{\alpha} _{{.}{%
\beta}{\tau}}+ {\sigma}_{({\beta}}F^{\alpha}_{{\tau})}, \quad {\hat {\gamma}}%
^{\alpha}_{{.}{\beta}{\lambda}}={\star {\gamma}}^{\alpha}_ {{.}{\beta}{%
\lambda}} + {\widetilde {\sigma}}_{({\beta}}{\delta}^{\alpha}_ {{\lambda})},%
$$
where ${\widetilde {\sigma}}_{\beta}={\sigma}_{\alpha}F^{\alpha}_{\beta}.$

\begin{theorem}
Four classes of reciprocal na--maps of la--spaces are characterized by
corresponding invariant criterions:

\begin{enumerate}
\item  for a--maps $^{(0)}T_{{.}{\alpha }{\beta }}^\mu =^{(0)}{\underline{T}}%
_{{.}{\alpha }{\beta }}^\mu ,$
$$
{}^{(0)}W_{{.}{\alpha }{\beta }{\gamma }}^\delta =^{(0)}{\underline{W}}_{{.}{%
\alpha }{\beta }{\gamma }}^\delta ;\eqno(4.12)
$$

\item  for $na_{(1)}$--maps
$$
3({^{({\gamma })}D}_\lambda P_{{.}{\alpha }{\beta }}^\delta +P_{{.}{\tau }{%
\lambda }}^\delta P_{{.}{\alpha }{\beta }}^\tau )=r_{({\alpha }{.}{\beta }){%
\lambda }}^{{.}{\delta }}-{\underline{r}}_{({\alpha }{.}{\beta }){\lambda }%
}^{{.}{\delta }}+\eqno(4.13)
$$
$$
[T_{{.}{\tau }({\alpha }}^\delta P_{{.}{\beta }{\lambda })}^\tau +Q_{{.}{%
\tau }({\alpha }}^\delta P_{{.}{\beta }{\lambda })}^\tau +b_{({\alpha }}P_{{.%
}{\beta }{\lambda })}^\delta +{\delta }_{({\alpha }}^\delta a_{{\beta }{%
\lambda })}];
$$

\item  for $na_{(2)}$--maps ${\hat T}_{{.}{\beta }{\tau }}^\alpha ={\star T}%
_{{.}{\beta }{\tau }}^\alpha ,$
$$
{\hat W}_{{.}{\alpha }{\beta }{\gamma }}^\delta ={\star W}_{{.}{\alpha }{%
\beta }{\gamma }}^\delta ;\eqno(4.14)
$$

\item  for $na_{(3)}$--maps $^{(3)}T_{{.}{\beta }{\gamma }}^\alpha =^{(3)}{%
\underline{T}}_{{.}{\beta }{\gamma }}^\alpha ,$
$$
{}^{(3)}W_{{.}{\beta }{\gamma }{\delta }}^\alpha =^{(3)}{\underline{W}}_{{.}{%
\beta }{\gamma }{\delta }}^\alpha .\eqno(4.15)
$$
\end{enumerate}
\end{theorem}

{\it Proof. }

\begin{enumerate}
\item  Let us prove that a--invariant conditions (4.12) hold. Deformations
of d--connections of type
$$
{}^{(0)}{\underline{\gamma }}_{\cdot \alpha \beta }^\mu ={{\gamma }^\mu }%
_{\alpha \beta }+{\psi }_{(\alpha }{\delta }_{\beta )}^\mu \eqno(4.16)
$$
define a--applications. Contracting indices $\mu $ and $\beta $ we can write
$$
{\psi }_\alpha ={\frac 1{m+n+1}}({{\underline{\gamma }}^\beta }_{\alpha
\beta }-{{\gamma }^\beta }_{\alpha \beta }).\eqno(4.17)
$$
Introducing d--vector ${\psi }_\alpha $ into previous relation and
expressing
$$
{{\gamma }^\alpha }_{\beta \tau }=-{T^\alpha }_{\beta \tau }+{{\Gamma }%
^\alpha }_{\beta \tau }
$$
and similarly for underlined values we obtain the first invariant conditions
from (4.12).

Putting deformation (4.16) into the formula for
$$
{\underline{r}}_{\alpha \cdot \beta \gamma }^{\cdot \tau }\quad \mbox{and}%
\quad {\underline{r}}_{\alpha \beta }={\underline{r}}_{\alpha \tau \beta
\tau }^{\cdot \tau }
$$
we obtain respectively relations
$$
{\underline{r}}_{\alpha \cdot \beta \gamma }^{\cdot \tau }-r_{\alpha \cdot
\beta \gamma }^{\cdot \tau }={\delta }_\alpha ^\tau {\psi }_{[\gamma \beta
]}+{\psi }_{\alpha [\beta }{\delta }_{\gamma ]}^\tau +{\delta }_{(\alpha
}^\tau {\psi }_{\varphi )}{w^\varphi }_{\beta \gamma }\eqno(4.18)
$$
and
$$
{\underline{r}}_{\alpha \beta }-r_{\alpha \beta }={\psi }_{[\alpha \beta
]}+(n+m-1){\psi }_{\alpha \beta }+{\psi }_\varphi {w^\varphi }_{\beta \alpha
}+{\psi }_\alpha {w^\varphi }_{\beta \varphi },\eqno(4.19)
$$
where
$$
{\psi }_{\alpha \beta }={}^{({\gamma })}D_\beta {\psi }_\alpha -{\psi }%
_\alpha {\psi }_\beta .
$$
Putting (4.16) into (4.19) we can express ${\psi }_{[\alpha \beta ]}$ as

$$
{\psi }_{[\alpha \beta ]}={\frac 1{n+m+1}}[{\underline{r}}_{[\alpha \beta ]}+%
{\frac 2{n+m+1}}{\underline{\gamma }}_{\cdot \varphi \tau }^\tau {w^\varphi }%
_{[\alpha \beta ]}-{\frac 1{n+m+1}}{\underline{\gamma }}_{\cdot \tau [\alpha
}^\tau {w^\varphi }_{\beta ]\varphi }]-
$$
$$
{\frac 1{n+m+1}}[r_{[\alpha \beta ]}+{\frac 2{n+m+1}}{{\gamma }^\tau }%
_{\varphi \tau }{w^\varphi }_{[\alpha \beta ]}-{\frac 1{n+m+1}}{{\gamma }%
^\tau }_{\tau [\alpha }{w^\varphi }_{\beta ]\varphi }].\eqno(4.20)
$$
To simplify our consideration we can choose an a--transform, parametrized by
corresponding $\psi $--vector from (4.16), (or fix a local coordinate cart)
the antisymmetrized relations (4.20) to be satisfied by d--tensor
$$
{\psi }_{\alpha \beta }={\frac 1{n+m+1}}[{\underline{r}}_{\alpha \beta }+{%
\frac 2{n+m+1}}{\underline{\gamma }}_{\cdot \varphi \tau }^\tau {w^\varphi }%
_{\alpha \beta }-{\frac 1{n+m+1}}{\underline{\gamma }}_{\cdot \alpha \tau
}^\tau {w^\varphi }_{\beta \varphi }-r_{\alpha \beta }-
$$
$$
{\frac 2{n+m+1}}{{\gamma }^\tau }_{\varphi \tau }{w^\varphi }_{\alpha \beta
}+{\frac 1{n+m+1}}{{\gamma }^\tau }_{\alpha \tau }{w^\varphi }_{\beta
\varphi }]\eqno(4.21)
$$
Introducing expressions (4.16),(4.20) and (4.21) into deformation of
curvature (4.17) we obtain the second conditions (4.12) of a-map invariance:
$$
^{(0)}W_{\alpha \cdot \beta \gamma }^{\cdot \delta }={}^{(0)}{\underline{W}}%
_{\alpha \cdot \beta \gamma }^{\cdot \delta },
$$
where the Weyl d--tensor on $\underline{\xi }$ (the extension of the usual
one for geodesic maps on (pseudo)--Riemannian spaces to the case of
v--bundles provided with N--connection structure) is defined as
$$
{}^{(0)}{\underline{W}}_{\alpha \cdot \beta \gamma }^{\cdot \tau }={%
\underline{r}}_{\alpha \cdot \beta \gamma }^{\cdot \tau }+{\frac 1{n+m+1}}[{%
\underline{\gamma }}_{\cdot \varphi \tau }^\tau {\delta }_{(\alpha }^\tau {%
w^\varphi }_{\beta )\gamma }-({\delta }_\alpha ^\tau {\underline{r}}%
_{[\gamma \beta ]}+{\delta }_\gamma ^\tau {\underline{r}}_{[\alpha \beta ]}-{%
\delta }_\beta ^\tau {\underline{r}}_{[\alpha \gamma ]})]-
$$
$$
{\frac 1{{(n+m+1)}^2}}[{\delta }_\alpha ^\tau (2{\underline{\gamma }}_{\cdot
\varphi \tau }^\tau {w^\varphi }_{[\gamma \beta ]}-{\underline{\gamma }}%
_{\cdot \tau [\gamma }^\tau {w^\varphi }_{\beta ]\varphi })+{\delta }_\gamma
^\tau (2{\underline{\gamma }}_{\cdot \varphi \tau }^\tau {w^\varphi }%
_{\alpha \beta }-{\underline{\gamma }}_{\cdot \alpha \tau }^\tau {w^\varphi }%
_{\beta \varphi })-
$$
$$
{\delta }_\beta ^\tau (2{\underline{\gamma }}_{\cdot \varphi \tau }^\tau {%
w^\varphi }_{\alpha \gamma }-{\underline{\gamma }}_{\cdot \alpha \tau }^\tau
{w^\varphi }_{\gamma \varphi })].
$$

\item  To obtain $na_{(1)}$--invariant conditions we rewrite $na_{(1)}$%
--equations (4.9) as to consider in explicit form covariant derivation $^{({%
\gamma })}D$ and deformations (4.2) and (4.3):
$$
2({}^{({\gamma })}D_\alpha {P^\delta }_{\beta \gamma }+{}^{({\gamma }%
)}D_\beta {P^\delta }_{\alpha \gamma }+{}^{({\gamma })}D_\gamma {P^\delta }%
_{\alpha \beta }+{P^\delta }_{\tau \alpha }{P^\tau }_{\beta \gamma }+
$$
$$
{P^\delta }_{\tau \beta }{P^\tau }_{\alpha \gamma }+{P^\delta }_{\tau \gamma
}{P^\tau }_{\alpha \beta })={T^\delta }_{\tau (\alpha }{P^\tau }_{\beta
\gamma )}+
$$
$$
{H^\delta }_{\tau (\alpha }{P^\tau }_{\beta \gamma )}+b_{(\alpha }{P^\delta }%
_{\beta \gamma )}+a_{(\alpha \beta }{\delta }_{\gamma )}^\delta .\eqno(4.22)
$$
Alternating the first two indices in (4.22) we have
$$
2({\underline{r}}_{(\alpha \cdot \beta )\gamma }^{\cdot \delta }-r_{(\alpha
\cdot \beta )\gamma }^{\cdot \delta })=2({}^{(\gamma )}D_\alpha {P^\delta }%
_{\beta \gamma }+
$$
$$
{}^{(\gamma )}D_\beta {P^\delta }_{\alpha \gamma }-2{}^{(\gamma )}D_\gamma {%
P^\delta }_{\alpha \beta }+{P^\delta }_{\tau \alpha }{P^\tau }_{\beta \gamma
}+{P^\delta }_{\tau \beta }{P^\tau }_{\alpha \gamma }-2{P^\delta }_{\tau
\gamma }{P^\tau }_{\alpha \beta }).
$$
Substituting the last expression from (4.22) and rescalling the deformation
parameters and d--tensors we obtain the conditions (4.9).

\item  Now we prove the invariant conditions for $na_{(0)}$--maps satisfying
conditions
$$
\epsilon \neq 0\quad \mbox{and}\quad \epsilon -F_\beta ^\alpha F_\alpha
^\beta \neq 0
$$
Let define the auxiliary d--connection
$$
{\tilde \gamma }_{\cdot \beta \tau }^\alpha ={\underline{\gamma }}_{\cdot
\beta \tau }^\alpha -{\psi }_{(\beta }{\delta }_{\tau )}^\alpha ={{\gamma }%
^\alpha }_{\beta \tau }+{\sigma }_{(\beta }F_{\tau )}^\alpha \eqno(4.23)
$$
and write
$$
{\tilde D}_\gamma ={}^{({\gamma })}D_\gamma F_\beta ^\alpha +{\tilde \sigma }%
_\gamma F_\beta ^\alpha -{\epsilon }{\sigma }_\beta {\delta }_\gamma ^\alpha
,
$$
where ${\tilde \sigma }_\beta ={\sigma }_\alpha F_\beta ^\alpha ,$ or, as a
consequence from the last equality,
$$
{\sigma }_{(\alpha }F_{\beta )}^\tau ={\epsilon }F_\lambda ^\tau ({}^{({%
\gamma })}D_{(\alpha }F_{\beta )}^\alpha -{\tilde D}_{(\alpha }F_{\beta
)}^\lambda )+{\tilde \sigma }_{(}{\alpha }{\delta }_{\beta )}^\tau .
$$
Introducing auxiliary connections
$$
{\star {\gamma }}_{\cdot \beta \lambda }^\alpha ={\gamma }_{\cdot \beta
\lambda }^\alpha +{\epsilon }F_\tau ^\alpha {}^{({\gamma })}D_{(\beta
}F_{\lambda )}^\tau
$$
and
$$
{\check \gamma }_{\cdot \beta \lambda }^\alpha ={\tilde \gamma }_{\cdot
\beta \lambda }^\alpha +{\epsilon }F_\tau ^\alpha {\tilde D}_{(\beta
}F_{\lambda )}^\tau
$$
we can express deformation (4.23) in a form characteristic for a--maps:
$$
{\hat \gamma }_{\cdot \beta \gamma }^\alpha ={\star {\gamma }}_{\cdot \beta
\gamma }^\alpha +{\tilde \sigma }_{(\beta }{\delta }_{\lambda )}^\alpha .%
\eqno(4.24)
$$
Now it's obvious that $na_{(2)}$--invariant conditions (4.24) are equivalent
with a--invariant conditions (4.12) written for d--connection (4.24). As a
matter of principle we can write formulas for such $na_{(2)}$--invariants in
terms of ''underlined'' and ''non--underlined'' values by expressing
consequently all used auxiliary connections as deformations of ''prime''
connections on $\xi $ and ''final'' connections on $\underline{\xi }.$ We
omit such tedious calculations in this work.

\item  Finally, we prove the last statement, for $na_{(3)}$--maps, of the
theorem IV.3. Let
$$
q_\alpha {\varphi }^\alpha =e=\pm 1,\eqno(4.25)
$$
where ${\varphi }^\alpha $ is contained in
$$
{\underline{\gamma }}_{\cdot \beta \gamma }^\alpha ={{\gamma }^\alpha }%
_{\beta \gamma }+{\psi }_{(\beta }{\delta }_{\gamma )}^\alpha +{\sigma }%
_{\beta \gamma }{\varphi }^\alpha .\eqno(4.26)
$$
Acting with operator $^{({\gamma })}{\underline{D}}_\beta $ on (4.25) we
write
$$
{}^{({\gamma })}{\underline{D}}_\beta q_\alpha ={}^{({\gamma })}D_\beta
q_\alpha -{\psi }_{(\alpha }q_{\beta )}-e{\sigma }_{\alpha \beta }.%
\eqno(4.27)
$$
Contracting (4.27) with ${\varphi }^\alpha $ we can express
$$
e{\varphi }^\alpha {\sigma }_{\alpha \beta }={\varphi }^\alpha ({}^{({\gamma
})}D_\beta q_\alpha -{}^{({\gamma })}{\underline{D}}_\beta q_\alpha )-{%
\varphi }_\alpha q^\alpha q_\beta -e{\psi }_\beta .
$$
Putting the last formula in (4.26) contracted on indices $\alpha $ and $%
\gamma $ we obtain
$$
(n+m){\psi }_\beta ={\underline{\gamma }}_{\cdot \alpha \beta }^\alpha -{{%
\gamma }^\alpha }_{\alpha \beta }+e{\psi }_\alpha {\varphi }^\alpha q_\beta
+e{\varphi }^\alpha {\varphi }^\beta ({}^{({\gamma })}{\underline{D}}_\beta
-{}^{({\gamma })}D_\beta ).\eqno(4.28)
$$
From these relations, taking into consideration (4.25), we have%
$$
(n+m-1){\psi }_\alpha {\varphi }^\alpha =
$$
$$
{\varphi }^\alpha ({\underline{\gamma }}_{\cdot \alpha \beta }^\alpha -{{%
\gamma }^\alpha }_{\alpha \beta })+e{\varphi }^\alpha {\varphi }^\beta ({}^{(%
{\gamma })}{\underline{D}}_\beta q_\alpha -{}^{({\gamma })}D_\beta q_\alpha
)
$$

Using the equalities and identities (4.27) and (4.28) we can express
deformations (4.26) as the first $na_{(3)}$--invariant conditions from
(4.15).

To prove the second class of $na_{(3)}$--invariant conditions we introduce
two additional d--tensors:
$$
{{\rho }^\alpha }_{\beta \gamma \delta }=r_{\beta \cdot \gamma \delta
}^{\cdot \alpha }+{\frac 12}({\psi }_{(\beta }{\delta }_{\varphi )}^\alpha +{%
\sigma }_{\beta \varphi }{\varphi }^\tau ){w^\varphi }_{\gamma \delta }{%
\quad }
$$
and
$$
{\underline{\rho }}_{\cdot \beta \gamma \delta }^\alpha ={\underline{r}}%
_{\beta \cdot \gamma \delta }^{\cdot \alpha }-{\frac 12}({\psi }_{(\beta }{%
\delta }_{{\varphi })}^\alpha -{\sigma }_{\beta \varphi }{\varphi }^\tau ){%
w^\varphi }_{\gamma \delta }.\eqno(4.29)
$$
Using deformation (4.26) and (4.29) we write relation
$$
{\tilde \sigma }_{\cdot \beta \gamma \delta }^\alpha ={\underline{\rho }}%
_{\cdot \beta \gamma \delta }^\alpha -{\rho }_{\cdot \beta \gamma \delta
}^\alpha ={\psi }_{\beta [\delta }{\delta }_{\gamma ]}^\alpha -{\psi }_{[{%
\gamma }{\delta }]}{\delta }_\beta ^\alpha -{\sigma }_{\beta \gamma \delta }{%
\varphi }^\alpha ,\eqno(4.30)
$$
where
$$
{\psi }_{\alpha \beta }={}^{({\gamma })}D_\beta {\psi }_\alpha +{\psi }%
_\alpha {\psi }_\beta -({\nu }+{\varphi }^\tau {\psi }_\tau ){\sigma }%
_{\alpha \beta },
$$
and
$$
{\sigma }_{\alpha \beta \gamma }={}^{({\gamma })}D_{[\gamma }{\sigma }_{{%
\beta }]{\alpha }}+{\mu }_{[\gamma }{\sigma }_{{\beta }]{\alpha }}-{\sigma }%
_{{\alpha }[{\gamma }}{\sigma }_{{\beta }]{\tau }}{\varphi }^\tau .
$$
Let multiply (4.30) on $q_\alpha $ and write (taking into account relations
(4.25)) the relation
$$
e{\sigma }_{\alpha \beta \gamma }=-q_\tau {\tilde \sigma }_{\cdot \alpha
\beta \delta }^\tau +{\psi }_{\alpha [\beta }q_{\gamma ]}-{\psi }_{[\beta
\gamma ]}q_\alpha .\eqno(4.31)
$$
The next step is to express ${\psi }_{\alpha \beta }$ trough d--objects on ${%
\xi }.$ To do this we contract indices $\alpha $ and $\beta $ in (4.30) and
obtain
$$
(n+m){\psi }_{[\alpha \beta ]}=-{\sigma }_{\cdot \tau \alpha \beta }^\tau
+eq_\tau {\varphi }^\lambda {\sigma }_{\cdot \lambda \alpha \beta }^\tau -e{%
\tilde \psi }_{[\alpha }{\tilde \psi }_{\beta ]}.
$$
Then contracting indices $\alpha $ and $\delta $ in (4.30) and using (4.31)
we write
$$
(n+m-2){\psi }_{\alpha \beta }={\tilde \sigma }_{\cdot \alpha \beta \tau
}^\tau -eq_\tau {\varphi }^\lambda {\tilde \sigma }_{\cdot \alpha \beta
\lambda }^\tau +{\psi }_{[\beta \alpha ]}+e({\tilde \psi }_\beta q_\alpha -{%
\hat \psi }_{(\alpha }q_{\beta )},\eqno(4.32)
$$
where ${\hat \psi }_\alpha ={\varphi }^\tau {\psi }_{\alpha \tau }.$ If the
both parts of (4.32) are contracted with ${\varphi }^\alpha ,$ it results
that
$$
(n+m-2){\tilde \psi }_\alpha ={\varphi }^\tau {\sigma }_{\cdot \tau \alpha
\lambda }^\lambda -eq_\tau {\varphi }^\lambda {\varphi }^\delta {\sigma }%
_{\lambda \alpha \delta }^\tau -eq_\alpha ,
$$
and, in consequence of ${\sigma }_{\beta (\gamma \delta )}^\alpha =0,$ we
have
$$
(n+m-1){\varphi }={\varphi }^\beta {\varphi }^\gamma {\sigma }_{\cdot \beta
\gamma \alpha }^\alpha .
$$
By using the last expressions we can write
$$
(n+m-2){\underline{\psi }}_\alpha ={\varphi }^\tau {\sigma }_{\cdot \tau
\alpha \lambda }^\lambda -eq_\tau {\varphi }^\lambda {\varphi }^\delta {%
\sigma }_{\cdot \lambda \alpha \delta }^\tau -e{(n+m-1)}^{-1}q_\alpha {%
\varphi }^\tau {\varphi }^\lambda {\sigma }_{\cdot \tau \lambda \delta
}^\delta .\eqno(4.33)
$$
Contracting (4.32) with ${\varphi }^\beta $ we have
$$
(n+m){\hat \psi }_\alpha ={\varphi }^\tau {\sigma }_{\cdot \alpha \tau
\lambda }^\lambda +{\tilde \psi }_\alpha
$$
and taking into consideration (4.33) we can express ${\hat \psi }_\alpha $
through ${\sigma }_{\cdot \beta \gamma \delta }^\alpha .$

As a consequence of (4.31)--(4.33) we obtain this formulas for d--tensor ${%
\psi }_{\alpha \beta }:$
$$
(n+m-2){\psi }_{\alpha \beta }={\sigma }_{\cdot \alpha \beta \tau }^\tau
-eq_\tau {\varphi }^\lambda {\sigma }_{\cdot \alpha \beta \lambda }^\tau +
$$
$$
{\frac 1{n+m}}\{-{\sigma }_{\cdot \tau \beta \alpha }^\tau +eq_\tau {\varphi
}^\lambda {\sigma }_{\cdot \lambda \beta \alpha }^\tau -q_\beta (e{\varphi }%
^\tau {\sigma }_{\cdot \alpha \tau \lambda }^\lambda -q_\tau {\varphi }%
^\lambda {\varphi }^\delta {\sigma }_{\cdot \alpha \lambda \delta }^\tau )+
$$
$$
eq_\alpha [{\varphi }^\lambda {\sigma }_{\cdot \tau \beta \lambda }^\tau
-eq_\tau {\varphi }^\lambda {\varphi }^\delta {\sigma }_{\cdot \lambda \beta
\delta }^\tau -{\frac e{n+m-1}}q_\beta ({\varphi }^\tau {\varphi }^\lambda {%
\sigma }_{\cdot \tau \gamma \delta }^\delta -eq_\tau {\varphi }^\lambda {%
\varphi }^\delta {\varphi }^\varepsilon {\sigma }_{\cdot \lambda \delta
\varepsilon }^\tau )]\}.
$$

Finally, putting the last formula and (4.31) into (4.30) and after a
rearrangement of terms we obtain the second group of $na_{(3)}$-invariant
conditions (4.15). If necessary we can rewrite these conditions in terms of
geometrical objects on $\xi $ and $\underline{\xi }.$ To do this we mast
introduce splittings (4.29) into (4.15). \qquad $\Box $
\end{enumerate}

For the particular case of $na_{(3)}$--maps when
$$
{\psi}_{\alpha}=0 , {\varphi}_{\alpha} = g_{\alpha \beta} {\varphi}^{\beta}
= {\frac{\delta }{\delta u^{\alpha}}} ( \ln {\Omega} ) , {\Omega}(u) > 0
$$
and
$$
{\sigma}_{\alpha \beta} = g_{\alpha \beta}%
$$
we define a subclass of conformal transforms ${\underline g}_{\alpha \beta}
(u) = {\Omega}^2 (u)  g_{\alpha \beta}$ which, in consequence of the fact
that d--vector ${\varphi}_{\alpha}$ must satisfy equations (4.11),
generalizes the class of concircular transforms (see \cite{sin}  for
references and details on concircular mappings of Riemannaian spaces) .

We emphasize that basic na--equations (4.9)--(4.11) are systems of first
order partial differential equations. The study of their geometrical
properties and definition of integral varieties, general and particular
solutions are possible by using the formalism of Pffaf systems \cite{vog}.
Here we point out that by using algebraic methods we can always verify if
systems of na--equations of type (4.9)--(4.11) are, or not, involute, even
to find their explicit solutions it is a difficult task (see more detailed
considerations for isotropic ng--maps in \cite{sin} and, on language of
Pffaf systems for na--maps, in \cite{vb13}). We can also formulate the
Cauchy problem for na--equations on $\xi $~ and choose deformation
parameters (4.7) as to make involute mentioned equations for the case of
maps to a given background space ${\underline{\xi }}$. If a solution, for
example, of $na_{(1)}$--map equations exists, we say that space $\xi $ is $%
na_{(1)}$--projective to space ${\underline{\xi }}.$ In general, we have to
introduce chains of na--maps in order to obtain involute systems of
equations for maps (superpositions of na-maps) from $\xi $ to ${\underline{%
\xi }}:$
$$U \buildrel {ng<i_{1}>} \over \longrightarrow  {U_{\underline 1}}
\buildrel ng<i_2> \over \longrightarrow \cdots
\buildrel ng<i_{k-1}> \over \longrightarrow U_{\underline {k-1}}
\buildrel ng<i_k> \over \longrightarrow {\underline U} $$
where $U\subset {\xi },U_{\underline{1}}\subset {\xi }_{\underline{1}%
},\ldots ,U_{k-1}\subset {\xi }_{k-1},{\underline{U}}\subset {\xi }_k$ with
corresponding splittings of auxiliary symmetric connections
$$
{\underline{\gamma }}_{.{\beta }{\gamma }}^\alpha =_{<i_1>}P_{.{\beta }{%
\gamma }}^\alpha +_{<i_2>}P_{.{\beta }{\gamma }}^\alpha +\cdots +_{<i_k>}P_{.%
{\beta }{\gamma }}^\alpha
$$
and torsion
$$
{\underline{T}}_{.{\beta }{\gamma }}^\alpha =T_{.{\beta }{\gamma }}^\alpha
+_{<i_1>}Q_{.{\beta }{\gamma }}^\alpha +_{<i_2>}Q_{.{\beta }{\gamma }%
}^\alpha +\cdots +_{<i_k>}Q_{.{\beta }{\gamma }}^\alpha
$$
where cumulative indices $<i_1>=0,1,2,3,$ denote possible types of na--maps.

\begin{definition}
Space $\xi $~ is nearly conformally projective to space ${\underline{\xi }},{%
\quad }nc:{\xi }{\to }{\underline{\xi }},$~ if there is a finite chain of
na--maps from $\xi $~ to ${\underline{\xi }}.$
\end{definition}

For nearly conformal maps we formulate :

\begin{theorem}
For every fixed triples $(N_j^a,{\Gamma }_{{.}{\beta }{\gamma }}^\alpha
,U\subset {\xi }$ and $(N_j^a,{\underline{\Gamma }}_{{.}{\beta }{\gamma }%
}^\alpha $, ${\underline{U}}\subset {\underline{\xi }})$, components of
nonlinear connection, d--connection and d--metric being of class $C^r(U),C^r(%
{\underline{U}})$, $r>3,$ there is a finite chain of na--maps $nc:U\to {%
\underline{U}}.$
\end{theorem}

Proof is similar to that for isotropic maps \cite{vk,vob12,vrjp} (we have to
introduce a finite number of na-maps with corresponding components of
deformation parameters and deformation tensors in order to transform step
by step coefficients of d-connection ${\Gamma}^{\alpha}_{\gamma \delta}$
into ${\underline {\Gamma}}^{\alpha}_{\beta \gamma} ).$

Now we introduce the concept of the Category of la--spaces, ${\cal C}({\xi}).
$ The elements of ${\cal C}({\xi})$ consist from $Ob{\cal C}({\xi})=\{{\xi},
{\xi}_{<i_{1}>}, {\xi}_{<i_{2}>},{\ldots}, \}$ being la--spaces, for
simplicity in this work, having common N--connection structures, and\\ $Mor
{\cal C}({\xi})=\{ nc ({\xi}_{<i_{1}>}, {\xi}_{<i_{2}>})\}$ being chains of
na--maps interrelating la--spaces. We point out that we can consider
equivalent models of physical theories on every object of ${\cal C}({\xi})$
(see details for isotropic gravitational models in \cite
{vk,vrjp,vog,vb12,vob12,vob13}  and anisotropic gravity in \cite
{vcl96,voa,vodg}). One of the main purposes of this section is to develop a
d--tensor and variational formalism on ${\cal C}({\xi}),$ i.e. on
la--multispaces, interrelated with nc--maps. Taking into account the
distinguished character of geometrical objects on la--spaces we call tensors
on ${\cal C}({\xi})$ as distinguished tensors on la--space Category, or
dc--tensors.

Finally, we emphasize that presented in this subsection definitions and
theorems can be generalized for v--bundles with arbitrary given structures
of nonlinear connection, linear d--connection and metric structures. Proofs
are similar to those from \cite{vth,sin}.

\subsection{Na-Tensor-Integral on La-Spaces}

The aim of this subsection is to define tensor integration not only for
bitensors, objects defined on the same curved space, but for dc--tensors,
defined on two spaces, $\xi$ and ${\underline {\xi}}$, even it is necessary
on la--multispaces. A. Mo\'or tensor--integral formalism having a lot of
applications in classical and quantum gravity \cite{syn,dewd,goz}  was
extended for locally isotropic multispaces in \cite{vog,vob12}. The
unispacial locally anisotropic version is given in \cite{v295a,gv}.

Let $T_{u}{\xi}$~ and $T_{\underline u}{\underline {\xi}}$ be tangent spaces
in corresponding points $u {\in} U {\subset} {\xi}$ and ${\underline u} {\in}
{\underline U} {\subset} {\underline {\xi}}$ and, respectively, $T^{\ast}_{u}%
{\xi}$ and $T^{\ast}_{\underline u}{\underline {\xi}} $ be their duals (in
general, in this subsection we shall not consider that a common
coordinatization is introduced for open regions $U$ and ${\underline U}$ ).
We call as the dc--tensors on the pair of spaces (${\xi}, {\underline {\xi}}$
) the elements of distinguished tensor algebra
$$
( {\otimes}_{\alpha} T_{u}{\xi}) {\otimes} ({\otimes}_{\beta} T^{\ast}_{u} {%
\xi}) {\otimes}({\otimes}_{\gamma} T_{\underline u}{\underline {\xi}}) {%
\otimes} ({\otimes}_{\delta} T^{\ast}_{\underline u} {\underline {\xi}})%
$$
defined over the space ${\xi}{\otimes} {\underline {\xi}}, $ for a given $nc
: {\xi} {\to} {\underline {\xi}} $.

We admit the convention that underlined and non--underlined indices refer,
respectively, to the points ${\underline u}$ and $u$. Thus $Q_{{.}{%
\underline {\alpha}}}^{\beta}, $ for instance, are the components of
dc--tensor $Q{\in} T_{u}{\xi} {\otimes} T_{\underline u}{\underline {\xi}}.$

Now, we define the transport dc--tensors. Let open regions $U$ and ${%
\underline U}$ be homeomorphic to sphere ${\cal R}^{2n}$ and introduce
isomorphism ${\mu}_{{u},{\underline u}}$ between $T_{u}{\xi}$ and $%
T_{\underline u}{\underline {\xi}}$ (given by map $nc : U {\to} {\underline U%
}).$ We consider that for every d--vector $v^{\alpha} {\in} T_{u}{\xi}$
corresponds the vector ${\mu}_{{u},{\underline u}}
(v^{\alpha})=v^{\underline {\alpha}} {\in} T_{\underline u}{\underline {\xi}}%
,$ with components $v^{\underline {\alpha}}$ being linear functions of $%
v^{\alpha}$:
$$
v^{\underline {\alpha}}=h^{\underline {\alpha}}_{\alpha}(u, {\underline u})
v^{\alpha}, \quad v_{\underline {\alpha}}= h^{\alpha}_{\underline {\alpha}}({%
\underline u}, u)v_{\alpha},%
$$
where $h^{\alpha}_{\underline {\alpha}}({\underline u}, u)$ are the
components of dc--tensor associated with ${\mu}^{-1}_{u,{\underline u}}$. In
a similar manner we have
$$
v^{\alpha}=h^{\alpha}_{\underline {\alpha}}({\underline u}, u) v^{\underline
{\alpha}}, \quad v_{\alpha}=h^{\underline {\alpha}}_{\alpha} (u, {\underline
u})v_{\underline {\alpha}}.%
$$

In order to reconcile just presented definitions and to assure the identity
for trivial maps ${\xi }{\to }{\xi },u={\underline{u}},$ the transport
dc-tensors must satisfy conditions :
$$
h_\alpha ^{\underline{\alpha }}(u,{\underline{u}})h_{\underline{\alpha }%
}^\beta ({\underline{u}},u)={\delta }_\alpha ^\beta ,h_\alpha ^{\underline{%
\alpha }}(u,{\underline{u}})h_{\underline{\beta }}^\alpha ({\underline{u}}%
,u)={\delta }_{\underline{\beta }}^{\underline{\alpha }}
$$
and ${\lim }_{{({\underline{u}}{\to }u})}h_\alpha ^{\underline{\alpha }}(u,{%
\underline{u}})={\delta }_\alpha ^{\underline{\alpha }},\quad {\lim }_{{({%
\underline{u}}{\to }u})}h_{\underline{\alpha }}^\alpha ({\underline{u}},u)={%
\delta }_{\underline{\alpha }}^\alpha .$

Let ${\overline S}_{p} {\subset} {\overline U} {\subset} {\overline {\xi}}$
is a homeomorphic to $p$-dimensional sphere and suggest that chains of
na--maps are used to connect regions :
$$ U \buildrel nc_{(1)} \over \longrightarrow {\overline S}_p
     \buildrel nc_{(2)} \over \longrightarrow {\underline U}.$$

\begin{definition}
The tensor integral in ${\overline{u}}{\in }{\overline{S}}_p$ of a
dc--tensor $N_{{\varphi }{.}{\underline{\tau }}{.}{\overline{\alpha }}_1{%
\cdots }{\overline{\alpha }}_p}^{{.}{\gamma }{.}{\underline{\kappa }}}$ $({%
\overline{u}},u),$ completely antisymmetric on the indices ${{\overline{%
\alpha }}_1},{\ldots },{\overline{\alpha }}_p,$ over domain ${\overline{S}}%
_p,$ is defined as
$$
N_{{\varphi }{.}{\underline{\tau }}}^{{.}{\gamma }{.}{\underline{\kappa }}}({%
\underline{u}},u)=I_{({\overline{S}}_p)}^{\underline{U}}N_{{\varphi }{.}{%
\overline{\tau }}{.}{\overline{\alpha }}_1{\ldots }{\overline{\alpha }}_p}^{{%
.}{\gamma }{.}{\overline{\kappa }}}({\overline{u}},{\underline{u}})dS^{{%
\overline{\alpha }}_1{\ldots }{\overline{\alpha }}_p}=
$$
$$
{\int }_{({\overline{S}}_p)}h_{\underline{\tau }}^{\overline{\tau }}({%
\underline{u}},{\overline{u}})h_{\overline{\kappa }}^{\underline{\kappa }}({%
\overline{u}},{\underline{u}})N_{{\varphi }{.}{\overline{\tau }}{.}{%
\overline{\alpha }}_1{\cdots }{\overline{\alpha }}_p}^{{.}{\gamma }{.}{%
\overline{\kappa }}}({\overline{u}},u)d{\overline{S}}^{{\overline{\alpha }}_1%
{\cdots }{\overline{\alpha }}_p},\eqno(4.34)
$$
where $dS^{{\overline{\alpha }}_1{\cdots }{\overline{\alpha }}_p}={\delta }%
u^{{\overline{\alpha }}_1}{\land }{\cdots }{\land }{\delta }u_p^{\overline{%
\alpha }}$.
\end{definition}

Let suppose that transport dc--tensors $h_\alpha ^{\underline{\alpha }}$~
and $h_{\underline{\alpha }}^\alpha $~ admit covariant derivations of
or\-der two and pos\-tu\-la\-te ex\-is\-ten\-ce of de\-for\-ma\-ti\-on
dc--ten\-sor\\ $B_{{\alpha }{\beta }}^{{..}{\gamma }}(u,{\underline{u}})$~
satisfying relations
$$
D_\alpha h_\beta ^{\underline{\beta }}(u,{\underline{u}})=B_{{\alpha }{\beta
}}^{{..}{\gamma }}(u,{\underline{u}})h_\gamma ^{\underline{\beta }}(u,{%
\underline{u}})\eqno(4.35)
$$
and, taking into account that $D_\alpha {\delta }_\gamma ^\beta =0,$

$$
D_\alpha h_{\underline{\beta }}^\beta ({\underline{u}},u)=-B_{{\alpha }{%
\gamma }}^{{..}{\beta }}(u,{\underline{u}})h_{\underline{\beta }}^\gamma ({%
\underline{u}},u).
$$
By using formulas  for torsion and, respectively,
curvature of connection ${\Gamma }_{{\beta }{\gamma }}^\alpha $~ we can
calculate next commutators:
$$
D_{[{\alpha }}D_{{\beta }]}h_\gamma ^{\underline{\gamma }}=-(R_{{\gamma }{.}{%
\alpha }{\beta }}^{{.}{\lambda }}+T_{{.}{\alpha }{\beta }}^\tau B_{{\tau }{%
\gamma }}^{{..}{\lambda }})h_\lambda ^{\underline{\gamma }}.\eqno(4.36)
$$
On the other hand from (4.35) one follows that
$$
D_{[{\alpha }}D_{{\beta }]}h_\gamma ^{\underline{\gamma }}=(D_{[{\alpha }}B_{%
{\beta }]{\gamma }}^{{..}{\lambda }}+B_{[{\alpha }{|}{\tau }{|}{.}}^{{..}{%
\lambda }}B_{{\beta }]{\gamma }{.}}^{{..}{\tau }})h_\lambda ^{\underline{%
\gamma }},\eqno(4.37)
$$
where ${|}{\tau }{|}$~ denotes that index ${\tau }$~ is excluded from the
action of antisymmetrization $[{\quad }]$. From (4.36) and (4.37) we obtain
$$
D_{[{\alpha }}B_{{\beta }]{\gamma }{.}}^{{..}{\lambda }}+B_{[{\beta }{|}{%
\gamma }{|}}B_{{\alpha }]{\tau }}^{{..}{\lambda }}=(R_{{\gamma }{.}{\alpha }{%
\beta }}^{{.}{\lambda }}+T_{{.}{\alpha }{\beta }}^\tau B_{{\tau }{\gamma }}^{%
{..}{\lambda }}).\eqno(4.38)
$$

Let ${\overline{S}}_p$~ be the boundary of ${\overline{S}}_{p-1}$. The
Stoke's type formula for tensor--integral (4.34) is defined as

$$
I_{{\overline{S}}_p}N_{{\varphi }{.}{\overline{\tau }}{.}{\overline{\alpha }}%
_1{\ldots }{\overline{\alpha }}_p}^{{.}{\gamma }{.}{\overline{\kappa }}}dS^{{%
\overline{\alpha }}_1{\ldots }{\overline{\alpha }}_p}=
$$
$$
I_{{\overline{S}}_{p+1}}{^{{\star }{(p)}}{\overline{D}}}_{[{\overline{\gamma
}}{|}}N_{{\varphi }{.}{\overline{\tau }}{.}{|}{\overline{\alpha }}_1{\ldots }%
{{\overline{\alpha }}_p]}}^{{.}{\gamma }{.}{\overline{\kappa }}}dS^{{%
\overline{\gamma }}{\overline{\alpha }}_1{\ldots }{\overline{\alpha }}_p},%
\eqno(4.39)
$$
where
$$
{^{{\star }{(p)}}D}_{[{\overline{\gamma }}{|}}N_{{\varphi }{.}{\overline{%
\tau }}{.}{|}{\overline{\alpha }}_1{\ldots }{\overline{\alpha }}_p]}^{{.}{%
\gamma }{.}{\overline{\kappa }}}=
$$
$$
D_{[{\overline{\gamma }}{|}}N_{{\varphi }{.}{\overline{\tau }}{.}{|}{%
\overline{\alpha }}_1{\ldots }{\overline{\alpha }}_p]}^{{.}{\gamma }{.}{%
\overline{\kappa }}}+pT_{{.}[{\overline{\gamma }}{\overline{\alpha }}_1{|}}^{%
\underline{\epsilon }}N_{{\varphi }{.}{\overline{\tau }}{.}{\overline{%
\epsilon }}{|}{\overline{\alpha }}_2{\ldots }{\overline{\alpha }}_p]}^{{.}{%
\gamma }{.}{\overline{\kappa }}}-B_{[{\overline{\gamma }}{|}{\overline{\tau }%
}}^{{..}{\overline{\epsilon }}}N_{{\varphi }{.}{\overline{\epsilon }}{.}{|}{%
\overline{\alpha }}_1{\ldots }{\overline{\alpha }}_p]}^{{.}{\gamma }{.}{%
\overline{\kappa }}}+B_{[{\overline{\gamma }}{|}{\overline{\epsilon }}}^{..{%
\overline{\kappa }}}N_{{\varphi }{.}{\overline{\tau }}{.}{|}{\overline{%
\alpha }}_1{\ldots }{\overline{\alpha }}_p]}^{{.}{\gamma }{.}{\overline{%
\epsilon }}}.\eqno(4.40)
$$
We define the dual element of the hypersurfaces element $dS^{{j}_1{\ldots }{j%
}_p}$ as
$$
d{\cal S}_{{\beta }_1{\ldots }{\beta }_{q-p}}={\frac 1{{p!}}}{\epsilon }_{{%
\beta }_1{\ldots }{\beta }_{k-p}{\alpha }_1{\ldots }{\alpha }_p}dS^{{\alpha }%
_1{\ldots }{\alpha }_p},\eqno(4.41)
$$
where ${\epsilon }_{{\gamma }_1{\ldots }{\gamma }_q}$ is completely
antisymmetric on its indices and
$$
{\epsilon }_{12{\ldots }(n+m)}=\sqrt{{|}G{|}},G=det{|}G_{{\alpha }{\beta }{|}%
},
$$
$G_{{\alpha }{\beta }}$ is taken from (2.12). The dual of dc--tensor $N_{{%
\varphi }{.}{\overline{\tau }}{.}{\overline{\alpha }}_1{\ldots }{\overline{%
\alpha }}_p}^{{.}{\gamma }{\overline{\kappa }}}$ is defined as the
dc--tensor \\ ${\cal N}_{{\varphi }{.}{\overline{\tau }}}^{{.}{\gamma }{.}{%
\overline{\kappa }}{\overline{\beta }}_1{\ldots }{\overline{\beta }}_{n+m-p}}
$ satisfying
$$
N_{{\varphi }{.}{\overline{\tau }}{.}{\overline{\alpha }}_1{\ldots }{%
\overline{\alpha }}_p}^{{.}{\gamma }{.}{\overline{\kappa }}}={\frac 1{{p!}}}%
{\cal N}_{{\varphi }{.}{\overline{\tau }}}^{{.}{\gamma }{.}{\overline{\kappa
}}{\overline{\beta }}_1{\ldots }{\overline{\beta }}_{n+m-p}}{\epsilon }_{{%
\overline{\beta }}_1{\ldots }{\overline{\beta }}_{n+m-p}{\overline{\alpha }}%
_1{\ldots }{\overline{\alpha }}_p}.\eqno(4.42)
$$
Using (4.16), (4.41) and (4.42) we can write
$$
I_{{\overline{S}}_p}N_{{\varphi }{.}{\overline{\tau }}{.}{\overline{\alpha }}%
_1{\ldots }{\overline{\alpha }}_p}^{{.}{\gamma }{.}{\overline{\kappa }}}dS^{{%
\overline{\alpha }}_1{\ldots }{\overline{\alpha }}_p}={\int }_{{\overline{S}}%
_{p+1}}{^{\overline{p}}D}_{\overline{\gamma }}{\cal N}_{{\varphi }{.}{%
\overline{\tau }}}^{{.}{\gamma }{.}{\overline{\kappa }}{\overline{\beta }}_1{%
\ldots }{\overline{\beta }}_{n+m-p-1}{\overline{\gamma }}}d{\cal S}_{{%
\overline{\beta }}_1{\ldots }{\overline{\beta }}_{n+m-p-1}},\eqno(4.43)
$$
where
$$
{^{\overline{p}}D}_{\overline{\gamma }}{\cal N}_{{\varphi }{.}{\overline{%
\tau }}}^{{.}{\gamma }{.}{\overline{\kappa }}{\overline{\beta }}_1{\ldots }{%
\overline{\beta }}_{n+m-p-1}{\overline{\gamma }}}=
$$
$$
{\overline{D}}_{\overline{\gamma }}{\cal N}_{{\varphi }{.}{\overline{\tau }}%
}^{{.}{\gamma }{.}{\overline{\kappa }}{\overline{\beta }}_1{\ldots }{%
\overline{\beta }}_{n+m-p-1}{\overline{\gamma }}}+(-1)^{(n+m-p)}(n+m-p+1)T_{{%
.}{\overline{\gamma }}{\overline{\epsilon }}}^{[{\overline{\epsilon }}}{\cal %
N}_{{\varphi }{.}{{\overline{\tau }}}}^{{.}{|}{\gamma }{.}{\overline{\kappa }%
}{|}{\overline{\beta }}_1{\ldots }{\overline{\beta }}_{n+m-p-1}]{\overline{%
\gamma }}}-
$$
$$
B_{{\overline{\gamma }}{\overline{\tau }}}^{{..}{\overline{\epsilon }}}{\cal %
N}_{{\varphi }{.}{\overline{\epsilon }}}^{{.}{\gamma }{.}{\overline{\kappa }}%
{\overline{\beta }}_1{\ldots }{\overline{\beta }}_{n+m-p-1}{\overline{\gamma
}}}+B_{{\overline{\gamma }}{\overline{\epsilon }}}^{{..}{\overline{\kappa }}}%
{\cal N}_{{\varphi }{.}{\overline{\tau }}}^{{.}{\gamma }{.}{\overline{%
\epsilon }}{\overline{\beta }}_1{\ldots }{\overline{\beta }}_{n+m-p-1}{%
\overline{\gamma }}}.
$$
To verify the equivalence of (4.42) and (4.43) we must take in consideration
that
$$
D_\gamma {\epsilon }_{{\alpha }_1{\ldots }{\alpha }_k}=0\ \mbox{and}\ {%
\epsilon }_{{\beta }_1{\ldots }{\beta }_{n+m-p}{\alpha }_1{\ldots }{\alpha }%
_p}{\epsilon }^{{\beta }_1{\ldots }{\beta }_{n+m-p}{\gamma }_1{\ldots }{%
\gamma }_p}=p!(n+m-p)!{\delta }_{{\alpha }_1}^{[{\gamma }_1}{\cdots }{\delta
}_{{\alpha }_p}^{{\gamma }_p]}.
$$
The developed tensor integration formalism will be used in
the next subsection for definition of conservation laws on spaces with local
anisotropy.

\subsection{On Conservation Laws on La--Spaces}

To define conservation laws on locally anisotropic spaces is a challenging
task because of absence of global and local groups of automorphisms of such
spaces. Our main idea is to use chains of na--maps from a given, called
hereafter as the fundamental la--space to an auxiliary one with trivial
curvatures and torsions admitting a global group of automorphisms. The aim
of this subsection is to  formulate conservation laws for la-gravitational
fields by using dc--objects and tensor--integral values, na--maps and
variational calculus on the Category of la--spaces.

\subsubsection{Nonzero divergence of the energy--momentum d--tensor}

R. Miron and M. Anastasiei \cite{ma87,ma94} pointed to this specific form of
conservation laws of matter on la--spaces: They calculated the divergence of
the energy--momentum d--tensor on la--space $\xi ,$%
$$
D_\alpha {E}_\beta ^\alpha ={\frac 1{\ {\kappa }_1}}U_\alpha ,\eqno(4.44)
$$
and concluded that d--vector
$$
U_\alpha ={\frac 12}(G^{\beta \delta }{{R_\delta }^\gamma }_{\phi \beta }{T}%
_{\cdot \alpha \gamma }^\phi -G^{\beta \delta }{{R_\delta }^\gamma }_{\phi
\alpha }{T}_{\cdot \beta \gamma }^\phi +{R_\phi ^\beta }{T}_{\cdot \beta
\alpha }^\phi )
$$
vanishes if and only if d--connection $D$ is without torsion.

No wonder that conservation laws, in usual physical theories being a
consequence of global (for usual gravity of local) automorphisms of the
fundamental space--time, are more sophisticate on the spaces with local
anisotropy. Here it is important to emphasize the multiconnection character
of la--spaces. For example, for a d--metric (2.12) on $\xi $ we can
equivalently introduce another (see (2.21)) metric linear connection $%
\tilde D$ The Einstein equations
$$
{\tilde R}_{\alpha \beta }-{\frac 12}G_{\alpha \beta }{\tilde R}={\kappa }_1{%
\tilde E}_{\alpha \beta }\eqno(4.45)
$$
constructed by using connection (4.23) have vanishing divergences
$$
{\tilde D}^\alpha ({{\tilde R}_{\alpha \beta }}-{\frac 12}G_{\alpha \beta }{%
\tilde R})=0\mbox{ and }{\tilde D}^\alpha {\tilde E}_{\alpha \beta }=0,
$$
similarly as those on (pseudo)Riemannian spaces. We conclude that by using
the connection (2.21) we construct a model of la--gravity which looks like
locally isotropic on the total space $E.$ More general gravitational models
with local anisotropy can be obtained by using deformations of connection ${%
\tilde \Gamma }_{\cdot \beta \gamma }^\alpha ,$
$$
{{\Gamma }^\alpha }_{\beta \gamma }={\tilde \Gamma }_{\cdot \beta \gamma
}^\alpha +{P^\alpha }_{\beta \gamma }+{Q^\alpha }_{\beta \gamma },
$$
were, for simplicity, ${{\Gamma }^\alpha }_{\beta \gamma }$ is chosen to be
also metric and satisfy Einstein equations (4.45). We can consider
deformation d--tensors ${P^\alpha }_{\beta \gamma }$ generated (or not) by
deformations of type (4.9)--(4.11) for na--maps. In
this case d--vector $U_\alpha $ can be interpreted as a generic source of
local anisotropy on $\xi $ satisfying generalized conservation laws (4.44).

\subsubsection{Deformation d--tensors and tensor--integral conservation laws}

From (4.34) we obtain a tensor integral on ${\cal C}({\xi})$ of a d--tensor:
$$
N_{{\underline {\tau}}}^{{.}{\underline {\kappa}}}(\underline u)= I_{{%
\overline S}_{p}}N^{{..}{\overline {\kappa}}}_ {{\overline {\tau}}{..}{%
\overline {\alpha}}_{1}{\ldots}{\overline {\alpha}}_{p}} ({\overline u})h^{{%
\overline {\tau}}}_{{\underline {\tau}}}({\underline u}, {\overline u})h^{{%
\underline {\kappa}}}_{{\overline {\kappa}}} ({\overline u}, {\underline u}%
)dS^{{\overline {\alpha}}_{1}{\ldots} {\overline {\alpha}}_{p}}.%
$$

We point out that tensor--integral can be defined not only for dc--tensors
but and for d--tensors on $\xi $. Really, suppressing indices ${\varphi }$~
and ${\gamma }$~ in (4.42) and (4.43), considering instead of a deformation
dc--tensor a deformation tensor
$$
B_{{\alpha }{\beta }}^{{..}{\gamma }}(u,{\underline{u}})=B_{{\alpha }{\beta }%
}^{{..}{\gamma }}(u)=P_{{.}{\alpha }{\beta }}^\gamma (u)\eqno(4.46)
$$
(we consider deformations induced by a nc--transform) and integration\\ $%
I_{S_p}{\ldots }dS^{{\alpha }_1{\ldots }{\alpha }_p}$ in la--space $\xi $ we
obtain from (4.34) a tensor--integral on ${\cal C}({\xi })$~ of a
d--tensor:
$$
N_{{\underline{\tau }}}^{{.}{\underline{\kappa }}}({\underline{u}}%
)=I_{S_p}N_{{\tau }{.}{\alpha }_1{\ldots }{\alpha }_p}^{.{\kappa }}(u)h_{{%
\underline{\tau }}}^\tau ({\underline{u}},u)h_\kappa ^{\underline{\kappa }%
}(u,{\underline{u}})dS^{{\alpha }_1{\ldots }{\alpha }_p}.
$$
Taking into account (4.38) we can calculate that curvature
$$
{\underline{R}}_{{\gamma }{.}{\alpha }{\beta }}^{.{\lambda }}=D_{[{\beta }%
}B_{{\alpha }]{\gamma }}^{{..}{\lambda }}+B_{[{\alpha }{|}{\gamma }{|}}^{{..}%
{\tau }}B_{{\beta }]{\tau }}^{{..}{\lambda }}+T_{{.}{\alpha }{\beta }}^{{%
\tau }{..}}B_{{\tau }{\gamma }}^{{..}{\lambda }}
$$
of connection ${\underline{\Gamma }}_{{.}{\alpha }{\beta }}^\gamma (u)={%
\Gamma }_{{.}{\alpha }{\beta }}^\gamma (u)+B_{{\alpha }{\beta }{.}}^{{..}{%
\gamma }}(u),$ with $B_{{\alpha }{\beta }}^{{..}{\gamma }}(u)$~ taken from
(4.46), vanishes, ${\underline{R}}_{{\gamma }{.}{\alpha }{\beta }}^{{.}{%
\lambda }}=0.$ So, we can conclude that la--space $\xi $ admits a tensor
integral structure on ${\cal {C}}({\xi })$ for d--tensors associated to
deformation tensor $B_{{\alpha }{\beta }}^{{..}{\gamma }}(u)$ if the
nc--image ${\underline{\xi }}$~ is locally parallelizable. That way we
generalize the one space tensor integral constructions in \cite{goz,gv,v295a}%
, were the possibility to introduce tensor integral structure on a curved
space was restricted by the condition that this space is locally
parallelizable. For $q=n+m$~ relations (4.43), written for d--tensor ${\cal N%
}_{\underline{\alpha }}^{{.}{\underline{\beta }}{\underline{\gamma }}}$ (we
change indices ${\overline{\alpha }},{\overline{\beta }},{\ldots }$ into ${%
\underline{\alpha }},{\underline{\beta }},{\ldots })$ extend the Gauss
formula on ${\cal {C}}({\xi })$:
$$
I_{S_{q-1}}{\cal N}_{\underline{\alpha }}^{{.}{\underline{\beta }}{%
\underline{\gamma }}}d{\cal S}_{\underline{\gamma }}=I_{{\underline{S}}_q}{^{%
\underline{q-1}}D}_{{\underline{\tau }}}{\cal N}_{{\underline{\alpha }}}^{{.}%
{\underline{\beta }}{\underline{\tau }}}d{\underline{V}},\eqno(4.47)
$$
where $d{\underline{V}}={\sqrt{{|}{\underline{G}}_{{\alpha }{\beta }}{|}}}d{%
\underline{u}}^1{\ldots }d{\underline{u}}^q$ and
$$
{^{\underline{q-1}}D}_{{\underline{\tau }}}{\cal N}_{\underline{\alpha }}^{{.%
}{\underline{\beta }}{\underline{\tau }}}=D_{{\underline{\tau }}}{\cal N}_{%
\underline{\alpha }}^{{.}{\underline{\beta }}{\underline{\tau }}}-T_{{.}{%
\underline{\tau }}{\underline{\epsilon }}}^{{\underline{\epsilon }}}{\cal N}%
_{{\underline{\alpha }}}^{{\underline{\beta }}{\underline{\tau }}}-B_{{%
\underline{\tau }}{\underline{\alpha }}}^{{..}{\underline{\epsilon }}}{\cal N%
}_{{\underline{\epsilon }}}^{{.}{\underline{\beta }}{\underline{\tau }}}+B_{{%
\underline{\tau }}{\underline{\epsilon }}}^{{..}{\underline{\beta }}}{\cal N}%
_{{\underline{\alpha }}}^{{.}{\underline{\epsilon }}{\underline{\tau }}}.%
\eqno(4.48)
$$

Let consider physical values $N_{{\underline{\alpha }}}^{{.}{\underline{%
\beta }}}$ on ${\underline{\xi }}$~ defined on its density ${\cal N}_{{%
\underline{\alpha }}}^{{.}{\underline{\beta }}{\underline{\gamma }}},$ i. e.
$$
N_{{\underline{\alpha }}}^{{.}{\underline{\beta }}}=I_{{\underline{S}}_{q-1}}%
{\cal N}_{{\underline{\alpha }}}^{{.}{\underline{\beta }}{\underline{\gamma }%
}}d{\cal S}_{{\underline{\gamma }}}\eqno(4.49)
$$
with this conservation law (due to (4.47)):%
$$
{^{\underline{q-1}}D}_{{\underline{\gamma }}}{\cal N}_{{\underline{\alpha }}%
}^{{.}{\underline{\beta }}{\underline{\gamma }}}=0.\eqno(4.50)
$$
We note that these conservation laws differ from covariant conservation laws
for well known physical values such as density of electric current or of
energy-- momentum tensor. For example, taking density ${E}_\beta ^{{.}{%
\gamma }},$ with corresponding to (4.48) and (4.50) conservation law,
$$
{^{\underline{q-1}}D}_{{\underline{\gamma }}}{E}_{{\underline{\beta }}}^{{%
\underline{\gamma }}}=D_{{\underline{\gamma }}}{E}_{{\underline{\beta }}}^{{%
\underline{\gamma }}}-T_{{.}{\underline{\epsilon }}{\underline{\tau }}}^{{%
\underline{\tau }}}{E}_{{\underline{\beta }}}^{{.}{\underline{\epsilon }}%
}-B_{{\underline{\tau }}{\underline{\beta }}}^{{..}{\underline{\epsilon }}}{E%
}_{\underline{\epsilon }}^{{\underline{\tau }}}=0,\eqno(4.51)
$$
we can define values (see (4.47) and (4.49))
$$
{\cal P}_\alpha =I_{{\underline{S}}_{q-1}}{E}_{{\underline{\alpha }}}^{{.}{%
\underline{\gamma }}}d{\cal S}_{{\underline{\gamma }}}.
$$
Defined conservation laws (4.51) for ${E}_{{\underline{\beta }}}^{{.}{%
\underline{\epsilon }}}$ have nothing to do with those for energy--momentum
tensor $E_\alpha ^{{.}{\gamma }}$ from Einstein equations for the almost
Hermitian gravity \cite{ma87,ma94} or with ${\tilde E}_{\alpha \beta }$ from
(4.45) with vanishing divergence $D_\gamma {\tilde E}_\alpha ^{{.}{\gamma }%
}=0.$ So ${\tilde E}_\alpha ^{{.}{\gamma }}{\neq }{E}_\alpha ^{{.}{\gamma }}.
$ A similar conclusion was made in \cite{goz} for unispacial locally
isotropic tensor integral. In the case of multispatial tensor integration we
have another possibility (firstly pointed in \cite{vog,v295a} for
Einstein-Cartan spaces), namely, to identify ${E}_{{\underline{\beta }}}^{{.}%
{\underline{\gamma }}}$ from (4.51) with the na-image of ${E}_\beta ^{{.}{%
\gamma }}$ on la--space $\xi .$ We shall consider this construction in the
next subsection.

\subsection{Na--Conservation Laws in La--Gravity}

It is well known that the standard pseudo--tensor description of the
energy--momen\-tum values for the Einstein gravitational fields is full of
ambiguities. Some light was shed by introducing additional geometrical
structures on curved space--time (bimetrics \cite{ros,koh}, biconnections
\cite{che}, by taking into account background spaces \cite{gri,dewd}, or
formulating variants of general relativity theory on flat space \cite
{log,gip}).  We emphasize here that rigorous mathematical investigations
based on two (fundamental and background) locally anisotropic, or isotropic,
spaces should use well--defined, motivated from physical point of view,
mappings of these spaces. Our na--model largely contains both attractive
features of the mentioned approaches; na--maps establish a local 1--1
correspondence between the fundamental la--space and auxiliary la--spaces on
which biconnection (or even multiconnection) structures are induced. But
these structures are not a priory postulated as in a lot of gravitational
theories, we tend to specify them to be locally reductible to the locally
isotropic Einstein theory \cite{gri,log}.

Let us consider a fixed background la--space $\underline {\xi}$ with given
metric ${\underline G}_{\alpha \beta} = ({\underline g}_{ij} , {\underline h}%
_{ab} )$ and d--connection ${\underline {\tilde {\Gamma}}}^{\alpha}_{\cdot
\beta\gamma}.$ For simplicity, we consider compatible
metric and connections  being torsionless and with vanishing curvatures.
Supposing that there is an nc--transform from the fundamental la--space $\xi$
to the auxiliary $\underline {\xi} .$ we are interested in the equivalents
of the Einstein equations (4.45) on $\underline {\xi} .$

We consider that a part of gravitational degrees of freedom is "pumped out"
into the dynamics of deformation d--tensors for d--connection, ${P^{\alpha}}%
_{\beta \gamma},$ and metric, $B^{\alpha \beta} =  ( b^{ij} , b^{ab} ) .$
The remained part of degrees of freedom is coded into the metric ${%
\underline G}_{\alpha \beta}$ and d--connection ${\underline {\tilde
{\Gamma}}}^{\alpha}_{\cdot \beta \gamma} .$

Following \cite{gri,vrjp} we apply the first order formalism and consider $%
B^{\alpha \beta }$ and ${P^\alpha }_{\beta \gamma }$ as independent
variables on $\underline{\xi }.$ Using notations
$$
P_\alpha ={P^\beta }_{\beta \alpha },\quad {\Gamma }_\alpha ={{\Gamma }%
^\beta }_{\beta \alpha },
$$
$$
{\hat B}^{\alpha \beta }=\sqrt{|G|}B^{\alpha \beta },{\hat G}^{\alpha \beta
}=\sqrt{|G|}G^{\alpha \beta },{\underline{\hat G}}^{\alpha \beta }=\sqrt{|%
\underline{G}|}{\underline{G}}^{\alpha \beta }
$$
and making identifications
$$
{\hat B}^{\alpha \beta }+{\underline{\hat G}}^{\alpha \beta }={\hat G}%
^{\alpha \beta },{\quad }{\underline{\Gamma }}_{\cdot \beta \gamma }^\alpha -%
{P^\alpha }_{\beta \gamma }={{\Gamma }^\alpha }_{\beta \gamma },
$$
we take the action of la--gravitational field on $\underline{\xi }$ in this
form:
$$
{\underline{{\cal S}}}^{(g)}=-{(2c{\kappa }_1)}^{-1}\int {\delta }^qu{}{%
\underline{{\cal L}}}^{(g)},\eqno(4.52)
$$
where
$$
{\underline{{\cal L}}}^{(g)}={\hat B}^{\alpha \beta }(D_\beta P_\alpha
-D_\tau {P^\tau }_{\alpha \beta })+({\underline{\hat G}}^{\alpha \beta }+{%
\hat B}^{\alpha \beta })(P_\tau {P^\tau }_{\alpha \beta }-{P^\alpha }%
_{\alpha \kappa }{P^\kappa }_{\beta \tau })
$$
and the interaction constant is taken ${\kappa }_1={\frac{4{\pi }}{{c^4}}}k,{%
\quad }(c$ is the light constant and $k$ is Newton constant) in order to
obtain concordance with the Einstein theory in the locally isotropic limit.

We construct on $\underline{\xi }$ a la--gravitational theory with matter
fields (denoted as ${\varphi }_A$ with $A$ being a general index)
interactions by postulating this Lagrangian density for matter fields
$$
{\underline{{\cal L}}}^{(m)}={\underline{{\cal L}}}^{(m)}[{\underline{\hat G}%
}^{\alpha \beta }+{\hat B}^{\alpha \beta };{\frac \delta {\delta u^\gamma }}(%
{\underline{\hat G}}^{\alpha \beta }+{\hat B}^{\alpha \beta });{\varphi }_A;{%
\frac{\delta {\varphi }_A}{\delta u^\tau }}].\eqno(4.53)
$$

Starting from (4.52) and (4.53) the total action of la--gravity on $%
\underline{\xi }$ is written as
$$
{\underline{{\cal S}}}={(2c{\kappa }_1)}^{-1}\int {\delta }^qu{\underline{%
{\cal L}}}^{(g)}+c^{-1}\int {\delta }^{(m)}{\underline{{\cal L}}}^{(m)}.%
\eqno(4.54)
$$
Applying variational procedure on $\underline{\xi },$ similar to that
presented in \cite{gri} but in our case adapted to N--connection by using
derivations (2.4) instead of partial derivations, we derive from (4.54) the
la--gravitational field equations
$$
{\bf {\Theta }}_{\alpha \beta }={{\kappa }_1}({\underline{{\bf t}}}_{\alpha
\beta }+{\underline{{\bf T}}}_{\alpha \beta })\eqno(4.55)
$$
and matter field equations
$$
{\frac{{\triangle }{\underline{{\cal L}}}^{(m)}}{\triangle {\varphi }_A}}=0,%
\eqno(4.56)
$$
where $\frac{\triangle }{\triangle {\varphi }_A}$ denotes the variational
derivation.

In (4.55) we have introduced these values: the energy--momentum d--tensor
for la--gravi\-ta\-ti\-on\-al field
$$
{\kappa }_1{\underline{{\bf t}}}_{\alpha \beta }=({\sqrt{|G|}})^{-1}{\frac{%
\triangle {\underline{{\cal L}}}^{(g)}}{\triangle G^{\alpha \beta }}}%
=K_{\alpha \beta }+{P^\gamma }_{\alpha \beta }P_\gamma -{P^\gamma }_{\alpha
\tau }{P^\tau }_{\beta \gamma }+
$$
$$
{\frac 12}{\underline{G}}_{\alpha \beta }{\underline{G}}^{\gamma \tau }({%
P^\phi }_{\gamma \tau }P_\phi -{P^\phi }_{\gamma \epsilon }{P^\epsilon }%
_{\phi \tau }),\eqno(4.57)
$$
(where
$$
K_{\alpha \beta }={\underline{D}}_\gamma K_{\alpha \beta }^\gamma ,
$$
$$
2K_{\alpha \beta }^\gamma =-B^{\tau \gamma }{P^\epsilon }_{\tau (\alpha }{%
\underline{G}}_{\beta )\epsilon }-B^{\tau \epsilon }{P^\gamma }_{\epsilon
(\alpha }{\underline{G}}_{\beta )\tau }+
$$
$$
{\underline{G}}^{\gamma \epsilon }h_{\epsilon (\alpha }P_{\beta )}+{%
\underline{G}}^{\gamma \tau }{\underline{G}}^{\epsilon \phi }{P^\varphi }%
_{\phi \tau }{\underline{G}}_{\varphi (\alpha }B_{\beta )\epsilon }+{%
\underline{G}}_{\alpha \beta }B^{\tau \epsilon }{P^\gamma }_{\tau \epsilon
}-B_{\alpha \beta }P^\gamma {\quad }),
$$
$$
2{\bf \Theta }={\underline{D}}^\tau {\underline{D}}_{tau}B_{\alpha \beta }+{%
\underline{G}}_{\alpha \beta }{\underline{D}}^\tau {\underline{D}}^\epsilon
B_{\tau \epsilon }-{\underline{G}}^{\tau \epsilon }{\underline{D}}_\epsilon {%
\underline{D}}_{(\alpha }B_{\beta )\tau }
$$
and the energy--momentum d--tensor of matter
$$
{\underline{{\bf T}}}_{\alpha \beta }=2{\frac{\triangle {\cal L}^{(m)}}{%
\triangle {\underline{\hat G}}^{\alpha \beta }}}-{\underline{G}}_{\alpha
\beta }{\underline{G}}^{\gamma \delta }{\frac{\triangle {\cal L}^{(m)}}{%
\triangle {\underline{\hat G}}^{\gamma \delta }}}.\eqno(4.58)
$$
As a consequence of (4.56)--(4.58) we obtain the d--covariant on $\underline{%
\xi }$ conservation laws
$$
{\underline{D}}_\alpha ({\underline{{\bf t}}}^{\alpha \beta }+{\underline{%
{\bf T}}}^{\alpha \beta })=0.\eqno(4.59)
$$
We have postulated the Lagrangian density of matter fields (4.53) in a form
as to treat ${\underline{{\bf t}}}^{\alpha \beta }+{\underline{{\bf T}}}%
^{\alpha \beta }$ as the source in (4.55).

Now we formulate the main results of this subsection:

\begin{proposition}
The dynamics of the Einstein la--gravitational fields, modeled as solutions
of equations (4.45) and matter fields on la--space $\xi ,$ can be
equivalently locally modeled on a background la--space $\underline{\xi }$
provided with a trivial d-connection and metric structures having zero
d--tensors of torsion and curvature by field equations (4.55) and (4.56) on
condition that deformation tensor ${P^\alpha }_{\beta \gamma }$ is a
solution of the Cauchy problem posed for basic equations for a chain of
na--maps from $\xi $ to $\underline{\xi }.$
\end{proposition}

\begin{proposition}
Local, d--tensor, conservation laws for Einstein la--gravita\-ti\-on\-al
fields can be written in form (4.59) for la--gravita\-ti\-on\-al (4.57) and
matter (4.58) energy--momentum d--tensors. These laws are d--covariant on
the background space $\underline{\xi }$ and must be completed with invariant
conditions of type (4.12)--(4.15) for every deformation parameters of a chain
of na--maps from $\xi $ to $\underline{\xi }.$
\end{proposition}

The above presented considerations consist proofs of both propositions.

We emphasize that nonlocalization of both locally anisotrop\-ic and
isotropic gravitational energy--momentum values on the fundamental (locally
anisotrop\-ic or isotropic) space $\xi $ is a consequence of the absence of
global group automorphisms for generic curved spaces. Considering
gravitational theories from view of multispaces and their mutual maps {\quad
} (directed by the basic geometric structures on $\xi \quad $ such as {\quad
} N--connection, d--connection, d--torsion and d--curvature components, see
coefficients for basic na--equations (4.9)--(4.11)), we can formulate local
d--tensor conservation laws on auxiliary globally automorphic spaces being
related with space $\xi $ by means of chains of na--maps. Finally, we remark
that as a matter of principle we can use d--connection deformations in order
to modelate the la--gravitational interactions with nonvanishing torsion and
nonmetricity. In this case we must introduce a corresponding source in
(4.59) and define generalized conservation laws as in (4.44) \quad (see
similar details for locally isotropic generalizations of the Einstein
gravity in Refs \cite{vob13,vog,vb12}.

\subsection{Na--maps in Einstein Gravity}

The nearly autoparallel map methods plays also an important role in
formulation of conversation laws for locally isotropic gravity \cite
{vk,vast,vob13} (Einstein  gravity), \cite{vrjp} (Einstein-Cartan theory),
\cite{vb12,vog} (gauge gravity  and gauge fields), and \cite{vob12,vth}
(gravitation with torsion and nonmetricity). We shall present some examples
of na-maps for solutions of Einstein equations (for trivial N-connection
structures the equations (4.45) are usual gravitational field equations in
general relativity) and analyze the  problem of formulation laws on curved
locally anisotropic spaces).

\subsubsection{Nc--flat solutions of Einstein equations}

In order to illustrate some applications of the na-map theory we consider
three particular solutions of Einstein equations.

\paragraph{Example 1.}

In works \cite{bon,bonp} one introduced te metric
$$
d s^2 = g_{ij} d x^i d x^j =exp (2 \Omega (q)) (d t^2 -d x^2) -%
$$
$$
q^2 (exp(2\beta (q)) dy^2 + exp(-2\beta (q)) d z^2), \eqno(4.60)
$$
where $q=t-x .$ If functions $\Omega (q)$ and $\beta (q) ,$ depending only
the variable $q,$ solve equations $2 d\Omega /dq =q (d\beta (q)/dq),$ the
metric (4.60) satisfies the vacuum Einstein equations. In general this
metric describes curved spaces, but if function $\beta (q)$ is a solution of
equations
$$
q(d^2 \beta /dq^2 ) + 2 d\beta /dq =q^2 (d\beta /dq)^3%
$$
we have a flat space, i. e. $R^{\ i}_{j\ kl} =0.$

1a)\ Let us show that metric (4.60) admits also $na_{(3)} + na_{(1)}$-maps
to Min\-kow\-ski space. If $q=0$, using conformal rescaling
$$
{\tilde g}_{ij} =q^{-2} exp(-2\beta ) {\tilde g}_{ij},%
$$
and introducing new variables $x^a = (p,v,y,z) (a=0,1,2,3),$ where
$$
v=t+p,\ p= \frac{1}{2} \int dq q^{-2} exp[2(\Omega - \beta )]%
$$
and $S(p) = -4 {\beta} (q)_{|q=q(p)},$ we obtain
$$
na_{(3)} : ds^2 \to d{\tilde s}^2 =2 dp dv -dy^2 - exp{S(p)} dz^2 \eqno(4.61)
$$
with one non-zero component of connection ${\tilde {\Gamma}}^3_{03} = \frac{1%
}{2}{(dS/dp)} = \frac{1}{2} {\dot S}.$ A next possible step will be a
na-map to the flat space with metric $ds^2 = dp^2 - dv^2 - dy^2 - dz^2. $
Really, the set of values ${\tilde P}^3_{03} = - \frac{1}{2} {\dot R},$ the
rest of components of the deformation tensor being equal to zero, and $b_0 =
[2 \ddot S - (\dot S)^2 ]/ S, b_1 = b_2 = 0, b_3 =-\dot S , a_{ij}=0,
Q^i_{jk}=0$ solves equations (4.45) for a map $na_{(1)}: d {\tilde s}^2 \to d%
{\underline s}^2.$

1b)The metric $d {\tilde s}$ from (4.61) also admits a $na_{(2)}$-map to the
flat space (as N. Sinyukov pointed out \cite{sin} intersections of type of
na-maps are possible). For example, the set of values $r=(r_0 (p),0,0,0), {%
\nu}_b =0, \sigma = (0,0,0, {\sigma}_3 ),$
$$
{\sigma}_3 = -(2 F^3_{0(0)})^{-1} {\dot S}exp [-\int r_0 (p) dp + S(p)],
F^3_0 =F^3_{0(0)} exp [\int r_0 (p)dp -  S(p)],%
$$
constant $F^3_{0(0)}\neq 0,$ the rest of the affinor components being  equal
to zero, and $r(p)$ is an arbitrary function on $p$ satisfies basic
equations with $e=0$ for a map $na_{(2)}: ds^2 \to d{\tilde s}^2.$

1c) Metric $d{\tilde s}^2$ from (4.61) does not admit $na_{(3)}$-maps to the
flat space because, in general, conditions $^{(3)}{\tilde W}^a_{bcd}=0,$ are
not satisfied, for example, $^{(3)}{\tilde W}^1_{001} = \frac{1}{4} ( \ddot
S +\frac{1}{2} \dot S ).$

\paragraph{Example 2.}

The Peres's solution of Einstein equations \cite{per} is given by metric
$$
d s^2 = -dx^2 -dy^2 + 2 d\eta d\lambda - Q(x,y) N(\zeta ) d {\zeta}^2 ,
\eqno(4.62)
$$
where $x=x^1 , y = x^2 , \lambda = x^3 , \zeta = x^4 .$ Parametrizing $%
Q(x,y)=\frac{1}{4} (x^2 - y^2 ), N(\zeta ) =sin \zeta $ we obtain plane wave
solution of Einstein equations. Choosing
$$
Q(x,y) = - \frac{xy}{2(x^2 +y^2)^2} \mbox{ and } N\left( \zeta \right) =\{
\begin{array}{c}
\exp \left[ \left( b^2-\zeta ^2\right) \right] ,
\mbox{ if }\left| \zeta \right| <b; \\ 0\qquad ,\qquad \mbox{if }\left|
\zeta \right| \geq 0 ,
\end{array}
$$
where $b$ is a real constant, we obtain a wave packed solutions of Einstein
equations. By straightforward calculations we can convince ourselves that
values
$$
{\sigma}_1 = \frac{1}{2} Q_1 N, {\sigma}_2 = \frac{1}{2} Q_2 N, {\sigma}_3
=0, {\sigma}_4 = \frac{1}{2} Q {\dot N}, r_b =0, {\nu}_b =0,
$$
$$
Q_1 = {\frac{\partial Q}{\partial x}},\qquad {\dot N} = dN/d\zeta
$$
solve basic equations for a map $na_{(2)}: d{\underline s}^2 \to d s^2, $
where\\ ${\underline g}_{ij} =diag(-1,-1,-1,1).$ So, metrics of type (4.62),
being of type II, according to Petrov's classification \cite{pet}, admits $%
na_{(2)}$-maps to the flat space.

\paragraph{Example 3.}

Now, we shall show that by using superpositions of na-maps and imbeddings
into pseudo-Riemannian spaces with dimension $d>2$ we can construct metrics
satisfying Einstein equations. Let consider a 2-dimensional metric
$$
d s^2_{(2)} = -d^2 (ln t) + (b-c)^{-2} d t^2 ,%
$$
where $b$ and $c$  are constant and $t>0.$
Firstly, we use the conformal rescaling:
$$
na_{(3)} : d s^2_{(2)} \to d {\tilde s}^2_{(2)}=e^{2x} t^{2(b+c-1)}
ds^2_{(2)}
$$
and the canonical imbedding into a 3-dimensional space,
$$
d {\tilde s}^2_{(2)} \subset d s^2_{(3)} = e^{2x} t^{2(b+c+1)} (-d^2(lnt) +
(b-c)^{-2} d x^2 ) + d y^2 .%
$$
Then we consider a conformal rescaling of this 3-dimensional space,
$$
na_{(3)}: d s^2_{(3)} \to d {\tilde s}^2_{(3)} = e^{-4x} d s^2_{(3)}%
$$
and, the next, the canonical imbedding into a 4 dimensional
pseudo-Riemannian space
$$
d {\tilde s}^2_{(3)} \subset d {\tilde s}^2_{(4)} =%
$$
$$
e^{-4x} [e^{2x} t^{2(b+c+1)}(-d^2(lnt)+(b-c)^{-2}d x^2 ) + d y^2] + d z^2 .%
$$
Finally, after a conformal mapping from $d {\tilde s}^2_{(4)},$
$$
na_{(3)}: d{\tilde s}^2_{(4)} \to d s^2 =%
$$
$$
e^{2x} t^{-2(b+c)} \{ e^{-4x}[t^{2(b+c+1)}(-d^2 lnt) + (b-c)^{-2} d x^2 ) +
dy^2 ] + d z^2 \},%
$$
we obtain a metric $d s^2 $ describing a class of space-times with ideal
liquid matter (as a rule, some relations between constants $n$ and $m$ and
the physical parameters of the liquid medium are introduced, see details in
\cite{kra}).

Similar considerations, but in general containing investigations of more
complicated systems of first order Pfaff equations, show that a very large
class of the Einstein equations solutions can be locally generated by using
imbeddings and chains of na-maps from flat auxiliary background spaces of
lower dimensions than that of the fundamental space-time.

\subsubsection{Na--conservation laws in general relativity}

We have considered two central topics in the section IV: the first was the
generalization of the nearly autoparallel map theory to the case of spaces
with local anisotropy and the second was the formulation of the
gravitational models on locally anisotropic (or in particular cases,
isotropic) na-backgrounds. For trivial N-connection and d-torsion structures
and pseudo-Riemannian metric the la-gravitational field equations (4.45)
became usual Einstein equations for general relativity. The results of
subsection IV.E hold good in simplified locally isotropic version. In this
subsection we pay  a special attention to the very important problem of
definition of conservation  laws for gravitational fields.

A brief historical note is in order. A. Z. Petrov at the end of 60th
initiated \cite{petk,petd,petd1,petk1} a programme of modeling field
interactions and modeling general relativity on arbitrary pseudo-Riemannian
spaces. He advanced an original approach to the gravitational
energy-momentum problem.  After Petrov's death (1972) N. S. Sinykov finally
elaborated \cite{sin} the  geometrical theory of nearly geodesics maps
(ng-maps; on (pseudo-)Riemannian spaces geodesics coincide with
autoparallels) of affine connected spaces as an  extension for the
(n-2)-projective space theory \cite{vra,moc}, concircular  geometry \cite{ya}%
, holomorphic projective mappings of Kahlerian spaces \cite{ots} and
holomorphic projective correspondences for almost complex spaces \cite
{ishih,tas}. A part of Petrov's geometric purposes have been achieved, but
Sinyukov's works practically do not contain investigations and applications
in gravitational theories. In addition to the canonical formulation of the
general relativity theory in the framework of the pseudo-Riemannian
geometry, we tried to elaborate well-defined criterions for  conditions when
for gravitational (locally isotropic or anisotropic) field theories
equivalent reformulations in arbitrary (curved or flat) spaces are possible
\cite{vk,vast,vob13,vrjp,vb12,vog,vob12,vth}.

The concept of spaces na-maps implies a kind of space-field Poincare
conventionality \cite{poi,poi1} which, in our case, states the possibility
to  formulate physical theories equivalently on arbitrary curved, or flat,
spaces (na-backgrounds, being the images of na-map chains) with a further
choice of one of them to be the space-time from a view of convenience or
simplicity. If the existence of ideal probing bodies (not destroying the
space-time structure by measurements) is postulated, the fundamental
pseudo-Riemannian, or a corresponding generalization, structure can be
established experimentally. Mappings to other curved or flat spaces would be
considered as mathematical "tricks" for illustrating some properties and
possible transformations of field equations. We remark a different situation
in quantum gravity where ideal probing bodies are not introduced and one has
to consider all na-backgrounds as equal in rights. Perhaps, it is preferred
to concern quantum gravitational problems on the space-time category ${\cal C%
}( \eta )$ (see Sec. IV.B, Theorem IV.4 ; to analyze the  renormalization
of field interactions we shall extend our considerations on  the tangent
bundle, or bundles of higher tangence) in order to develop quantum field
theories with gravitational interactions on multispaces interrelated with
"quantized" na-maps, on the category theory see \cite{bucur}.

In spite of some common features between our approach and those based on
background methods \cite{dewd,gri,gip} the na-method gives rise to a
different class of gravitational theories. As a rule, background
investigations  are carried out in the linear approximation and do not take
into consideration the "back reaction" of perturbations. Sometimes one uses
successive approximations. In the framework of the na-map theory we do not
postulate the existence of any {\it a priori} given background space-time.
The auxiliary spaces (curved, or flat) being considered by us, should belong
to the set of spaces which are images of na-map chains from the fundamental
space-time. As a matter of principle such nc-transforms can be constructed
as to be directed by solutions of the Einstein equations; if the Cauchy
problem is posed for both field and basic na-maps equations the set of
possible na-backgrounds can be defined exactly and without any
approximations and additional suppositions. Physical processes can be
modeled locally equivalently on every admissible na-background space.
According to A. Z. Petrov \cite{petk,petd,petd1,petk1}, if we tend to model
the gravitational  interactions, for example, in the Minkowski space, the
space-time curvature  should be "pumped out" into a "gravitational force".
Free gravitating ideal  probing bodies do not "fall" along geodesics in a
such auxiliary flat space.  We have to generalize the concept of geodesics
by introducing nearly geodesics or, more generally, nearly autoparallels in
order to describe equivalently trajectories of point probing mass as well
field interactions on  the fundamental space-time and auxiliary
na-backgrounds.

So, we can conclude that to obtain an equivalent formulation of the general
relativity theory on na-backgrounds one has to complete the gravitational
field equations (4.55) with equations for na-map chains (minimal chains are
preferred).\\ Of course, one also has to consider two types of conservation
laws: for values of energy-momentum type, for gravitational fields, see
(4.59), and conditions of type (4.12),(4.13),(4.14) and (4.15) for na-maps
invariants.

We point out the solitarity of mappings to the Minkowski space-time. This
space  is not only a usual and convenient arena for developing new variants
of theoretical models and investigating conservation laws in already known
manners. Perhaps, the "simplest" pseudo-Euclidean background should be
considered as a primary essence from which, by using different types of
maps, more generalized curved spaces (auxiliary, or being fundamental
Einstein space-times) can be constructed. Here the question arises:\ May be
the  supplementary relations to the background field equations and
conservation  laws are motivated only by our requirement that mappings from
the curved space-time to a flat space to be obtained only by the mean of
na-maps and do not reflect any, additional to the background formulations,
properties of  the Einstein gravitational fields? The answer is related with
the problem of  concordance of geometrical structures on all spaces taken
into consideration.  Really, on backgrounds we have biconnection (or even
multi--connection structures). So, various types of geodesics, nearly
geodesics, motion and field equations, and a lot of other properties, being
characteristic for such spaces, can be defined. That is why, we need a
principle to base and describe  the splitting of geometrical structures by
1-1 local mappings from one space  to another (which we have advocated to be
the transformation of geodesics into nearly autoparallels). Of course, other
types of space mappings can be advanced. But always equations defined for
maps splittings of geometrical structures and corresponding invariants must
be introduced. This is the price  we shall pay if the general relativity is
considered out of its natural pseudo-Riemannian geometry. Sometimes such
digressions from the fundamental space-time geometry are very useful and, as
it was illustrated, gives us the possibility to formulate, in a well-known
for flat spaces manner, the conservation laws for gravitational fields.
However, all introduced ma-map  equations and invariant conditions are not
only motivated by a chosen type of  mappings' and auxiliary spaces'
artifacts. Being determined by the fundamental objects on curved space-time,
such as metric, connection and curvature, these na-relations pick up a part
of formerly unknown properties and symmetries of gravitational fields an can
be used (see Sec. IV.D) for a new type  (with respect to na-maps
flexibility) classification of curved spaces.

Finally, we remark that because na-maps consists a class of nearest
extensions of conformal rescalings we suppose them to play a preferred role
among others being more sophisticate and, at present time, less justified
from physical point of view.

\section{Higher Order Anisotropic Strings}

The relationship between two dimensional $\sigma $-models and strings has
been considered \cite{lov,fts,cfmp,sen,alw} in order to discuss the
effective low energy field equations for the massless models of strings. In
this section we shall study some of the problems associated with the theory
of higher order anisotropic strings being a natural generalization to higher
order anisotropic ) backgrounds (we shall write in brief ha-backgrounds,
ha-spaces and ha-geometry) of the Polyakov's covariant functional-integral
approach to string theory \cite{poly}. Our aim is to show that a
corresponding low-energy string dynamics contains the motion equations for
field equations on ha-spaces; models of ha-gravity could be more adequate
for investigation of quantum gravitational and Early Universe cosmology.

\subsection{Higher Order Anisotropic Sigma Models}

In this subsection we present a generalization of some necessary results on
nonlinear $\sigma$--model and string propagation to the case of
ha--backgrounds. Calculations on both type of locally anisotropic and
isotropic spaces are rather similar if we accept the Miron and Anastasiei
\cite{ma87,ma94} geometric formalism. We emphasize that on ha-backgrounds we
have to take into account the distinguished character, by N-connection, of
geometric objects.

\subsubsection{Action of nonlinear $\sigma$--model and torsion of ha-space}

Let a map of a two-dimensional (2d), for simplicity, flat space $M^2$ into
ha-space $\xi $ defines a $\sigma $ -model field $$u^{<\mu >}\left( z\right)
=\left( x^i\left( z\right) ,y^{<a>}\left( z\right) \right) =\left( x^i\left(
z\right) ,y^{a_1}\left( z\right) ,...,y^{a_z}\left( z\right) \right) ,$$
where $z=\left\{ z^A,A=0,1\right\} $ are two-dimensional complex coordinates
on $M^2.$ The moving of a bosonic string in ha--space is governed by the
nonlinear $\sigma $-model action (see, for instance, \cite{lov,fts,cfmp,sen}
for details on locally isotropic spaces):
$$
I=\frac 1{\lambda ^2}\int d^2z[\frac 12\sqrt{\gamma }\gamma ^{AB}\partial
_Au^{<\mu >}\left( z\right) \partial _Bu^{<\nu >}\left( z\right) G_{<\mu
><\nu >}\left( u\right) +
$$
$$
\frac{\widetilde{n}}3\epsilon ^{AB}\partial _Au^{<\mu >}\partial _Bu^{<\nu
>}b_{<\mu ><\nu >}\left( u\right) +\frac{\lambda ^2}{4\pi }\sqrt{\gamma }%
R^{(2)}\Phi \left( u\right) ],\eqno(5.1)
$$
where $\lambda ^2$ and $\widetilde{n}\,$ are interaction constants, $\Phi
\left( u\right) $ is the dilaton field, $R^{(2)}$ is the curvature of the 2d
world sheet provided with metric $\gamma _{AB},\gamma =\det \left( \gamma
_{AB}\right) $ and $\partial _A=\frac \partial {\partial z^A},$ tensor $%
\epsilon ^{AB}$ and d-tensor $b_{<\mu ><\nu >}$ are antisymmetric.

From the viewpoint of string theory we can interpret $b_{<\alpha ><\beta >}$
as the vacuum expectation of the antisymmetric, in our case locally
anisotropic, d-tensor gauge field $B_{<\alpha ><\beta ><\gamma >}$ (see
considerations for locally isotropic models in \cite{ts89,hul} and the
Wess-Zumino-Witten model \cite{wit,wesz}, which lead to the conclusion \cite
{bra} that $\widetilde{n}$ takes only integer values and that in the
perturbative quantum field theory the effective quantum action depends only
on B$_{...}$ and does not depend on b$_{...}$ ).

In order to obtain compatible with N-connection motions of ha-strings we
consider these relations between d-tensor $b_{<\alpha ><\beta >},$ strength $%
B_{<\alpha ><\beta ><\gamma >}=\delta _{[<\alpha >}b_{<\beta ><\gamma >]}$
and torsion $T_{.<\beta ><\gamma >}^{<\alpha >}:$%
$$
\delta _{<\alpha >}b_{<\beta ><\gamma >}=T_{<\alpha ><\beta ><\gamma >},%
\eqno(5.2)
$$
with the integrability conditions
$$
\Omega _{a_fb_f}^{<a_p>}\delta _{<a_p>}b_{<\beta ><\gamma >}=\delta
_{[a_f}T_{a_f]<\beta ><\gamma >},(0\leq f<p\leq z),\eqno(5.3)
$$
where $\Omega _{a_fb_f}^{<a_p>}$ are the coefficients of the N-connection
curvature. In this case we can express $B_{<\alpha ><\beta ><\gamma
>}=T_{[<\alpha ><\beta ><\gamma >]}.$ Conditions (5.2) and (5.3) define a
simplified model of ha--strings when the $\sigma $-model antisymmetric
strength is induced from the ha--background torsion. More general
constructions are possible by using normal coordinates adapted to both
N-connection and torsion structures on ha--background space. For simplicity,
we omit such considerations in this work.

Choosing the complex (conformal) coordinates $z=\iota ^0+i\iota ^1,\overline{%
z}=\iota ^0-i\iota ^1,$ where $\iota ^A,A=0,1$ are real coordinates, on the
world sheet we can represent the two-dimensional metric in the conformally
flat form:
$$
ds^2=e^{2\varphi }dzd\overline{z},\eqno(5.4)
$$
where $\gamma _{z\overline{z}}=\frac 12e^{2\varphi }$ and $\gamma
_{zz}=\gamma _{\overline{z}\overline{z}}=0.$

Let us consider an ha-field $U\left( u\right) ,u\in {\cal E}^{<z>}$ taking
values in ${\cal G}$ being the Lie algebra of a compact and semi simple Lie
group,%
$$
U\left( u\right) =\exp [i\varphi \left( u\right) ],\varphi \left( u\right)
=\varphi ^{\underline{\alpha }}\left( u\right) q^{\underline{\alpha }},
$$
where $q^{\underline{\alpha }}$ are generators of the Lie algebra with
antisymmetric structural constants $f^{\underline{\alpha }\underline{\beta }%
\underline{\gamma }}$ satisfying conditions%
$$
[q^{\underline{\alpha }},q^{\underline{\beta }}]=2if^{\underline{\alpha }%
\underline{\beta }\underline{\gamma }}q^{\underline{\gamma }},\quad tr(q^{%
\underline{\alpha }}q^{\underline{\beta }})=2\delta ^{\underline{\alpha }%
\underline{\beta }}.
$$

The action of the Wess-Zumino-Witten type ha--model should be written as
$$
I\left( U\right) =\frac 1{4\lambda ^2}\int d^2z\ tr\left( \partial
_AU\partial ^AU^{-1}\right) +\widetilde{n}\Gamma \left[ U\right] ,\eqno(5.5)
$$
where $\Gamma \left[ U\right] $ is the standard topologically invariant
functional \cite{bra}. For perturbative calculations in the framework of the
model (5.1) it is enough to know that as a matter of principle we can
represent the action of our theory as (5.5) and to use d-curvature $%
r_{<\beta >.<\gamma ><\delta >}^{.<\alpha >}$ for a torsionless d-connection
$\tau _{.<\beta ><\gamma >}^{<\alpha >},$ and strength $B_{<\alpha ><\beta
><\gamma >}$ respectively expressed as
$$
R_{<\alpha ><\beta ><\gamma ><\delta >}=f_{\underline{\alpha }\underline{%
\beta }\underline{\tau }}f_{\underline{\gamma }\underline{\delta }\underline{%
\tau }}V_{<\alpha >}^{\underline{\alpha }}V_{<\beta >}^{\underline{\beta }%
}V_{<\gamma >}^{\underline{\gamma }}V_{<\delta >}^{\underline{\delta }}
$$
and%
$$
B_{<\alpha ><\beta ><\gamma >}=\eta f_{\underline{\alpha }\underline{\beta }%
\underline{\tau }}V_{<\alpha >}^{\underline{\alpha }}V_{<\beta >}^{%
\underline{\beta }}V_\tau ^{\underline{\tau }},
$$
where a new interaction constant $\eta \equiv \frac{\widetilde{n}\lambda ^2}{%
2\pi }$ is used and $V_{<\alpha >}^{\underline{\alpha }}$ is a locally
adapted vielbein associated to the metric (2.12):%
$$
G_{<\alpha ><\beta >}=V_{<\alpha >}^{\underline{\alpha }}V_{<\beta >}^{%
\underline{\beta }}\delta ^{\underline{\alpha }\underline{\beta }}
$$
and
$$
G^{<\alpha ><\beta >}V_{<\alpha >}^{\underline{\alpha }}V_{<\beta >}^{%
\underline{\beta }}=\delta ^{\underline{\alpha }\underline{\beta }}.%
\eqno(5.6)
$$
For simplicity, we shall omit underlining of indices if this will not give
rise to ambiguities.

Finally,  we remark that for $\eta =1$ we obtain a
conformally invariant two-dimensional quantum field theory (being similar to
those developed in \cite{bel} ).

\subsubsection{The d--covariant method of ha--background field and $\sigma$%
--models}

Suggesting the compensation of all anomalies we can fix the gauge for the
two--dimensional metric when action (5.1) is written as
$$
I\left[ u\right] =\frac 1{2\lambda ^2}\int d^2z\{G_{<\alpha ><\beta >}\eta
^{AB}+\frac 23b_{<\alpha ><\beta >}\epsilon ^{AB}\}\partial _Au^{<\alpha
>}\partial _Bu^{<\beta >},\eqno(5.1a)
$$
where $\eta ^{AB}$ and $\epsilon ^{AB}$ are, respectively, constant
two-dimensional metric and antisymmetric tensor. The covariant method of
background field, as general references see \cite{ket,ketn,dew,agfm}, can be
extended for ha--spaces. Let consider a curve in ${\cal E}^{<z>}$
parameterized as $\rho ^{<\alpha >}\left( z,s\right) ,s\in [0,1],$
satisfying autoparallel equations%
$$
\frac{d^2\rho ^{<\alpha >}\left( z,s\right) }{ds^2}+\Gamma _{.<\beta
><\gamma >}^{<\alpha >}\left[ \rho \right] \frac{d\rho ^{<\beta >}}{ds}\frac{%
d\rho ^{<\gamma >}}{ds}=
$$
$$
\frac{d^2\rho ^{<\alpha >}\left( z,s\right) }{ds^2}+\tau _{.<\beta ><\gamma
>}^{<\alpha >}\left[ \rho \right] \frac{d\rho ^{<\beta >}}{ds}\frac{d\rho
^{<\gamma >}}{ds}=0,
$$
with boundary con\-di\-tions
$$
\rho \left( z,s=0\right) =u\left( z\right) \ \mbox{and}\ \rho \left(
z,s=1\right) u\left( z\right) +v\left( z\right) .
$$
For simplicity, hereafter we shall consider that d-con\-nec\-ti\-on ${{%
\Gamma }^{<\alpha >}}_{<\beta ><\gamma >}$ is defined by d--metric (2.12) and
N-con\-nec\-ti\-on structures, i.e.
$$
{{\Gamma }^{<\alpha >}}_{<\beta ><\gamma >}={}^{\circ }{{\Gamma }^{<\alpha >}%
}_{<\beta ><\gamma >}.
$$
The tangent d-vector $\zeta ^{<\alpha >}=\frac d{ds}\rho ^{<\alpha >},$
where $\zeta ^{<\alpha >}\mid _{s=0}=\zeta _{\left( 0\right) }^{<\alpha >}$
is chosen as the quantum d-field. Then the expansion of action $I[u+v\left(
\zeta \right) ],$ see (5.1a), as power series on $\zeta ,$
$$
I[u+v\left( \zeta \right) ]=\sum_{k=0}^\infty I_k,
$$
where
$$
I_k=\frac 1{k!}\frac{d^k}{ds^k}I\left[ \rho \left( s\right) \right] \mid
_{s=0},
$$
defines d-covariant, de\-pend\-ing on the background d-field,
in\-ter\-ac\-tion vortexes of locally an\-isot\-rop\-ic $\sigma $-model.

In order to compute $I_k\,$ it is useful to consider relations%
$$
\frac d{ds}\partial _A\rho ^{<\alpha >}=\partial _A\zeta ^{<\alpha >},\frac
d{ds}G_{<\alpha ><\beta >}=\zeta ^{<\tau >}\delta _{<\tau >}G_{<\alpha
><\beta >},
$$
$$
\partial _AG_{<\alpha ><\beta >}=\partial _A\rho ^{<\tau >}\delta _{<\tau
>}G_{<\alpha ><\beta >},
$$
to introduce auxiliary operators%
$$
\left( \widehat{\nabla }_A\zeta \right) ^{<\alpha >}=\left( \nabla _A\zeta
\right) ^{<\alpha >}-G^{<\alpha ><\tau >}T_{\left[ <\alpha ><\beta ><\gamma
>\right] }\left[ \rho \right] \epsilon _{AB}\partial ^B\rho ^{<\gamma
>}\zeta ^{<\beta >},
$$
$$
\left( \nabla _A\zeta \right) ^{<\alpha >}=\left[ \delta _{<\beta
>}^{<\alpha >}\partial _A+\tau _{.<\beta ><\gamma >}^{<\alpha >}\partial
_A\rho ^{<\gamma >}\right] \zeta ^{<\beta >},
$$
$$
\nabla \left( s\right) \xi ^{<\lambda >}=\zeta ^{<\alpha >}\nabla _{<\alpha
>}\xi ^{<\lambda >}=\frac d{ds}\xi ^{<\lambda >}+\tau _{.<\beta ><\gamma
>}^{<\lambda >}\left[ \rho \left( s\right) \right] \zeta ^{<\beta >}\xi
^{<\gamma >},\eqno(5.7)
$$
having properties%
$$
\nabla \left( s\right) \zeta ^{<\alpha >}=0,\nabla \left( s\right) \partial
_A\rho ^{<\alpha >}=\left( \nabla _A\zeta \right) ^{<\alpha >},
$$
$$
\nabla ^2\left( s\right) \partial _A\rho ^{<\alpha >}=r_{<\beta >.<\gamma
><\delta >}^{.<\alpha >}\zeta ^{<\beta >}\zeta ^{<\gamma >}\partial _A\rho
^{<\delta >},
$$
and to use the curvature d-tensor of d-connection (5.7),
$$
\widehat{r}_{<\beta ><\alpha ><\gamma ><\delta >}=r_{<\beta ><\alpha
><\gamma ><\delta >}-\nabla _{<\gamma >}T_{\left[ <\alpha ><\beta ><\delta
>\right] }+\nabla _{<\delta >}T_{\left[ <\alpha ><\beta ><\gamma >\right] }-
$$
$$
T_{\left[ <\tau ><\alpha ><\gamma >\right] }G^{<\tau ><\lambda >}T_{\left[
<\lambda ><\delta ><\beta >\right] }+T_{\left[ <\tau ><\alpha ><\delta
>\right] }G^{<\tau ><\lambda >}T_{\left[ <\lambda ><\gamma ><\beta >\right]
}.
$$

Values $I_k$ can be computed in a similar manner as in \cite{hul,bra,bel},
but in our case by using corresponding d-connections and d-objects. Here we
present the first four terms in explicit form:%
$$
I_1=\frac 1{2\lambda ^2}\int d^2z2G_{<\alpha ><\beta >}\left( \widehat{%
\nabla }_A\zeta \right) ^{<\alpha >}\partial ^Au^{<\beta >},\eqno(5.8)
$$
$$
I_2=\frac 1{2\lambda ^2}\int d^2z\{\left( \widehat{\nabla }_A\zeta \right)
^2+
$$
$$
\widehat{r}_{<\beta ><\alpha ><\gamma ><\delta >}\zeta ^{<\beta >}\zeta
^{<\gamma >}\left( \eta ^{AB}-\epsilon ^{AB}\right) \partial _Au^{<\alpha
>}\partial _Bu^{<\beta >}\},
$$
$$
I_3=\frac 1{2\lambda ^2}\int d^2z\{\frac 43(r_{<\beta ><\alpha ><\gamma
><\delta >}-
$$
$$
G^{<\tau ><\epsilon >}T_{\left[ <\epsilon ><\alpha ><\beta >\right]
}T_{\left[ <\tau ><\gamma ><\delta >\right] })\partial _Au^{<\alpha >}(%
\widehat{\nabla }^A\zeta ^{<\delta >})\zeta ^{<\beta >}\zeta ^{<\gamma >}+
$$
$$
\frac 43\nabla _{<\alpha >}T_{\left[ <\delta ><\beta ><\gamma >\right]
}\partial _Au^{<\beta >}\epsilon ^{AB}(\widehat{\nabla }_B\zeta ^{<\gamma
>})\zeta ^{<\alpha >}\zeta ^{<\delta >}+
$$
$$
\frac 23T_{\left[ <\alpha ><\beta ><\gamma >\right] }(\widehat{\nabla }%
_A\zeta ^{<\alpha >})\epsilon ^{AB}(\widehat{\nabla }\zeta ^{<\beta >})\zeta
^{<\gamma >}+
$$
$$
\frac 13\left( \nabla _{<\lambda >}r_{<\beta ><\alpha ><\gamma ><\delta
>}+4G^{<\tau ><\epsilon >}T_{\left[ <\epsilon ><\lambda ><\alpha >\right]
}\nabla _{<\beta >}T_{\left[ <\gamma ><\delta ><\tau >\right] }\right)
\times
$$
$$
\partial _Au^{<\alpha >}\partial ^Au^{<\delta >}\zeta ^{<\gamma >}\zeta
^{<\lambda >}+
$$
$$
\frac 13(\nabla _{<\alpha >}\nabla _{<\beta >}T_{\left[ <\tau ><\gamma
><\delta >\right] }+
$$
$$
2G^{<\lambda ><\epsilon >}G^{<\varphi ><\phi >}T_{\left[ <\alpha ><\lambda
><\varphi >\right] }T_{[<\epsilon ><\beta ><\delta >]}T_{\left[ <\phi ><\tau
><\gamma >\right] }+
$$
$$
2r_{<\alpha >.<\beta ><\gamma >}^{.<\lambda >}T_{\left[ <\alpha ><\tau
><\delta >\right] })\partial _Au^{<\gamma >}\epsilon ^{AB}\left( \partial
_B\zeta ^{<\tau >}\right) \zeta ^{<\alpha >}\zeta ^{<\beta >}\},
$$
$$
I_4=\frac 1{4\lambda ^2}\int d^2z\{(\frac 12\nabla _{<\alpha >}r_{<\gamma
><\beta ><\delta ><\tau >}-
$$
$$
G^{<\lambda ><\epsilon >}T_{\left[ <\epsilon ><\beta ><\gamma >\right]
}\nabla _{<\alpha >}T_{\left[ <\lambda ><\delta ><\tau >\right] })\times
$$
$$
\partial _Au^{<\beta >}(\widehat{\nabla }^A\zeta ^{<\tau >})\zeta ^{<\alpha
>}\zeta ^{<\gamma >}\zeta ^{<\delta >}+
$$
$$
\frac 13r_{<\beta ><\alpha ><\gamma ><\delta >}(\widehat{\nabla }_A\zeta
^{<\alpha >})(\widehat{\nabla }^A\zeta ^{<\delta >})\zeta ^{<\beta >}\zeta
^{<\gamma >}+
$$
$$
(\frac 1{12}\nabla _{<\alpha >}\nabla _{<\beta >}r_{<\delta ><\gamma ><\tau
><\lambda >}+\frac 13r_{<\delta >.<\tau ><\gamma >}^{.<\kappa >}r_{<\beta
><\kappa ><\alpha ><\lambda >}-
$$
$$
\frac 12(\nabla _{<\alpha >}\nabla _{<\beta >}T_{\left[ <\gamma ><\tau
><\epsilon >\right] })G^{<\epsilon ><\pi >}T_{\left[ <\pi ><\delta ><\lambda
>\right] }-
$$
$$
\frac 12r_{<\alpha >.<\beta ><\gamma >}^{.<\kappa >}T_{\left[ <\kappa ><\tau
><\epsilon >\right] }G^{<\epsilon ><\pi >}T_{\left[ <\pi ><\delta ><\lambda
>\right] }+
$$
$$
\frac 16r_{<\alpha >.<\beta ><\epsilon >}^{.<\kappa >}T_{\left[ <\kappa
><\gamma ><\tau >\right] }G^{<\epsilon ><\pi >}T_{\left[ <\epsilon ><\delta
><\lambda >\right] })\times
$$
$$
\partial _Au^{<\gamma >}\partial ^Au^{<\lambda >}\zeta ^{<\alpha >}\zeta
^{<\beta >}\zeta ^{<\delta >}\zeta ^{<\tau >}+
$$
$$
[\frac 1{12}\nabla _{<\alpha >}\nabla _{<\beta >}\nabla _{<\gamma
>}T_{\left[ <\lambda ><\delta ><\tau >\right] }+
$$
$$
\frac 12\nabla _{<\alpha >}\left( G^{<\kappa ><\epsilon >}T_{\left[
<\epsilon ><\lambda ><\delta >\right] }\right) r_{<\beta ><\kappa ><\gamma
><\tau >}+
$$
$$
\frac 12\left( \nabla _{<\alpha >}T_{\left[ <\pi ><\beta ><\kappa >\right]
}\right) G^{<\pi ><\epsilon >}G^{<\kappa ><\nu >}T_{\left[ <\epsilon
><\gamma ><\delta >\right] }T_{\left[ <\nu ><\lambda ><\tau >\right] }-
$$
$$
\frac 13G^{<\kappa ><\epsilon >}T_{\left[ <\epsilon ><\lambda ><\delta
>\right] }\nabla _{<\alpha >}r_{<\beta ><\kappa ><\delta ><\tau >}]\times
$$
$$
\partial _Au^{<\delta >}\epsilon ^{AB}\partial _Bu^{<\tau >}\zeta ^{<\lambda
>}\zeta ^{<\alpha >}\zeta ^{<\beta >}\zeta ^{<\gamma >}+
$$
$$
\frac 12[\nabla _{<\alpha >}\nabla _{<\beta >}T_{[<\tau ><\gamma ><\delta
>]}+T_{[<\kappa ><\delta ><\tau >]}r_{<\alpha >.<\beta ><\gamma >}^{.<\kappa
>}+
$$
$$
r_{<\alpha >.<\beta ><\delta >}^{.<\kappa >}T_{\left[ <\kappa ><\gamma
><\tau >\right] }]
$$
$$
\times \partial _Au^{<\gamma >}\epsilon ^{AB}(\widehat{\nabla }_B\zeta
^{<\delta >})\zeta ^{<\alpha >}\zeta ^{<\beta >}\zeta ^{<\tau >}+
$$
$$
\frac 12\nabla _{<\alpha >}T_{\left[ <\delta ><\beta ><\gamma >\right] }(%
\widehat{\nabla }_A\zeta ^{<\beta >})\epsilon ^{AB}(\widehat{\nabla }_B\zeta
^{<\gamma >})\zeta ^{<\alpha >}\zeta ^{<\delta >}\}.
$$

Now we con\-struct the d-covariant ha-background fun\-cti\-o\-nal (we use
meth\-ods, in our case cor\-res\-pond\-ing\-ly adapted to the
N-con\-nect\-i\-on structure, developed in \cite{hul,ketn,hs}). The standard
quantization technique is based on the functional integral
$$
Z\left[ J\right] =\exp \left( iW\left[ J\right] \right) =\int d[u]\exp
\{i\left( I+uJ\right) \},\eqno(5.9)
$$
with source $J^{<\alpha >}$ (we use condensed denotations and consider that
computations are made in the Euclidean space). The generation functional $%
\Gamma $ of one-particle irreducible (1PI) Green functions is defined as
$$
\Gamma \left[ \overline{u}\right] =W\left[ J\left( \overline{u}\right)
\right] -\overline{u}\cdot J\left[ \overline{u}\right] ,
$$
where $\overline{u}=\frac{\Delta W}{\delta J}$ is the mean field. For
explicit perturbative calculations it is useful to connect the source only
with the covariant quantum d-field $\zeta $ and to use instead of (5.9) the
new functional
$$
\exp \left( iW\left[ \overline{u},J\right] \right) =\int \left[ d\zeta
\right] \exp \{i\left( I\left[ \overline{u}+v\left( \zeta \right) \right]
+J\cdot \zeta \right) \}.\eqno(5.10)
$$

It is clear that Feynman diagrams obtained from this functional are
d-covariant.

Defining the mean d-field $\overline{\zeta }\left( u\right) =\frac{\Delta W}{%
\delta J\left( u\right) }$ and introducing the auxiliary d-field $\zeta
^{\prime }=\zeta -\overline{\zeta }$ we obtain from (2.9) a double expansion
on both classical and quantum ha--backgrounds:%
$$
\exp \left( i\overline{\Gamma [}\overline{u},\overline{\zeta ]}\right) =
$$
$$
\int \left[ d\zeta ^{\prime }\right] \exp \{i\left( I\left[ \overline{u}%
+v\left( \zeta ^{\prime }+\overline{\zeta }\right) \right] -\zeta ^{\prime }%
\frac{\Delta \Gamma }{\delta \overline{\zeta }}\right) \}.\eqno(5.11)
$$

The manner of fixing the measure in the functional (5.10) (and as a
consequence in (5.11) ) is obvious :
$$
[d\zeta ]=\prod_u\sqrt{\mid G\left( u\right) \mid }\prod_{<\alpha
>=0}^{n_E-1}d\zeta ^{<\alpha >}\left( u\right) .\eqno(5.12)
$$

Using vielbein fields (5.6) we can rewrite the measure (5.12) in the form
$$
\left[ d\zeta \right] =\prod_u\prod_{\alpha =0}^{n_E-1}d\zeta ^{\underline{%
\alpha }}\left( u\right) .
$$
The structure of renormalization of $\sigma $-models of type (5.10) (or
(5.11)) is analyzed, for instance, in \cite{hul,ketn,hs}. For ha--spaces we
must take into account the N--connection structure.

\subsection{Regularization and $\beta $--Funct\-i\-ons of HAS\--mo\-dels}

The aim of this subsection is to study the problem of regularization and
quantum ambiguities in $\beta$--functions of the renormalization group and
to present the results on one- and two-loop calculus for the
ha--$\sigma$--model ( HAS--model ).

\subsubsection{Renormalization group beta func\-ti\-ons}

Because our $\sigma $-model is a two-dimensional and massless locally
anisotropic theory we have to con\-sider both types of in\-frared and
ul\-tra\-vio\-let reg\-u\-lar\-iza\-tions (in brief, IR- and
UV-reg\-u\-lar\-iza\-tion). In order to regularize IR-divergences and
distinguish them from UV-divergences we can use a standard mass term in the
action (5.1) of the HAS--model%
$$
I_{\left( m\right) }=-\frac{\tilde m^2}{2\lambda ^2}\int d^2zG_{<\alpha
><\beta >}u^{<\alpha >}u^{<\beta >}.
$$
For regularization of UF-divergences it is convenient to use the dimensional
regularization. For instance, the regularized propagator of quantum d--fields
$\zeta $ looks like%
$$
<\zeta ^{<\alpha >}\left( u_1\right) \zeta ^{<\beta >}\left( u_2\right)
>=\delta ^{<\alpha ><\beta >}G\left( u_1-u_2\right) =
$$
$$
i\lambda ^2\delta ^{<\alpha ><\beta >}\int \frac{d^qp}{\left( 2\pi \right) ^q%
}\frac{\exp \left[ -ip\left( u_1-u_2\right) \right] }{\left( p^2-\tilde
m^2+i0\right) },
$$
where $q=2-2\epsilon .$

The d-covariant dimensional regularization of UF-divergences is complicated
because of existence of the antisymmetric symbol $\epsilon ^{AB}.$ One
introduces \cite{ketnp,ketn} this general prescription:%
$$
\epsilon ^{LN}\eta _{NM}\epsilon ^{MR}=\psi \left( \epsilon \right) \eta
^{LR}
$$
and
$$
\epsilon ^{MN}\epsilon ^{RS}=\omega \left( \epsilon \right) \left[ \eta
^{MS}\eta ^{NR}-\eta ^{MR}\eta ^{NS}\right] ,
$$
where $\eta ^{MN}$ is the q-dimensional Minkowski metric, and $\psi \left(
\epsilon \right) $ and $\omega \left( \epsilon \right) $ are arbitrary
d-functions satisfying conditions $\psi \left( 0\right) =\omega \left(
0\right) =1$ and depending on the type of renormalization.

We use the standard dimensional regularization , with dimensionless scalar
d-field $u^{<\alpha >}\left( z\right) ,$ when expressions for unrenormalized
$G_{<\alpha ><\beta >}^{\left( ur\right) }$ and $B_{<\alpha ><\beta
>}^{(k,l)}$ have a d-tensor character, i.e. they are polynoms on d-tensors
of curvature and torsion and theirs d-covariant derivations (for simplicity
  we consider $\lambda ^2=1;$ in general one-loop
1PI-diagrams must be proportional to $\left( \lambda ^2\right) ^{l-1}$).

RG $\beta $-functions are defined by relations (for simplicity we shall omit
index R for renormalized values)
$$
\mu \frac d{d\mu }G_{<\alpha ><\beta >}=\beta _{\left( <\alpha ><\beta
>\right) }^G\left( G,B\right) ,\mu \frac d{d\mu }B_{\left[ <\alpha ><\beta
>\right] }=\beta _{\left[ <\alpha ><\beta >\right] }^B\left( G,B\right) ,
$$
$$
\beta _{<\alpha ><\beta >}=\beta _{\left( <\alpha ><\beta >\right) }^G+\beta
_{\left[ <\alpha ><\beta >\right] }^B.
$$

By using the scaling property of the one--loop counter--term under global
conformal transforms%
$$
G_{<\alpha ><\beta >}^G\rightarrow \Lambda ^{\left( l-1\right) }G_{<\alpha
><\beta >}^{\left( k,l\right) },B_{<\alpha ><\beta >}^{\left( k,l\right)
}\rightarrow \Lambda ^{\left( l-1\right) }B_{<\alpha ><\beta >}^{\left(
k,l\right) }
$$
we obtain%
$$
\beta _{\left( <\alpha ><\beta >\right) }^G=-\sum_{l=1}^{(1,l)}lG_{(<\alpha
><\beta >)}^{(1,l)},\beta _{[<\alpha ><\beta >]}^B=-\sum_{l=1}^\infty
lB_{[<\alpha ><\beta >]}^{(1,l)}
$$
in the leading order on $\epsilon $ (compare with the usual perturbative
calculus from \cite{thoo}).

The d-covariant one-loop counter-term is taken as%
$$
\Delta I^{\left( l\right) }=\frac 12\int d^2zT_{<\alpha ><\beta >}^{\left(
l\right) }\left( \eta ^{AB}-\epsilon ^{AB}\right) \partial _Au^{<\alpha
>}\partial _Bu^{<\beta >},
$$
where
$$
T_{<\alpha ><\beta >}^{\left( l\right) }=\sum_{k=1}^l\frac 1{\left(
2\epsilon ^k\right) }T_{<\alpha ><\beta >}^{\left( k,l\right) }\left(
G,B\right) .\eqno(5.13)
$$
For instance, in the three-loop approximation we have
$$
\beta _{<\alpha ><\beta >}=T_{<\alpha ><\beta >}^{(1,1)}+2T_{<\alpha ><\beta
>}^{(1,2)}+3T_{<\alpha ><\beta >}^{(1,3)}.\eqno(5.14)
$$

The next step is to consider constraints on the structure
of $\beta $-functions connected with conditions of integrability (caused by
conformal invariance of the two-dimensional world-sheet).

\subsubsection{One--loop divergences and ha--renorm gro\-up equ\-ations}

We generalize the one-loop results \cite{cz} to the case of ha-backgrounds.
If in locally isotropic models one considers an one-loop diagram, for the
HAS-model the distinguished by N-connection character of ha-interactions
leads to the necessity to consider one-loop diagrams (see Fig. 1).To these
diagrams one corresponds counter-terms:%
$$
I_1^{\left( c\right)
}=I_1^{(c,x^2)}+I_1^{(c,y^2)}+I_1^{(c,xy)}+I_1^{(c,yx)}=
$$
$$
-\frac 12I_1\int d^2z\widehat{r}_{ij}\left( \eta ^{AB}-\epsilon ^{AB}\right)
\partial _Ax^i\partial _Bx^j-
$$
$$
\frac 12I_1\int d^2z\widehat{r}_{<a><b>}\left( \eta ^{AB}-\epsilon
^{AB}\right) \partial _Ay^{<a>}\partial _By^{<b>}-
$$
$$
\frac 12I_1\int d^2z\widehat{r}_{i<a>}\left( \eta ^{AB}-\epsilon
^{AB}\right) \partial _Ax^i\partial _By^{<a>}-
$$
$$
\frac 12I_1\int d^2z\widehat{r}_{<a>i}\left( \eta ^{AB}-\epsilon
^{AB}\right) \partial _Ay^{<a>}\partial _Bx^i,
$$
where $I_1$ is the standard integral%
$$
I_1=\frac{G\left( 0\right) }{\lambda ^2}=i\int \frac{d^qp}{(2\pi )^2}\frac
1{p^2-\tilde m^2}=\frac{\Gamma \left( \epsilon \right) }{4\pi ^{\frac
q2}(\tilde m^2)^\epsilon }=
$$
$$
\frac 1{4\pi \epsilon }-\frac 1\pi \ln \tilde m+\mbox{finite counter terms}.
$$

There are one--loops on the base and fiber spaces or describing quantum
interactions between fiber and base components of d-fields. If the
ha-background d-connection is of distinguished Levi-Civita type we obtain
only two one-loop diagrams (on the base and in the fiber) because in this
case the Ricci d-tensor is symmetric. It is clear that this four-multiplying
(doubling for the Levi-Civita d-connection) of the number of one-loop
diagrams is caused by the ''indirect'' interactions with the N-connection
field. Hereafter, for simplicity, we shall use a compactified
(non-distinguished on x- and y-components) form of writing out diagrams and
corresponding formulas and emphasize that really all expressions containing
components of d-torsion generate irreducible types of diagrams (with
respective interaction constants) and that all expressions containing
components of d-curvature give rise in a similar manner to irreducible types
of diagrams. We shall take into consideration these details in the
Sec. V.B.3 where we shall write the two-loop effective action.

Subtracting in a trivial manner $I_1,$
$$
I_1+subtractions=\frac 1{4\pi \epsilon },
$$
we can write the one-loop $\beta $-function in the form:%
$$
\beta _{<\alpha ><\beta >}^{\left( 1\right) }=\frac 1{2\pi }\widehat{r}%
_{<\alpha ><\beta >}=\frac 1{2\pi }(r_{<\alpha ><\beta >}-
$$
$$
G_{<\alpha ><\tau >}T^{\left[ <\tau ><\gamma ><\phi >\right] }T_{\left[
<\beta ><\gamma ><\phi >\right] }+G^{<\tau ><\mu >}\nabla _{<\mu >}T_{\left[
<\alpha ><\beta ><\tau >\right] }).
$$

We also note that the mass term in the action generates the mass one-loop
counter-term
$$
\Delta I_1^{(m)}=\frac{\tilde m^2}2I_1\int d^2z\{\frac 13r_{<\alpha ><\beta
>}u^{<\alpha >}u^{<\beta >}-u_{<\alpha >}\tau _{.<\beta ><\gamma >}^{<\alpha
>}G^{<\beta ><\gamma >}\}.
$$

The last two formulas can be used for a study of effective charges as in
\cite{bra} where some solutions of RG-equations are analyzed. We shall not
consider in this work such methods connected with the theory of differential
equations.

\subsubsection{Two-loop $\beta $-functions for the HAS-model}

In order to obtain two--loops of the HAS--model we add to the list (5.9) the
expansion
$$
\Delta I_{1\mid 2}^{\left( c\right) }=-\frac 12I_1\int d^2z\{\widehat{r}%
_{<\alpha ><\beta >}\left( \eta ^{AB}-\epsilon ^{AB}\right) \left( \widehat{%
\nabla }_A\zeta ^{<\alpha >}\right) \left( \widehat{\nabla }_B\zeta ^{<\beta
>}\right) +
$$
$$
(\nabla _\tau \widehat{r}_{<\alpha ><\beta >}+\widehat{r}_{<\alpha ><\gamma
>}G^{<\gamma ><\delta >}T_{\left[ <\delta ><\beta ><\tau >\right] })\times
$$
$$
\left( \eta ^{AB}-\epsilon ^{AB}\right) \zeta ^{<\tau >}(\widehat{\nabla }%
_A\zeta ^{<\alpha >})\partial _Bu^{<\beta >}+
$$
$$
\left( \nabla _{<\tau >}\widehat{r}_{<\alpha ><\beta >}-T_{\left[ <\alpha
><\tau ><\gamma >\right] }\widehat{r}_{.<\beta >}^{<\gamma >}\right) (\eta
^{AB}-\epsilon ^{AB})\partial _Au^{<\alpha >}\left( \widehat{\nabla }_B\zeta
^{<\beta >}\right) \zeta ^{<\tau >}+
$$
$$
\left( \eta ^{AB}-\epsilon ^{AB}\right) (\frac 12\nabla _{<\gamma >}\nabla
_{<\tau >}\widehat{r}_{<\alpha ><\beta >}+\frac 12\widehat{r}_{<\epsilon
><\beta >}r_{<\gamma >.<\tau ><\alpha >}^{.<\epsilon >}+
$$
$$
\frac 12\widehat{r}_{<\alpha ><\epsilon >}r_{<\gamma >.<\tau ><\beta
>}^{.<\epsilon >}+
$$
$$
T_{\left[ <\alpha ><\tau ><\epsilon >\right] }\widehat{r}^{<\epsilon
><\delta >}T_{\left[ <\delta ><\gamma ><\beta >\right] }+G^{<\mu ><\nu
>}T_{\left[ <\nu ><\beta ><\gamma >\right] }\nabla _{<\tau >}\widehat{r}%
_{<\alpha ><\mu >}-
$$
$$
G^{<\mu ><\nu >}T_{\left[ <\alpha ><\gamma ><\nu >\right] }\nabla _{<\tau >}%
\widehat{r}_{<\mu ><\beta >})\partial _Au^{<\alpha >}\partial _Bu^{<\beta
>}\zeta ^{<\tau >}\zeta ^{<\gamma >}\}
$$
and the d--covariant part of the expansion for the one--loop mass
counter-term%
$$
\Delta I_{1\mid 2}^{\left( m\right) }=\frac{\left( \tilde m\right) ^2}%
2I_1\int d^2z\frac 13r_{<\alpha ><\beta >}\zeta ^{<\alpha >}\zeta ^{<\beta
>}.
$$

The non-distinguished diagrams defining two--loop divergences are illustrated
in Fig.2. We present the explicit form of corresponding counter-terms
computed by using, in our case adapted to ha-backgrounds, methods developed
in \cite{ketn,ketnp}:

For counter-term of the diagram $(\alpha )$ we obtain
$$
(\alpha )=-\frac 12\lambda ^2(I_i)^2\int d^2z\{(\frac 14\Delta r_{<\delta
><\varphi >}-\frac 1{12}\nabla _{<\delta >}\nabla _{<\varphi >}%
\overleftarrow{r}+
$$
$$
\frac 12r_{<\delta ><\alpha >}r_{<\varphi >}^{<\alpha >}-\frac 16r_{.<\delta
>.<\psi >}^{<\alpha >.<\beta >}r_{<\alpha ><\beta >+}\frac 12r_{.<\delta
>}^{<\alpha >.<\beta ><\gamma >}r_{<\alpha ><\varphi ><\beta ><\gamma >}+
$$
$$
\frac 12G_{<\delta ><\tau >}T^{\left[ <\tau ><\alpha ><\beta >\right]
}\Delta T_{\left[ <\varphi ><\alpha ><\beta >\right] }+
$$
$$
\frac 12G_{<\varphi ><\tau >}r_{<\delta >}^{<\alpha >}T_{\left[ <\alpha
><\beta ><\gamma >\right] }T^{\left[ <\tau ><\beta ><\gamma >\right] }-
$$
$$
\frac 16G^{<\beta ><\tau >}T_{\left[ <\delta ><\alpha ><\beta >\right]
}T_{\left[ <\varphi ><\gamma ><\tau >\right] }r^{<\alpha ><\gamma >}+
$$
$$
G^{<\gamma ><\tau >}T_{[<\delta ><\alpha ><\tau >]}\nabla ^{(<\alpha
>}\nabla ^{<\beta >)}T_{\left[ <\varphi ><\beta ><\gamma >\right] }+
$$
$$
\frac 34G^{<\kappa ><\tau >}r_{.<\delta >}^{<\alpha >.<\beta ><\gamma
>}T_{\left[ <\beta ><\gamma ><\kappa >\right] }T_{\left[ <\alpha ><\varphi
><\tau >\right] }-
$$
$$
\frac 14r^{<\kappa ><\alpha ><\beta ><\gamma >}T_{\left[ <\delta ><\beta
><\gamma >\right] }T_{\left[ <\kappa ><\alpha ><\varphi >\right] })\partial
_Au^{<\delta >}\partial ^Au^{<\varphi >}+
$$
$$
\frac 14[\nabla ^{<\beta >}\Delta T_{\left[ <\delta ><\varphi ><\beta
>\right] }-3r_{...<\delta >}^{<\gamma ><\beta ><\alpha >}\nabla _{<\alpha
>}T_{\left[ <\beta ><\gamma ><\varphi >\right] }-
$$
$$
3T_{\left[ <\alpha ><\beta ><\delta >\right] }\nabla ^{<\gamma
>}r_{..<\gamma ><\psi >}^{<\beta ><\alpha >}+
$$
$$
\frac 14r^{<\alpha ><\gamma >}\nabla _{<\alpha >}T_{\left[ <\gamma ><\delta
><\varphi >\right] }+\frac 16T_{\left[ <\delta ><\varphi ><\alpha >\right]
}\nabla ^{<\alpha >}r-
$$
$$
4G^{<\gamma ><\tau >}T_{\left[ <\tau ><\beta ><\delta >\right] }T_{\left[
<\alpha ><\kappa ><\varphi >\right] }\nabla ^{<\beta >}(G_{<\gamma
><\epsilon >}T^{\left[ <\alpha ><\kappa ><\epsilon >\right] })+
$$
$$
2G_{<\delta ><\tau >}G^{<\beta ><\epsilon >}\nabla _{<\alpha >}(G^{<\alpha
><\nu >}T_{\left[ <\nu ><\beta ><\gamma >\right] })\times
$$
$$
T^{\left[ <\gamma ><\kappa ><\tau >\right] }T_{\left[ <\epsilon ><\kappa
><\varphi >\right] }]\epsilon ^{AB}\partial _Au^{<\delta >}\partial
_Bu^{<\varphi >}\}.
$$

In order to computer the counter-term for diagram $(\beta )$ we use
integrals:%
$$
\lim \limits_{u\rightarrow v}i<\partial _A\zeta \left( u\right) \partial
^A\zeta \left( v\right) >=i\int \frac{d^qp}{(2\pi )^q}\frac{p^2}{p^2-\tilde
m^2}=\tilde m^2I_1
$$
(containing only a IR-divergence) and
$$
J\equiv i\int \frac{d^2p}{(2\pi )^2}\frac 1{(p^2-\tilde m^2)^2}=-\frac
1{(2\pi )^2}\int d^2k_E\frac 1{(k_E^2+\tilde m^2)^2}.
$$
(being convergent). In result we can express%
$$
(\beta )=\frac 16\lambda ^2\left( I_1^2+2\tilde mI_1J\right) \times
$$
$$
\int d^2z\widehat{r}_{<\beta ><\alpha ><\gamma ><\delta >}r^{<\beta ><\gamma
>}\left( \eta ^{AB}-\epsilon ^{AB}\right) \partial _Au^{<\alpha >}\partial
_Bu^{<\beta >}.
$$
In our further considerations we shall use identities (we can verify them by
straightforward calculations):%
$$
\widehat{r}_{(<\alpha >.<\beta >)}^{.[<\gamma >.<\delta >]}=-\nabla
_{(<\alpha >}(G_{<\beta >)<\tau >}T^{[<\tau ><\gamma ><\delta >]}),
$$
$$
\widehat{r}_{\left[ <\beta ><\alpha ><\gamma ><\delta >\right] }=2G^{<\kappa
><\tau >}T_{\left[ <\tau >[<\alpha ><\beta >\right] }T_{\left[ <\gamma
><\delta >]<\kappa >\right] ,}
$$
in the last expression we have three type of antisymmetrizations on indices,
$\left[ <\tau ><\alpha ><\beta >\right] ,$ $\left[ <\gamma ><\delta ><\kappa
>\right] $ and $\left[ <\alpha ><\beta ><\gamma ><\delta >\right] ,$
$$
\widehat{\nabla }_{<\delta >}T_{\left[ <\alpha ><\beta ><\gamma >\right] }%
\widehat{\nabla }_{<\varphi >}T^{\left[ <\alpha ><\beta ><\gamma >\right]
}=\frac 9{16}\left( \widehat{r}_{\left[ \beta \alpha \gamma \right] \delta }-%
\widehat{r}_{\delta \left[ \alpha \beta \gamma \right] }\right) \times
\eqno(5.15)
$$
$$
\left( \widehat{r}_{...........<\varphi >}^{\left[ <\beta ><\alpha ><\gamma
>\right] }-\widehat{r}_{...........<\varphi >}^{.\left[ <\alpha ><\beta
><\gamma >\right] }\right) -\frac 94\widehat{r}_{\left[ <\alpha ><\beta
><\gamma ><\delta >\right] }\widehat{r}_{.........<\varphi >]}^{[<\alpha
><\beta ><\gamma >}+
$$
$$
\frac 94\widehat{r}_{......[<\delta >}^{<\alpha ><\beta ><\gamma >}\widehat{r%
}_{\left[ <\varphi >]<\alpha ><\beta ><\gamma >\right] }+\frac 94\widehat{r}%
_{.[<\delta >}^{<\alpha >.<\beta ><\gamma >}\widehat{r}_{\left[ <\varphi
>]<\alpha ><\beta ><\gamma >\right] }.
$$

The momentum integral for the first of diagrams $(\gamma )$
$$
\int \frac{d^qpd^qp^{\prime }}{(2\pi )^{2q}}\frac{p_Ap_B}{(p^2-\tilde
m^2)([k+q]^2-\tilde m^2)([p+q]^2-\tilde m^2)}
$$
diverges for a vanishing exterior momenta $k_{<\mu >}.$The explicit calculus
of the corresponding counter-term results in
$$
\gamma _1=-\frac{2\lambda ^2}{3q}I_1^2\int d^2z\{(r_{<\alpha >(<\beta
><\gamma >)<\delta >}+\eqno(5.16)
$$
$$
G^{<\varphi ><\tau >}T_{[<\tau ><\alpha >(<\beta >]}T_{[<\gamma >)<\delta
><\varphi >]})\times
$$
$$
\left( r_{.<\mu >}^{<\beta >.<\gamma ><\delta >}-G_{<\mu ><\tau >}G_{<\kappa
><\epsilon >}T^{[<\tau ><\beta ><\kappa >]}T^{\left[ <\gamma ><\delta
><\epsilon >\right] }\right) \partial ^Au^{<\alpha >}\partial _Au^{<\mu >}+
$$
$$
(\nabla _{(<\beta >}T_{[<\delta >)<\alpha ><\gamma >]})\nabla ^{<\beta
>}(G_{<\mu ><\tau >}T^{[<\tau ><\gamma ><\delta >]})$$
$$\epsilon ^{LN}\eta
_{NM}\epsilon ^{MR}\partial _Lu^{<\alpha >}\partial _Ru^{<\mu >}-
$$
$$
2(r_{<\alpha >(<\beta ><\gamma >)<\delta >}+G^{<\varphi ><\tau >}T_{[<\alpha
><\tau >(<\beta >]}T_{[<\gamma >)<\delta ><\varphi >]})\times
$$
$$
\nabla ^{<\beta >}(G_{<\mu ><\epsilon >}T^{[<\epsilon ><\delta ><\gamma
>]})\epsilon ^{MR}\partial _Mu^{<\alpha >}\partial _Ru^{<\mu >}\}.
$$
The counter-term of the sum of next two $(\gamma )$-diagrams is chosen to be
the ha-extension of that introduced in \cite{ketn,ketnp}%
$$
\gamma _2+\gamma _3=-\lambda ^2\omega \left( \epsilon \right) \frac{10-7q}{%
18q}I_1^2\times \eqno(5.17)
$$
$$
\int d^2z\{\widehat{\nabla }_AT_{[<\alpha ><\beta ><\gamma >]}\widehat{%
\nabla }^AT^{[<\alpha ><\beta ><\gamma >]}+
$$
$$
6G^{<\tau ><\epsilon >}T_{[<\delta ><\alpha ><\tau >]}T_{[<\epsilon ><\beta
><\gamma >]}\widehat{\nabla }_{<\varphi >}T^{[<\alpha ><\beta ><\gamma
>]}\times
$$
$$
\left( \eta ^{AB}-\epsilon ^{AB}\right) \partial _Au^{<\delta >}\partial
_Bu^{<\varphi >}\}.
$$
In a similar manner we can computer the rest part of counter-terms:%
$$
\delta =\frac 12\lambda ^2(I_1^2+m^2I_1J)\times
$$
$$
\int d^2z\widehat{r}_{<\alpha >(<\beta ><\gamma >)<\delta >}\widehat{r}%
^{<\beta ><\gamma >}\left( \eta ^{AB}-\epsilon ^{AB}\right) \partial
_Au^{<\alpha >}\partial _Bu^{<\delta >},
$$
$$
\epsilon =\frac 14\lambda ^2I_1^2\int d^2z\left( \eta ^{AB}-\epsilon
^{AB}\right) \times
$$
$$
[\Delta \widehat{r}_{<\delta ><\varphi >}+r_{<\delta >}^{<\alpha >}\widehat{r%
}_{<\alpha ><\varphi >}+r_{<\varphi >}^{<\alpha >}\widehat{r}_{<\delta
><\alpha >}
$$
$$
-2(G^{<\alpha ><\tau >}T_{\left[ <\delta ><\beta ><\alpha >\right]
}T_{\left[ <\varphi ><\gamma ><\tau >\right] }\widehat{r}^{<\beta ><\gamma
>}-
$$
$$
G_{<\varphi ><\tau >}T^{\left[ <\tau ><\alpha ><\beta >\right] }\nabla
_{<\alpha >}\widehat{r}_{<\delta ><\beta >}+
$$
$$
G_{<\delta ><\tau >}T^{[<\tau ><\alpha ><\beta >]}\nabla _{<\alpha >}%
\widehat{r}_{<\beta ><\varphi >}]\partial _Au^{<\delta >}\partial
_Bu^{<\varphi >},
$$
$$
\iota =\frac 16\lambda ^2\tilde m^2I_1J\times
$$
$$
\int d^2z\widehat{r}_{<\beta ><\alpha ><\gamma ><\delta >}r^{<\beta ><\gamma
>}\left( \eta ^{AB}-\epsilon ^{AB}\right) \partial _Au^{<\alpha >}\partial
_Bu^{<\delta >},
$$
$$
\eta =\frac 14\lambda ^2\omega \left( \epsilon \right) \left( I_1^2+2\tilde
m^2I_1J\right) \times
$$
$$
\int d^2z\widehat{r}_{.<\alpha ><\gamma ><\delta >}^{<\beta >}T_{\left[
<\beta ><\varphi ><\tau >\right] }T^{\left[ <\gamma ><\varphi ><\tau
>\right] }\left( \eta ^{AB}-\epsilon ^{AB}\right) \partial _Au^{<\alpha
>}\partial _Bu^{<\delta >}.
$$

By using relations (5.5) we can represent terms (5.6) and (5.7) in the
canonical form (5.1) from which we find the contributions in the $\beta
_{<\delta ><\varphi >}$-function (5.2):%
$$
\gamma _1:-\frac 2{3(2\pi )^2}\widehat{r}_{<\delta >(<\alpha ><\beta
>)<\gamma >}\widehat{r}_{........<\varphi >}^{<\gamma >(<\alpha ><\beta >)}-%
\eqno(5.18)
$$
$$
\frac{(\omega _1-1)}{(2\pi )^2}\{\frac 43\widehat{r}_{[<\gamma >(<\alpha
><\beta >)<\delta >]}\widehat{r}_{........<\varphi >]}^{[<\gamma >(<\alpha
><\beta >)}+
$$
$$
\widehat{r}_{[<\alpha ><\beta ><\gamma ><\delta >]}\widehat{r}%
_{.......<\varphi >]}^{[<\alpha ><\beta ><\gamma >}\},
$$
$$
\gamma _2+\gamma _3:\frac{(4\omega _1-5)}{9(2\pi )^2}\{\widehat{\nabla }%
_{<\delta >}T_{\left[ <\alpha ><\beta ><\gamma >\right] }\widehat{\nabla }%
_{<\varphi >}T^{\left[ <\alpha ><\beta ><\gamma >\right] }+
$$
$$
6G^{<\tau ><\epsilon >}T_{\left[ <\delta ><\alpha ><\tau >\right] }T_{\left[
<\epsilon ><\beta ><\gamma >\right] }\widehat{\nabla }_{<\varphi >}T^{\left[
<\alpha ><\beta ><\gamma >\right] }-
$$
$$
\frac{(\omega _1-1)}{(2\pi )^2}\widehat{r}_{\left[ <\alpha ><\beta ><\gamma
><\delta >\right] }\widehat{r}_{.......<\varphi >]}^{[<\alpha ><\beta
><\gamma >}\},
$$
$$
\eta :\frac{\omega _1}{(2\pi )^2}\widehat{r}^\alpha ._{\delta \beta \varphi
}T_{\left[ \alpha \tau \epsilon \right] }T^{\left[ \beta \tau \epsilon
\right] }.
$$

Finally,  we remark that two-loop $\beta $-function can
not be written only in terms of curvature $\widehat{r}_{<\alpha ><\beta
><\gamma ><\delta >}$ and d-derivation $\widehat{\nabla }_{<\alpha >}$
(similarly as in the locally isotropic case \cite{ketn,ketnp} ).

\subsubsection{Low--energy effective action for locally anisotrop\-ic strings}

The conditions of vanishing of $\beta $-functions describe the propagation
of string in the background of ha-fields $G_{<\alpha ><\beta >}$ and $%
b_{<\alpha ><\beta >}.$ (in this subsection we chose the canonic d-connection 
$%
^{\circ }{\Gamma }_{\cdot <\beta ><\gamma >}^{<\alpha >}$ on ${\cal E}^{<z>}%
{\cal ).}$ The $\beta $-functions are proportional to d-field equations
obtained from the on--shell string effective action
$$
I_{eff}=\int du\sqrt{|\gamma |}L_{eff}\left( \gamma ,b\right) .\eqno(5.19)
$$
The adapted to N-connection variations of (5.17) with respect to $G^{<\mu
><\nu >}$ and $b^{<\mu ><\nu >}$ can be written as
$$
\frac{\Delta I_{eff}}{\delta G^{<\alpha ><\beta >}}=W_{<\alpha ><\beta
>}+\frac 12G_{<\alpha ><\beta >}(L_{eff}+{\mbox{complete derivation)})},
$$
$$
\frac{\Delta I_{eff}}{\delta b^{<\alpha ><\beta >}}=0.
$$

The invariance of action (5.17) with respect to N-adapted diffeomorfisms
gives rise to the identity%
$$
\nabla _{<\beta >}W^{<\alpha ><\beta >}-T^{\left[ <\alpha ><\beta ><\gamma
>\right] }\frac{\Delta I_{eff}}{\delta b^{<\beta ><\gamma >}}=
$$
$$
-\frac 12\nabla ^{<\alpha >}(L_{eff}+\mbox{complete derivation)}
$$
(in the locally isotropic limit we obtain the well-known results from \cite
{zan,ckp}). This points to the possibility to write out the integrability
conditions as
$$
\nabla ^{<\beta >}\beta _{(<\alpha ><\beta >)}-G_{<\alpha ><\tau >}T^{\left[
<\tau ><\beta ><\gamma >\right] }\beta _{[<\beta ><\gamma >]}=-\frac
12\nabla _{<\alpha >}L_{eff}.\eqno(5.20)$$

For one-loop $\beta $-function, $\beta _{<\alpha ><\beta >}^{\left( 1\right)
}=\frac 1{2\pi }\widehat{r}_{<\alpha ><\beta >},$ we find from the last
equations%
$$
\nabla ^{<\beta >}\beta _{(<\delta ><\beta >)}^{(1)}-G_{<\delta ><\tau
>}T^{[<\tau ><\beta ><\gamma >]}\beta _{[<\beta ><\gamma >]}^{(1)}=
$$
$$
\frac 1{4\pi }\nabla _{<\delta >}\left( \overleftarrow{R}+\frac 13T_{\left[
<\alpha ><\beta ><\gamma >\right] }T^{\left[ <\alpha ><\beta ><\gamma
>\right] }\right)
$$
We can take into account two-loop $\beta$--functions by fixing an explicit
form of
$$
\omega (\epsilon )=1+2\omega _1\epsilon +4\omega _2\epsilon ^2+...
$$
when
$$
\omega _{HVB}\left( \epsilon \right) =\frac 1{(1-\epsilon )^2},\omega
_1^{HVB}=1,\omega _2^{HVB}=\frac 34
$$
(the t'Hooft-Veltman-Bos prescription \cite{thoov}). Putting values (5.18)
into (5.20) we obtain the two--loop approximation for ha--field equations
$$
\nabla ^{<\beta >}\beta _{(<\delta ><\beta >)}^{(2)}-G_{<\delta ><\tau
>}T^{[<\tau ><\beta ><\gamma >]}\beta _{[<\beta ><\gamma >]}^{(2)}=
$$
$$
\frac 1{2(2\pi )^2}\nabla _\delta [-\frac 18r_{<\alpha ><\beta ><\gamma
><\delta >}r^{<\alpha ><\beta ><\gamma ><\delta >}+
$$
$$
\frac 14r_{<\alpha ><\beta ><\gamma ><\delta >}G_{<\tau ><\epsilon
>}T^{[<\alpha ><\beta ><\tau >]}T^{[<\epsilon ><\gamma ><\delta >]}+
$$
$$
\frac 14G^{<\beta ><\epsilon >}G_{<\alpha ><\kappa >}T^{[<\alpha ><\tau
><\sigma >]}T_{[<\epsilon ><\tau ><\sigma >]}T^{[<\kappa ><\mu ><\nu
>]}T_{[<\beta ><\mu ><\nu >]}-
$$
$$
\frac 1{12}G^{<\beta ><\epsilon >}G_{<\gamma ><\lambda >}T_{[<\alpha ><\beta
><\tau >]}T^{[<\alpha ><\gamma ><\varphi >]}T_{[<\epsilon ><\varphi ><\kappa
>]}T^{[<\lambda ><\tau ><\kappa >]}],
$$
which can be obtained from effective action%
$$
I^{eff}\sim \int \delta ^{n_E}u\sqrt{|\gamma |}[-\overleftarrow{r}+\frac
13T_{[<\alpha ><\beta ><\gamma >]}T^{[<\alpha ><\beta ><\gamma >]}-
$$
$$
\frac{\alpha ^{\prime }}4(\frac 12r_{<\alpha ><\beta ><\gamma >\delta
}r^{<\alpha ><\beta ><\gamma ><\delta >}-
$$
$$
G_{<\tau ><\epsilon >}r_{<\alpha ><\beta ><\gamma ><\delta >}T^{[<\alpha
><\beta ><\tau >]}T^{[<\epsilon ><\gamma ><\delta >]}-
$$
$$
G^{<\beta ><\epsilon >}G_{<\alpha ><\upsilon >}T^{[<\alpha ><\gamma ><\kappa
>]}T_{[<\epsilon ><\gamma ><\kappa >]}T^{[<\upsilon ><\sigma ><\varsigma
>]}T_{[<\beta ><\sigma ><\varsigma >]}+
$$
$$
\frac 13G^{<\beta ><\epsilon >}G_{<\delta ><\kappa >}T_{[<\alpha ><\beta
><\gamma >]}T^{[<\alpha ><\delta ><\varphi >]}T_{[<\varphi ><\upsilon
><\epsilon >]}T^{[<\gamma ><\upsilon ><\kappa >]})]\eqno(5.21)
$$

The action (5.21) (for 2$\pi \alpha ^{\prime }=1\,$ and in locally isotropic
limit) is in good concordance with the similar ones on usual closed strings
\cite{gs,ketn}.

We note that the existence of an effective action is assured by the
Zamolodchikov c-theorem \cite{zam} which was generalized \cite{ts87} for the
case of bosonic nonlinear $\sigma $-model with dilaton connection. In a
similar manner we can prove that such results hold good for ha-backgrounds.

\subsection{Scattering of Ha--Gravitons and Duality of Models}

The quantum theory of ha-strings can be naturally considered by using the
formalism of functional integrals on ''hypersurfaces'' (see Polyakov's works
\cite{poly}). In this subsection we study the structure of scattering
amplitudes of ha-gravitons. Questions on duality of ha-string theories will
be also analyzed.

\subsubsection{Ha--string amplitudes for ha-gravitons scattering}

We introduce the Green function of ha--tachyons, the fundamental state of
ha--string, as an integral ( after Weeck rotation in the Euclidean space )%
$$
G_t(p_{1,...,}p_N)=$$
$$\int [D\gamma _{AB}(\zeta )][Du^{<\alpha >}(z)]\exp
(-\frac 1{4\pi \alpha ^{\prime }}\int d^2z\sqrt{|\gamma |}\gamma
^{AB}\partial _Au^{<\alpha >}\partial _Bu^{<\alpha >})
$$
$$
\int \left[ \prod_AD^2z_A\right] \sqrt{|\gamma (z_B)|}\exp (ip_B^{<\alpha
>}u(z_B)),\eqno(5.22)
$$
where the integration measure on $\gamma _{AB}$ includes the standard ghost
Fadeev-Popov determinant corresponding to the fixation of the
reparametrization invariance and%
$$
\left[ \prod_AD^2z_A\right] =\prod_{A\neq M,N,K}d^2z_A|z_M-z_{N\,}|^2\left|
z_K-z_M\right| ^2\left| z_K-z_N\right| ^2.
$$
Because the string quantum field theory can be uncontradictory formulated
for spaces of dimension d=26 we consider that in formula (5.22) $\alpha $
takes values from 0 to 25. Formula (5.22) leads to dual amplitudes for
ha-tachyon scatterings for $p^2=\frac 4{\alpha ^{\prime }}$ (see \cite{sch}
for details and references on usual locally isotropic tachyon scattering).

The generating functional of Green functions (5.22) in the coordinate
u-repre\-sen\-ta\-ti\-on can be formally written as a hyper surface mean
value
$$
\Gamma ^0\left[ \Phi \right] =<\exp \left( -\frac 1{2\pi \alpha ^{\prime
}}\int d^2z\sqrt{|\gamma |}\Phi \left[ u\left( z\right) \right] \right) >.
$$
In order to conserve the reparametrization invariance we define the
ha-graviton source as
$$
\Gamma ^0\left[ G\right] =<\exp \left( -\frac 1{4\pi \alpha ^{\prime }}\int
d^2z\sqrt{|\gamma |}\gamma ^{AB}\partial _Au^{<\alpha >}\partial _Bu^{<\beta
>}G_{<\alpha ><\beta >}[u\left( z\right) ]\right) >
$$
from which we obtain the Green function of a number of $K$ elementary
perturbations of the closed ha-string ($K$ ha--gravitons)
$$
G_g(u_1,...,u_K)=
<\frac 12\int [\prod_{[j]}D^2z_{[j]}]\sqrt{|\gamma (z)|}G^{AB\;}(z_{[j]})%
$$
$$\partial _Au^{<\alpha >}(z_{[j]})\partial _Bu^{<\beta >}(z_{[j]\;})\chi
_{<\alpha ><\beta >}^{\left( j\right) }\delta ^{(d)}(u_{[j]}-u(z_{[j]}))>,
$$
where $\chi _{<\alpha ><\beta >}^{\left( j\right) }$ are polarization
d-tensors of exterior ha-gravitons and $[j]=1,2,...K.$ Applying the Fourier
transform we obtain
$$
G_g(p_1,...,p_K)=
$$
$$
\int [D\gamma _{AB}\left( z\right) ][Du^{<\alpha >}\left( z\right) ]\exp
\left( -\frac 1{2\pi \alpha ^{\prime }}\int d^2z\sqrt{|\gamma |}\frac
12G^{AB}\partial _Au^{<\alpha >}\partial _Bu^{<\alpha >}\right)
$$
$$
\int \left[ \prod_{[j]}D^2z_{[j]}\right] \frac 12\sqrt{|\gamma (z_{[j])}|}%
\gamma ^{AB}(z_{[j]})\partial _Au^{<\beta >}(z_{[j]})\partial _Bu^{<\gamma
>}(z_{[j]})\chi _{<\alpha ><\beta >}^{(j)}$$
$$\exp [ip_{[j]}^\delta u^\delta
(z)].
$$
Integrating the last expression on $G_{<\mu ><\nu >}\,$ and $u^{<\alpha >}$
for d=26, when there are not conformal anomalies, we have%
$$
G_g(p_1,...,p_K)=
$$
$$
\int \left[ \prod_{[j]}D^2z_{[j]}\right] (\partial _A\frac \partial
{\partial p_{[j]}}\cdot \chi ^{(j)}.\partial ^A\frac \partial {\partial
p_{[j]}})\times
$$
$$
\int [D\sigma (z)]\exp [-\pi \alpha ^{\prime
}\sum_{[i,j]}p_{[i]}p_{[j]}V(z_{[i]},z_{[j]},\sigma )],\eqno(5.23)
$$
where V is the Green function of the Laplacian for the conformally-flat
metric $G_{AB}=e^\sigma \delta _{AB}:$%
$$
\partial _A(\sqrt{|\gamma |}\gamma ^{AB}\partial _A)V=-\delta
^2(z_{[i]}-z_{[j]})
$$
which can be represented as
$$
V(z_{[i]},z_{[j]},\sigma )=-\frac 1{4\pi }\ln
|z_{[i]}-z_{[j]}|^2,z_{[i]}\neq z_{[j]},
$$
$$
V\left( z_{[k]},z_{[k]}^p,\sigma \right) =\frac 1{4\pi }(\sigma
(z_{[k]})-\ln (\frac 1\epsilon )),
$$
for $\epsilon \,$ being the cutting parameter. \ Putting the last expression
into (5.23) we compute the Green function of ha-gravitons:%
$$
G_g(p_1,...p_K)=
$$
$$
\int \prod_{[j]\neq
[p],[q],[s]}d^2z_{[j]}|z_{[q]}-z_{[p]}|^2|z_{[s]}-z_{[q]}|^2|z_{[s]}-
z_{[p]}|^2(\partial _A\frac \partial {\partial p_{[j]}}\cdot \chi ^{(j)}
\cdot \partial ^A\frac \partial {\partial p_{[j]}})
$$
$$
\int d\sigma (z)\prod_{[i]<[m]}|z_{[i]}-
$$
$$
z_{[m]}|^{\alpha ^{\prime }p_{[i]}p_{[j]}}\exp \left[ -\frac 14\alpha
^{\prime }\sum_{[k]}p_{[k]}^2\sigma (z_{[k]})\right] \exp \left[ \frac{%
\alpha ^{\prime }}4\ln (\frac 1\epsilon )\sum_{[k]}p_{[k]}^2\right] ,
$$
where the definition of integration on $\sigma (z)$ is extended as%
$$
\int d\sigma (z)\equiv
$$
$$
\lim \limits_{\sigma _d(z_{[j]})\rightarrow -\infty }^{\sigma
_s(z_{[j]})\rightarrow +\infty }\prod_{[j]}\int_{\sigma _d(z_{[j]})}^{\sigma
_s(z_{[j]})}d\sigma (z_{[j]}).
$$
So the scattering amplitude
$$
A_g(p_1,...,p_K)=
$$
$$
\lim \limits_{p_{[j]}^2\rightarrow 0}\prod_{[j]}p_{[j]}^2G_g(p_{1,}...,p_K)
$$
is finite if%
$$
\lim \limits_{\sigma _d(z_{[j]})\to -\infty ,p_{[j]}^2\to 0}|p_{[j]}^2\sigma
(z_{[j]})|=const<\infty .
$$
$G_g(p_1,...,p_{K)}\,$ has poles on exterior momenta corresponding to
massless higher order  anisotropic modes of spin 2 (ha-gravitons). The final
result for the scattering amplitute of ha-gravitons is of the form
$$
A_g(p_1,...,p_K)\sim \int \prod_{[j]\neq
[p],[q],[s]}d^2z_{[j]}|z_{[p]}-z_{[q]}|^2|z_{[s]}-z_{[q]}|^2|z_{[s]}-z_{[p]}|^2
$$
$$
\partial _A\frac \partial {\partial p_{[j]}^{<\alpha >}}\chi _{<\alpha
><\beta >}^{(j)}\partial ^A\frac \partial {\partial p_{[j]}^{<\beta
>}}\prod_{[m]<[n]}|z_{[m]}-z_{[n]}|^{\alpha ^{\prime }p_{[m]}\cdot p_{[n]}}.
$$
If instead of polarization d-tensor $\chi _{<\alpha ><\beta }^{(j)}$ the
graviton polarization tensor $\chi _{ik}^{(j)}$ is taken we obtain the well
known results on scattering of gravitons in the framework of the first
quantization of the string theory \cite{fts,keti}.

\subsubsection{Duality of Ha--$\sigma$--mo\-dels}

Two theories are dual if theirs non-equivalent second order actions can be
generated by the same first order action. The action principle assures the
equivalence of the classical dual theories. But, in general, the duality
transforms affects the quantum conformal properties \cite{bus}. In this
subsection we shall prove this for the ha-$\sigma $-model (5.1) when metric $%
\gamma $ and the torsion potential b on ha-background ${\cal E}^{<z>}$ do
not depend on coordinate u$^0.$ If such conditions are satisfied we can
write for (5.1) the first order action
$$
I=\frac 1{4\pi \alpha ^{\prime }}\int d^2z\{\sum_{<\alpha >,<\beta
>=1}^{n_E-1}[\sqrt{|\gamma |}\gamma ^{AB}(G_{00}V_AV_B+2G_{0\alpha
}V_A\left( \partial _Bu^{<\alpha >}\right) +\eqno(5.24)
$$
$$
G_{<\alpha ><\beta >}(\partial _Au^{<\alpha >})\left( \partial _Bu^{<\beta
>}\right) )+\epsilon ^{AB}(b_{0<\alpha >}V_B(\partial _Au^{<\alpha >})+
$$
$$
b_{<\alpha ><\beta >}(\partial _Au^{<\alpha >})(\partial _Bu^{<\beta >}))]+
$$
$$
\epsilon ^{AB}\widehat{u}^0(\partial _AV_B)+\alpha ^{\prime }\sqrt{|\gamma |}%
R^{(2)}\Phi (u)\},
$$
where string interaction constants from (5.1) and (5.24) are related as $%
\lambda ^2=2\pi \alpha ^{\prime }.$

This action will generate an action of type (5.1) if we shall exclude the
Lagrange multiplier $\widehat{u}^0.$ The dual to (5.24) action can be
constructed by substituting \ V$_A$ expressed from the motion equations for
fields V$_A$ (also obtained from action (5.23)):%
$$
\widehat{I}=\frac 1{4\pi \alpha ^{\prime }}\int d^2z\{\sqrt{|\gamma |}\gamma
^{AB}\widehat{G}_{<\alpha ><\beta >}(\partial _A\widehat{u}^{<\alpha
>})(\partial _B\widehat{u}^{<\beta >})+
$$
$$
\epsilon ^{AB}\widehat{b}_{<\alpha ><\beta >}(\partial _A\widehat{u}%
^{<\alpha >})(\partial _B\widehat{u}^{<\beta >})+\alpha ^{\prime }\sqrt{%
|\gamma |}R^{(2)}\Phi (u)\},
$$
where the knew metric and torsion potential are introduced respectively as
$$
\widehat{G}_{00}=\frac 1{G_{00}},\widehat{G}_{0<\alpha >}=\frac{b_{0<\alpha
>}}{G_{00}},
$$
$$
\widehat{G}_{<\alpha ><\beta >}=G_{<\alpha ><\beta >}-\frac{G_{0<\alpha
>}G_{0<\beta >}-b_{0<\alpha >}b_{0<\beta >}}{G_{00}}
$$
and
$$
\widehat{b}_{0<\alpha >}=-\widehat{b}_{<\alpha >0}=\frac{G_{0<\alpha >}}{%
G_{00}},
$$
$$
\widehat{b}_{<\alpha ><\beta >}=b_{<\alpha ><\beta >}+\frac{G_{0<\alpha
>}b_{0<\beta >}-b_{0<\alpha >}G_{0<\beta >}}{G_{00}}
$$
(in the formulas for the new metric and torsion potential indices $\alpha $
and $\beta $ take values $1,2,...n_E-1$).

If the model (5.1) satisfies the conditions of one--loop conformal invariance
(see details for locally isotropic backgrounds in \cite{cfmp}, one holds
these ha-field equations%
$$
\frac 1{\alpha ^{\prime }}\frac{n_E-25}3+
$$
$$
[4(\nabla \Phi )^2-4\nabla ^2\Phi -r-\frac 13T_{[<\alpha ><\beta ><\gamma
>]}T^{[<\alpha ><\beta ><\gamma >]}]=0,
$$
$$
\widehat{r}_{(<\alpha ><\beta >)}+2\nabla _{(<\alpha >}\nabla _{<\beta
>)}\Phi =0,
$$
$$
\widehat{r}_{[<\alpha ><\beta >]}+2T_{[<\alpha ><\beta ><\gamma >]}\nabla
^{<\gamma >}\Phi =0.\eqno(5.25)
$$
By straightforward calculations we can show that the dual theory has the
same conformal properties and satisfies the conditions (5.25) if the dual
transform is completed by the shift of dilaton field%
$$
\widehat{\Phi }=\Phi -\frac 12\log G_{00}.
$$

The system of ha--field equations (5.25), obtained as a low-energy limit of
the ha-string theory, is similar to Einstein--Cartan equations (2.34) and
(2.36). We note that the explicit form of locally anisotropic
energy-momentum source in (5.25) is defined from well defined principles and
symmetries of string interactions and this form is not postulated, as in
usual locally isotropic field models, from some general considerations in
order to satisfy the necessary conservation laws on ha-space whose
formulation is very sophisticated because of nonexistence of global and even
local group of symmetries of such type of spaces. Here we also remark that
the higher order anisotropic model with dilaton field interactions does not
generate in the low-energy limit the Einstein-Cartan ha-theory because the
first system of equations from (5.25) represents some constraints (being a
consequence of the two-dimensional symmetry of the model) on torsion and
scalar curvature which can not be interpreted as some algebraic relations of
type (2.36) between locally anisotropic spin-matter source and torsion. As
a matter of principle we can generalize our constructions by introducing
interactions with gauge ha-fields and considering a variant of locally
anisotropic chiral $\sigma $-model \cite{jac} in order to get a system of
equations quite similar to (2.36). However, there are not exhaustive
arguments for favoring the Einstein-Cartan theory and we shall not try in
this work to generate it necessarily from ha--strings.

\section{Summary and outlook}

We have developed the spinor differential geometry of distinguished vector
bundles provided with nonlinear and distinguished connections and metric
structures and shown in detail the way of formulation the theory of
fundamental field (gravitational, gauge and spinor) interactions on generic
locally an\-isot\-rop\-ic spaces.

We investigated the problem of definition of spinors on spaces with higher
order anisotropy. Our approach is based on the formalism of Clifford
d-algebras. We introduced spinor structures on ha-spaces as Clifford
d-module structures on dv-bundles. We also proposed the second definition,
as distinguished spinor structures, by using Clifford fibrations. It was
shown that $H^{2n}$-spaces admit as a proper characteristic the almost
complex spinor structures. We argued that one of the most important
properties of spinors in both dv-bundles with compatible N-connection,
d-connection and metric and in $H^{2n}$-spaces is the periodicity 8 on the
dimension of the base and on the dimension of the typical fiber spaces.

It should be noted that we introduced \cite{vjmp,vb295,vod} d-spinor
structures in an algebraic topological manner, and that in our
considerations the compatibility of d-connection and metric, adapted to a
given N-connection, plays a crucial role. The Yano and Ishihara method of
lifting of geometrical objects in the total spaces of tangent bundles \cite
{yan} and the general formalism for vector bundles of Miron and Anastasiei
\cite{ma87,ma94} clearing up the possibility and way of definition of
spinors on higher order anisotropic spaces. Even a straightforward
definition of spinors on Finsler and Lagrange spaces, and, of course, on
various theirs extensions, with general noncompatible connection and metric
structures, is practically impossible (if spinors are introduced locally
with respect to a given metric quadratic form, the spinor constructions will
not be invariant on parallel transports), we can avoid this difficulty by
lifting in a convenient manner the geometric objects and physical values
from the base of a la-space on the tangent bundles of v- and t-bundles under
consideration. We shall introduce corresponding discordance laws and values
and define nonstandard spinor structures by using nonmetrical d-tensors (see
such constructions for locally isotropic curved spaces with torsion and
nonmetricity in \cite{lue}).

The distinguishing by a N-connection structure of the multidimensional space
into horizontal and vertical subbundles points out to the necessity to start
up the spinor constructions for la-spaces with a study of distinguished
Clifford algebras for vector spaces split into h- and v-subspaces. The
d-spinor objects exhibit a eight-fold periodicity on dimensions of the
mentioned subspaces. As it was shown in \cite{vjmp,vb295}, see also the
 Secs. II.B--II.D of this work,
a corresponding d-spinor technique can be
developed, which is a generalization for higher dimensional with
N-connection structure of that proposed by Penrose and Rindler \cite
{pen,penr1,penr2} for locally isotropic curved spaces, if the locally
adapted to the N-connection structures d-spinor and d-vector frames are
used. It is clear the d-spinor calculus is more tedious than the 2-spinor
one for Einstein spaces because of multidimensional and multiconnection
character of generic ha-spaces.
The d-spinor differential geometry formulated in Sec. II.F can be
considered as a branch of the geometry of Clifford fibrations for v-bundles
provided with N-connection, d-connection and metric structures. We have
emphasized only the features containing d-spinor torsions and curvatures
which are necessary for a d-spinor formulation of la-gravity. To develop a
conformally invariant d-spinor calculus is possible only for a particular
class of ha-spaces when the Weyl d-tensor (2.33) is defined by the
N-connection and d-metric structures. In general, we have to extend the
class of conformal transforms to that of nearly autoparallel maps of
ha-spaces (see \cite{vod,voa,vodg,vcl96} and Sec. IV).

Having fixed compatible N-connection, d-connection and metric structures on
a dv--bundle  ${\cal E}^{<z>}$ we can develop physical models on this space
by using a covariant variational d-tensor calculus as on Riemann-Cartan
spaces (really there are specific complexities because, in general, the
Ricci d-tensor is not symmetric and the locally anisotropic frames are
nonholonomic). The systems of basic field equations for fundamental matter
(scalar, Proca and Dirac) fields and gauge and gravitational fields have
been introduced in a geometric manner by using d-covariant operators and
la-frame decompositions of d-metric. These equations and expressions for
energy-momentum d-tensors and d-vector currents can be established by using
the standard variational procedure, but correspondingly adapted to the
N-connection structure if we work by using la-bases.

In this section we have reformulated the fiber bundle formalism for both
Yang-Mills and gravitational fields in order to include into consideration
space-times with higher order anisotropy.
We have argued that our approach has the
advantage of making manifest the relevant structures of the theories with
local anisotropy and putting greater emphasis on the analogy with
anisotropic models than the standard coordinate formulation in Finsler
geometry.

Our geometrical approach to higher order anisotropic gauge and gravitational
interactions are refined in such a way as to shed light on some of the more
specific properties and common and distinguishing features of the Yang-Mills
and Einstein fields. As we have shown, it is possible to make a gauge like
treatment for both models with local anisotropy (by using correspondingly
defined linear connections in bundle spaces with semisimple structural
groups, with variants of nonlinear realization and extension to semisimple
structural groups, for gravitational fields).

We have extended our models proposed in \cite{vg,v295a,v295b} for
generalized Lagrange spaces to the case on ha-spaces modeled as
 higher order vector
bundles provided with N-connection, d-connection and d-metric structures.
The same geometric machinery can be used for developing of gauge field
theories on spaces with higher order anisotropy \cite{mirata,yan} or on Jet
bundles \cite{asa89}.

Let us try to summarize our results, discuss their possible implications and
make the basic conclusions. Firstly, we have shown that the Einstein-Cartan
theory has a natural extension for a various class of ha-spaces. Following
the R. Miron and M. Anastesiei approach \cite{ma87,ma94} to the geometry of
la-spaces it becomes evident the possibility and manner of formulation of
classical and quantum field theories on such spaces. Here we note that in
la-theories we have an additional geometric structure, the N-connection.
From physical point of view it can be interpreted, for instance, as a
fundamental field managing the dynamics of splitting of high-dimensional
space-time into the four-dimensional and compactified ones. We can also
consider the N-connection as a generalized type of gauge field which
reflects some specifics of ha-field interactions and possible intrinsic
structure of ha-spaces. It was convenient to analyze the geometric structure
of different variants of ha-spaces (for instance, Finsler, Lagrange and
generalized Lagrange spaces) in order to make obvious physical properties
and compare theirs perspectives in developing of new physical models.

According to modern-day views the theories of fundamental field interactions
should be a low energy limit of the string theory. One of the main results
of this work is the proof of the fact that in the framework of ha-string
theory is contained a more general, locally anisotropic, gravitational
physics. To do this we have developed the locally anisotropic nonlinear
sigma model and studied it general properties. We shown that the condition
of self consistent propagation of string on ha-background impose
corresponding constraints on the N-connection curvature, ha-space torsion
and antisymmetric d-tensor. Our extension of background field method for
ha-spaces has a distinguished by N-connection character and the main
advantage of this formalism is doubtlessly its universality for all types of
locally isotropic or anisotropic spaces.

The presented one- and two-loop calculus for the higher order anisotropic
string model and used in this work d-covariant dimensional regularization
are developed for ha-background spaces modelled as vector bundles provided
with compatible N-connection, d-connection and metric structures. In the
locally isotropic limit we obtain the corresponding formulas for the usual
nonlinear sigma model.

Finally, it should be stressed that we firstly calculated the amplitudes for
scattering of ha-gravitons and that the duality properties of the formulated
in this work higher order anisotropic models are similar to those of models
considered for locally isotropic strings
\newpage

\newpage
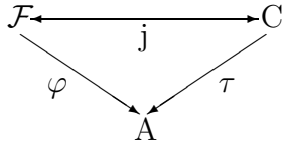
\begin{figure}[htbp]
\begin{center}
\begin{picture}(100,50) \setlength{\unitlength}{1pt}
\thinlines
\put(0,45){${\cal F}$}
\put(96,45){C}
\put(48,2){A}
\put(50,38){j}
\put(50,48){ \vector(-1,0){45}}
\put(50,48){\vector(1,0){45}}
\put(5,42){\vector(3,-2){45}}
\put(98,42){\vector(-3,-2){45}}
\put(15,20){$\varphi$}
\put(80,20){$\tau$}
\end{picture}
\end{center}
\caption{Diagram 1}
\end{figure}
\vskip20pt
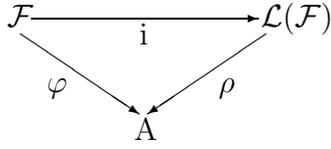
\begin{figure}[htbp]
\begin{center}
\begin{picture}(100,50) \setlength{\unitlength}{1pt}
\thinlines
\put(0,45){${\cal F}$}
\put(96,45){${\cal L}({\cal F})$}
\put(48,2){A}
\put(50,38){i}
\put(50,48){ \line(-1,0){45}}
\put(50,48){\vector(1,0){45}}
\put(5,42){\vector(3,-2){45}}
\put(98,42){\vector(-3,-2){45}}
\put(15,20){$\varphi$}
\put(80,20){$\rho$}
\end{picture}
\end{center}
\caption{Diagram 2}
\end{figure}

\vskip20pt
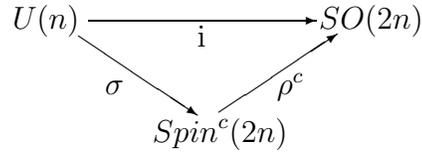
\begin{figure}[htbp]
\begin{center}
\begin{picture}(255,50) \setlength{\unitlength}{1pt}
\thinlines
\put(50,45){$U(n)$}
\put(166,45){$SO(2n)$}
\put(103,2){${Spin}^c (2n)$}
\put(120,38){i}
\put(120,48){ \line(-1,0){45}}
\put(120,48){\vector(1,0){45}}
\put(75,42){\vector(3,-2){45}}
\put(128,13) {\vector(3,2){45}}
\put(85,20){$\sigma$}
\put(150,20){${\rho}^c$}
\end{picture}
\end{center}
\caption{Diagram 3}
\end{figure}
\vskip20pt
\begin{figure}[htbp]
\begin{center}
\begin{picture}(190,50) \setlength{\unitlength}{1pt}
\thinlines
\put(0,45){$Spin(n)\times Spin(n')$}
\put(160,45){$Spin(n+n')$}
\put(15,2){$O(n)\times O(n')$}
\put(168,02){$O(n+n')$}
\put(100,45){ \vector(1,0){55}}
\put(48,40){\vector(0,-1){25}}
\put(90,5){\vector(1,0){70}}
\put(185,40) {\vector(0,-1){25}}
\end{picture}
\end{center}
\caption{Diagram 4}
\end{figure}
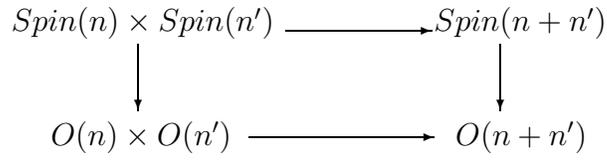
\vskip20pt

\begin{figure}[htbp]
\begin{center}
\begin{picture}(150,50) \setlength{\unitlength}{1pt}
\thinlines
\put(0,45){${\cal E}_{\cal C}^{0,2n} (M^{2n})$}
\put(145,45){${\cal E}_{\cal C}^{{\lambda}^c \oplus {\lambda}^c } (M^n )$}
\put(81,0){${\cal E}_{\cal C}^W (M^n )$}
\put(99,38){${\epsilon}^c$}
\put(99,48){ \line(-1,0){45}}
\put(99,48){\vector(1,0){45}}
\put(54,42){\vector(3,-2){45}}
\put(147,42){\vector(-3,-2){45}}
\put(64,20){${\tilde{\varepsilon}}^c$}
\put(129,20){$\cal H$}
\end{picture}
\end{center}
\caption{Diagram 5}
\end{figure}
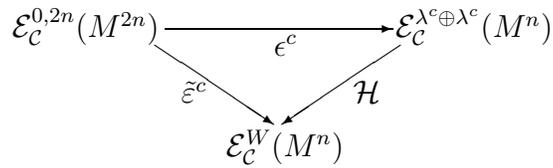
\vskip20pt

\begin{figure}[htbp]
\begin{center}
\begin{picture}(140,70) \setlength{\unitlength}{1pt}
\thinlines
\put(0,60){$Spin(2n)$}
\put(100,60){${Spin}^c (2n)$}
\put(6,0){$SO(n)$}
\put(106,0){$SO(2n)$}
\put(36,2){ \vector(1,0){65}}
\put(20,13){\vector(0,1){40}}
\put(126,13) {\vector(0,1){40}}
\put(8,30){$\beta$}
\put(70,60){$\subset$}
\end{picture}
\end{center}
\caption{Diagram 6}
\end{figure}
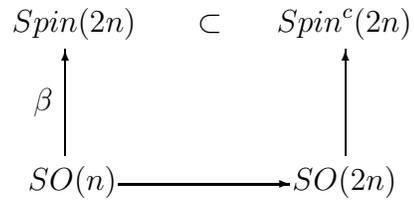
\vskip20pt

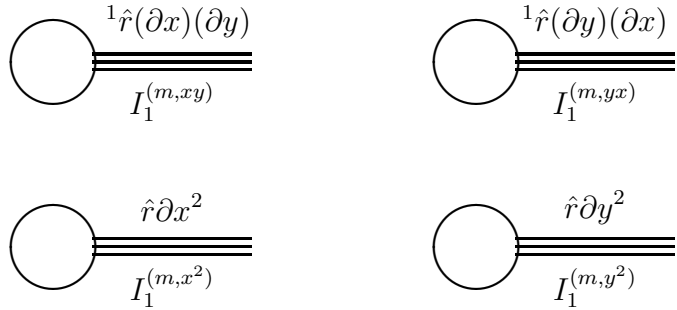
\begin{figure}[htbp]
\begin{picture}(360,150) \setlength{\unitlength}{1pt}
\thicklines
\put(105,35){\circle{30}}
\put(120,35){\line(1,0){60}}
\put(120,32){\line(1,0){60}}
\put(120,38){\line(1,0){60}}
\put(130,10){\makebox(40,20){$I_1^{(m,x^2)}$}}
\put(125,40){\makebox(50,20){$\hat r {\partial x}^2$}}

\put(265,35){\circle{30}}
\put(280,35){\line(1,0){60}}
\put(280,32){\line(1,0){60}}
\put(280,38){\line(1,0){60}}
\put(290,10){\makebox(40,20){$I_1^{(m,y^2)}$}}
\put(285,40){\makebox(50,20){$\hat r {\partial y}^2$}}

\put(105,105){\circle{30}}
\put(120,105){\line(1,0){60}}
\put(120,102){\line(1,0){60}}
\put(120,108){\line(1,0){60}}
\put(130,80){\makebox(40,20){$I_1^{(m,xy)}$}}
\put(125,110){\makebox(55,20){$^1{\hat r}(\partial x)(\partial y)$}}

\put(265,105){\circle{30}}
\put(280,105){\line(1,0){60}}
\put(280,102){\line(1,0){60}}
\put(280,108){\line(1,0){60}}
\put(290,80){\makebox(40,20){$I_1^{(m,yx)}$}}
\put(285,110){\makebox(50,20){$^1{\hat r}(\partial y)(\partial x)$}}

\end{picture}
\caption{Feyn\-man di\-a\-grams for  one-loop $\beta$-functions
of HAS--models}
\end{figure}

\begin{figure}[htbp]
\begin{picture}(360,500) \setlength{\unitlength}{1pt}
\thicklines

\put(91,55){\circle{30}}
\put(67,55){\circle{16}}
\put(67,47){\line(0,1){16}}
\put(19,55){\line(1,0){40}}
\put(19,52){\line(1,0){40}}
\put(19,58){\line(1,0){40}}
\put(107,52){\line(1,0){60}}
\put(107,55){\line(1,0){60}}
\put(107,58){\line(1,0){60}}
\put(20,60){\makebox(40,20){$m^2 \hat r$}}
\put(111,60){\makebox(60,20){$\hat r (\eta - \varepsilon ) {\partial u}^2$}}
\put(50,00){\makebox(70,20){Diagram $(\iota )$}}

\put(130,150){\line(1,0){40}}
\put(130,147){\line(1,0){40}}
\put(130,153){\line(1,0){40}}
\put(115,150){\circle{30}}
\put(92,150){\circle{16}}
\put(14,150){\line(1,0){70}}
\put(14,147){\line(1,0){70}}
\put(14,153){\line(1,0){70}}
\put(86,144){\line(1,1){12}}
\put(86,155){\line(1,-1){12}}
\put(50,110){\makebox(70,20){Diagram $(\varepsilon )$}}
\put(14,154){\makebox(70,20){$\hat r (\eta -\varepsilon )^2(\partial u)^2$}}
\put(131,154){\makebox(40,20){$\hat r(\eta -\varepsilon )$}}

\put(365,150){\circle{30}}
\put(341,150){\circle{16}}
\put(202,150){\line(1,0){130}}
\put(202,147){\line(1,0){130}}
\put(202,153){\line(1,0){130}}
\put(335,144){\line(1,1){12}}
\put(335,155){\line(1,-1){12}}
\put(290,100){\makebox(70,15){Diagram $(\delta )$}}
\put(202,155){\makebox(150,15){$(\nabla ^2\hat r +r\hat r + T^2\hat r +
T\nabla \hat r)$}}
\put(205,130){\makebox(100,15){$(\eta - \varepsilon )(\partial u)^2$}}

\put(305,55){\circle{30}}
\put(250,55){\line(1,0){110}}
\put(320,52){\line(1,0){40}}
\put(320,58){\line(1,0){40}}
\put(250,58){\line(1,0){40}}
\put(250,52){\line(1,0){40}}
\put(305,25){\line(0,1){15}}
\put(308,25){\line(0,1){15}}
\put(302,25){\line(0,1){15}}
\put(250,60){\makebox(40,20){$\varepsilon T$}}
\put(320,60){\makebox(50,20){$\varepsilon T$}}
\put(310,26){\makebox(70,15){$(\eta - \varepsilon ) (\partial u)^2$}}
\put(270,00){\makebox(70,15){Diagram $(\eta )$}}

\put(350,260){\circle{30}}
\put(365,260){\line(1,0){15}}
\put(365,257){\line(1,0){15}}
\put(365,263){\line(1,0){15}}
\put(320,260){\line(1,0){15}}
\put(320,257){\line(1,0){15}}
\put(320,263){\line(1,0){15}}
\put(350,230){\line(0,1){15}}
\put(347,230){\line(0,1){15}}
\put(353,230){\line(0,1){15}}
\put(330,210){\makebox(40,15){$\varepsilon T(\partial u)$}}
\put(365,265){\makebox(15,15){$\varepsilon T$}}
\put(320,265){\makebox(15,15){$\varepsilon T$}}
\put(343,296){\makebox(5,10){+}}

\put(300,300){\circle{30}}
\put(270,300){\line(1,0){60}}
\put(270,303){\line(1,0){15}}
\put(270,297){\line(1,0){15}}
\put(315,303){\line(1,0){15}}
\put(315,297){\line(1,0){15}}
\put(316,305){\makebox(15,15){$\varepsilon T$}}
\put(270,305){\makebox(15,15){$\varepsilon T$}}
\put(253,296){\makebox(5,10){+}}

\put(100,225){\makebox(70,15){Diagrams $(\gamma )$}}

\put(00,300){\line(1,0){240}}
\put(120,300){\circle{30}}
\put(00,303){\line(1,0){105}}
\put(00,297){\line(1,0){105}}
\put(135,303){\line(1,0){105}}
\put(135,297){\line(1,0){105}}
\put(00,305){\makebox(105,20){$(r+T^2+\varepsilon \nabla T)\partial u$}}
\put(136,305){\makebox(105,20){$(r+T^2+\varepsilon \nabla T)\partial u$}}

\put(00,470){\line(1,0){230}}
\put(230,485){\circle{28}}
\put(230,455){\circle{28}}
\put(00,467){\line(1,0){230}}
\put(00,473){\line(1,0){230}}
\put(05,450){\makebox(200,15){$(\nabla ^2r+r^2+T\nabla ^2T=rT^2)
(\partial u)^2+$}}
\put(05,430){\makebox(200,15){$\varepsilon (\nabla ^3T+r\nabla T+T^2\nabla T+
T\nabla r)(\partial u)^2$}}
\put(80,390){\makebox(70,15){Diagram $(\alpha )$}}

\put(305,485){\circle{28}}
\put(305,455){\circle{28}}
\put(270,470){\line(1,0){30}}
\put(270,473){\line(1,0){30}}
\put(270,467){\line(1,0){30}}
\put(260,390){\makebox(70,15){Diagram $(\beta )$}}
\put(320,485){\line(1,0){65}}
\put(320,482){\line(1,0){65}}
\put(320,488){\line(1,0){65}}
\put(275,475){\makebox(5,10){$r$}}
\put(320,465){\makebox(64,15){$\hat r(\eta -\varepsilon )(\partial u)^2$}}

\end{picture}
\caption{Two-loops diagrams for  HAS--models}
\end{figure}


\begin{references}



\bibitem{ald}  R. Aldovandi and E. Stedile, {\sl Int. J. Theor. Phys.}\ {\bf %
23} (1984) 301


\bibitem{agfm}  L. Alvarez-Gaume, D. Z. Freedman and S. Mukhi, {\sl Ann.
Phys.}\ {\bf 134} (1981) 85


\bibitem{alw}  S. P. de Alwis, {\sl Phys. Rev.} {\bf D34} (1986) 3760

\bibitem{ana77}  M. Anastasiei, {\sl C. R. Acad. Sci. Paris} {\bf A284,}
(1977) 943

\bibitem{ana86}  M. Anastasiei,{\sl \ An. \c st. Univ. Ia\c si, s I a} {\bf %
32 }(1986) 17

\bibitem{ana87}  M. Anastasiei, {\sl Tensor, N. S.} {\bf 46} (1987) 323

\bibitem{ana94}  M. Anastasiei, in {\it Coloquium on Differential Geometry,}
25-30 July 1994 (Lajos Kossuth University, Debrecen, Hungary, 1994) p.1

\bibitem{anask}  M. Anastasiei and H. Kawaguchi, {\sl Tensor, N. S.} {\bf 49}
(1990) 296

\bibitem{am}  P. L. Antonelli and R. Miron (eds), {\it Lagrange and Finsler
Geometry, Applications to Physics and Biology} (Kluwer Academic Publishers,
Dordrecht, Boston, London, 1996)

\bibitem{az94}  P. L. Antonelli and T. J. Zastavniak (guest eds), {\it %
Lagrange Geometry, Finsler Spaces and Noise Applied in Biology and Physics.}%
, in {\it Mathematical and Computer Modelling,} ed. Rolin, E. Y. 20. N415
(Plenum Press, 1994)

\bibitem{ari}  A. K. Aringazin and G. S. Asanov, {\sl Rep. Math. Phys.} {\bf %
25} (1988) 35


\bibitem{asa}  G. S. Asanov, {\it Finsler Geometry, Relativity and Gauge
Theories} (Reidel, Boston, 1985)

\bibitem{asa88}  G. S. Asanov and S. F. Ponomarenko, {\it Finsler Bun\-dle
on Space--Time. Associated Gauge Fields and Con\-nec\-tions} (\c Stiin\c ta,
Chi\c sin\u au, 1988) [in Russian]

\bibitem{asa89}  G. S. Asanov, {\it Fibered Generalization of the Gauge
Field Theory. Finslerian and Jet Gauge Fields} (Moscow University Press,
Moscow, 1989) [in Russian]

\bibitem{ash}  A. Ashtekar, {\sl Phys. Rev.} {\bf D36} (1987) 1587

\bibitem{atick}  J. Atick, A. Dhar and B. Ratra, {\sl Phys. Rev.} {\bf D33}
(1986) 2824

\bibitem{ati}  M. F. Atiyah, R. Bott and A. Shapiro, {\sl Topology} {\bf 3}
(1964) 3





\bibitem{bei}  G. G. Beil, {\sl Int. J. Theor. Phys.} {\bf 31} (1992) 1025

\bibitem{bej}  A. Bejancu, {\it Finsler Geometry and Applications} (Ellis
Horwood, Chichester, England, 1990)

\bibitem{bej89h}  A. Bejancu, {\it Generalized Gauge Theories,} in {\sl %
Colloquia Mathematica Societatis J\'anos Bolyai.} {\bf 56}. Differential
geometry (EGER, Hungary, 1989) p. 101


\bibitem{bej91a}  A. Bejancu, {\sl An. \c St. Univ. ''Al. I. Cuza'' Ia\c si
(new ser., fasc.2)} {\bf 36} (1991) 123


\bibitem{bel}  A. A. Belavin, A. M. Polyakov and A. B. Zamolodchikov, {\sl %
Nucl. Phys.} {\bf B241} (1984) 333





\bibitem{berw}  L. Berwald, {\sl Math. Z.} {\bf 25} (1926) 40; Correction,
{\sl Math. Z.} {\bf 26} (1927) 176

\bibitem{bis}  R. D. Bishop and R. J. Crittenden, {\it Geometry of Manifolds}
(Academic Press, 1964)

\bibitem{bog}  G. Yu. Bogoslovsky, {\it Theory of Locally-Anisotropic
Space-Time} (Izdatel'stvo Moscovskogo Universiteta, Moscow, 1992) [in
Russian]

\bibitem{bon}  H. Bondi, {\sl Nature} (1979) 1072

\bibitem{bonp}  H. Bondi, F. Pirani and I. Robinson, {\sl Proc. Roy. Soc.
London.} {\bf A251} (259) (1959) 519

\bibitem{bra}  E. Braaten, T. L. Curtright, C. K. and Zachos, {\sl Nucl.
Phys.} {\bf B260} (1985) 630







\bibitem{bucur}  I. Bucur and A. Deleanu, {\it Introduction to the Theory of
Categories and Functors} (John Wiley and Sons LTD, London, New York, Sydney,
1969); (Mir, Moscow, 1972) [in Russian]

\bibitem{bus}  T. H. Busher, {\sl Phys. Lett.} {\bf B201} (1988) 466

\bibitem{cfmp}  C. G. Callan, D. Friedan, E. J. Martinec and M. J. Perry,
{\sl Nucl. Phys.} {\bf B262} (1985) 593

\bibitem{ckp}  C. G. Callan, I. R. Klebanov, and M. J. Perry, {\sl Nucl.
Phys.} {\bf B278} (1986) 78

\bibitem{car35}  E. Cartan, {\it Les Espaces de Finsler} (Hermann, Paris,
1935)




\bibitem{che}  N. A. Chernikov, {\it Preprint JINR} P2-88-27, Dubna, 1988
[in Russian]

\bibitem{chev} C. Chevalley, {\it The Construction and Study of Certain
 Important Algebras,} (Publications of Mathematical Society, Tokyo, 1955)





\bibitem{cru1}  A. Crumeyrolle, {\sl Ann. Inst. H. Poincare} {\bf 11} (1969)
19

\bibitem{cru2}  A. Crumeyrolle, {\sl Ann. Inst. H. Poincare} {\bf 14} (1971)
309


\bibitem{cz}  T. L. Curtright and C. K. Zachos, {\sl Phys. Rev. Lett.} {\bf %
53} (1984) 1799










\bibitem{fin}  P. Finsler, $\ddot U${\it ber Kur\-ven und Fl}$\ddot a${\it %
chen in All\-gemeiner R}$\ddot a${\it men} (Disser\-ta\-ti\-on, G$\ddot o$%
ttingen, 1918); reprinted (Birkh$\ddot a$user, Basel, 1951).

\bibitem{foc}  V. Fock, {\sl Zs. f. Phys.} {\bf 57} (1929) 261

\bibitem{fts}  S. Fradkin and A. A. Tseytlin, {\sl Phys. Lett.} {\bf B158}
(1985) 316; {\sl Nucl. Phys.} {\bf B261} (1985) 1











\bibitem{gs}  M. B. Green and J. H. Schwarz, {\sl Nucl. Phys.} {\bf B243}
(1984) 285

\bibitem{gsw}  M. B. Green, J. H. Schwarz and E. Witten, {\it Superstring
Theory,} vol. 1 and 2 (Cambridge University Press, Cambridge, 1987)


\bibitem{gri}  L. P. Grishchuk, A. N. Petrov and A. D. Popova, {\sl Commun.
Math. Phys.} {\bf 94} (1984) 379





\bibitem{god}  A. Godbillon, {\it Elements de Topologie Algebraique}
(Herman, Paris, 1971)

\bibitem{gip}  I. Gottlieb and N. Ionescu--Pallas, {\sl Rev. Roum. Phys.}
{\bf 35} (1990) 25

\bibitem{goz}  J. Gottlieb, V. Oproiu and G. Zet, {\sl An. \c Sti. Univ.
''Al. I. Cuza'' Iasi, Sect. Ia. Mat. (N.S.)} {\bf 20} (1974) 123

\bibitem{gvd}  I. Gottlieb and S. Vacaru,
 in {\it Colloquium on Differential Geometry,} 25-30
July, 1994, Debrecen, Hungary (Lajos Kossuth University, Debrecen, Hungary,
1994), p. 9

\bibitem{gv}  I. Gottlieb and S. Vacaru,
in {\it  Lagrange and Finsler
Geometry, Applications to Physics and Biology,} eds. P. L. Antonelli and
Radu Miron (Kluwer Academic Publishers, Dordrecht, Boston, London, 1996), p.
209





\bibitem{thoo}  G. t'Hooft, {\sl Nucl. Phys.} {\bf B61} (1973) 455

\bibitem{thoov}  G. t'Hooft and M. Veltman, {\sl Nucl. Phys.} {\bf B44}
(1971) 189; M. Bos, {\sl Ann. Phys. (USA)} {\bf 181} (1988) 177




\bibitem{hs}  R. S. Howe and K. S. Stelle, {\it The Background Field Method
and the Nonlinear }$\sigma ${\it -model} (Preprint IAS, 1986)



\bibitem{hul}  C. M. Hull, {\it Lectures on Non-linear Sigma Models and
Strings} (DAMPT, Cambridge, England, 1987)



\bibitem{ikes}  S. Ikeda, {\sl An. Phys.}{\bf \ 47} (1990) 1

\bibitem{in}  R. S. Ingarden, {\sl Tensor N.S.} {\bf 30} (1976) 201

\bibitem{ing}  R. S. Ingarden, {\sl Tensor N. S.} {\bf 46} (1987) 354

\bibitem{ishih}  S. Ishihara, {\sl Thoku Math. J.} {\bf 9} (1957) 273

\bibitem{ish}  H.\ Ishikawa, {\sl J. Math. Phys.} {\bf 22} (1981) 995




\bibitem{jac}  I. Jack, {\sl Nucl. Phys.} {\bf B234} (1984) 365

\bibitem{jad}  A. Jadczic and K. Pilch, {\sl Commun. Math. Phys.} {\bf 78}
(1981) 373




\bibitem{kar}  M. Karoubi, {\it K-Theory} (Springer-Verlag, Berlin, 1978)



\bibitem{ker}  J. Kern, {\sl Arch. Math.} {\bf 25} (1974) 438

\bibitem{ketnp}  S. V. Ketov, {\sl Nucl. Phys.} {\bf 294} (1987) 813

\bibitem{keti}  S. V. Ketov, {\sl Izvestya Vuzov. Fizica} {\bf 31} (1988) 17
[in Russian]

\bibitem{ket}  S. V. Ketov,{\it \ Introduction into Quantum Theory of
Strings and Superstrings} (Nauka, Novosibirsk, 1990) [in Russian]

\bibitem{ketn}  S. V. Ketov, {\it Nonlinear Sigma-Models in Quantum Field
Theory and String Theory} (Nauka, Novosibirsk, 1992) [in Russian]


\bibitem{koh}  Z. Kohler, {\sl Z. Physik} {\bf 134} (1954) 286


\bibitem{kra}  D. Kramer, H. Stephani, M. McCalum and E. Hert, {\it Exact
Solutions of the Einstein's Field Equations,} \ E. Schmutzer, ed., (Berlin,
1980)






\bibitem{log}  A. A. Logunov and M. A. Mestvirishvily, {\it Relativistic
Theory of Gravitation} (Nauka, Moscow, 1989) [in Russian]

\bibitem{lov}  C. Lovelace, {\sl Phys. Lett.} {\bf B35} (1984) 75

\bibitem{lue}  C. P. Luehr and M. Rosenbaum, {\sl J. Math. Phys.} {\bf 15}
(1974) 1120

\bibitem{lue80}  C. P. Luehr and M. Rosenbaum, {\sl J. Math. Phys.} {\bf 21}
(1980) 1432




\bibitem{mat}  M. Matsumoto, {\it Foundations of Finsler Geometry and
Special Finsler Spaces} (Kaisisha, Shigaken, 1986)

\bibitem{ma87}  R. Miron and M. Anastasiei, {\it Vector Bundles. Lagrange
Spaces. Application in Relativity} (Academiei, Romania, 1987) [in Romanian]

\bibitem{ma94}  R. Miron and M. Anastasiei, {\it The Geometry of Lagrange
Spaces: Theory and Applications} (Kluwer Academic Publishers, Dordrecht,
Boston, London, 1994)

\bibitem{mirata}  R. Miron and Gh. Atanasiu, {\it Compendium sur les Espaces
Lagrange D'ordre Sup\'erieur, Seminarul de Mecanic\v a. Universitatea din
Timi\c soara. Facultatea de Matematic\v a,} 1994 40 p.

\bibitem{mk}  R. Miron and T. Kavaguchi, {\sl Int. J. Theor. Phys.} {\bf 30}
(1991) 1521

\bibitem{mirbal}  R. Miron, R. K. Tavakol, V. Balan and I. Roxburgh, {\it %
Geometry of Space-Time and Generalized Lagrange Gauge Theory,} {\sl %
Publicationes Mathematicae,} Debrecen, Hungary {\bf 42} (1993) 215

\bibitem{moc}  P. Mocanu, {\it Spa\c tii Par\c tial Proiective}, v. {\bf 6},
N 3-4, Academia R. P. Rom\^ane, (Bucure\c sti, 1955)

\bibitem{moo}  A. Mo\'or, {\sl Acta Math.} {\bf 86} (1951) 71










\bibitem{ono}  T. Ono and Y. Takano, {\sl Tensor N. S.} {\bf 52} (1993) 56


\bibitem{ots}  T. Otsuki and Y. Tashiro, {\sl Math. J. Okayama Univ.} {\bf 4}
(1954) 57

\bibitem{pen}  R. Penrose, {\it Structure of Space-time,} in: {\sl Battelle
Rencontres, Lectures in Mathematics and Physics,} eds. C. M. DeWitt and J.
A. Wheeler (Benjamin, New York, 1967)

\bibitem{penr1}  R. Penrose and W. Rindler, {\it Spinors and Space-Time,}
vol. {\bf 1}, {\it Two-Spinor Calculus and Relativistic Fields} (Cambridge
University Press, Cambridge, 1984)

\bibitem{penr2}  R. Penrose and W. Rindler, {\it Spinors and Space-Time,}
vol. {\bf 2}, {\it Spinor and Twistor Methods in Space-Time Geometry}
(Cambridge University Press, Cambridge, 1986)

\bibitem{per}  A. Peres, {\sl Phys. Rev. Lett.} {\bf 3} (1959) 571

\bibitem{petk}  A. Z. Petrov, (ed) {\it Modeling of physical fields
Gravitatsia i teoria otnositelinosti,}\ vip. {\bf 4-5} (Izdatelistvo
Kazanskogo Univeriteta, 1968) p. 7 \ [in Russian]

\bibitem{pet}  A. Z. Petrov, {\it Einstein Spaces} (Pergamon Press, Oxford,
1969)

\bibitem{petd}  A. Z. Petrov, {\sl Doklady Academii Nauk SSSR} {\bf 186}
(1969) 1302 [in Russian]

\bibitem{petd1}  A. Z. Petrov, {\sl Doklady Academii Nauk SSSR} {\bf 190}
(1970) 305 [in Russian]

\bibitem{petk1}  A. Z. Petrov, {\it On modelling of gravitational fields.}
Preprint ITF-71-14R, Kiev, 1971 [in Russian]



\bibitem{poi}  H. Poincare, {\it Science and Hypothesis} (Walter Scott,
London, 1905); (Dover, New York, 1952)

\bibitem{poi1}  H. Poincare, {\it Oeuvres de Henry Poincare,} vol. {\bf 9},
(Gauther-Villars, Paris, 1954)

\bibitem{poly}  A. M. Polyakov, {\sl Phys. Lett.} {\bf B103} (1981) 207;
211;\ {\it Gauge Fields and Strings, Contemporary Concepts in Physics}
(Harwood Acad. Publ., Chur e. a., Switzerland, 1987)

\bibitem{pon}  V. N. Ponomarev, A. O. Barvinsky and Yu. N. Obukhov, \ {\it %
Geometrodynamical Methods and Gauge Approach to Gravity Theory}
(Energoatomizdat, Moscow, 1985) [in Russian]

\bibitem{p}  D. A. Popov, {\sl Theor. Math. Phys.} {\bf 24 }(1975) 347 [in
Russian]

\bibitem{pd}  D. A. Popov and L. I. Dikhin, {\sl Doklady Academii Nauk SSSR}
{\bf 225} (1975) 347 [in Russian]












\bibitem{ros}  N. Rosen, {\sl Phys. Rev.} {\bf 57} (1940) 147

\bibitem{rox}  I. W. Roxburgh, {\sl Tensor N. S.,} {\bf 51} (1992) 59

\bibitem{run}  H. Rund, {\it The Differential Geometry of Finsler Spaces}
(Springer--Verlag, Berlin, 1959)






\bibitem{sch}  J. H. Schwarz (ed), {\it Superstrings: the first 15 years of
superstring theory: reprints \& commentary,} (World Scientific, Singapore e.
a., 1985)

\bibitem{sen}  A. Sen, {\sl Phys. Rev. Lett.} {\bf 55} (1986) 1846

\bibitem{sin}  N. S. Sinyukov, {\it Geodesic Maps of Riemannian Spaces}
(Nauka, Moscow, 1979) [in Russian]

\bibitem{shore}  G. M. Shore, {\it Nucl. Phys.} {\bf B286} (1987) 349

\bibitem{vast}  Dorina Stancu and S. Vacaru,  in {\it Quantum Field Methods
in Investigations of High Temperature Superconductivity and Unordered
Systems,} ed. K. K. Gudima and all ( \c Stiin\c ta, Chi\c sin\v au, 1992) p.
74

\bibitem{syn}  J. L. Synge, {\it Relativity: The General Theory}
(North-Holland Publishing Company, Amsterdam, 1960)

\bibitem{takp}  Y. Takano, {\sl Progr. Theor. Phys.} {\bf 40} (1968), 1159

\bibitem{takn}  Y. Takano, {\sl Nuovo Cimento} {\bf 10} (1974) 747

\bibitem{tas}  Y. Tashiro, {\sl Math. J. Okayama Univ.} {\bf 6} (1957) 147

\bibitem{tav}  R. K. Tavakol, N. Van der Bergh, {\sl Gen. Rel. Grav.} {\bf 18%
} (1986) 849

\bibitem{ts}  A. A. Tseytlin, {\sl Phys. Rev.} {\bf D26} (1982) 3327


\bibitem{ts87}  A. A. Tseytlin, {\sl Phys. Lett.} {\bf B194} (1987) 63


\bibitem{ts89}  A. A. Tseytlin, {\sl Int. J. Mod. Phys.} {\bf A4} (1989) 1257

\bibitem{tur}  A. Turtoi, {\it Applications of Algebra and Geometry in
Spinors Theory} (Editura Tehnoc\v a, Bucure\c sti, 1989) [in Romanian]

\bibitem{v87}  S. Vacaru, {\sl Vestnik Moscovskogo Universiteta, Fizica i
Astronomia} {\bf 28} (1987) 5 [in Russian]

\bibitem{vb13}  S. Vacaru,
{\sl Studia Uni\-ver\-si\-ta\-tis Ba\-bes--Bo\-ly\-ai,} Cluj--Na\-po\-ca,
Romania {\bf XXXV,} 2
(1990) 36

\bibitem{v91}  S. Vacaru,  in {\it Methods of Quantum Field Theory in
 Condenced Matter Physics,} ed. P.I. Hadji (Stiin\c
ta, Chi\c sinau, Moldova, 1991) p.98 [in Russian]

\bibitem{vk}  S. Vacaru,
 in {\sl Contr. Int. Conf. ''
Lobachevski and Modern Geometry '',} Part {\bf II, }ed. V. Bajanov et all
(Kazani, University Press, 1992) p. 64

\bibitem{vb12}  S. Vacaru,
 {\sl Buletinul Academiei de \c Stiin\c te a
Republicii Moldova, Fizica \c si Tehnica} {\bf 3}(12) (1993) 17

\bibitem{vth}  S. Vacaru, {\it Applications of Nearly Autoparallel Maps and
Twistor--Gauge Methods} {\it in Gravity and Condensed States,} Ph D Thesis
(''Al. I. Cuza'' University, Ia\c si, Rom\^ania, 1993) [in Romanian]

\bibitem{vrjp}  S. Vacaru,
 {\sl Romanian Journal of Physics} {\bf 39} (1994) 37



\bibitem{vjmp}  S. Vacaru,  {\sl J. Math.
Phys} {\bf 37} (1996) 508

\bibitem{vb295}  S. Vacaru,
 {\sl Buletinul Academiei de \c Stiin\c te a
Republicii Moldova, Fizica \c si Tehnica} {\bf 3} (1995) 53

\bibitem{v295a}  S. Vacaru,
 {\sl Buletinul Academiei de \c Stiin\c te a
Republicii Moldova, Fizica \c si Tehnica} {\bf 1} (1995) p. 54

\bibitem{v295b}  S. Vacaru,
 {\sl Buletinul Academiei de \c Stiin\c te a Republicii Moldova,
Fizica \c si Tehnica} {\bf 1} (1996) p. 62


\bibitem{vlasg}  S. Vacaru, {\it Generalized Lagrange and Finsler
Supergravity;} E-print: gr-qc/9604016.

\bibitem{vlags}  S. Vacaru, {\it Locally anisotropic gravity and strings,}\\
E-print: gr-qc/9604013.


\bibitem{vst96}  S. Vacaru, {\it Stochas\-tic pro\-cesses and dif\-fu\-sion
on spaces with lo\-cal an\-i\-sot\-ropy}; E-print: gr-qc/9604014.

\bibitem{vcl96}  S. Vacaru, {\it Nearly autoparallel maps, tensor integral
and conservation laws on locally anisotropic spaces;} E--print:
gr-qc/9604017.

\bibitem{vsp96}  S. Vacaru, {\it Spinors in higher dimensional and locally
anisotropic spaces;} E-print: gr-qc/9604015.

\bibitem{v96jpa1}  S. Vacaru, {\it Locally anisotropic interactions: I.
Nonlinear connections in higher order anisotropic superspaces;}
 E--print: hep--th/9607196.

\bibitem{v96jpa2}  S. Vacaru, {\it Locally anisotropic interactions: II.
Torsions and curvatures of higher order anisotropic superspaces;}
 E--print: hep--th/9607195.

\bibitem{v96jpa3}  S. Vacaru, {\it Locally anisotropic interactions: III.
Higher order anisotropic supergravity;}
 E--print: hep--th/9607194.


\bibitem{vg}  S. Vacaru and Yu. Goncharenko,  {\sl Int. J.
Theor. Phys.} {\bf 34} (1995) 1955

\bibitem{vob12}  S. Vacaru and S. Ostaf,  {\sl Buletinul
Academiei de \c Stiin\c te a Republicii Moldova, Fizica \c si Tehnica} {\bf 3%
} (1993) 4

\bibitem{vob13}  S. Vacaru and S. Ostaf,
{\sl Buletinul Academiei de \c Stiin\c te a Republicii Moldova, Fizica \c si
Tehnica} {\bf 1} (1994) 64

\bibitem{vod}  S. Vacaru and S. Ostaf, in
{\sl Coloquium on Differential Geometry,} 25-30 July 1994 (Lajos
Kossuth University, Debrecen, Hungary) p.56

\bibitem{voa}  S. Vacaru and S. Ostaf,
 in {\sl Lagrange and Finsler Geometry,} eds.
P. L. Antonelli and R. Miron, ( Kluwer Academic Publishers, Dordrecht,
Boston, London, 1996) 241

\bibitem{voarm}  S. Vacaru and S. Ostaf,
 {\sl Rep. Math. Phys.} {\bf 37} (1996) 309; E-print:\
gr-qc/9602010

\bibitem{vog}  S. Vacaru, S. Ostaf and Yu. Goncharenko,
 {\sl Romanian J. Physics} {\bf 39} (1994) 199

\bibitem{vodg}  S. Vacaru, S. Ostaf, Yu. Goncharenko and A. Doina,
 {\sl Buletinul Academiei de \c Stiin\c te a Republicii Moldova,
Fizica \c si Tehnica} {\bf 3} (1994) 42

\bibitem{var}  J. Vargas and D. G. Torr,{\sl \ J. Math. Phys.} {\bf 34}
(1993) 4898.


\bibitem{vlas}  A. A. Vlasov,{\it \ Statistical Distribution Functions}
(Nauka, Moscow, 1966) [in Russian]


\bibitem{vra}  G. Vr\v anceanu, {\it Lec\c tii de Geometrie Differential\v a,%
} vol. {\bf 2} (Ed. Didactic\v a \c si Pedagogic\v a, Bucure\c sti, 1977)
[in Romanian]

\bibitem{wal}  R. P. Walner, {\sl General Relativity and Gravitation} {\bf 17%
} (1985) 1081


\bibitem{wesz}  J. Wess and B. Zumino, {\sl Phys. Lett.} {\bf 37} (1971) 95


\bibitem{dewd}  B. S. DeWitt, {\it Dynamical Theory of Groups and Fields}
(Gordon and Breach, New York, 1965)

\bibitem{dew}  B. DeWitt, {\it Supermanifolds} (Cambridge University Press,
1984)


\bibitem{wit}  E. Witten, {\sl Commun. Math. Phys.} {\bf 92} (1984) 455


\bibitem{ya}  K. Yano, {\sl Proc. Imp. Acad. Tokyo, }{\bf 16} (1940) 195;
354; 442; 505

\bibitem{yan}  K. Yano and S. I. Ishihara, {\it Tangent and Cotangent
Bundles. Differential Geometry} (Marcel Dekker, New York, 1973)

\bibitem{zam}  A. B. Zamolodchikov, {\sl Pis'ma J. T. P.} {\bf 43} (1986)
565 [in Russian]

\bibitem{zan}  D. Zanon, {\sl Phys. Lett.} {\bf B191} (1987) 363

\end{references}
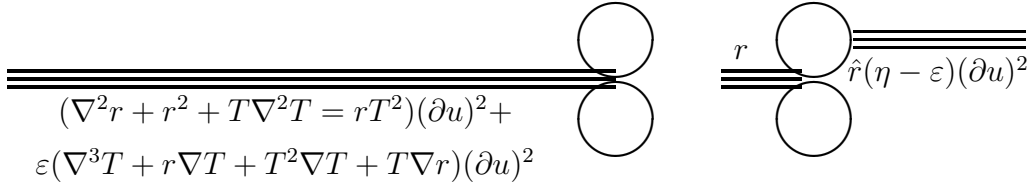
\end{document}